\documentclass[twocolumn,prd,floatfix,preprintnumbers,nofootinbib,superscriptaddress, aps]{revtex4-1}
        
\usepackage{amssymb,amsmath,verbatim,mathtools,needspace,enumitem,etoolbox,graphicx,physics,microtype,afterpage,bm,soul}
\usepackage[dvipsnames]{xcolor}
\definecolor{linkcolor}{rgb}{0.0,0.3,0.5}
\usepackage[unicode, colorlinks=true, linkcolor=linkcolor, citecolor=linkcolor, filecolor=linkcolor,urlcolor=linkcolor, pdfusetitle]{hyperref}
\usepackage{orcidlink}
\usepackage[all]{hypcap}
\usepackage[T1]{fontenc}
\usepackage[utf8]{inputenc}
\usepackage{tabularx}
\usepackage{cancel}
\graphicspath{{Plots/}}

\usepackage{lmodern}
\usepackage{ragged2e}
\allowdisplaybreaks
\usepackage{tikz}
\usepackage{color}
\usepackage{nicefrac}
\usepackage{framed}
\usepackage{hyperref}
\hypersetup{colorlinks, citecolor=darkblue, linkcolor=black, urlcolor=darkblue}
\definecolor{rossos}{cmyk}{0,1,1,0.55}
\definecolor{bluscuro}{rgb}{0.15, 0.2, .85}
\definecolor{bluchiaro}{cmyk}{1,.3,0.,0.1}
\definecolor{ForestGreen}{rgb}{0.13, 0.55, 0.13}
\definecolor{darkblue}{rgb}{0,0, 1.39}
\newcommand{\msun}{M_{\odot}}

\newcommand{\be}{\begin{equation}}

\newcommand{\ee}{\end{equation}}
\renewcommand{\d}{{\rm d}}
\def\BH{\text{\tiny BH}}

\newcommand{\PBH}{\text{\tiny PBH}}
\newcommand{\NS}{\text{\tiny NS}}
\newcommand{\BNS}{\text{\tiny BNS}}
\newcommand{\GW}{\text{\tiny GW}}

\newcommand{\orb}{\text{\tiny orb}}

\def\lsim{\mathrel{\rlap{\lower4pt\hbox{\hskip0.5pt$\sim$}}
    \raise1pt\hbox{$<$}}}         
\def\gsim{\mathrel{\rlap{\lower4pt\hbox{\hskip0.5pt$\sim$}}
    \raise1pt\hbox{$>$}}}         

\newcommand{\jhu}{William H.\ Miller III Department of Physics and Astronomy, Johns Hopkins University, \\ 3400 North Charles Street, Baltimore, Maryland, 21218, USA}
\newcommand{\columbia}{Department of Physics and Columbia Astrophysics Laboratory, Columbia University, New York, NY 10027, USA}

\begin{document}

\title{Gravitational wave signatures of subsolar neutron star formation in collapsar disks}

\author{Vishal Baibhav\orcidlink{0000-0002-2536-7752}}
\email{vb2630@columbia.edu}\thanks{NASA Einstein Fellow}
\affiliation{\columbia}

\author{Valerio De Luca\orcidlink{0000-0002-1444-5372}}
\email{vdeluca2@jh.edu}
\affiliation{\jhu}

\author{Loris Del Grosso\orcidlink{0000-0002-6722-4629}}
\email{ldelgro1@jh.edu}
\affiliation{\jhu}

\author{Emanuele Berti\orcidlink{0000-0003-0751-5130}}
\email{berti@jhu.edu}
\affiliation{\jhu}

\author{Brian D.~Metzger\orcidlink{0000-0002-4670-7509}}
\email{bdm2129@columbia.edu}
\affiliation{\columbia}
\affiliation{Center for Computational Astrophysics, Flatiron Institute, 162 5th Avenue, New York, NY 10010, USA}

\author{Lam Hui\orcidlink{0000-0001-7003-4132}}
\email{lh399@columbia.edu}
\affiliation{\columbia}

\begin{abstract}
\medskip
\noindent
Standard models for stellar core collapse predict the birth of neutron stars with masses exceeding a solar mass. In the collapsar model, a rapidly rotating massive star collapses to a black hole surrounded by a dense, neutrino-cooled accretion disk, whose outer regions may in some cases fragment into neutron-rich clumps that collapse to form subsolar-mass neutron stars. These disk-formed objects can then merge either with one another or, following possibly hierarchical evolution, with the central black hole. In this work, we characterize the gravitational-wave signatures of these binaries, focusing on distinctive features of the collapsar environment: Doppler and other phase modulations from orbital motion about the central black hole, gas-driven migration in the disk, and tidal interactions during inspiral. We show that the combination of anomalously low component masses with these environmental imprints produces unique waveform signatures that can distinguish the collapsar channel from other subsolar merger scenarios. Current and next-generation detectors have good prospects for identifying the intrinsic properties of such objects and, if subsolar mergers are discovered, for testing and constraining the collapsar formation pathway.
\end{abstract}

\preprint{ET-0728A-26}

\maketitle


\tableofcontents


\clearpage

\section{Introduction}
\label{sec:intro}

\noindent
The direct detection of gravitational waves (GWs) by the LIGO-Virgo-KAGRA (LVK)
Collaboration has opened a new observational window onto the Universe, revealing a
population of compact binary mergers that was previously inaccessible to electromagnetic (EM)
astronomy~\cite{LIGOScientific:2016vlm}.
Four observing runs have cataloged almost four hundred
binary black hole (BH), binary neutron star (NS), and NS-BH
coalescences, transforming
compact-object astrophysics, nuclear physics, and tests of general
relativity into precision sciences~\cite{LIGOScientific:2018mvr,LIGOScientific:2020ibl, LIGOScientific:2021usb,KAGRA:2021vkt, LIGOScientific:2025hdt,LIGOScientific:2025slb, LIGOScientific:2026wfs,LIGOScientific:2026wxz}.  Next-generation ground-based interferometers, including the Einstein
Telescope (ET)~\cite{Punturo:2010zz,Hild:2010id,Branchesi:2023mws, ET:2025xjr}, Cosmic
Explorer (CE)~\cite{Reitze:2019iox,Evans:2021gyd,Evans:2023euw}, and the
space-borne Laser Interferometer Space Antenna (LISA) mission~\cite{LISA:2017pwj,Colpi:2024xhw}, promise
a leap of one to two orders of magnitude in sensitivity and reach.
These future detectors will probe the GW sky to cosmological distances,
resolve thousands of sources per year, and access frequency bands
presently beyond reach~\cite{Branchesi:2023mws, ET:2025xjr}.

Among the most compelling targets for current and future GW searches
are compact objects with masses well below the Chandrasekhar
limit $\sim 1\,M_\odot$ --- the so-called subsolar-mass
regime.  Conventional stellar evolution produces neither stellar-remnant BHs nor ordinary core-collapse NSs in this mass range~\cite{Chandrasekhar:1931ih,Shapiro:1983du, Burrows:2020qrp, Lattimer:2021emm, Janka:2025tvf, Karim:2026bva}, so
a confirmed detection would constitute unambiguous evidence for
new physics or exotic formation pathways~\cite{Shandera:2018xkn, Singh:2020wiq, Essick:2024olf, Krnjaic:2026sgl, Gao:2026zwc}. The LVK Collaboration has performed
dedicated searches for subsolar-mass compact binary coalescences in
its observing runs, placing upper limits on their merger rate but finding no
confident detections~\cite{Abbott:2018oah,Authors:2019qbw,Nitz:2020bdb,Nitz:2021mzz,Nitz:2022ltl,
LIGOScientific:2021job,Phukon:2021cus,LVK:2022ydq, Kacanja:2024hme, Soni:2024cdb,  Kacanja:2026byy, LIGOScientific:2026wxz}.  Third-generation detectors will dramatically lower current sensitivity limits: ET and CE are expected to detect or tightly constrain subsolar merger rates out to cosmological distances, enabling not only discovery but also precise measurements of masses, tidal deformabilities, and environmental effects. Spanning the interface of fundamental physics, nuclear physics, and astrophysics, searches for subsolar-mass compact objects represent one of the highest-priority targets for next-generation GW observatories~\cite{ET:2025xjr}.

The prospect of a GW detection in the subsolar range is further
enriched by the possibility of associated EM counterparts. The LVK Collaboration has already demonstrated the power of
multi-messenger observations through  the binary NS event GW170817,
accompanied by a short gamma-ray burst and a kilonova~\cite{LIGOScientific:2017vwq, Goldstein:2017mmi,Savchenko:2017ffs, Alexander:2017aly, LIGOScientific:2017ync, Coulter:2017wya, Evans:2017mmy, Troja:2017nqp, Pan:2017jem, Arcavi:2017xiz, Drout:2017ijr, Hallinan:2017woc, Kasliwal:2017ngb}. During the ongoing run, the LVK Collaboration has issued public low-latency alerts for two subsolar event candidates, S250818k and S251112cm~\cite{2025GCN.41437....1L, ligo_scientific_collaboration_ligovirgokagra_2025}, that triggered
wide-field follow-up campaigns across the EM
spectrum~\cite{Kasliwal:2025keb, Franz:2025uan, Gillanders:2025fwf, Yang:2025dtj, Ackley:2026zlf, Vieira:2026eof, ODwyer:2026caq, Hall:2026gov,Tejera:2026dpr, Paek:2026kat}. 

\begin{figure*}
    \centering
    \includegraphics[width=1\linewidth]{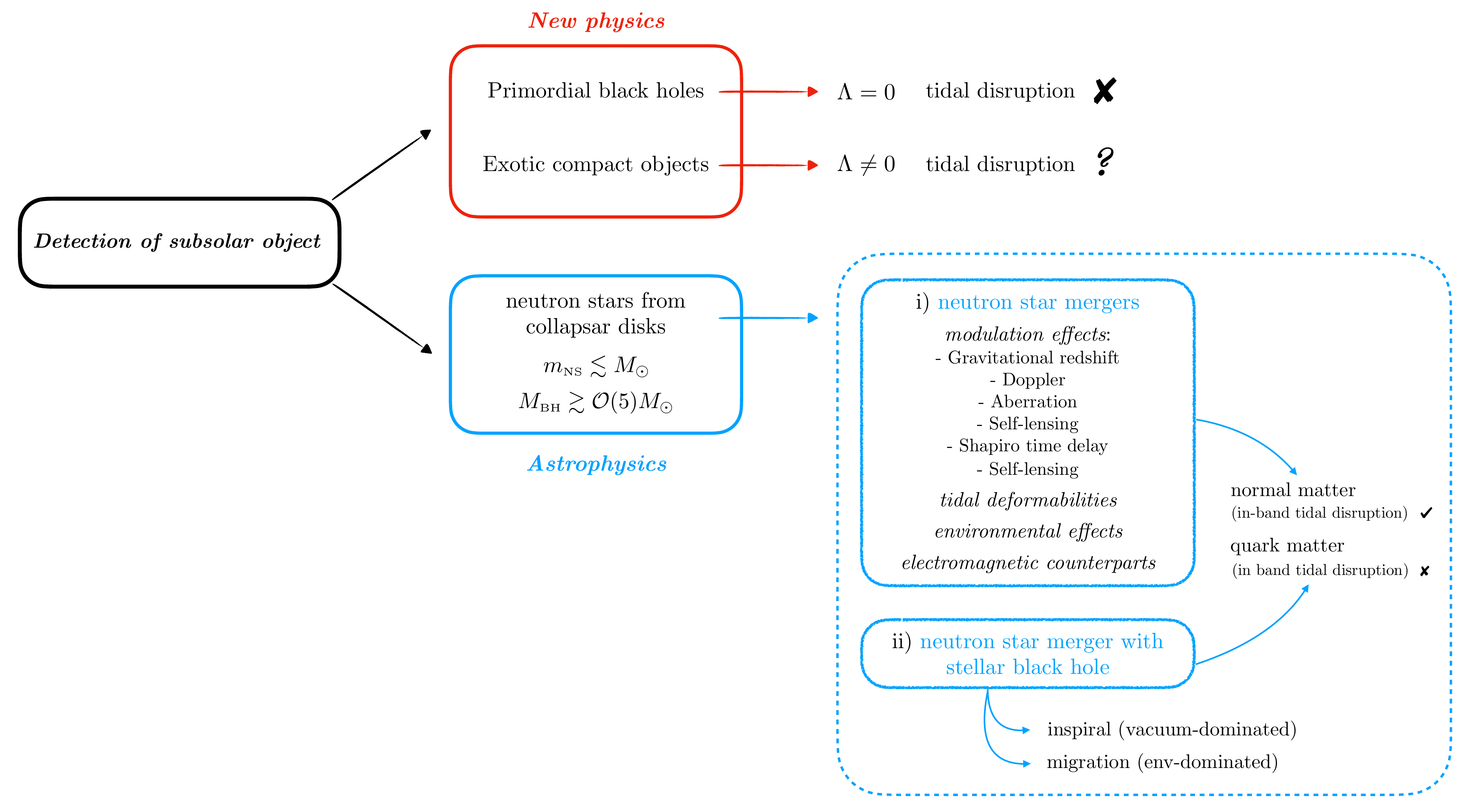}
    \caption{Schematic flowchart illustrating possible interpretations of a subsolar object detection, distinguishing between scenarios involving new physics (PBHs and exotic compact objects) and an astrophysical origin (NSs from collapsar disks), and listing their main associated observational signatures.
    } 
    \label{fig:chart}
\end{figure*}

\begin{figure*}
    \centering
    \includegraphics[width=0.98\linewidth]{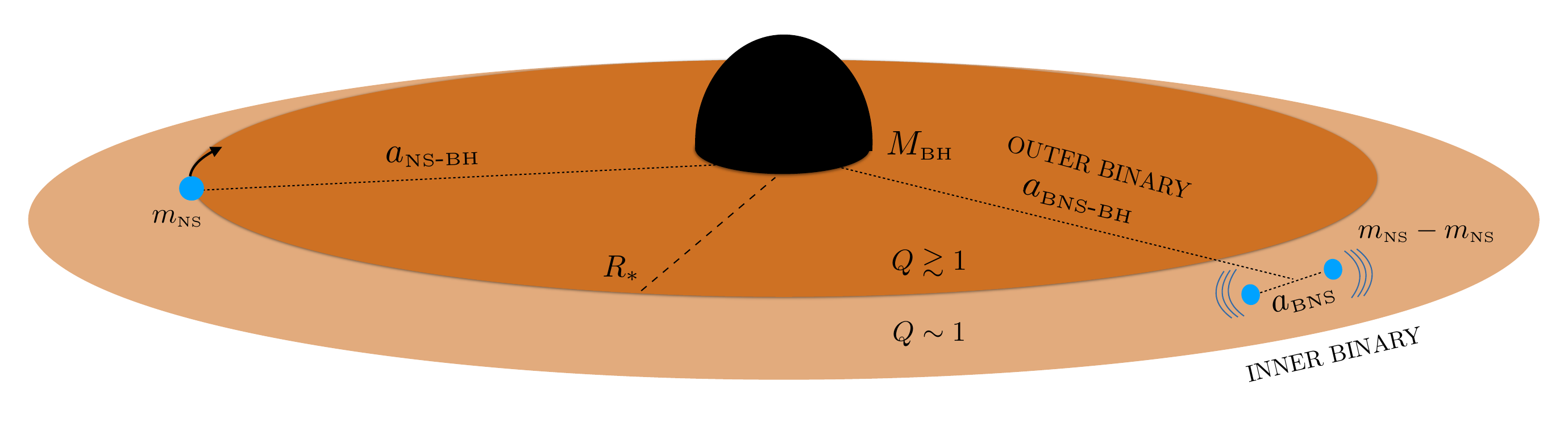}
    \caption{Schematic illustration of binary mergers within the collapsar disk scenario.  The outer regions (light orange) of the accretion disk encircling the central stellar-mass BH ($M_\BH$) are gravitationally unstable (i.e., they have Toomre parameter $Q \sim 1$) and hence may host (potentially hierarchical) mergers among subsolar NSs ($m_\NS$) produced by disk fragmentation. Following the hierarchical-triple nomenclature, we refer to such a BNS system as the \emph{inner binary} and to the orbit of this pair about the central BH as the \emph{outer binary}. In the inner parts of the disk (dark orange), where fragmentation has not occurred $(Q \gtrsim 1)$, an individual (potentially remnant) NS merges with the central BH, starting from the fragmentation distance $R_*$. We use this inner/outer terminology throughout, referring to the orbital hierarchy of the triple (only while the inner binary remains intact) and not to the radial location within the disk.
    }
    \label{fig:cartoon}
\end{figure*}

\begin{figure*}[t]
    \centering    \includegraphics[width=0.94\linewidth]{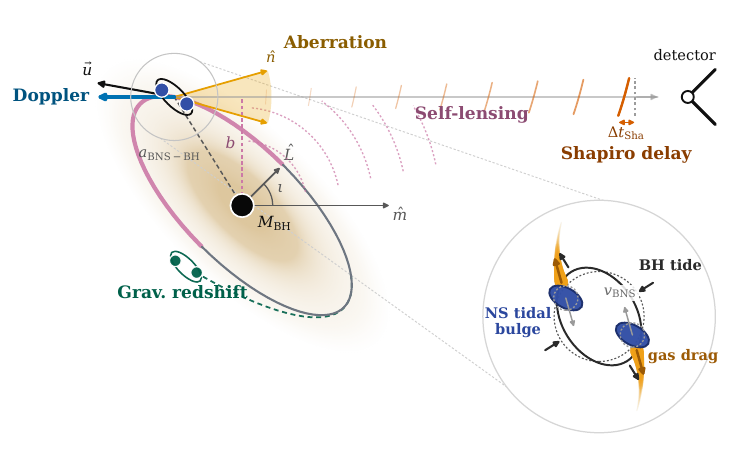}
    \caption{Schematic representation of the imprints on the GW signal emitted by a subsolar BNS merging inside a collapsar disk, where the binary orbits the central BH as part of a hierarchical triple. The binary's orbital motion around the BH modulates the waveform through a Doppler shift and, transversely, through relativistic aberration of the emission direction $\hat n$. The BH potential further redshifts the emitted frequency, while propagation near the BH produces self-lensing and a Shapiro delay. These five propagation and kinematic modulations are quantified in Sec.~\ref{sec:modeffects}. Additional effects arise from the local environment and internal structure: residual disk gas exerts drag on the two stars, removing orbital energy; each NS raises a tidal bulge on its companion (Sec.~\ref{sec:envtidalNSNS}); stellar rotation produces a spin-induced quadrupole (Appendix~\ref{sec:spin_quadrupole}); and the tidal field of the central BH modifies the conservative dynamics of the inner binary (Appendix~\ref{sec:tidalfield}).}
    \label{fig:scene}
\end{figure*}

One class of subsolar compact object that has attracted sustained theoretical
attention are primordial black holes (PBHs), i.e., BHs formed in the
radiation-dominated era from the gravitational collapse of
primordial density perturbations, well before any star
formation~\cite{Zeldovich:1967lct,Hawking:1971ei,Carr:1974nx,Carr:1975qj} (see~\cite{Sasaki:2018dmp, Carr:2020gox, Green:2020jor,Byrnes:2025tji,LISACosmologyWorkingGroup:2023njw} for recent reviews).
PBHs with masses $M_\PBH \lesssim M_\odot$ are viable subsolar
GW sources if they form binaries in the early Universe or through
dynamical capture~\cite{Cholis:2016kqi,Bird:2016dcv,Clesse:2016vqa, Sasaki:2016jop,Raidal:2018bbj,Raidal:2024bmm}, and are currently constrained by a broad array of astrophysical and cosmological
observations, including microlensing surveys and GW data~\cite{Carr:2020gox,Green:2020jor, Andres-Carcasona:2024wqk, Carr:2026hot} (see Refs.~\cite{Baumgarte:2025syh, Baumgarte:2026azn} for alternative subsolar formation channels based on PBHs). In particular, in the 
mass range $\sim (0.1 - 1)\,M_\odot$, PBHs could constitute at most 1\% of the dark matter, while potentially leaving a detectable
imprint in subsolar GW searches~\cite{Prunier:2023uoo, Markin:2023fxx, Yuan:2024yyo, Pujolas:2021yaw, DeLuca:2021hde, Franciolini:2023opt, Miller:2024rca,  Magaraggia:2026jhk, Christensen:2026qfx}. Their waveforms are essentially those of point-mass compact binaries in vacuum, making the astrophysical interpretation of any subsolar
GW event inherently ambiguous unless additional physical imprints --- tidal deformability, environmental effects, or EM counterparts --- can
break the degeneracy between a PBH binary and a compact object of
astrophysical origin~\cite{Cardoso:2017cfl,Cardoso:2019rvt,Cardoso:2019upw,Barsanti:2021ydd,Franciolini:2021tla,Franciolini:2021xbq,Cole:2022fir,Coogan:2021uqv,Crescimbeni:2024cwh,Crescimbeni:2024qrq, DeLuca:2024uju, Russo:2025ivk, DeLuca:2025bph, Begnoni:2025aqc, Corman:2026lbt}.

A physically motivated and observationally testable astrophysical
alternative to PBHs in the subsolar GW window is provided by NSs formed in the outer regions of collapsar
disks~\cite{Metzger:2024ujc,Lerner:2025dkd, Chen:2025uwd}.  Collapsars
are the end state of rapidly rotating massive stars, in which
core collapse produces a stellar-mass BH surrounded by a dense,
neutrino-cooled accretion disk, and are
thought to power long gamma-ray
bursts~\cite{MacFadyen:1998vz,Woosley:2006fn}.  If the disk contains sufficient mass at large radii, it can become gravitationally unstable, fragmenting into self-bound clumps which, in the
neutron-rich rapidly cooling environment of the outer disk, contract to form
subsolar-mass NSs~\cite{Lattimer:2004pg,Metzger:2024ujc}. The
resulting population of subsolar NSs orbiting
within the disk undergoes a rich dynamical evolution, leading to binary NS (BNS) mergers within the outer
disk, and subsequently to the
inspiral of any surviving NS remnant into the central BH~\cite{Piro:2006ja,Metzger:2024ujc,Chen:2025uwd,Wu:2023qeh}. Together, these stages produce a distinctive combination of GW signatures absent from isolated PBH or ordinary BNSs, that are potentially within reach of current and next-generation GW detectors.
The number of clumps depends on the disk conditions: while fragmentation typically yields $\mathcal{O}(10)$ subsolar NSs, a disk that produces a single clump --- or one whose fragments promptly coalesce into a single remnant --- would instead leave a lone subsolar NS that could eventually merge with the central BH.

\subsection{Executive Summary}
\noindent
A compact-object merger in the subsolar-mass regime would signal either new physics or an unconventional astrophysical formation channel. Mass alone, however, would not determine its origin, because PBHs, exotic compact objects, and disk-formed NSs can occupy overlapping mass ranges. As illustrated in Fig.~\ref{fig:chart}, the central challenge is to distinguish between these possibilities using information beyond the masses. In this work, we identify waveform signatures that provide the necessary observational discriminants.

We develop a framework for addressing this question in the context of subsolar NSs produced by the fragmentation of collapsar accretion disks, as illustrated schematically in Fig.~\ref{fig:cartoon}. At distances of roughly $\mathcal{O}(10^2 - 10^3)$ gravitational radii the outer disk reaches a Toomre parameter of order unity ($Q \sim 1$) and fragments into compact, neutron-rich clumps (Sec.~\ref{sec:model}). These cool and contract into NSs of mass $\sim 0.01 - 1\,\msun$, well below the conventional core-collapse floor, yet stable under realistic cold equations of state (EoSs), and with highly unusual structure: a $0.1\,\msun$ NS made of nuclear matter (NM), for example, has a radius of several tens of km and a dimensionless tidal deformability of order $10^9$. We evaluate these properties for three representative cold EoSs: an ideal non-relativistic neutron-gas polytrope (Poly), which provides a simple analytical benchmark spanning the entire subsolar range, and NM and quark-matter (QM) EoSs, whose predictions differ at potentially observable levels. The same disk environment that produces these objects also governs their subsequent evolution: characteristic orbital separations, gas densities, and migration timescales set the dynamics of the newly formed NSs, which interact with one another and with the central BH to generate a sequence of GW sources carrying the imprint of their gaseous birth environment.

\begin{figure*}
    \centering
    \includegraphics[width=\linewidth]{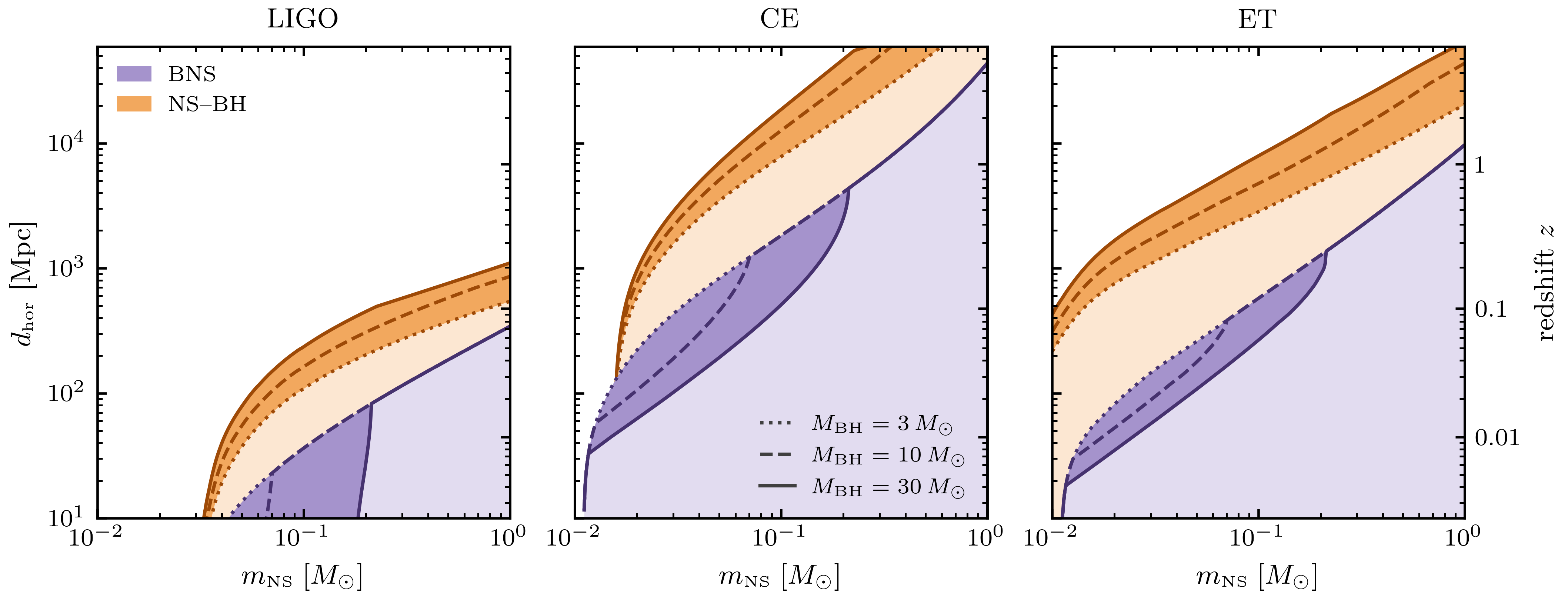}   
    \caption{Horizon distance $d_\text{\tiny hor}$ as a function of $m_\NS \in[0.01,1]\,M_\odot$ for NS-BH binaries (orange) and equal-mass BNS mergers (purple). The NS mass-radius relation follows the polytropic model of Eq.~\eqref{eq:Rpoly}. BNSs are assumed to form at a separation equal to half of Hill's radius, $f_\text{\tiny H}=0.5$.  The three columns correspond to LIGO (left), CE (middle), and ET (right). Curves are shown for $M_\BH=3$, $10$, and $30\,M_\odot$, indicated by dotted, dashed, and solid lines, respectively. 
   For NS-BH mergers, the inspiral starts from $R_*$ (given by $\rho (R_*)= 10^8\,{\rm g\,cm^{-3}}$) and terminates at the tidal-disruption (Roche) radius~$\sim 2R_\NS(M_\BH/m_\NS)^{1/3}$ or at the innermost stable circular orbit frequency, and when necessary 
    it is truncated at each detector's low-frequency cutoff. For each channel the darker band spans the horizon over $M_\BH\in[3,30]\,\msun$ (bounded by the dotted and solid curves), while the lighter shading below marks the region in which a source would be detectable, $d_\text{\tiny L}<d_\text{\tiny hor}$.
    }
    \label{fig:nsbh_bns_horizon}
\end{figure*}

Building on this framework, we map the parameter space and identify the distinct evolutionary pathways available to the disk-born NSs (Sec.~\ref{sec:param_space}). Two merger channels emerge naturally: pairs of subsolar NSs form BNS systems that merge within the outer disk as a hierarchical triple about the central BH (Sec.~\ref{sec:param_NSNS}), while individual NSs or the remnants of hierarchical mergers migrate inward and ultimately merge with the central BH itself, forming a NS-BH system (Sec.~\ref{sec:param_NSBH}). The imprints that the collapsar environment leaves on the BNS signals are summarized in Fig.~\ref{fig:scene}.

These two channels differ dramatically in loudness. The NS-BH merger is the loudest event of the scenario --- its signal-to-noise ratio (SNR) is roughly an order of magnitude above the BNS stage, since the chirp mass is mostly set by the $3 - 30\,\msun$ central BH --- and is therefore the most likely to be detected (see Fig.~\ref{fig:nsbh_bns_horizon} and Sec.~\ref{Sec:detNSBH} below). For a fiducial $0.1\,\msun$ NS, the NS-BH horizon reaches a few hundred Mpc in LIGO, a few Gpc in ET, and up to $\sim 20$~Gpc in CE.  This is three to ten times larger than the typical BNS horizon, which depends on the BNS separation at formation, on the distance from the central BH, and on the component masses, including a broad range of possibilities that go from ``clean'' mergers all the way to configurations in which the BH unbinds the pair before contact (see Sec.~\ref{sec:detection}). 

\begin{figure*}
    \centering
\includegraphics[width=0.49\linewidth]{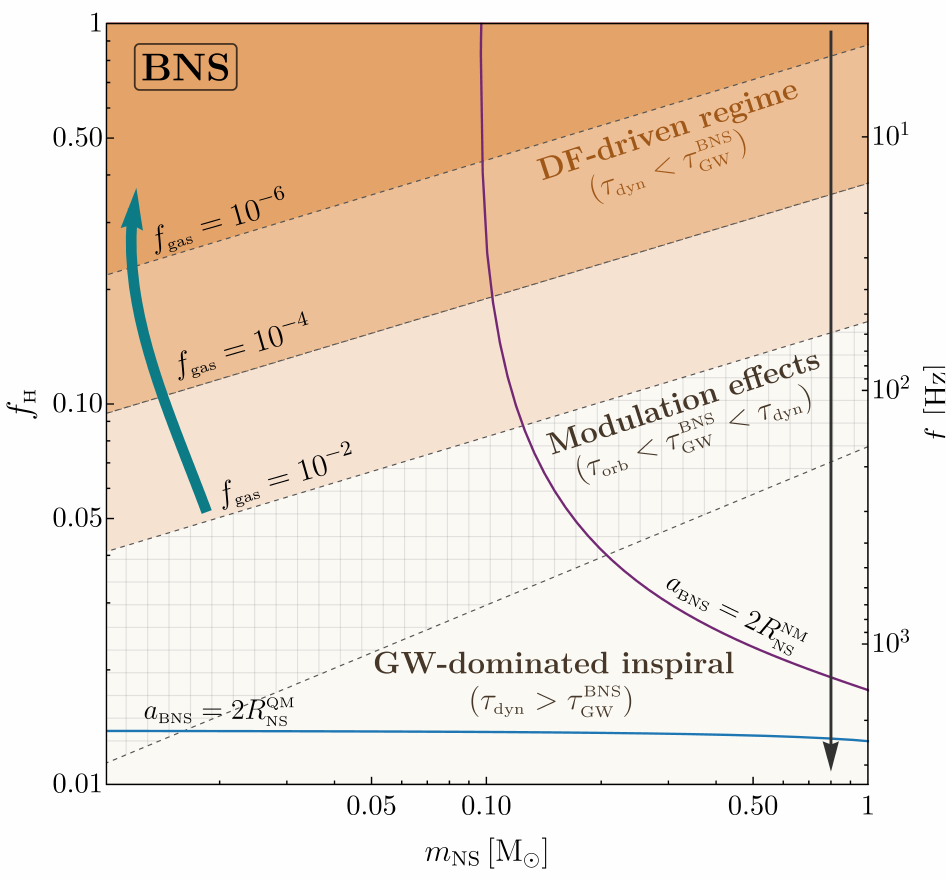}
\includegraphics[width=0.49\linewidth]{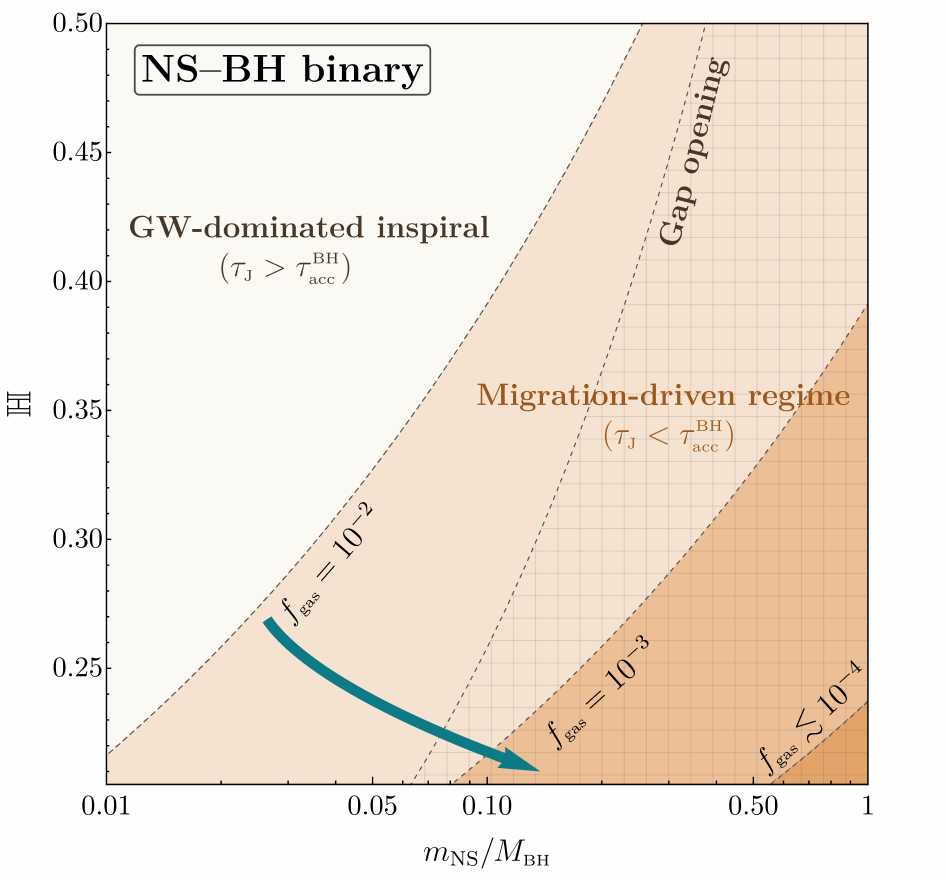}
 \caption{Dynamical regimes for BNS (left panel) and NS-BH
(right panel) mergers in a collapsar disk.
Left panel: Regimes in the plane of component NS mass $m_\NS$ and initial binary separation, expressed as the dimensionless Hill fraction $f_\text{\tiny H}$ defined in Eq.~\eqref{eq:fH} below, for an equal-mass BNS. Orange shading denotes dynamical-friction driven evolution ($\tau_\text{\tiny dyn} < \tau_\GW^\BNS$); increasing the residual post-fragmentation gas fraction $f_\text{\tiny gas}$ enlarges this regime ($f_\text{\tiny gas}$ is decreasing in the direction of the teal arrow). In the beige region (the ``GW-dominated inspiral''), the binary evolves primarily through GW emission. The cross-hatched subset satisfies $\tau_\GW^\BNS > \tau_\orb$, allowing the outer orbit about the central BH to show an imprint of the modulation effects discussed in Sec.~\ref{sec:modeffects}. The magenta and blue curves mark the contact boundaries for NM and QM EoSs, respectively. We adopt a pre-fragmentation density $\rho(R_*) = 10^{8}\,{\rm g\,cm^{-3}}$.
Right panel: Regimes for a NS of mass $m_\NS$ migrating toward and merging with a central BH of mass $M_\BH$, shown in the plane of mass ratio $m_\NS/M_\BH$ and vertical disk aspect ratio $\mathbb{H}$.
In the top-left beige region, the BH consumes the disk before the NS reaches the stable inner disk at $r < R_*$, and the subsequent inspiral proceeds in vacuum. The orange colored regions instead correspond to migration through a gas-rich disk, with boundaries shown for different residual gas fractions $f_\text{\tiny gas}$, which  is decreasing in the direction of the teal arrow. In the cross-hatched region, the NS may open a gap in the disk.}
    \label{fig:param}
\end{figure*}

The GW signals emitted by these systems can vary significantly, because their dynamics belong to several physically distinct regimes, as illustrated in Fig.~\ref{fig:param}. The BNS channel (left panel) is mainly set by the initial separation --- bounded from below by the NS contact condition, whose floor sits lower for QM than for NM, opening a narrow strip accessible only to the former (Sec.~\ref{sec:param_NSNS}). The residual post-fragmentation gas, at a fraction $f_\text{\tiny gas}  \approx 10^{-6} - 10^{-2}$ of the original disk density, then selects the evolution: dense gas drives dynamical-friction hardening of the BNS, while near-vacuum gives a clean GW-driven inspiral, and an intermediate regime lets the orbit about the central BH modulate the waveform. The phase space of the NS-BH channel (right panel), mainly driven by the vertical aspect ratio of the disk (Sec.~\ref{sec:param_NSBH}), can be partitioned into a vacuum GW-dominated inspiral or a gas-rich, migration-assisted one (Type~I, the planetary-migration analogue), with gap opening (Type~II) only once hierarchical growth pushes the remnant towards large masses $\gtrsim 1 - 10\,\msun$.

The BNS channel gives rise to the richest signatures (Sec.~\ref{sec:NS-NS}), summarized in Fig.~\ref{fig:scene}. Because these binaries evolve while orbiting the central BH and while embedded in residual disk gas, their waveforms carry imprints unavailable to an isolated, point-like binary. The hierarchical-triple geometry adds to the signal Doppler modulations, relativistic aberration, gravitational redshift, Shapiro delay, and self-lensing, several of which are distinguishable from a vacuum waveform for any detectable source (Sec.~\ref{sec:modeffects}). In particular, the Doppler modulation and gravitational redshift are detectable across much of the accessible parameter space. The surrounding gas imprints a dynamical-friction phase shift entering at $-5.5\,$ post-Newtonian (PN) order, so the binary accumulates fewer cycles and coalesces sooner than an equal-mass point-particle system (Sec.~\ref{sec:envtidalNSNS}). At the same time, the large tidal deformabilities of subsolar NSs contribute a $+5$ PN correction. These tidal effects are potentially measurable with current and next-generation detectors and provide a direct probe of the NS EoS. Eccentricity is a further tell (Sec.~\ref{sec-eccentricity}): disk formation is generically eccentric, and because subsolar binaries accumulate $\sim 10^{5}$ in-band cycles in next-generation detectors, even residual eccentricities of order $10^{-3}$ may be measurable, providing a sensitive probe of the binary assembly channel~\cite{Breivik:2016ddj,Nishizawa:2016eza,Zevin:2021rtf,Romero-Shaw:2022xko} (see also Ref.~\cite{Wu:2026hth} for a recent analysis). The simultaneous presence of anomalously low masses and these environmental and tidal signatures would be a powerful indicator of an astrophysical origin (see Fig.~\ref{fig:chart}).

The inspiral of a NS into the central BH is an independent probe of the collapsar scenario (Sec.~\ref{sec:NS-BH}). Here, the measurability prospects are inverted: the tidal deformability is suppressed by the large mass ratio and is essentially unmeasurable through the inspiral (Sec.~\ref{sec:FisherII}), so the EoS discriminant becomes a spectral one: the Roche tidal-disruption frequency at which the central BH shreds the NS 
represents a sharp cutoff encoding the stellar radius, and again separating NM from QM. Depending on the disk state and the efficiency of gas-driven migration, the system evolves through a combination of environmental torques and GW emission before entering the band, and any residual eccentricity (Sec.~\ref{sec:NS-BH_ecc}) further distinguishes it from a vacuum inspiral. Together with the BNS channel, these systems establish a multi-stage GW picture of collapsar evolution that can be tested observationally.

Although PBH binaries occupy the same mass range, they evolve as effectively point-like systems in vacuum and lack these environmental and tidal effects, so the two scenarios can be separated by waveform rather than by mass alone (Sec.~\ref{sec:PBHs}). We find that a PBH binary with the same chirp mass and symmetric mass ratio can be distinguished from a BNS merger in both ET and LIGO, even when the residual gas density is very low, thanks to the enhanced tidal effects characterizing the astrophysical objects. For the NS-BH merger, the signals remain distinguishable over the entire mass range in ET; in LIGO, the result depends on the masses of the central BH and of the NS. 

Taken together, anomalously low masses, in-band tidal disruption, environmental phase shifts, hierarchical-triple modulation, and resolvable eccentricity yield waveform fingerprints with no analog in PBH binaries or ``standard'' BNSs (Sec.~\ref{sec:conclusions}). EM corroboration may come as a coincident long or short gamma-ray burst~\cite{MacFadyen:1998vz, Woosley:2006fn}, a r-process ``fragmented superkilonova''~\cite{Siegel:2018zxq, Metzger:2024ujc}, a stripped-envelope supernova~\cite{Galama:1998ea, Cano:2016ccp}, or a fast blue optical transient~\cite{Drout:2014dma, Prentice:2018qxn, Margutti:2018rri, Yao:2021qqz} (see specifically Refs.~\cite{Metzger:2024ujc,Chen:2025uwd,Kasliwal:2025keb,Baibhav:2026zuy} for discussions regarding EM signatures associated with the collapsar scenario); but the most conclusive evidence, independent of any counterpart, is a spatially and temporally coincident subsolar BNS merger and NS-BH merger, a particularly distinctive double-merger signature of the collapsar channel. More broadly, the framework developed here shows how next-generation GW observations can connect compact-object populations, accretion physics, stellar evolution, and fundamental physics within a single program --- turning an ambiguous subsolar trigger into a measurement of masses, tidal deformabilities, and formation environment out to cosmological distances, and into a direct test of the collapsar pathway.  

\subsection{Outline}
\noindent
The paper is organized as follows. In Sec.~\ref{sec:model} we review the collapsar disk model, the
conditions for fragmentation and NS formation, and the properties of
the resulting compact remnants.  Sec.~\ref{sec:param_space} maps out
the parameter space and dynamical phases governing the two merger
stages.  Secs.~\ref{sec:NS-NS} and~\ref{sec:NS-BH} analyze the GW
signals from the BNS  and NS-BH mergers,
respectively, covering detection prospects and parameter
measurability with current and future GW detectors, while Sec.~\ref{sec:PBHs} addresses the interplay with PBHs.
We summarize our conclusions in Sec.~\ref{sec:conclusions}. 

Technical details are collected in various appendices that focus on the spins of the compact objects (Appendix~\ref{app:NSspin}), binary assembly induced by environmental effects (Appendix~\ref{app:capturerate}),  further results on BNS mergers  (Appendix~\ref{app:NSNSdetails}), accretion  onto the compact objects (Appendix~\ref{app:accretion}),  tidal field of the central BH on
the inner binary (Appendix~\ref{sec:tidalfield}), EoS dependence of
GW observables (Appendix~\ref{app:EoS}), detailed derivations of the modulation effects (Appendix~\ref{app:modulations}), and post-merger emission of the subsolar
remnant (Appendix~\ref{sec:postmerger}). 
The notation used throughout the paper is summarized in Table~\ref{tab:nomenclature}. 
Henceforth, we adopt geometrical units ($G = c = 1$).

\section{Subsolar NSs from collapsar disks}
\label{sec:model}
\noindent
In this section we review the collapsar-disk model that underlies our analysis, and we characterize the subsolar objects it produces. In Sec.~\ref{sec:coll_overview} we recall why such low masses lie beyond the reach of standard stellar evolution, and then describe the structure of the neutrino-cooled accretion disk and the conditions under which its outer regions become gravitationally unstable and fragment --- fixing the characteristic mass, number, and birth environment of the resulting clumps, together with the disk profiles used throughout the rest of the paper. In Sec.~\ref{sec:properties} we turn to the microphysical and structural properties of the compact remnants themselves (their mass-radius relation and tidal deformability), which follow from the cold EoS prevailing after collapse, and directly shape the GW signatures studied in Secs.~\ref{sec:NS-NS} and~\ref{sec:NS-BH}. The spins of the NSs and of the central BH are discussed in Appendix~\ref{app:NSspin}.

\subsection{Overview}
\label{sec:coll_overview}

\noindent
A central prediction of massive-star evolution is the existence of a characteristic lower mass scale for NSs, inherited from the Chandrasekhar mass of the collapsing iron core~\cite{Chandrasekhar:1931ih},
\begin{equation}
    M_\mathrm{Ch} \simeq 1.45\,M_\odot
    \left(\frac{Y_e}{0.5}\right)^2.
    \label{eq:MCh}
\end{equation} 
For values of the electron fractions $Y_e \simeq 0.4 - 0.5$, typical of pre-supernova iron cores, Eq.~\eqref{eq:MCh} yields the familiar lower bound $m_\NS \gtrsim 1.2\,M_\odot$ found in simulations of core-collapse supernovae~\cite{Sukhbold:2015wba,Burrows:2019rtd,Ertl:2019zks,Woosley:2020mze}.

However, if the collapsing material is substantially more neutron rich ($Y_e<0.5$), the Chandrasekhar mass decreases, allowing stable NSs with significantly lower masses~\cite{Lattimer:2004pg}. Although such conditions are uncommon in standard stellar evolution, they arise naturally in the collapsar scenario. The collapse of a rapidly rotating massive star produces a stellar-mass BH (with masses $M_\BH \lesssim \mathcal{O}(30)\,M_\odot$, consistent with stripped-envelope supernova models and observations~\cite{Schneider:2020vvh}) surrounded by a massive, hot, neutrino-cooled accretion disk~\cite{MacFadyen:1998vz,Woosley:2006fn,Gottlieb:2023cgm}. A substantial fraction of the stellar envelope fails to accrete directly onto the BH and instead circularizes into this disk, driving accretion rates of $\dot{M}_\BH \sim (0.1-1)\,M_\odot\,\mathrm{s}^{-1}$. Although these rates are sustained for only the first few seconds of the collapse, they are sufficient to establish a quasi-steady disk.

If the disk is sufficiently large, extending to radii $r \gtrsim \mathcal{O}(100)\,R_g$, where $R_g \equiv GM_\BH/c^2$, the outer disk can become self-gravitating and susceptible to fragmentation. The resulting bound clumps provide a way to form NSs with masses well below those produced by the core-collapse of more slowly rotating stars. The condition for gravitational instability is described by the Toomre parameter~\cite{Toomre:1964zx,Lodato:2004wf}
\begin{equation}
    Q \equiv \frac{c_s \kappa}{\pi \Sigma} \lesssim Q_0 \simeq \mathcal{O}(1)\,,
\end{equation}
where $c_s$ is the midplane sound speed, $\kappa \simeq \Omega \equiv \sqrt{M_\BH/r^3}$ is the epicyclic (orbital) frequency for a Keplerian disk, and $\Sigma$ the surface density. In the unstable regime, perturbations on the scale of the disk thickness ($h \simeq c_s / \Omega$) can collapse and fragment.  

To estimate the characteristic masses and formation sites of these fragments, we model the outer collapsar disk using a steady-state $\alpha$-disk prescription \cite{Shakura:1972te}.  We assume an effective viscosity $\alpha \simeq 0.03$, close to that predicted by numerical simulations of the magnetorotational instability~\cite{Siegel:2018zxq}, which we shall fix from now on. The kinematic viscosity is then $\nu = \alpha c_s h \simeq \alpha c_s^2/\Omega$. The accretion rate onto the central BH, assumed to be radially constant, satisfies~\cite{Shakura:1972te}
\begin{equation}
    \dot{M}_\BH \simeq 3\pi \nu \Sigma \simeq 3\pi \alpha \frac{c_s^2}{\Omega} \Sigma\,.
\end{equation}
If we parametrize the scale height as $h = \mathbb{H} r$ in terms of the disk's vertical aspect ratio $\mathbb{H}$, the sound speed becomes $c_s =  \mathbb{H} v$, in terms of the local Keplerian velocity $v = \Omega \, r$. 

Combining these relations gives the surface density profile
\begin{equation}
    \Sigma (r) = \frac{\dot{M}_\BH \Omega}{3\pi \alpha c_s^2} = \frac{\dot{M}_\BH}{3\pi \alpha \mathbb{H}^2} \frac{1}{\sqrt{M_\BH r}} \,,
\end{equation}
which exhibits an explicit radial dependence $\Sigma \propto r^{-1/2}$ for a fixed accretion rate. Similarly, the disk midplane density, crucial for the dynamical friction and accretion processes to be discussed below, follows as
\begin{equation}
\label{densityprofile}
    \rho (r) = \frac{\Sigma}{2 h} = \frac{\dot{M}_\BH}{6\pi \alpha \mathbb{H}^3} \frac{1}{\sqrt{M_\BH r^3}}\,.
\end{equation}
The midplane gas density is therefore completely determined by the accretion rate and disk aspect ratio, and the total enclosed disk mass is given by
\begin{align}
\label{eq:mdisk}
M_\text{\tiny disk} &\simeq \int_{R_g}^{R_\text{\tiny disk}} 
   \Sigma(r)\, 2\pi r\, {\rm d}r  \nonumber \\
   & \simeq 0.8 \, M_\BH \left(\frac{\mathbb{H}}{0.3}\right)^{-2} \left(\frac{\dot{M}_\BH}{M_\odot/\rm s}\right) \left(\frac{R_\text{\tiny disk}}{10^2 R_g}\right)^{3/2}
   \,,
\end{align}
where, in the second equality, we have assumed that the disk size is much larger than the BH gravitational radius, $R_\text{\tiny disk} \gg R_g$.  

The Toomre parameter, in turn, can be rewritten by substituting the above expressions; for a disk at radius $r$,
\begin{equation}
Q = \frac{3 \alpha c_s^3}{\dot{M}_\BH} = \frac{3\alpha \mathbb{H}^3}{\dot{M}_\BH} \left(\frac{M_\BH}{ r}\right)^{3/2}\,.
\end{equation}
Requiring that fragmentation and hence NS formation be possible at least as far out as the characteristic disk size $R_\text{\tiny disk} \sim \mathcal{O}(100) R_g$, one obtains a lower bound on the accretion rate,
\begin{equation}
\dot{M}_\BH \gtrsim \frac{0.5}{Q_0}\, M_\odot\,{\rm s}^{-1}\left(\frac{\mathbb{H}}{0.3}\right)^3 \left(\frac{R_\text{\tiny disk}}{10^{2} R_g}\right)^{-3/2}\,,
\end{equation}
ensuring $Q \lesssim Q_0$ in the outer disk. Fragmentation therefore requires accretion rates of order $\sim M_{\odot}$ s$^{-1}$ or larger, consistent with the early phases of collapsar evolution. 

The same high accretion rates that trigger fragmentation also imply rapid depletion of the disk, limiting the time available for fragment evolution. This occurs on the characteristic ``disk-depletion'' timescale
\begin{equation}
    \tau_\text{\tiny acc}^\BH
    = \frac{M_\text{\tiny disk}}{\dot{M}_\BH}
    \simeq 10\, \text{s}\,
    \left(\frac{M_\text{\tiny disk}}{10 \,M_\odot}\right)
    \left(\frac{\dot{M}_\BH}{M_\odot \, {\rm s}^{-1}}\right)^{-1}\,.
    \label{eq:taccBH}
\end{equation}
Since $M_\text{\tiny disk}$ is comparable to the central mass
[Eq.~\eqref{eq:mdisk}], this depletion time is of order seconds, and will
play an important role in the subsequent dynamics of the NS.

Once $Q\lesssim Q_0 \approx 1$, fragmentation first occurs near the characteristic fragmentation radius $R_*$, given by
\begin{align}
\label{eq:Rstar}
    R_{*} & =  \left(\frac{M_\BH}{2 \pi\, Q_0\, \rho (R_*)} \right)^{1/3} = R_g \left(\frac{3 \mathbb{H}^3 \alpha}
    {\dot{M}_\BH Q_0}\right)^{2/3} \nonumber \\
    &\simeq \frac{62}{Q_0^{2/3}} R_g
    \left(\frac{\mathbb{H}}{0.3}\right)^{2} 
    \left(\frac{\dot{M}_\BH}{M_\odot/\rm s}\right)^{-2/3}\,.
\end{align}
At this radius, overdense regions collapse on a timescale $\sim Q/\Omega$, comparable to the local orbital period and therefore much shorter than the viscous evolution time of the disk~\cite{Lerner:2025dkd}. Fragmentation is possible only when the disk extends beyond this radius, i.e., when $R_\text{\tiny disk}\gtrsim R_*$.
 
The minimum mass participating in fragmentation can be 
estimated as
\begin{align}
   M_\text{\tiny frag} & \simeq \int_{R_*}^{R_*+h} 
   \Sigma(r)\, 2\pi r\, {\rm d}r \simeq 
   \frac{2 \mathbb{H}^2}{Q_0} M_\BH \nonumber \\
   & \simeq 0.2 \, M_\BH \left(\frac{\mathbb{H}}{0.3}\right)^{2} \ll M_\text{\tiny disk}
   \,,
\end{align}
giving rise to about $N_\NS \simeq 
2/\mathbb{H} \sim \mathcal{O}(10)$ clumps, assuming $\mathbb{H}\sim 0.1 - 0.3$ as seen in the simulations of Ref.~\cite{Chen:2025uwd}, with 
individual masses
\begin{align}
\label{clumpmass}
    m_\text{\tiny clump} &\simeq \pi h^2 \Sigma (R_*)
    \simeq \frac{\mathbb{H}^3}{Q_0} M_\BH \nonumber \\
    &\simeq 0.1 M_\odot \left(\frac{\mathbb{H}}{0.3}\right)^{3} 
    \left(\frac{M_\BH}{5 M_\odot}\right)
    \,.
\end{align}
These masses, while well below the canonical core-collapse NS mass scale, are nevertheless sufficient to form stable NSs once the fragments cool and relax to the hydrostatic equilibrium described by a cold NS EoS~\cite{Lattimer:2004pg,Lerner:2025dkd}. Before collapsing, the clumps form with initial densities
\begin{align}
\rho (R_*)
& \simeq  10^{10} \frac{\rm g}{\rm cm^3} \, Q_0 
\left( \frac{\dot{M}_\BH}{M_\odot\, {\rm s}^{-1}}\right)^2 
\left( \frac{5\,M_\odot}{M_\BH}\right)^2 
\left( \frac{0.3}{\mathbb{H}}\right)^6\,,
\end{align}
intermediate between white dwarf and NS 
densities~\cite{Lerner:2025dkd}.

A second key criterion for disk fragmentation is the 
requirement of rapid cooling: for self-gravitating 
overdensities to successfully collapse and decouple 
from the disk, the gas must be able to radiate away 
the heat generated during compression faster than it 
can be stabilized by thermal pressure. If cooling is 
inefficient, thermal support can halt the collapse and 
prevent the formation of gravitationally bound clumps. 
This is quantified by the dimensionless cooling 
parameter $\beta \equiv \tau_\text{\tiny cool}\,\Omega$, 
defined as the ratio of the internal energy to the 
local cooling rate times the orbital frequency, where 
the cooling is dominated by neutrino and 
antineutrino emission and photodisintegration 
processes~\cite{Gammie:2001bw, Lerner:2025dkd}. 
Fragmentation requires $\beta \lesssim \beta_c$, 
where  $\beta_c \approx 10$ for the 
radiation-dominated, neutrino-cooled regime of 
collapsar disks for a stiff EoS~\cite{2021MNRAS.503.4192B}. 
The extreme accretion rates $\dot{M}_\BH \gtrsim \mathcal{O}(0.1)\,M_\odot\,{\rm s}^{-1}$ in collapsar disks 
drive mid-plane temperatures to $T \sim {\rm MeV}$, 
partially lifting electron degeneracy and powering 
copious neutrino emission, especially through 
electron and positron capture on 
nucleons~\cite{Beloborodov:2002af}. The dissociation of alpha particles at similar temperatures is also a significant coolant~\cite{Piro:2006ja}. At such accretion 
rates, it is possible to achieve $\beta \ll \beta_c$ 
in collapsar disks~\cite{Lerner:2025dkd}, making 
fragmentation plausible over a broad parameter space.

\begin{figure*}[t!] 
    \centering
\includegraphics[width=0.49\linewidth]{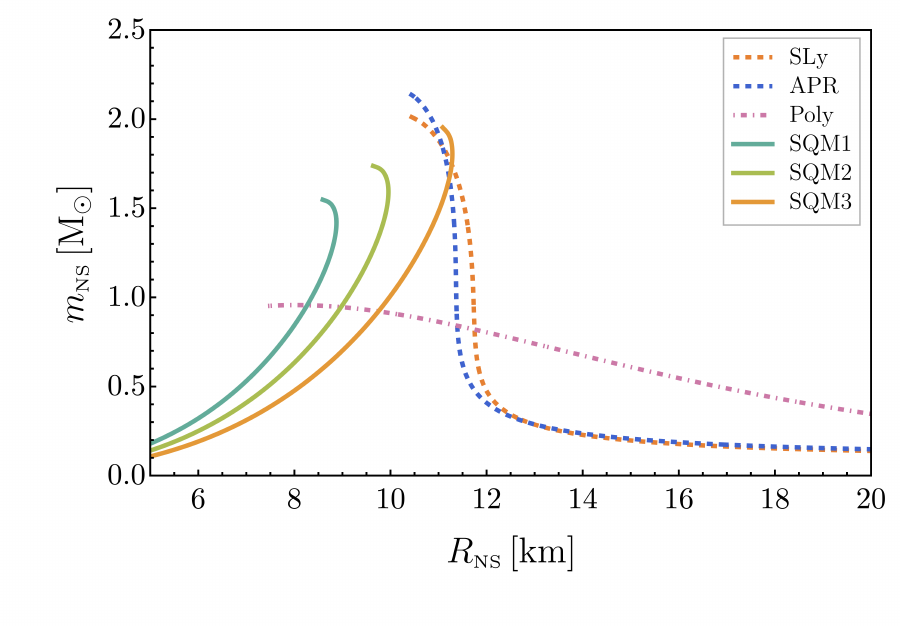}
\includegraphics[width=0.49\linewidth]{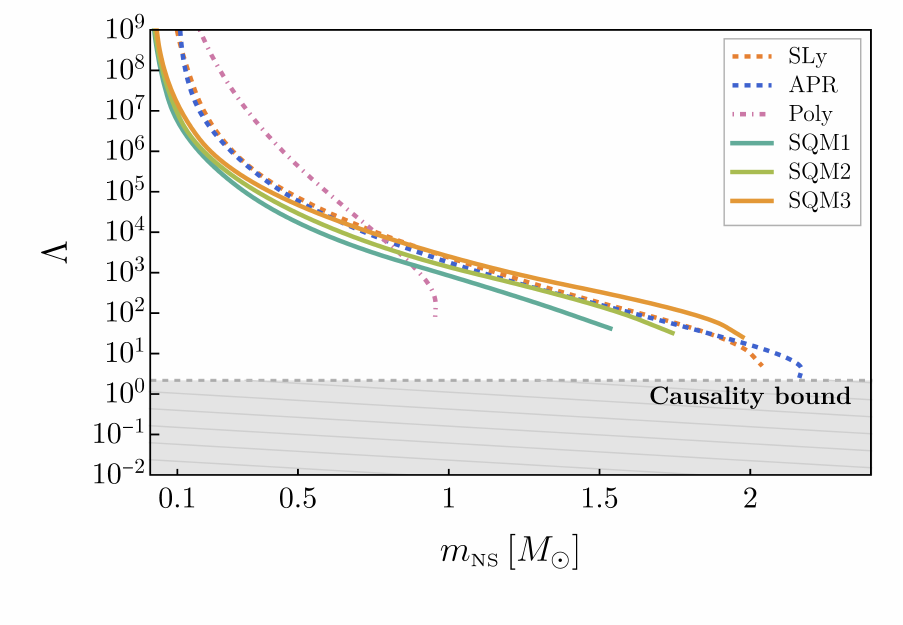}
    \caption{Left panel: Mass-radius relation for NSs with various EoSs. Dashed lines refer to NM EoSs, solid lines to QM EoSs, and the dot-dashed line shows the non-relativistic neutron polytrope of Eq.~\eqref{eq:Rpoly} with $Y_e = 0$. Right panel: NS tidal deformability as a function of mass. The gray region marks values below the causality bound for tidal interactions of Ref.~\cite{Russo:2025ivk}.
 }
    \label{fig:mr-TLN}
\end{figure*}

Fragmentation therefore requires a combination of sufficiently high accretion rates, large outer disk radii, moderately thin disks, and efficient cooling. Under these conditions, gravitationally unstable overdensities can collapse into self-bound clumps that subsequently evolve into subsolar-mass NSs. Because the same weak interactions which cool the disk gas act to neutronize the material and reduce its electron fraction, the local Chandrasekhar mass $M_\text{\tiny Ch} \propto Y_e^{2}$ in the disk is $\ll M_{\odot}$ [Eq.~\eqref{eq:MCh}]. Numerical simulations~\cite{Chen:2025uwd} confirm that most of the formed bound clumps are above the Chandrasekhar mass, and thus likely to continue to collapse to form subsolar NSs on a timescale much shorter than the dynamical timescale. Fragments that form below the local Chandrasekhar mass do not collapse to NSs: being pressure-supported, they either persist as low-mass degenerate objects or are tidally sheared and reincorporated into the disk flow, and thus do not contribute to the subsolar NS population.

Clump formation consumes most of the mass in the outer disk, $r\gtrsim R_*$, leaving a much lower residual gas density. Motivated by the large uncertainty in the degree of gas depletion, we consider suppression factors spanning $f_\text{\tiny gas} \approx 10^{-6}-10^{-2}$ relative to the initial density $\rho(R_*)$~\cite{Chen:2025uwd}. This residual gas density plays a crucial role in the subsequent dynamical evolution by setting the strength of dynamical friction acting on the BNSs, and determining whether their hardening is gas-driven or GW-driven (Sec.~\ref{sec:param_space}).  Accretion feedback --- energy released as gas falls onto the NS (see also Appendix~\ref{app:accretion} for further details) --- may also push gas away from the binary, reducing the effective $f_\text{\tiny gas}$ entering the drag force. In what follows we treat $f_\text{\tiny gas}$ as a free parameter to encapsulate these significant uncertainties.

\subsection{NS properties}
\label{sec:properties}
\noindent
The compact remnants formed in fragmenting collapsar disks are subsolar-mass NSs with microphysical properties that differ markedly from those of standard core-collapse NSs. At the densities and temperatures $T\sim{\rm MeV}$ characteristic of collapsar disk fragments, nucleons are non-relativistic and the clump is predominantly baryonic, with an electron fraction $Y_e \lesssim 0.1$~\cite{Chen:2025uwd}. 
After clump collapse, efficient neutrino cooling increases the core density by orders of magnitude on a timescale typically much shorter than subsequent merger or inspiral timescales (less than seconds)~\cite{Roberts:2012}, so the clump has time to contract and settle to a cold EoS before binary interactions become relevant. Only in rare cases of prompt, repeated mergers are finite-temperature effects likely to be important.

The left panel of Fig.~\ref{fig:mr-TLN} shows the mass-radius relations for NSs computed with a representative set of EoSs, including an idealized analytical benchmark and two physically distinct families of realistic cold matter.

The simplest description, which we refer to as the polytropic (Poly) EoS, is
an ideal gas of free, non-relativistic neutrons. The central densities reached
by subsolar configurations (in absence of QM) lie far below the scale $\sim 6\times 10^{15}\,
{\rm g\,cm^{-3}}$ at which neutrons become relativistic, so that degeneracy
pressure alone gives the polytropic relation~\cite{Shapiro:1983du}
\begin{equation}
    P = K \rho^{5/3}\,, \quad
    K = \frac{3^{2/3}\pi^{4/3}}{5}\frac{\hbar^2}{m_n^{8/3}}\,,
\label{eq:polyEoS}
\end{equation}
in terms of the rest-mass density $\rho$ and of the neutron mass $m_n$, i.e.,
a  polytrope of index $n = 3/2$. Its Lane-Emden solution fixes the
mass-radius relation to $R_\NS \propto m_\NS^{-1/3}$, with a prefactor set by
$K$ and by the electron fraction $Y_e$ of the material~\cite{Chen:2025uwd} (here $Y_e$ is treated as spatially uniform for simplicity; in reality weak interactions imprint a $Y_e$ profile across the star, so the value entering below should be understood as a representative average),
\begin{equation}
    R_\NS \simeq 33\,{\rm km}\,(1-Y_e)^{-1/3}
    \left(\frac{m_\NS}{0.1\,\msun}\right)^{-1/3}\,,
\label{eq:Rpoly}
\end{equation}
so that a $0.1\,\msun$ object reaches tens of km. Crucially for our purposes, this sequence admits no minimum mass: $R_\NS \to \infty$ as $m_\NS \to 0$, allowing the relation to be continued below the NM floor discussed below. This continuation is idealized in isolation, since at sufficiently low densities neutron-rich nuclei become unstable to $\beta$-decay and the composition, rather than the mechanical structure, sets the true lower limit~\cite{Shapiro:1983du}. In the collapsar environment, however, such an extrapolation may be partly justified if the dynamical time is shorter than the weak-interaction time, preventing $\beta$-decays from appreciably raising $Y_e$. We therefore use the polytrope as an analytical benchmark spanning the full subsolar range, rather than as a realistic model of the stellar interior.

The first realistic family we adopt assumes cold $\beta$-equilibrium and corresponds to NM EoSs, such as APR~\cite{Akmal:1998cf} and SLy~\cite{Douchin:2001sv}, which predict NS radii $R_\NS \sim 10 - 10^2$ km and a minimum cold stable mass of $m_\NS \gtrsim 0.1\,M_\odot$~\cite{Gondek:1997fd}. Although the lightest fragments can be born below this threshold, they could grow by gas accretion as they migrate toward the central BH (by up to two to three orders of magnitude in mass~\cite{Lerner:2025dkd}) so that a sub-threshold fragment crosses the minimum stable mass and settles onto the stable nuclear-matter branch rather than failing to form a NS.

The second family corresponds to QM EoSs, such as SQM1~\cite{Haensel:1986qb, Prakash:1995uw, Prakash:1996xs}, in 
which strange QM is absolutely stable. 
Quark stars lie on a qualitatively different
mass-radius sequence: they are self-bound rather
than gravity-bound, so arbitrarily low-mass
configurations exist in principle, while reaching radii as large as $10$ km when $m_\NS \gtrsim 1\, M_\odot$. This behavior is
captured by the bag model underlying the SQM1 EoS~\cite{Chodos:1974je, Johnson:1975zp},
in which the light quarks form a
nearly massless, deconfined Fermi gas held together by a
constant vacuum energy density $B$ (the bag constant),
giving an approximately linear EoS. Because the pressure vanishes at a finite surface energy density, the
star is bound by the strong interaction rather than by
gravity, and in the low-mass limit its nearly uniform
interior gives $m_\NS \propto R_\NS^3$, so that the
radius shrinks to zero as $m_\NS \to 0$.
Therefore, the window of 
masses below the nuclear-matter minimum, 
$m_\NS \lesssim 0.1\,M_\odot$, is accessible only if 
the NS interior contains QM. As we discuss in 
Sec.~\ref{sec:detection}, this mass window 
therefore provides a direct observational 
discriminant between the two EoS families.
Regarding the possibility of QM, note 
that along the quark star mass-radius relation, 
the energy density is set by the bag constant 
($B\sim 10^{14}-10^{15}\,{\rm g\,cm^{-3}}$), 
which is much larger than typical fragmentation 
densities, so QM is excluded at 
formation. However, collapse may bring the core into the density regime where a transition to deconfined QM becomes possible, provided strange matter is sufficiently stable and conversion can nucleate.

Aside from their mass-radius relations, NSs are characterized by a nonzero tidal deformability, usually quantified in terms of the (dimensionless)  response parameter
\begin{equation}\label{eq:Lambda}
    \Lambda = \frac{2}{3}\,k_2
    \left(\frac{R_\NS}{m_\NS}\right)^5\,,
\end{equation}
where $k_2$ is the $\ell=2$ tidal Love number (TLN)~\cite{Hinderer:2007mb} and
$R_\NS$ is the stellar radius. The TLN encodes the
quadrupolar polarizability of the star and depends
sensitively on its EoS, which characterizes its interior properties.  Throughout, $k_2$ (and hence $\Lambda$) is computed for nonrotating, unmagnetized equilibrium configurations. The fragments considered here may nonetheless carry sizable spin (Appendix~\ref{app:NSspin}), which modifies the stellar tidal response: rotation couples the spin to the external tidal field and introduces spin-dependent corrections to the TLNs~\cite{Pani:2015nua,Pani:2015hfa,Landry:2015zfa}. These corrections scale with the spin and vanish in the static limit; we neglect them here, retaining spin only through the spin-induced quadrupole and the spin-orbit resonance discussed in Appendix~\ref{app:NSspin}, and defer a self-consistent treatment of spin-dependent tides for subsolar NSs to future work.

The right panel of Fig.~\ref{fig:mr-TLN} shows the 
dimensionless tidal deformability $\Lambda$ as a 
function of NS mass for the same set of EoSs. 
Subsolar NSs are characterized by extremely large 
tidal deformabilities: for an NM EoS, 
$\Lambda$ ranges from $\mathcal{O}(10^3)$ at 
$m_\NS\sim 1\,M_\odot$ to $\mathcal{O}(10^9)$ at 
$m_\NS\sim 0.1\,M_\odot$ because heavier configurations are more compact and therefore less tidally deformable,  
while quark stars generically have smaller $\Lambda$
at fixed mass due to their larger compactness. 
For the polytropic EoS below masses $m_\NS\lesssim 0.4\,M_\odot$, the TLN is nearly mass-independent, $k_2\simeq0.14$, as obtained by solving the standard $\ell=2$ tidal perturbation equation on the $n=3/2$ Lane--Emden profile~\cite{Hinderer:2007mb}, so that Eq.~\eqref{eq:Lambda} can be evaluated in closed form:
\begin{equation}
   {\rm Poly}: \quad  \Lambda \simeq  4\times10^{10}\,(1-Y_e)^{-5/3}
    \left(\frac{m_\NS}{0.1\,\msun}\right)^{-20/3}\,,
\label{eq:LambdaPoly}
\end{equation}
the steep scaling following directly from $R_\NS \propto m_\NS^{-1/3}$ through
the compactness $m_\NS/R_\NS \propto m_\NS^{4/3}$. Over the plotted range, the polytropic model provides an approximate upper envelope to more realistic tidal deformabilities.

These large tidal deformabilities have two 
important consequences. First, they leave 
a significant imprint on the GW inspiral 
waveform~\cite{Franciolini:2021xbq, Chia:2023tle, Crescimbeni:2024cwh,Crescimbeni:2024qrq, DeLuca:2024uju, Russo:2025ivk, Corman:2026lbt}, with subsolar 
NSs being tidally disrupted at much lower 
frequencies than their standard-mass counterparts, 
potentially within the sensitive band of current 
and future detectors (see Sec.~\ref{sec:NS-BH}). 
Second, a confident measurement of a large 
tidal deformability would be incompatible with a 
BH interpretation, providing clear 
evidence for a NS and breaking the 
degeneracy with PBH 
scenarios~\cite{Franciolini:2021xbq, Crescimbeni:2024cwh,Crescimbeni:2024qrq, DeLuca:2024uju, Russo:2025ivk, Corman:2026lbt}. 
The gray shaded region in the right panel marks 
values below the causality bound on tidal 
interactions derived in Ref.~\cite{Russo:2025ivk}: 
tidal deformabilities in this region would 
require superluminal sound speeds, and are 
therefore excluded for any physical EoS.

In summary, the disk structure and stability analysis provide the expected density and surface-density profiles at each radius, constraints on the accretion rates required for fragmentation, and natural scales for the masses and number of clumps formed. The assumed cold EoS then fixes the radii and tidal deformabilities of the NS clumps. These ingredients are crucial for analyzing the subsequent orbital evolution, which we split into two parts:

\noindent
\emph{(i)~the inner merger phase}, involving interactions, hardening, and eventual coalescence of the subsolar-mass NS clumps formed through fragmentation within the disk, and 

\noindent
\emph{(ii)~the outer merger phase}, in which NS clumps (or their hierarchical merger product) inspiral and merge with the central BH. 

Each phase, summarized schematically in Fig.~\ref{fig:cartoon}, has distinctive GW  signatures that we will explore in the following sections.

\section{Parameter space and dynamical phases}
\label{sec:param_space}
\noindent
The two merger phases introduced at the end of Sec.~\ref{sec:model} offer distinctive GW signatures that we analyze in Secs.~\ref{sec:NS-NS} and~\ref{sec:NS-BH}, respectively. 
Before doing so, we map out the parameter space and the dynamical timescales that govern each phase, as summarized in Fig.~\ref{fig:param}.

\subsection{BNS}
\label{sec:param_NSNS}
\noindent
Predicting the dynamics of the NS clumps formed by disk fragmentation is, in general, a complicated $N$-body problem that will involve hierarchical mergers~\cite{Metzger:2024ujc,Tagawa:2019osr} (and the corresponding gravitational recoil kicks that could unbind some clumps from the disk~\cite{Varma:2020nbm, Lerner:2025dkd}) as well as torques from the surrounding gas disk.
The collapsing clumps typically acquire large intrinsic angular momentum due to the velocity shear across them, which is generally above the mass-shedding limit for NSs~\cite{Alexander:2008wv}; the excess angular momentum must therefore be redistributed, either through fission into one or more sub-bodies or through strong gravitational scattering among clumps.
Furthermore, the final outcome is subject to statistical fluctuations due to the intrinsically stochastic nature of the fragmentation process~\cite{Chen:2025uwd}.

Rather than attempting to model the full N-body evolution, we isolate a representative subprocess: the merger of two subsolar-mass NSs of mass $m_\NS$ and initial separation $a_{\BNS,i}$,
driven by GW emission.  Two channels can produce such a
binary, leading to qualitatively different initial orbital
configurations: 

\noindent
{\it (i)}  \emph{Fission} of an over-spinning clump could deposit excess angular momentum into the relative orbit of two proto-NSs, plausibly producing a close binary with modest eccentricity~\cite{Alexander:2008wv} (see also Appendix~\ref{app:NSspin} for a discussion of the NS spins); 

\noindent
{\it (ii)} \emph{Capture} of two independently formed NSs, either in vacuum via GW energy loss, or in the gas-rich environment where dynamical friction enhances the capture cross section and accelerates coalescence once a binary forms, generally results in an eccentric binary. We estimate capture rates induced by disk migration in Appendix~\ref{app:capturerate}. 

Both channels can operate simultaneously, and multi-body encounters among the $N_\NS\sim\mathcal{O}(10)$ fragments add further stochasticity. The initial eccentricity is an important observational discriminant that we quantify in Sec.~\ref{sec-eccentricity}.

In what follows, we adopt an agnostic approach, treating $m_\NS$ and $a_{\BNS,i}$ as free parameters that span the uncertain outcomes of the underlying N-body dynamics. An upper bound on $a_{\BNS,i}$ is set by the Hill radius of the NS binary
relative to the central BH, located at an initial distance $a_{\BNS-\BH,i}$.  Assuming for simplicity an equal-mass NS system, we have
\begin{align}
    r_\text{\tiny Hill} &= a_{\BNS-\BH,i} \left(\frac{2m_\NS}{3 M_\BH}\right)^{1/3} \nonumber \\
    & \simeq 
    R_*
    \left(\frac{2m_\NS}{3 M_\BH}\right)^{1/3}
    \simeq 0.2 \,R_*
    \left(\frac{m_\NS/M_\BH}{10^{-2}}\right)^{1/3}\,,
    \label{eq:rHill_outer}
\end{align}
so that configurations with $a_{\BNS,i} > r_\text{\tiny Hill}$ cannot form a bound binary (in the second equality, we have approximated the initial BNS-BH distance to be of the order of the fragmentation radius $R_*$).  We parametrize the initial separation through the dimensionless Hill fraction,
\begin{equation}
    f_\text{\tiny H} \equiv \frac{a_{\BNS,i}}{r_\text{\tiny Hill}}\,,
    \qquad f_\text{\tiny H} \in (0,1)\,,
    \label{eq:fH}
\end{equation}
which serves as the central free parameter of the NS-merger analysis. 
Not all values of $f_\text{\tiny H}$ are dynamically stable, however: the periodic tidal perturbations exerted by the central BH on the inner NS binary can drive it into chaotic instability whenever the two orbits are not sufficiently hierarchical.
This requirement places an upper bound on $f_\text{\tiny H}$ that depends on the mass ratio $m_\NS/M_\BH$ and, for an eccentric inner binary, weakens as $1/(1+e_{\BNS,i})$~\cite{Mardling2001}. At $m_\NS=0.1\,M_\odot$ and $M_\BH=5\,M_\odot$ the criterion gives $f_\text{\tiny H} \lesssim 0.4$ and, more generally, it permits $f_\text{\tiny H}\lesssim 0.5$ for prograde triples, while allowing  fractions as large as $f_\text{\tiny H}\sim 1$ for retrograde configurations~\cite{Grishin2017}. In what follows we therefore let $f_\text{\tiny H}$ span its whole range, keeping in mind that values $f_\text{\tiny H}\gtrsim 0.5$ are generally unstable for prograde configurations, and therefore preferentially require retrograde or otherwise unusually stable geometries.

The initial separation is also bounded from below: the BNS cannot start with the two NSs already in contact, so
$a_{\BNS,i}(1-e_{\BNS,i})>2R_{\NS}$, where
\begin{align}
{\rm Poly}:& \quad  R_{\NS} \approx 34 \, {\rm km} \, \left(\frac{1-Y_e}{0.9}\right)^{-1/3}  \left(\frac{m_\NS}{0.1\,M_\odot}\right)^{-1/3}\nonumber\,,
\\
{\rm NM}:& \quad  R_{\NS} \approx 11 \, {\rm km}\left[1+
    \!3\left(\frac{m_\NS}{0.1\,M_\odot}\right)^{-3}\right] \nonumber\,,
\\
{\rm QM}:& \quad R_{\text{\tiny NS}} \approx 4 \, {\rm km}\left(\frac{m_\NS}{0.1\,M_\odot}\right)^{1/3} \,.
\label{eq:R_NS}
\end{align}
The first line is the analytical polytropic result of Eqs.~\eqref{eq:polyEoS}
and~\eqref{eq:Rpoly}, which we evaluate throughout at the fiducial disk
composition $Y_e \simeq 0.1$; the remaining two are calibrated against the APR
EoS for NM and the SQM1 EoS for QM, and capture the qualitative behavior of
the $m_\NS - R_\NS$ relation of either class of matter, largely independently
of the specific EoS within that class. The three expressions have complementary ranges of validity. We adopt the polytrope as the fiducial prescription across the full range $m_\NS \in [0.01,1]\,\msun$, though at the lowest densities it ceases to describe a $\beta$-stable interior; all baseline detection prospects and modulation results of Secs.~\ref{sec:detection}, \ref{sec:modeffects}, \ref{sec-eccentricity}, \ref{Sec:detNSBH} and~\ref{sec:NS-BH_ecc} refer to it. The self-bound QM branch extends to arbitrarily low masses, whereas the NM fit is restricted to $m_\NS \gtrsim 0.1\,\msun$, its minimum stable mass. The EoS enters the detectability through the mass-radius relation --- via the formation and termination frequencies --- and we discuss the corresponding NM and QM modifications at the end of each of those sections. Wherever the EoS is itself the observable at stake (e.g. in the environmental and tidal analysis of Sec.~\ref{sec:envtidalNSNS}, and in the measurability study of Sec.~\ref{sec:FisherII}) we work directly with the NM and QM pair, whose contrasting predictions provide the discriminant.

The radius of the NS ultimately translates into a lower bound on the Hill fraction of the form
\begin{equation}
    {\rm no \,\, initial \,\, contact:}  \quad   f_\text{\tiny H} \gtrsim
        \frac{2R_\NS}{R_*\,(2m_\NS/3M_\BH)^{1/3}}\,.
        \label{eq:nstouch}
    \end{equation}
Numerically, this translates to
\begin{align}
    f_\text{\tiny H} &\gtrsim 0.12 \left(\frac{R_\NS}{34\,{\rm km}}\right)
    \left(\frac{m_\NS}{0.1\,\msun}\right)^{-1/3} \nonumber \\
    &\times \left(\frac{M_\BH}{5\,\msun}\right)^{-2/3}
    \left(\frac{R_*}{340\,R_g}\right)^{-1},
    \label{eq:nstouch_num}
\end{align}
so that inserting the mass-radius relations of Eq.~\eqref{eq:R_NS} gives $f_\text{\tiny H} \gtrsim 0.12\,(m_\NS/0.1\,\msun)^{-2/3}$ for the polytrope, and a mass-independent bound $f_\text{\tiny H} \gtrsim 0.014$ for QM.
The blue and magenta lines in Fig.~\ref{fig:param} (left panel) show this contact boundary for an NM and a QM EoS, respectively; the strip between the two lines is accessible only if the NS interior contains QM. The different trends in the magenta and blue lines depicted in Fig.~\ref{fig:param} ultimately reflect the fundamental difference between the QM and NM EoSs. In the former case, decreasing the mass also leads to a decrease in the radius, such that the NS density remains approximately constant, and therefore so does the dimensionless Hill fraction, since $f_{\text{\tiny H}} \sim R_{\NS} / m_{\NS}^{1/3}$. Conversely, the density of standard NM decreases as the mass decreases, leading to a rapidly growing
$f_{\text{\tiny H}}$ as $m_\NS \to 0.1 \, M_\odot$.

As the subsolar-mass binary evolves in the residual gas environment
after fragmentation, it will experience dynamical friction~\cite{Ostriker:1998fa,Kim:2007zb,Kim:2008ab}.
Following disk fragmentation, the clumps consume most of the outer disk
mass, leaving a suppressed residual density
$\rho_\text{\tiny env} \equiv f_\text{\tiny gas} \, \rho(R_*) \approx (10^{-6} - 10^{-2})\,\rho(R_*)$~\cite{Chen:2025uwd}. On scales $a_{\BNS,i}\ll h$, the binary is immersed in an approximately homogeneous medium, making the standard Ostriker drag formula a suitable local approximation~\cite{Chandrasekhar:1943ys,Ostriker:1998fa} (see Ref.~\cite{Vicente:2019ilr} for extensions beyond the homogeneous assumption).
The gas exerts a retarding force on each NS
proportional to $\rho_\text{\tiny env}$, extracting orbital energy at a rate
$\dot{E}_\text{\tiny dyn} \propto \rho_\text{\tiny env}\,m_\NS^2/v_\BNS$, $v_\BNS$ being the orbital velocity of the NS binary,
and giving rise to a dynamical-friction hardening timescale
\begin{align}
\tau_\text{\tiny dyn}
    & = \frac{E_\text{\tiny orb}}{|\dot{E}_\text{\tiny dyn}|}
    = \frac{E_\text{\tiny orb}}{\left(4\pi \rho_\text{\tiny env} m_\NS^2 / v_{\BNS}\right)\mathcal{I}} \nonumber \\
    & \simeq 0.3 \,  {\rm s} \left( \frac{M_\BH}{5\,M_\odot}\right) 
\left( \frac{\mathbb{H}}{0.3}\right)^3
\left( \frac{f_\text{\tiny H}}{0.5}\right)^{-3/2} \nonumber \\
& \times \left(\frac{\dot{M}_\BH}{M_\odot\, {\rm s}^{-1}}\right)^{-1} \left( \frac{f_\text{\tiny gas}}{10^{-2}}\right)^{-1}
    \,,
    \label{eq:dyntimescaleSSNSs}
\end{align}
where $E_\text{\tiny orb}=m_\NS^2/(2a_\BNS)$ is the orbital energy, and $\mathcal{I} \approx \mathcal{O}(5)$
is a Coulomb logarithm. The latter depends on the object's Mach number $\sim 1/\mathbb{H}$ and the binary's initial separation $f_\text{\tiny H}$ through the expression~\cite{Ostriker:1998fa}
\begin{align}
\label{calIDF}
\mathcal{I} \simeq \frac{1}{2}\log \left(1 - \frac{c_s^2}{v_\BNS^2} \right) + \log \left(\frac{v_\BNS \, \tau_\text{\tiny orb}}{m_\NS} \right)\,,
\end{align}
in terms of the outer orbital period
$\tau_\text{\tiny orb}=2\pi/\Omega$. This estimate neglects disk shear, wake overlap, and the backreaction of the binary on the local gas, and should therefore be regarded as an order-of-magnitude hardening timescale~\cite{Derdzinski:2020wlw,Zwick:2021dlg}.

Because $\tau_\text{\tiny dyn}
\propto \rho_\text{\tiny env}^{-1}$, the role of the gas depends critically on
the post-fragmentation density suppression factor: high $\rho_\text{\tiny env}$
drives the binary by gas drag until GW emission becomes
significant (orange region in Fig.~\ref{fig:param}, left panel),
while low $\rho_\text{\tiny env}$ leaves GW emission as the dominant channel. Since the boundaries of these regions depend on the assumed disk density, fixed to $\rho(R_*) = 10^8\,{\rm g\,cm^{-3}}$ in Fig.~\ref{fig:param}, in Appendix~\ref{app:NSNSdetails} we show the corresponding parameter space for a higher value, $\rho(R_*) = 10^{10}\,{\rm g\,cm^{-3}}$ (Fig.~\ref{fig:param_app}). There we also discuss how the contact boundary, the entry frequency, and the density suppression factors that determine the dynamical-friction-driven regions scale with $\rho(R_*)$.

The other relevant timescale is the Peters GW
inspiral time~\cite{Peters:1963ux,Peters:1964zz}
\begin{equation}
    \tau_\GW^{\BNS}
    = \frac{5}{512}\frac{a_\BNS^4}{m_\NS^3}\,,
    \label{eq:GWtimescaleSSNSs}
\end{equation}
which is the time for GW emission to remove an $\mathcal{O}(1)$ fraction of the orbital energy.  

When $\tau_\text{\tiny dyn}\gtrsim \tau_\GW^{\BNS}$ (beige region in Fig.~\ref{fig:param}), the inspiral is driven primarily by GW emission and the BNS survives for many outer orbits. If, in addition, $\tau_\GW^{\BNS} \gtrsim \tau_\text{\tiny orb}$ (cross-hatched beige region in Fig.~\ref{fig:param}), the hierarchical triple geometry can imprint characteristic modulations on the GW signal, including Doppler phase shifts, relativistic aberration, gravitational redshift, Shapiro time delay, and self-lensing; these effects are analyzed in Sec.~\ref{sec:modeffects}. By contrast, when $\tau_\GW^{\BNS} \ll \tau_\text{\tiny dyn},\,\tau_\text{\tiny orb}$, these modulation effects are suppressed, while GW emission brings the NSs to contact, potentially at frequencies inside the band of ground-based GW detectors. This regime is the main focus of Sec.~\ref{sec:envtidalNSNS}.

As shown in Appendix~\ref{app:accretion}, both GW emission and dynamical friction harden the binary on timescales much shorter than the NS mass-doubling time by Bondi accretion. Accretion-driven mass growth can therefore be neglected in the waveform model of Sec.~\ref{sec:NS-NS}. In the same Appendix, we compare these hardening timescales with the disk-depletion time $\tau_\text{\tiny acc}^\BH$ of the central BH. Across most of the parameter space, the inner binary hardens before the gas is consumed --- by GW emission at small $f_\text{\tiny H}$ and by dynamical friction at large $f_\text{\tiny H}$ --- so the residual gas persists through merger and can imprint the environmental effects discussed below.

\subsection{NS-BH binary}
\label{sec:param_NSBH}
\noindent 
A second channel is the inward migration and eventual merger of a NS with the central BH~\cite{Metzger:2024ujc}. The NS may form directly from the collapse of a single clump, or as the remnant of a hierarchical BNS merger chain like that described in Sec.~\ref{sec:NS-NS}; in the latter case, the merger of two subsolar NSs is likely to leave a NS remnant rather than a BH~\cite{Corman:2026lbt}. We therefore treat $m_\NS$ as a free parameter to cover both possibilities.

Once formed near the fragmentation radius $R_*$, the clump migrates inward
toward the central BH through the surrounding gas disk.  Its inward drift
is driven primarily by the asymmetric gravitational torque it exerts on
the gas, in analogy with Type~I planetary migration in
protoplanetary disks~\cite{Ward:1997di,Tanaka2002,Lin1986,Kley2012}.
The physical mechanism is the following: as the NS orbits, it launches
spiral density waves at the Lindblad resonances ahead of and behind
its orbit. The wave trailing the NS carries more angular momentum than
the leading one, producing a net negative torque that extracts orbital
angular momentum and drives the NS inward.  The angular momentum of
the NS at $R_*$ is $J_\NS = m_\NS \sqrt{M_\BH R_*}$, and the
net torque can be written as~\cite{Ward:1997di,Tanaka2002, 2010MNRAS.401.1950P}
\begin{equation}
    \Gamma_\text{\tiny mig,I}
    \simeq \frac{C_{\rm I}\,\dot{M}_\BH\,m_\NS^2}{3\pi\alpha\,\mathbb{H}^4\,M_\BH^2}
    \sqrt{M_\BH R_*}\,,
    \label{eq:Gamma_mig}
\end{equation}
where $C_{\rm I}$ encodes the torque efficiency, which depends primarily on the local logarithmic slopes of the disk temperature and surface density~\cite{2017MNRAS.471.4917J}. For the collapsar disk profiles of Sec.~\ref{sec:model}, these gradients are set by the accretion rate and aspect ratio, giving $C_{\rm I} \sim \mathcal{O}(1)$ across the parameter space of interest. We note that the Type~I torque formula strictly applies in the linear regime, i.e., before the fragment opens a gap in the disk (which we discuss below). The corresponding migration timescale is then
\begin{align}
    \tau_\text{\tiny mig}
    & = \frac{J_\NS}{2\,\Gamma_\text{\tiny mig,I}}
    = \frac{3\pi\alpha\,\mathbb{H}^4\,M_\BH^2}
           {2\,C_{\rm I}\,\dot{M}_\BH\,m_\NS} \nonumber \\
    & \simeq 5 \times 10^{-2} \,  {\rm s} \left( \frac{M_\BH}{5\,M_\odot}\right) 
\left( \frac{\mathbb{H}}{0.3}\right)^4 \nonumber \\
& \times \left(\frac{m_\NS/M_\BH}{10^{-2}}\right)^{-1} \left( \frac{\dot{M}_\BH}{M_\odot\, {\rm s}^{-1}}\right)^{-1}
           \,.
    \label{eq:tau_mig}
\end{align}
We set $C_{\rm I}=2.1$ in the numerical estimates; variations by factors of a few shift the regime boundaries correspondingly.

Equating the rate of change of the orbital energy to the combined losses from GW emission and migration torques gives
\begin{equation}
\dot{E}_\text{\tiny orb} = - \dot{E}_\GW - \dot{E}_\text{\tiny mig} \,,
\end{equation}
or, equivalently,
\begin{equation}
    \dot{v} = \frac{v}{2}
    \left(\frac{1}{\tau_\text{\tiny mig}}
    + \frac{v^8}{4 \, v_*^8\,\tau_{\GW}^{\NS-\BH} }\right),
    \label{eq:eom}
\end{equation}
where $v_* = \sqrt{M_\BH/R_*}$ is the orbital velocity at $R_*$, and $\tau_\GW^{\NS-\BH}$ is the Peters inspiral time for the NS-BH binary evaluated at $R_*$,
\begin{equation}
\label{tau_GW_BH}
\tau_{\GW}^{\NS-\BH} =  \frac{5}{256}\frac{M_\BH^2}{v_*^8 m_{\text{\tiny NS}}}\,.
\end{equation}
Approximating $\tau_\text{\tiny mig}$ as constant during the inspiral, separation of variables in Eq.~\eqref{eq:eom} gives
\begin{equation}
    \tau_\text{\tiny merger}
    \approx \frac{1}{4}\tau_\text{\tiny mig}
      \log\!\left(1 + \frac{4\, \tau_{\GW}^{\NS-\BH}}{\tau_\text{\tiny mig}}\right).
    \label{eq:tmerger}
\end{equation}
In the migration-dominated limit ($\tau_\text{\tiny mig}\ll\tau_{\GW}^{\NS-\BH}$),
the inspiral proceeds exponentially fast and
$\tau_\text{\tiny merger}\approx\tau_\text{\tiny mig}$; in the GW-dominated limit
($\tau_{\GW}^{\NS-\BH} \ll\tau_\text{\tiny mig}$), one recovers the standard Peters
result $\tau_\text{\tiny merger}\approx\tau_{\GW}^{\NS-\BH}$. Even in the latter regime, the residual disk torque leaves a perturbative
imprint on the waveform. The relative correction to the frequency evolution is
the ratio of the migration to the GW term in Eq.~\eqref{eq:eom},
\begin{equation}
    \varepsilon_\text{\tiny mig}
    = \frac{4\,v_*^{8}\,\tau_\GW^{\NS-\BH}}{\tau_\text{\tiny mig}\,v^{8}}
    = \frac{5\,C_{\rm I}\,\dot{M}_\BH}{96\pi\,\alpha\,\mathbb{H}^{4}\,v^{8}}\,.
    \label{eq:eps_mig}
\end{equation}
The associated dephasing is given by
\begin{equation}
    \delta\psi_\text{\tiny mig}
    = -\frac{5}{26}\,\varepsilon_\text{\tiny mig}\,\psi_\text{\tiny PP}
    \propto f^{-13/3}\,,
    \label{eq:dpsi_mig}
\end{equation}
with $\psi_\text{\tiny PP}\propto f^{-5/3}$ the leading-order vacuum phase:
a $-4$\,PN correction, the ground-based analog of the disk-migration
imprints studied for extreme mass-ratio
inspirals~\cite{Kocsis:2011dr,Yunes:2011ws,Derdzinski:2018qzv,Zwick:2021dlg,Speri:2022upm}.

This inspiral, however, requires disk gas to survive until the NS reaches the inner disk. Since the clump forms near $R_*$, where $Q\simeq Q_0$, it must first migrate inward by a radial distance of order the disk scale height, $h=\mathbb{H}R_*$, before entering the stable inner disk, $r\lesssim R_*$, where $Q>1$ and the gas density is no longer suppressed by fragmentation. Repeating the integration of Eq.~\eqref{eq:eom}, but stopping when the orbit has shrunk from $R_*$ to $(1-\mathbb{H})R_*$ rather than continuing to coalescence, gives the crossing timescale
\begin{equation}
    \tau_\text{\tiny J} = \frac{\tau_\text{\tiny mig}}{4}\,
    \log\!\left[\frac{\tau_\text{\tiny mig}+4\,\tau_\GW^{\NS-\BH}}
    {\tau_\text{\tiny mig}+4\,(1-\mathbb{H})^4\,\tau_\GW^{\NS-\BH}}\right]\,,
    \label{eq:tauJ}
\end{equation}
which reduces to $\tau_\text{\tiny mig}\,\log[1/(1-\mathbb{H})]$ in the
migration-dominated limit ($\tau_\text{\tiny mig}\ll\tau_\GW^{\NS-\BH}$) and to
$\big[1-(1-\mathbb{H})^4\big]\,\tau_\GW^{\NS-\BH}$ in the GW-dominated one
($\tau_\GW^{\NS-\BH}\ll\tau_\text{\tiny mig}$).

The critical comparison is between $\tau_\text{\tiny J}$ and the disk-depletion time,
$\tau_\text{\tiny acc}^\BH=M_\text{\tiny disk}/\dot{M}_\BH$, over which accretion onto the central BH consumes the disk [see Eq.~\eqref{eq:taccBH}]. If $\tau_\text{\tiny J} > \tau_\text{\tiny acc}^\BH$, the disk has already been consumed by the time the NS reaches the inner disk, and the binary subsequently evolves in vacuum. This corresponds to the beige region in the right panel of Fig.~\ref{fig:param}; the resulting GW-dominated inspiral is studied in Sec.~\ref{sec:NS-BH}. Conversely, if $\tau_\text{\tiny J} < \tau_\text{\tiny acc}^\BH$ (orange regions), the NS enters a gas-rich environment, with the different colors indicating different values of the density suppression factor $f_\text{\tiny gas}$. As expected, smaller values of $f_\text{\tiny gas}$ enlarge the parameter space in which the inspiral is GW-dominated.

In the gas-rich region, accretion onto the NS as it moves through the disk may also become relevant, as discussed in Appendix~\ref{app:accretion}. However, wherever the disk survives until the NS arrives, the inward evolution is initially migration-dominated, $\tau_\GW^{\NS-\BH} \gg \tau_\text{\tiny mig}$. If the remaining lifetime of the gas, $\tau_\text{\tiny acc}^\BH-\tau_\text{\tiny J}$, exceeds $\tau_\text{\tiny mig}$, the NS is delivered to the central BH primarily by disk torques rather than by GW emission, and the merger time approaches the migration limit, $\tau_\text{\tiny merger}\approx\tau_\text{\tiny mig}$, of Eq.~\eqref{eq:tmerger}. If instead $\tau_\text{\tiny acc}^\BH-\tau_\text{\tiny J}<\tau_\text{\tiny mig}$, the inspiral should transition from a migration-dominated phase to a GW-driven phase as the environment fades. Whenever the dynamics is migration-dominated, most of the orbital energy is deposited into the gas rather than radiated as GWs, suppressing the GW signal.

For sufficiently massive NSs or sufficiently thin disks, the tidal torque exerted by the NS can overcome viscous diffusion, opening a gap around its orbit and driving a transition from Type~I to Type~II migration. In the Type~II regime, the NS is coupled to the viscous evolution of the disk and drifts inward on approximately the viscous timescale~\cite{Lin1986,Kley2012}. Following Refs.~\cite{Metzger:2024ujc,Crida:2005zp}, this transition occurs above a critical mass
\begin{equation}
    M_\text{\tiny gap}
    \simeq \frac{50\,\nu\,M_\BH}{R_*^2\,\Omega} \approx  \mathbb{H}^2 M_\BH  \,,
    \label{eq:Mgap}
\end{equation}
which, using the disk profiles of Sec.~\ref{sec:model}, can be evaluated to be $M_\text{\tiny gap}\sim 1 - 10\,M_\odot$ across the range of $R_*$ of interest~\cite{Metzger:2024ujc}, well above the typical clump mass $m_\text{\tiny clump}\simeq\mathbb{H}^3 M_\BH/Q_0$ [see Eq.~\eqref{clumpmass}].  Gap opening is therefore negligible for freshly formed clumps, but becomes relevant once hierarchical mergers (or gas accretion) have grown the NS to $m_\NS\gtrsim M_\text{\tiny gap}$; this corresponds to the cross-hatched orange region in the right panel of Fig.~\ref{fig:param}. Once this threshold is crossed, the migration transitions from Type~I to Type~II, with the  torque given by~\cite{Kanagawa2018}
\begin{equation}
    \Gamma_\text{\tiny mig,II} = 
    \frac{1}{1+0.04 \, \kappa_\text{\tiny gap}}\Gamma_\text{\tiny mig,I}\,,
    \label{eq:GammaII}
\end{equation}
in terms of the dimensionless parameter
\begin{equation}
    \kappa_\text{\tiny gap} \equiv 
    \left(\frac{m_\NS}{M_\BH}\right)^2 
    \frac{1}{\alpha \, \mathbb{H}^5} 
    \,.
    \label{eq:kappa_gap}
\end{equation}
Values $\kappa_\text{\tiny gap} \gtrsim 20$ generally imply gap opening, corresponding to $m_\NS \gtrsim M_\text{\tiny gap}$. We show the resulting parameter space in the $(m_\NS/M_\BH,\,\mathbb{H})$ plane, taking $\mathbb{H}\in[0.2,0.5]$ so that the initial clumps remain subsolar according to Eq.~\eqref{clumpmass}. We extend the mass ratio up to order unity to include the limiting case in which the NS is the remnant of a hierarchical merger chain involving $\sim 10$ clumps, so that $m_\NS \sim 10\,m_\text{\tiny clump}\sim M_\BH$.

\section{BNS mergers}
\label{sec:NS-NS}
\noindent
The BNSs formed in the collapsar disk (as described in Sec.~\ref{sec:param_space}) inspiral as hierarchical triples, with the inner BNS embedded in a dense gaseous environment while orbiting the central BH. This configuration can imprint several signatures on the GW signal, including Doppler phase shifts from the outer orbit, eccentricity of the inner binary, gas-driven hardening through dynamical friction, and tidal deformability effects near merger. In this section we quantify these effects in turn, beginning with the waveform tools used to assess their detectability.

\subsection{Inspiral waveform and data analysis}
\label{sec:waveform:tf2}
\noindent
We model the GW signal from compact binary coalescences using the frequency-domain, \texttt{TaylorF2} PN waveform template in the stationary-phase approximation~\cite{Buonanno:2009zt, Blanchet:2013haa, Damour:2000zb}. This template describes the inspiral phase of the coalescence and, in the Fourier domain, takes the form
\begin{equation}
    \tilde h (f; \bm{\theta}) = C_{\Omega}
    (\bm{\theta}) \, \mathcal{A}(f; \bm{\theta}) 
    \, \exp\Big[{\rm i}\, \psi_\text{\tiny PP} 
    (f; \bm{\theta}) \Big]\,,
\end{equation}
where $\bm{\theta}$ denotes the set of intrinsic and extrinsic source parameters. The amplitude $\mathcal{A}(f; \bm{\theta})$ and point-particle phase $\psi_\text{\tiny PP}$ depend on the binary masses and aligned spin components $\chi_{1,z}$ and $\chi_{2,z}$, which in this analysis are assumed to be sizable, given the discussion in Appendix~\ref{app:NSspin}. The leading Newtonian amplitude reads
\begin{equation}
\label{GWamplitude}
    \mathcal{A} (f ; \bm{\theta}) = 
    \sqrt{\frac{5}{24}} \frac{\mathcal{M}_c^{5/6}}
    {\pi^{2/3} d_\text{\tiny L} f^{7/6}} \,,
\end{equation}
where $\mathcal{M}_c$ is the chirp mass of the 
binary and $d_\text{\tiny L}$ its luminosity distance. 
The geometric coefficient
\begin{equation}
C_{\Omega} (\bm{\theta}) = \left[F_+^2 
(1+\cos^2 \iota)^2 + 4 F_\times^2 \cos^2 
\iota\right]^{1/2}
\end{equation}
encodes the detector response, where $\iota$ is the inclination angle and the antenna pattern functions $F_{+,\times}$ depend on the sky location and polarization angle~\cite{Jaranowski:1994xd}. We consider an optimally oriented binary, with sky location and polarization chosen to maximize the detector response,  so that $F_+ = 1$, $F_\times = 0$, and $C_\Omega$ attains its maximum value $C_\Omega = 2$.

The various physical phenomena arising from the collapsar environment (modulation effects from orbital motion around the central BH, gas-driven migration and eccentricity, dynamical friction, and tidal interactions) enter the waveform as perturbative corrections to the point-particle phase. We capture all such effects through a perturbed waveform of the form
\begin{equation}
    \label{eq:perturbed_waveform}
    \tilde h_\text{\tiny pert} (f; \bm{\theta}) 
    = \tilde h (f; \bm{\theta}) \, \mathcal{S}_\text{\tiny TD} (f;\bm{\theta} ) \, \exp\Big[
    {\rm i}\, \delta\psi (f; \bm{\theta}) \Big]\,,
\end{equation}
where $\mathcal{S}_\text{\tiny TD} (f;\bm{\theta} )$ and  $\delta\psi (f; \bm{\theta})$ encode the amplitude and
phase corrections, respectively, induced by orbital, environmental and tidal effects, which we specify individually 
in the following subsections. 

To quantify the observational impact of these 
corrections, we introduce two complementary 
statistics. First, we assess the 
\emph{distinguishability} of a signal 
$\tilde{h}_\text{\tiny pert}$ from the 
best-fitting vacuum template $\tilde{h}$ 
through the noise-weighted scalar product
\begin{equation}
\label{eq:scalarproduct}
    \langle h_1 | h_2 \rangle = 4\,{\rm Re}
    \int_{f_\text{\tiny min}}^{f_\text{\tiny max}} 
    \frac{\tilde{h}_1^*(f)\, \tilde{h}_2(f)}
    {S_n(f)}\,{\rm d}f\,,
\end{equation}
where $S_n(f)$ is the one-sided power spectral 
density of the detector noise, while $f_\text{\tiny min}$ and $f_\text{\tiny max}$ denote the minimum and maximum frequency of integration, respectively. Throughout this work we adopt
the O4 sensitivity curve for LIGO~\cite{TheLIGOScientific:2014jea,Aasi:2013wya}, the $40$~km configuration for
CE~\cite{Evans:2021gyd,Evans:2023euw}, and the ET-D design curve for the
triangular ET~\cite{Punturo:2010zz,Branchesi:2023mws,Hild:2010id}. Accordingly, the minimum
frequencies allowed by the GW detectors are given by $f_\text{\tiny min} = 20,
10, 2$~{\rm Hz} for LIGO, CE and ET, respectively. The upper integration frequency is taken to be the minimum between the detector’s high-frequency limit, the unbound frequency $f_{\text{\tiny unb}}$ (see Sec.~\ref{sec:detection}), and the ISCO frequency for point-particle systems (see
Appendix~\ref{app:NSspin}).

The optimal (face-on, overhead) SNR associated to the point-particle waveform  reads 
\begin{align}
\rho_0^2 = \langle h|h\rangle &= 4\int_{f_\text{\tiny min}}^{f_\text{\tiny max}}
\frac{|\tilde h(f)|^2}{S_n(f)}\,{\rm d}f \notag\\[2pt]
&= \frac{5}{6\,\pi^{4/3}}\,\frac{\mathcal{M}_c^{5/3}}{d_\text{\tiny L}^{2}}
\int_{f_\text{\tiny min}}^{f_\text{\tiny max}}\frac{f^{-7/3}}{S_n(f)}\,{\rm d}f\,.
\label{eq:rho0}
\end{align}
The detection horizon $d_\text{\tiny hor}$ is the luminosity distance $d_\text{\tiny L}$
at which $\rho_0 = 8$.
The number of GW cycles accumulated in band is obtained by integrating the
instantaneous rate of change of the GW frequency $\dot{f}$ over the in-band inspiral,
\begin{align}
    \mathcal{N}_\text{\tiny cyc} &= \int_{f_\text{\tiny min}}^{f_\text{\tiny max}} \frac{{\rm d}f}{\dot f}\,f \simeq \mathcal{N}_\text{\tiny GW} + \delta \mathcal{N}_\text{\tiny env} \,,
    \label{eq:Ncyc}
\end{align}
between the effective in-band limits $f_\text{\tiny min}$ and
$f_\text{\tiny max}$, where the first term on the right-hand side is the number of cycles associated only to GW emission 
\begin{align}
    \mathcal{N}_\GW &= \int_{f_\text{\tiny min}}^{f_\text{\tiny max}} \frac{{\rm d}f}{\dot f_\GW}\,f \notag\\[2pt]
    &= \frac{1}{32\,\pi^{8/3} \mathcal{M}_c^{5/3}}
    \left(f_\text{\tiny min}^{-5/3} - f_\text{\tiny max}^{-5/3}\right)\,,
    \label{eq:NGW}
\end{align}
and the second term is the correction $\delta \mathcal{N}_\text{\tiny env}$ induced by environmental effects in the binary evolution, which give rise to an effective orbital dephasing.
Because the subsolar chirp mass makes $\dot f_\GW \propto \mathcal{M}_c^{5/3}$ small, $\mathcal{N}_\GW$ is large and grows steeply at low masses; together with the chirp-mass amplitude, it controls both the SNR accumulated in band and the secular phase budget available to environmental and tidal imprints.

The mismatch  between the perturbed signal and the 
best-fitting unmodulated template is 
defined as~\cite{Owen:1995tm, Lindblom:2008cm}
\begin{equation}\label{eq:mismatch}
\mathcal{M} = 1 - \frac{\langle h_\text{\tiny 
pert}|h\rangle}{\sqrt{\langle h_\text{\tiny 
pert}|h_\text{\tiny pert}\rangle\langle 
h|h\rangle}}\,.
\end{equation}
 The \emph{distinguishability} 
SNR then reads~\cite{Lindblom:2008cm,Baird:2012cu,Toubiana:2024car}
\begin{equation}
    \delta\rho \equiv \lVert h_{\rm pert} - h \rVert
    \approx \rho_0\,\sqrt{2\mathcal{M}}\,,
    \label{eq:rho_diff}
\end{equation}
where the mismatch $\mathcal{M}$ is minimized over the relative time shift, 
phase, and intrinsic template parameters 
$\bm{\theta}$, so that what remains is the part of the
signal $h_{\rm pert}$ \emph{not} reabsorbable by a small redefinition of the intrinsic and
extrinsic parameters. The last equality assumes $\langle h_\text{\tiny pert}|h_\text{\tiny pert}\rangle\simeq\langle h|h\rangle$, i.e., that the modulation leaves the total SNR essentially unchanged and only redistributes signal power, so that $\delta\rho^2=\langle h_\text{\tiny pert}-h|h_\text{\tiny pert}-h\rangle\simeq2\rho_0^2\,\mathcal{M}$ follows from Eq.~\eqref{eq:mismatch}.
We adopt the threshold $\delta\rho > 3$ for waveform distinguishability. For a residual of unknown amplitude and phase, $\delta\rho^2$ is a central $\chi^2$ variable with two degrees of freedom, so $\delta\rho=3$ corresponds to a false-alarm probability $\mathrm{FAP}=e^{-\delta\rho^2/2}\sim 1\%$. Since
$\delta\rho \propto \rho_0$, it is convenient to quote its value at the detection
horizon $d_\text{\tiny hor}$, where $\rho_0 = 8$ by definition. We define this horizon value so that
\begin{equation}
    \delta\rho_h \equiv \delta\rho  \, (d_\text{\tiny hor}) =  8\sqrt{2\mathcal{M}}\,.
    \label{eq:rho_diff_hor}
\end{equation}
A horizon value $\delta\rho_h > 3$ corresponds to a residual mismatch
$\mathcal{M} > 9/128 \approx 0.07$. The maximum distinguishability distance
$d_\text{\tiny max}$ is the source distance at which $\delta\rho = 3$.

Second, we assess \emph{parameter measurability} using the Fisher information matrix formalism~\cite{Cutler:1994ys, Poisson:1995ef, Berti:2004bd, Vallisneri:2007ev}. At high SNR, the posterior for the source parameters $\bm{\theta}$ is well approximated by a Gaussian centered at the true values $\hat{\bm{\theta}}$, with covariance matrix $\Sigma_{ij} = (\Gamma_{ij})^{-1}$, where
\begin{equation}
  \Gamma_{ij} = \left\langle 
  \frac{\partial h}{\partial \theta_i} 
  \bigg| \frac{\partial h}{\partial \theta_j} 
  \right\rangle_{\bm{\theta}=\hat{\bm{\theta}}}
\end{equation}
is the Fisher information matrix. The $1\sigma$ statistical uncertainty on parameter $\theta_i$ is then $\sigma_i = \sqrt{\Sigma_{ii}}$. Together, the mismatch and Fisher analysis provide complementary diagnostics: the former quantifies whether orbital, environmental and tidal effects are detectable at all, while the latter quantifies how precisely the underlying binary parameters can be inferred once a signal is detected.

\subsection{In-band evolution and detection prospects}
\label{sec:detection}
\noindent
Having set up the waveform model and the data-analysis statistics in the previous section, we now follow the BNS through the detector band and quantify its detectability across the parameter space. Two frequencies bracket the in-band signal: {\it (i)} the frequency at which the GW-driven inspiral begins, fixed by the initial separation $a_{\BNS,i} = f_\text{\tiny H}\,r_\text{\tiny Hill}$ through 
{\begin{align}
f_i &= \frac{1}{\pi} \sqrt{\frac{2m_\NS}{a_{\BNS,i}^3}} \notag\\ 
&\simeq 8\left(\frac{5\,\msun}{M_\BH}\right)
   \left(\frac{f_\text{\tiny H}\,a_{\BNS-\BH,i}}{200\,R_g}\right)^{\!-3/2}{\rm Hz} \notag\\[2pt]
&\simeq 3.6\,f_\text{\tiny H}^{-3/2}
   \left(\frac{\rho(R_*)}{10^{8}\,{\rm g\,cm^{-3}}}\right)^{\!1/2}{\rm Hz}\,,
\label{eq:fi}
\end{align}
where in the second line we have expressed the Hill radius in terms of the initial BNS-BH separation $a_{\BNS-\BH,i}$, while the last line assumes $a_{\BNS-\BH,i}=R_*$, as shown in Eq.~\eqref{eq:rHill_outer}; and {\it (ii)} the frequency $f_\text{\tiny TD}$ at which the inspiral terminates, which we model through the amplitude factor $\mathcal{S}_\text{\tiny TD}$ introduced below.

The amplitude factor $\mathcal{S}_\text{\tiny TD}(f;\bm{\theta})$ accounts for the
end stage of the inspiral, where two distinct processes can truncate the
signal: the two NSs may tidally disrupt each other as they reach contact, or
the binary as a whole may be tidally \emph{unbound} by the central BH before
the components ever touch. We model this end stage as a cutoff of the form~\cite{DeLuca:2022xlz}
\begin{equation}
\label{eq:taperingfunction}
    \mathcal{S}_\text{\tiny TD} (f;\bm{\theta}) =\frac{1+e^{-f_{\text{\tiny TD}}/f_{\text{\tiny slope}}}}{1+e^{(f-f_{\text{\tiny TD}})/f_{\text{\tiny slope}}}}\,,
\end{equation}
which ensures that the inspiral terminates smoothly around the tidal-disruption frequency
\begin{equation}
    f_\text{\tiny TD}  = {\rm min}\,(f_\text{\tiny Roche}, f_\text{\tiny unb})
    \label{eq:fTD}
\end{equation}
set by the smaller of the two terminators, which we discuss in turn below. The rate at which this transition occurs is controlled by the characteristic slope $f_{\text{\tiny slope}}$, which we set to $f_{\text{\tiny slope}} = f_{\text{\tiny TD}}/10$, but we have verified that its specific value does not significantly affect our analysis.

The first terminator is the Roche frequency at which the tidal field of the companion disrupts the NS~\cite{GomezLopez:2025aiy},
\begin{align}\label{eq:fRoche}
    f_\text{\tiny Roche} = \frac{1}{\pi} \sqrt{\dfrac{m_\text{\tiny comp} + m_\NS}{d_{\text{\tiny Roche}}^3}}\,,
\end{align}
where the Roche radius reads
\begin{equation}
\label{Rocheradius}
    d_{\text{\tiny Roche}} = \gamma_{\text{\tiny Roche}} \, R_\NS (m_\NS) \left( \dfrac{m_\text{\tiny comp} }{m_\NS} \right)^{1/3} \,,
\end{equation}
in terms of the companion mass $m_\text{\tiny comp}$, the NS mass-radius relation
$R_\NS (m_\NS)$ (see Fig.~\ref{fig:mr-TLN} for details), and the
 coefficient $\gamma_{\text{\tiny Roche}}$, which takes values
ranging from 1.26 for rigid bodies to 2.44 for fluid
bodies~\cite{Shapiro:1983du}. For the symmetric
BNS considered here ($m_\text{\tiny comp} = m_\NS$) the Roche frequency
essentially coincides with the contact frequency of the pair, and marks the
natural end of a full inspiral that proceeds all the way to merger. For the
subsolar masses of interest, and taking $\gamma_\text{\tiny Roche} \simeq 2$
so that disruption coincides with contact, this takes the EoS-agnostic form
\begin{align}
    f_\text{\tiny Roche}^\BNS &= \frac{1}{\pi}\sqrt{\frac{m_\NS}{4 R_\NS^3}}\nonumber\\
    &\simeq 92\,{\rm Hz}\left(\frac{m_\NS}{0.1\,\msun}\right)^{1/2}
    \left(\frac{R_\NS}{34\,{\rm km}}\right)^{-3/2},
    \label{eq:fRocheBNS_gen}
\end{align}
which, upon inserting the mass-radius relations of Eq.~\eqref{eq:R_NS},
evaluates to
\begin{align}
{\rm Poly}:& \quad f_\text{\tiny Roche}^\BNS \simeq 91 \, {\rm Hz} \left(\frac{m_\NS}{0.1\,\msun}\right)\,,
\nonumber \\
{\rm NM}:& \quad f_\text{\tiny Roche}^\BNS \simeq  0.5\,{\rm kHz}\left(\frac{m_\NS}{0.1\,\msun}\right)^{1/2} \nonumber \\
& \qquad \quad \,\,\, \times \left[1+3\left(\frac{m_\NS}{0.1\,\msun}\right)^{-3}\right]^{-3/2}\,,
\nonumber \\
{\rm QM}:& \quad f_\text{\tiny Roche}^\BNS \simeq 2.3\,{\rm kHz}\,.
\label{eq:fRoche_num}
\end{align}
The EoS enters entirely through $R_\NS(m_\NS)$, and it does so sharply: since
$f_\text{\tiny Roche}^\BNS \propto (m_\NS/R_\NS^3)^{1/2}$, the three
mass-radius relations of Eq.~\eqref{eq:R_NS} give three qualitatively
different terminations. The polytrope, with $R_\NS \propto m_\NS^{-1/3}$,
yields a contact frequency growing linearly with the NS mass across the whole
subsolar range, and sitting close to the peak sensitivity of ground-based
detectors at $m_\NS \simeq 0.1\,\msun$. Quark stars, whose nearly constant
density gives $R_\NS \propto m_\NS^{1/3}$, terminate at an essentially
mass-independent frequency of a few kHz, at the very top of the detector band~\cite{Ackley:2020atn,Srivastava:2022slt}.
NM interpolates between the two, the contact frequency climbing
steeply as the radius collapses and dropping
below the LIGO floor as $m_\NS$ approaches the minimum stable mass.

The second terminator, the unbound frequency $f_\text{\tiny unb}$, accounts for the
disruption of the inner binary as a whole by the tidal field of the central
BH (see also Appendix~\ref{sec:tidalfield} for a dedicated discussion of the effects of the BH tidal field on the inner NS binary). Because the pair orbits the BH in a hierarchical triple, its center of mass
also spirals inward under GW emission while the inner orbit shrinks, so that
both the BNS separation $a_\BNS$ and the BNS-BH binary separation $a_{\BNS-\BH}$ decay through
radiation reaction. The BNS pair is unbound once the outer
separation has shrunk enough that the BH tidal field overwhelms the pair's
self-gravity, i.e., once the inner-binary Hill radius falls below the inner
separation 
\begin{equation}
    a_{\BNS-\BH} (f_\text{\tiny unb})\, \left(\frac{2 m_\NS}{3 M_\BH}\right)^{1/3} =  a_{\BNS} (f_\text{\tiny unb})\,.
    \label{eq:tidal_condition}
\end{equation}
Since both separations
follow the Peters power law $a^4 \propto 1 - t/\tau_\GW$, in terms of the corresponding GW timescales [see Eqs.~\eqref{eq:GWtimescaleSSNSs} and \eqref{tau_GW_BH}], the crossing
of the two curves yields the unbinding time 
\begin{align}
    \tau_\text{\tiny unb} &=
     \left(1-f_\text{\tiny H}^{4}\right)
    \left(\frac{1}{\tau_{\GW}^{\BNS-\BH}} - \frac{f_\text{\tiny H}^{4}}{\tau_{\GW}^{\BNS}}\right)^{\!-1} \notag\\[4pt]
    &= \frac{3^{4/3}\left(2q\right)^{2/3}\left(1-f_\text{\tiny H}^{4}\right)}
    {f_\text{\tiny H}^{4}\left[4 - 3^{4/3}\left(2q\right)^{2/3} + 8q\right]}\;\tau_{\GW}^{\BNS}\,,
    \label{eq:t_unb}
\end{align}
in terms of the compact objects mass ratio $q = m_\NS/M_\BH$.
Mapping this time to the instantaneous inner-binary GW frequency through the Keplerian relation $f \propto a_{\BNS}^{-3/2}$, together with the Peters decay for the inner BNS system $a_{\BNS}(t) = a_{\BNS,i}\,(1 - t/\tau_{\GW}^{\BNS})^{1/4}$, finally gives the unbinding frequency
\begin{equation}
    f_\text{\tiny unb} = f_i
    \left(1 - \frac{\tau_\text{\tiny unb}}{\tau_{\GW}^{\BNS}}\right)^{\!-3/8} \,.
    \label{eq:funb}
\end{equation}
The pair is then stripped before merger
once $\tau_\text{\tiny unb} < \tau_\text{\tiny GW}^\BNS$, i.e. 
\begin{equation}
{\rm unbinding \,\, before \,\, merger:} \quad f_\text{\tiny H} > \frac{3^{1/3}q^{1/6}}{2^{1/3}(1+2q)^{1/4}}\,.
\label{eq:fH_unb}
\end{equation}
Equations~\eqref{eq:t_unb} and~\eqref{eq:fH_unb} assume a wide inspiral, $f_i\ll f_\text{\tiny Roche}$, in vacuum. They break down either for light NSs at small $a_{\BNS-\BH,i}$, where the pair is born close to contact and merges before the BH can unbind it, or for values of $f_\text{\tiny H}$ above the critical threshold beyond which the gas dominates the BNS dynamics (see the orange regions in the left panel of Fig.~\ref{fig:param}).
Because this condition typically requires $f_\text{\tiny H}\gtrsim0.5$ (for example, $f_\text{\tiny H}>0.59$ at $q=0.02$), in-band unbinding occurs only for retrograde inner binaries, which can remain stable up to $f_\text{\tiny H}\sim1$. Prograde binaries are limited to $f_\text{\tiny H}\lesssim0.5$, so their inner inspiral reaches contact before the shrinking Hill radius can unbind them.

A bound inner binary thus has two possible outcomes: if it is tight enough, it spirals all the way to contact and terminates at $f_\text{\tiny Roche}$, so that the full inspiral is observed; if it is too wide, the central BH strips it before contact and the signal is truncated at the lower frequency $f_\text{\tiny unb}$. This competition determines the detectable parameter region.  Even if the inner BNS is tidally disrupted, or ionized, its two NSs remain on independent bound orbits around the central BH, effectively forming two NS-BH systems. Although no longer bound to each other, repeated close encounters and gravitational scattering between the NSs can generate distinctive, potentially detectable GW signatures beyond the standard binary-inspiral waveform~\cite{Baibhav:2026zuy}.

These two terminators are radiation-reaction deadlines: they fix which process ends the inspiral first, and hence the upper edge of the in-band signal.
Whether a wide pair is bound at all is a separate, secular question.
As discussed in Sec.~\ref{sec:param_space}, the periodic tidal perturbation from the central BH accumulates over many outer orbits and, above a critical Hill fraction, drives the inner binary into chaotic instability before it ever enters the GW-driven phase. Long-term stability confines prograde BNSs to $f_\text{\tiny H}\lesssim 0.5$, while the Coriolis stabilization of retrograde orbits extends their survival out to $f_\text{\tiny H}\sim 1$~\cite{Mardling2001,Grishin2017}.

\begin{figure*}[!t]
\centering
\includegraphics[width=\textwidth]{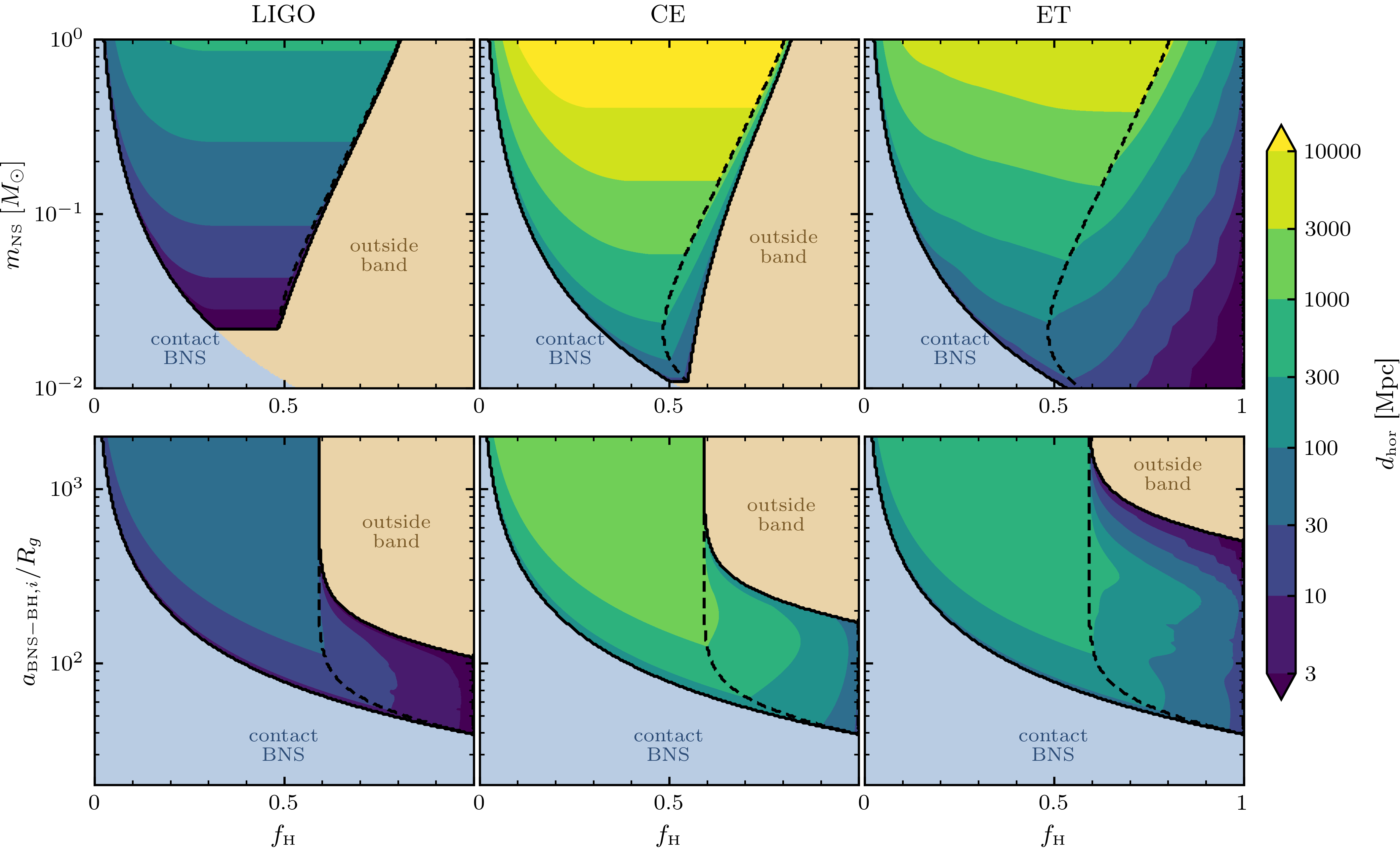}
\caption{Contours of the detection horizon, $d_\text{\tiny hor}$, for the inner BNS as a function of the initial Hill-radius filling fraction,
$f_\text{\tiny H}$, assuming polytropic EoS of Eq.~\eqref{eq:Rpoly}. The top row explores the $(f_\text{\tiny H},m_{\NS})$ parameter
space for a fixed central BH mass, $M_\BH=5\,M_\odot$, assuming that
the outer BNS-BH system begins its inspiral at the fragmentation radius,
$a_{\BNS-\BH,i}=R_*$. The bottom rows instead fix
$m_\NS=0.1\,M_\odot$ and $M_\BH=5\,M_\odot$, and show the
corresponding dependence on the initial outer-orbit separation,
$a_{\BNS-\BH,i}/R_g$. The columns correspond to LIGO, CE, and ET, respectively.
The shaded regions without contours mark configurations with no detectable inner-binary
signal. At low $f_\text{\tiny H}$, the NSs are already in contact at formation
(``contact BNS''). At high $f_\text{\tiny H}$, the central BH unbinds the binary
before it enters the detector band (``outside band''); at
sufficiently low $m_{\NS}$, the inner merger may likewise occur entirely
outside the band. The solid black curve bounds the detectable region. The
dashed black curve marks the boundary where the central BH tidally
unbinds the wider inner binary before the NSs reach contact. Beyond this boundary,
$d_\text{\tiny hor}$ decreases with increasing $f_\text{\tiny H}$, and eventually vanishes
once the signal falls below the detector band.}
\label{fig:hor_fH}
\end{figure*}

The behavior of the detection horizon $d_\text{\tiny hor}$ for the BNS channel is schematically summarized  in Fig.~\ref{fig:nsbh_bns_horizon}, for a fixed initial NS separation $f_\text{\tiny H}$, and examined in greater detail in Fig.~\ref{fig:hor_fH}, where we illustrate its dependence on the initial outer distance with the central BH $a_{\BNS-\BH,i}$ and on the initial BNS separation. 
  
The purple bands in Fig.~\ref{fig:nsbh_bns_horizon} show the BNS horizon $d_\text{\tiny hor}$ as a function of the NS mass $m_\NS$ for LIGO, CE and ET. The horizon increases with $m_\NS$: at $m_\NS=0.1\,\msun$ it reaches $\approx36\,$Mpc in LIGO, $\approx0.57\,$Gpc in ET and $\approx1.8\,$Gpc ($z\approx0.34$) in CE, and at $m_\NS=1\,\msun$ it reaches $\approx340\,$Mpc in LIGO, $z\approx4.7$ in CE, and $z\approx1.3$ in ET; the lightest pairs ($m_\NS\lesssim0.02\,\msun$) terminate below the detector noise floors and leave the band altogether. Because the inner-BNS chirp mass is set by the NS mass alone, $\mathcal{M}_c=2^{-1/5}m_\NS$, independently of the central BH mass $M_\BH$, the band spanning $M_\BH \in[3,\,30]\,\msun$ is narrow and collapses to a single line above $m_\NS\approx0.2\,\msun$; below that, $M_\BH$ enters only by tidally truncating the inner inspiral earlier through the unbounding frequency, so a lighter central mass reaches farther (for $m_\NS=0.1\,\msun$, the horizons associated with $M_\BH=3$ and $10\,\msun$  coincide near $36\,$Mpc, while for $M_\BH=30\,\msun$ the truncated signal falls entirely below the LIGO band and the source is undetectable there, with CE and ET still reaching $\approx0.5\,$Gpc and $\approx0.27\,$Gpc). 

In Fig.~\ref{fig:hor_fH} we plot the detection horizon as a function of the initial Hill fraction $f_\text{\tiny H}$ for the three detectors. The detectable binary configurations and the horizon reach are determined by the interplay of the two band edges introduced above --- the entry frequency $f_i$ [Eq.~\eqref{eq:fi}], where the pair becomes GW-driven, and the terminating frequency $f_\text{\tiny TD}=\min(f_\text{\tiny Roche},f_\text{\tiny unb})$ [Eq.~\eqref{eq:fTD}], where the inspiral ends --- together with the detector floor $f_\text{\tiny min}$. These three frequencies divide the parameter space into four regimes, which we describe in order of increasing $f_\text{\tiny H}$, together with their dependence on the main model parameters.

\noindent
\emph{(i) Contact.} At the lowest $f_\text{\tiny H}$ the inner orbit is so
tight that the two NSs already touch at formation: the initial separation
$a_{\BNS,i}=f_\text{\tiny H}\,r_\text{\tiny Hill}$ falls below $2R_\NS$
[Eq.~\eqref{eq:nstouch}], and there is no observable inspiral. Since
$R_\NS\propto m_\NS^{-1/3}$ [see Eq.~\eqref{eq:R_NS}], this region widens for light NSs. It corresponds to the light blue
region labeled as ``contact BNS'' in the bottom left corner of all panels in Fig.~\ref{fig:hor_fH}.

\noindent
\emph{(ii) Inner merger.} At intermediate $f_\text{\tiny H}$ the pair is wide enough to avoid contact at birth, yet tight enough to inspiral all the way to merger under GW emission. The signal then evolves across the full band from $f_i$ up to the contact frequency $f_\text{\tiny Roche}$, and the horizon is largest. This is the only regime that yields a complete, detectable inspiral.

\noindent
\emph{(iii) Unbinding.} As $f_\text{\tiny H}$ grows, the inner orbit becomes a larger fraction of the Hill radius, and tidal forces from the central BH unbind the pair before it reaches contact [Eq.~\eqref{eq:fH_unb}]. The inspiral is truncated at $f_\text{\tiny unb}<f_\text{\tiny Roche}$: the signal is still in band, but the observable frequency band is narrower, so the horizon shrinks as $f_\text{\tiny H}$ increases.

\noindent
\emph{(iv) Disruption before the band.} At the largest $f_\text{\tiny H}$ the orbit is so wide that unbinding occurs very early and the truncation frequency itself falls below the detector floor, $f_\text{\tiny unb}<f_\text{\tiny min}$. The truncated signal never enters the sensitive band and the source is undetectable. This is the brown region labeled as ``outside band'' at high $f_\text{\tiny H}$ in all panels of Fig.~\ref{fig:hor_fH}.

\noindent
\textbf{Hill-fraction dependence.} Moving horizontally across Fig.~\ref{fig:hor_fH}, the horizon occupies a band in $f_\text{\tiny H}$ that is bounded below by contact (regime~i) and above by the tidal-disruption boundary (regimes~iii--iv). Inside this band the behavior is controlled by the entry frequency. Larger values of $f_\text{\tiny H}$ increase the initial separation $a_{\BNS,i}$ and decrease $f_i\propto f_\text{\tiny H}^{-3/2}$. A lower entry frequency means that the signal begins earlier in the sensitive band of the detector, the inspiral is longer, and the horizon is larger --- but only until $f_i$ falls below the detector floor $f_\text{\tiny min}$. Beyond that point the band starts at $f_\text{\tiny min}$ irrespective of $f_\text{\tiny H}$, and the horizon saturates. This saturation is detector-specific: the entry frequency falls below the floor for $f_\text{\tiny H} \gtrsim (3.6\,{\rm Hz}/f_\text{\tiny min})^{2/3}$ [Eq.~\eqref{eq:fi}], i.e., near $f_\text{\tiny H}\approx0.3$ for LIGO ($f_\text{\tiny min}=20\,$Hz) and $f_\text{\tiny H}\approx0.5$ for CE ($10\,$Hz), whereas for ET the entry frequency stays above its $2\,$Hz floor over the whole allowed range, so its horizon continues to increase with $f_\text{\tiny H}$ up to the disruption boundary: the lower the floor, the larger the $f_\text{\tiny H}$ reached before saturation. At even larger values of $f_\text{\tiny H}$ (regime~iii) the trend reverses: the BH unbinds the pair before contact, the ending frequency $f_\text{\tiny unb}$ drops with $f_\text{\tiny H}$, the observable band shrinks, and the horizon falls steeply --- until the truncated signal decreases below $f_\text{\tiny min}$ (regime~iv) and the binary becomes unobservable.

\noindent
\textbf{Mass dependence.} In the upper panels of Fig.~\ref{fig:hor_fH} we show the dependence of the horizon on the NS mass $m_\NS$, at fixed $a_{\BNS-\BH,i}=R_*$ and $M_\BH=5\,\msun$. Two effects set its steep rise with mass. The first is the GW amplitude, because $d_\text{\tiny hor}\propto\mathcal{M}_c^{5/6}\propto m_\NS^{5/6}$ if the observable band were fixed. 
The second, and more important, is that the frequency band itself widens with mass. Because the NS radius shrinks with mass for a polytropic EoS, the contact frequency increases rapidly, $f_\text{\tiny Roche}\propto m_\NS$ [Eq.~\eqref{eq:fRoche_num}], while the entry frequency $f_i$ is essentially mass-independent [Eq.~\eqref{eq:fi}]. As $m_\NS$ grows, the in-band interval $[f_i,\,f_\text{\tiny Roche}]$ broadens, the cycle count $\mathcal{N}_\GW$ increases, and the horizon grows faster than the simple amplitude scaling $m_\NS^{5/6}$. At low masses, the same mechanism works in reverse: the band contracts to a narrow frequency interval just above $f_\text{\tiny min}$, producing the rapid low-mass decrease seen in the figure. Below $m_\NS\approx0.02\,\msun$ the pair is born in contact (regime~i) and becomes unobservable.

\noindent
\textbf{Outer-separation dependence.} In the lower panels of Fig.~\ref{fig:hor_fH} we instead vary the outer separation $a_{\BNS-\BH,i}$, at fixed $m_\NS=0.1\,\msun$ and $M_\BH=5\,\msun$. Because the Hill radius, and therefore $a_{\BNS,i}$, scales linearly with $a_{\BNS-\BH,i}$, the entry frequency obeys $f_i\propto a_{\BNS-\BH,i}^{-3/2}$, exactly as for $f_\text{\tiny H}$. A wider outer orbit thus lowers $f_i$, widens the band, and yields a deeper horizon, until $f_i$ once again reaches the detector floor and the horizon saturates. The crossover separation grows as the floor is lowered: it lies near $a_{\BNS-\BH,i}\approx190\,R_g$ for LIGO, $\approx270\,R_g$ for CE, and $\approx520\,R_g$ for ET, so that all three horizons plateau within the plotted range. At the opposite extreme, $a_{\BNS-\BH,i}\lesssim80\,R_g$ (at $f_\text{\tiny H}=0.5$; the bound scales as $1/f_\text{\tiny H}$), the Hill radius is too small to hold a detached pair and the two NSs are again in contact at formation (regime~i).

\noindent
\textbf{EoS dependence.} The horizons above are computed for a single polytropic NS mass-radius
relation. The EoS enters the inner-BNS signal only through $R_\NS(m_\NS)$, and
only at the two edges of the in-band window: the formation-contact bound of
Eq.~\eqref{eq:nstouch_num}, which determines whether a detached pair exists,
and the contact frequency of Eq.~\eqref{eq:fRocheBNS_gen} at which the inspiral
terminates; the entry frequency is set by the Hill geometry alone and carries no
EoS information. Because the noise-weighted signal power falls steeply with
frequency, most of the SNR is accumulated in the low-frequency portion of the band, so the
horizon depends only weakly on where the inspiral ends: a more compact star
reaches a higher contact frequency and deepens the horizon.
The horizons converge toward larger masses, becoming indistinguishable above
$m_\NS \approx 0.2\,\msun$, where every EoS terminates above the frequencies that
carry appreciable SNR weight. The EoS information contained in the detectability is therefore confined to the
lightest sources, which are the most EoS-sensitive: there the
formation-contact bound determines whether a given pair is born detached and
inspirals into band, or is already in contact at formation. The corresponding detectability curves for the
different EoSs are discussed in Appendix~\ref{app:EoS}.

\subsection{Modulation effects}
\label{sec:modeffects}
\noindent
The triple geometry imprints several distinct relativistic modulations on the BNS (the ``inner binary'') GW signal~\cite{Meiron:2016ipr,Chen:2018axp,Chen:2020iky,Toubiana:2020drf,Cardoso:2021vjq,Sberna:2022qbn,Yin:2024nyz,Santos:2025ass,Santos:2026lzq,Cardoso:2026ugm}.  In this section, we study the physics of each of these effects, that can take place in the cross-hatched region in the left panel of Fig.~\ref{fig:param}. 
Every modulation originates in the slow orbit of the inner BNS about the central BH.  The outer-orbit motion produces time dependence in the line-of-sight velocity, gravitational potential, and wave-propagation geometry, each of which perturbs the phase of the otherwise GW-driven inner-binary chirp.  We model every effect as a small phase correction to the inner carrier and then map it to the frequency domain, where it is measured against the inspiral template. 

Throughout this section $f$ denotes the GW frequency of the inner BNS (the carrier signal), with $\psi(f)$ its frequency-domain GW phase and $\phi(t)$ the corresponding time-domain phase. Modulation effects enter as small time-domain corrections $\delta\phi(t)$ to $\phi(t)$.  The time-domain phase correction $\delta\phi(t)$ maps to a frequency-domain perturbation $\delta\psi(f)$ of $\psi(f)$ in Eq.~\eqref{eq:perturbed_waveform} via the stationary-phase bijection $t \leftrightarrow f$:
\begin{equation}
    \delta\psi(f) = \delta\phi\bigl(t(f)\bigr) \,,
    \label{eq:dpsi_from_dphi}
\end{equation}
where $t(f) = \frac{1}{2\pi}\frac{\mathrm{d}\psi}{\mathrm{d} f}$ is the
stationary-phase time, that assigns to each carrier frequency $f$ the instant $t$ at which the inner binary radiates it.

The same outer orbit fixes the geometry shared by all five effects.
We place the observer in the $x$--$z$ plane along the unit vector
\begin{equation}
    \hat{m} = (\sin\iota,\; 0,\; \cos\iota)\,,
    \label{eq:mhat}
\end{equation}
where $\iota$ is the inclination between the line of sight and the outer orbital angular momentum (we take the outer orbital plane to coincide with the $(x,y)$ plane, so that the outer orbital angular momentum points along $\hat z$).
The inner binary sits at position $\vec{r}(t)$ on this outer orbit, at separation $a_{\BNS-\BH}(t)$ from the central BH, with outer orbital phase $\varphi_{\BNS-\BH}(t)$ and outer (GW) orbital frequency $f_{\BNS-\BH}(t)$.

Both binaries are evolved simultaneously on a single time coordinate $t$ spanning the in-band lifetime: each inspirals under its own GW emission, so as the inner binary evolves across the detector band, the outer separation decreases simultaneously, with $a_{\BNS-\BH}(t)$ shrinking and $f_{\BNS-\BH}(t)$ and $\varphi_{\BNS-\BH}(t)$ rising through $\sim\!10^{4}$--$10^{5}$ orbits. Each modulation is computed from the outer-orbit state at time $t$ and mapped to the inner frequency through the inner chirp $t(f)$. Eliminating $t$ between the two inspirals gives the outer separation at inner frequency $f$,
\begin{equation}
    a_{\BNS-\BH}(f) = a_{\BNS-\BH, i} \left[1 - R_\tau\left(1 -
    \left(\frac{f_i}{f}\right)^{8/3}\right)\right]^{1/4} ,
    \label{eq:r_of_f}
\end{equation}
where $R_\tau$ is the ratio of the initial outer-to-inner inspiral rates. Since the Peters
time of a circular binary scales as
$\tau_\GW \propto a^4/[m_1 m_2 (m_1+m_2)]$, and the outer orbit carries the inner
pair as a single body of mass $2 m_\NS$ about the central BH, one finds
\begin{align}
     R_\tau \equiv \frac{\tau_\GW^{\BNS}}{\tau_\GW^{\BNS-\BH}}
    &= \left(\frac{a_{\BNS,i}}{a_{\BNS-\BH,i}}\right)^{\!4}
       \frac{M_\BH\left(M_\BH + 2 m_\NS\right)}{m_\NS^2} \notag\\[4pt]
    &= \left(\frac{2}{3}\right)^{4/3} f_\text{\tiny H}^{4}\,
       \frac{1 + 2q}{q^{2/3}} \,,
    \label{eq:tR}
\end{align}
where the second equality follows from $a_{\BNS,i} = f_\text{\tiny H}\,(2q/3)^{1/3}\,a_{\BNS-\BH,i}$ [Eqs.~\eqref{eq:rHill_outer} and~\eqref{eq:fH}], in terms of the mass ratio $q = m_\NS/M_\BH$. Thus $R_\tau$ measures how far the outer orbit moves while the inner binary is in band: $R_\tau\ll1$ leaves it essentially static, whereas $R_\tau\geq1$ means that the outer orbit shrinks faster than the inner binary, so the modulation drifts as the geometry evolves underneath it.

The five GW modulations fall into two families characterized by their time signature. The Doppler shift, relativistic aberration, the oscillatory part of the Shapiro delay, and the self-lensing fringes are oscillatory, repeating once per outer orbit; the gravitational redshift and the secular part of the Shapiro delay accumulate monotonically as the outer orbit shrinks. For each effect we proceed in three steps: (i) we derive the time-domain phase correction $\delta\phi(t)$ from the geometry of the outer orbit; (ii) we map it to the frequency-domain residual $\delta\psi(f)$ through the stationary-phase relation of Eq.~\eqref{eq:dpsi_from_dphi}; and (iii) we evaluate $\delta\psi$ using a single detection statistic that quantifies how distinguishable the modulated waveform $h_\text{\tiny pert}$ is from the best-fitting ordinary inspiral template $h$.

The last of these steps is common to all five imprints. The same expression applies to all five effects. For a pure phase perturbation, using $|\tilde h|^2/S_n\propto f^{-7/3}/S_n$, the normalized overlap between the perturbed and the vacuum waveform is
\begin{equation}
  \frac{\langle h_\text{\tiny pert}|h\rangle}{\rho_0^2}
  =\mathrm{Re}\,\big\langle e^{i\delta\psi}\big\rangle_W\,,
  \label{eq:overlap}
\end{equation}
so the residual is weighted by where the inspiral SNR lies, through
the noise-weighted band average
\begin{equation}
    W(f)=\frac{f^{-7/3}}{S_n(f)}\,,\qquad
    \langle X\rangle_W=\frac{\int X\,W(f)\,\d f}{\int W(f)\,\d f}\,,
    \label{eq:weight}
\end{equation}
where the power law $f^{-7/3}$ characterizes the $|\tilde h|^2$ dependence of the inspiral [see Eq.~\eqref{eq:rho0}].
A real search re-fits the binary parameters, so any part of $\delta\psi$ that mimics a shift of those parameters is not an observable modulation and must be marginalized over in the mismatch computation [see Eq.~\eqref{eq:mismatch}]: a constant phase is degenerate with the coalescence phase, a phase $\propto f$ with the coalescence time, and a phase $\propto f^{-5/3}$ with the chirp mass $\mathcal{M}_c$~\cite{Chen:2017xbi,Tamanini:2019usx}. The oscillatory modulations all enter the frequency-domain phase in the same form:
\begin{equation}
    \delta\psi(f) = -\Theta(f)\,\cos\varphi_{\BNS-\BH} (t(f))\,,
    \label{eq:dpsi_osc}
\end{equation} 
with a slowly varying modulation index $\Theta(f)$ measuring the radians of unabsorbable phase contributed by the effect, in terms of the associated orbital phase $\varphi_{\BNS-\BH}$. Because the inner binary completes $\sim\!10^4$ outer orbits while it crosses the band, $e^{i\delta\psi}$ may be averaged over one outer orbit at fixed $\Theta$. As we show in Appendix~\ref{app:modulations}, this average reduces the horizon distinguishability of Eq.~\eqref{eq:rho_diff_hor} to
\begin{equation}
    \delta\rho_h\simeq8\sqrt{2\bigl[1-\langle J_0(\Theta)\rangle_W\bigr]}\,,
    \label{eq:rho_J0}
\end{equation}
in terms of the Bessel function of the first kind $J_0$. The modulation therefore enters the distinguishability through the single number $\Theta$, and two regimes follow:
\begin{equation}
  \delta\rho_h\simeq
  \begin{cases}
    \dfrac{8}{\sqrt2}\,\langle\Theta^2\rangle_W^{1/2}\,, & \Theta\ll1\,,\\[1.1ex]
    8\sqrt2\approx11.3\,, & \Theta\gtrsim1\,.
  \end{cases}
  \label{eq:rho_J0_limits}
\end{equation}
Weakly modulated effects ($\Theta\ll1$) remain in the linear regime, where $\delta\rho_h\simeq(8/\sqrt{2})\,\langle\Theta^2\rangle_W^{1/2}$ tracks the amplitude of the perturbation; effects reaching $\Theta\gtrsim1$ instead decorrelate the modulated and unmodulated waveforms completely, so that no template refit restores the match, $\delta\rho_h$ approaches the saturation limit $8\sqrt{2}\approx11.3$, and depends primarily on the noise-weighted frequency support rather than the amplitude of the imprint. The gravitational redshift and the secular part of the Shapiro delay, being monotonic rather than oscillatory, fall outside this scheme and are discussed separately below.

All quantitative results that follow --- in particular, the horizon SNR $\delta\rho_h$ maps shown in Figs.~\ref{fig:rho_diff_iota} and~\ref{fig:rho_diff_fH} --- are obtained with the full numerical prescription: both inspirals are evolved simultaneously, each phase correction $\delta\psi(f)$ is computed from the exact outer-orbit state along the chirp, and the distinguishability of Eqs.~\eqref{eq:rho_diff}--\eqref{eq:rho_diff_hor} is evaluated numerically, with no orbit averaging or weak-modulation expansion.
The analytic results --- the orbit-averaged Eq.~\eqref{eq:rho_J0}, its limits, and the closed forms derived in Appendix~\ref{app:modulations} --- provide order-of-magnitude estimates of the parameter dependence: they can omit subdominant ingredients, such as the orthogonal projection of the residual onto the re-fitted template parameters, and none of them is used in producing the figures below.

\subsubsection*{Doppler phase shift}
\label{sec:doppler}
\noindent
The Doppler modulation arises as the inner binary orbits the central BH, producing a time-dependent shift of the GW frequency~\cite{Bonvin:2016qxr,Meiron:2016ipr,Inayoshi:2017hgw,Robson:2018svj,Wong:2019hsq,Tamanini:2019usx,Toubiana:2020drf,Yu:2020dlm,Strokov:2021mkv,Xuan:2022qkw,Vijaykumar:2023tjg,Metzger:2024ujc,Lazarow:2024gdn,Stegmann:2023wzy,Samsing:2024syt,Metzger:2024ujc,Pompili:2026tdf,Tiwari:2026lvw,Hendriks:2026kys}.

The line-of-sight distance from the system's center of mass to the inner binary, projected along $\hat{m}$ [Eq.~\eqref{eq:mhat}], is
\begin{equation}
    d(t) = \vec{r}(t) \cdot \hat{m}
         = a_{\BNS-\BH}(t)\,\cos\varphi_{\BNS-\BH}(t)\,\sin\iota\,,
    \label{eq:d_Los}
\end{equation}
in terms of the outer separation $a_{\BNS-\BH}(t)$ and outer orbital phase $\varphi_{\BNS-\BH}(t)$ defined above.

The Doppler phase correction then reads
\begin{align}
    \delta\phi_\text{\tiny Dop}(t) & = -2\pi\,f(t)\,d(t)\,,
    \label{eq:dphi_doppler_explicit}
\end{align}
with the outer separation $a_{\BNS-\BH}(f)$ evolving as given by
Eq.~\eqref{eq:r_of_f} and the outer orbital phase
$\varphi_{\BNS-\BH}(t) = \pi\int_0^t f_{\BNS-\BH}(t')\,\mathrm{d} t'$ obtained
by integrating the outer GW frequency. Here $f(t)$ is the GW frequency of the inner BNS carrier [defined below Eq.~\eqref{eq:dpsi_from_dphi}], to be distinguished from the outer-orbit frequency $f_{\BNS-\BH}$ that enters the outer orbital phase $\varphi_{\BNS-\BH}(t)=\pi\int_0^t f_{\BNS-\BH}\,\mathrm{d}t'$.
When mapped to the frequency domain through
Eq.~\eqref{eq:dpsi_from_dphi} and inserted in the perturbed waveform of
Eq.~\eqref{eq:perturbed_waveform}, this correction is all that enters the full
numerical evaluation of the horizon SNR $\delta\rho_h$ shown in
Figs.~\ref{fig:rho_diff_iota} and~\ref{fig:rho_diff_fH}.
The Doppler effect is
oscillatory, repeating once per outer orbit, and comparison with
Eq.~\eqref{eq:dpsi_osc} identifies its modulation index directly from the
oscillation envelope, 
\begin{equation}
    \Theta_\text{\tiny Dop}(f) = 2\pi\,f\,a_{\BNS-\BH}(f)\,\sin\iota\,.
    \label{eq:Theta_dop}
\end{equation}
This is essentially the phase accumulated by the carrier over the light-crossing time of the outer orbit, and it grows across the band $\propto f$, only mildly offset by the slow shrinkage of $a_{\BNS-\BH}$. It scales linearly with $\sin\iota$, vanishing for face-on systems ($\iota=0$, no line-of-sight velocity) and being maximal for edge-on systems. For the reference triple ($m_\NS = 0.1\,\msun$, $M_\BH = 5\,\msun$, $a_{\BNS-\BH,i} = 340\,R_g$) the light-crossing time is $a_{\BNS-\BH,i}/c = 8.4$~ms, and scanning the accessible $(m_\NS, f_\text{\tiny H}, a_{\BNS-\BH})$ space at $\iota = 45^\circ$ gives $\Theta_\text{\tiny Dop} \sim 1 - 40$\,rad. The Doppler modulation is thus in the saturation regime of Eq.~\eqref{eq:rho_J0_limits} over most of the parameter space, and its distinguishability approaches the saturation limit $8\sqrt{2}$ rather than tracking the amplitude of the imprint, as we explain in more detail in Appendix~\ref{app:mod_doppler}.

\subsubsection*{Relativistic aberration}
\label{sec:aberration}
\noindent
Relativistic aberration shifts the apparent direction of GW propagation when the source moves, modifying the $(2,2)$ mode phase \cite{Torres-Orjuela:2018ejx,Torres-Orjuela:2020cly,Torres-Orjuela:2020oxq,Torres-Orjuela:2020dhw}.  An observer at rest sees the wave arriving from a fixed line-of-sight direction $-\hat{m}$ in the center-of-mass frame.  Because the inner binary moves on the outer orbit with velocity $\vec{u}(t) = v_{\BNS-\BH}(-\sin\varphi_{\BNS-\BH},\cos\varphi_{\BNS-\BH},0)$, where $v_{\BNS-\BH}$ is the outer  orbital speed, the propagation direction in the binary's instantaneous rest frame is obtained by Lorentz-boosting $-\hat{m}$:
\begin{equation}
    \hat{n} =
      \frac{-\hat{m} - \gamma\,\vec{u}
        - (\gamma - 1)\,\dfrac{(\hat{m}\!\cdot\!\vec{u})}{u^2}\,\vec{u}}
           {\gamma\,(1 + \hat{m}\!\cdot\!\vec{u})}\,,
    \label{eq:lorentz_aberration}
\end{equation}
where $\gamma = (1 - v_{\BNS-\BH}^2)^{-1/2}$.  The dominant $(2,2)$ GW mode carries an azimuthal phase $e^{-2\mathrm{i}\vartheta_n}$, where $\vartheta_n(t) = \mathrm{arctan2}(n_y, n_x)$ is the in-plane azimuthal angle of $\hat{n}$. As the binary orbits, $\hat{n}$ precesses about $-\hat{m}$ and $\vartheta_n(t)$ oscillates at the outer orbital frequency, shifting the mode phase by $\delta\phi_\text{\tiny abe}(t) = -2\,\vartheta_n(t)$. This exact phase, evaluated using Eqs.~\eqref{eq:dpsi_from_dphi} and~\eqref{eq:perturbed_waveform}, enters the full numerical evaluation of the horizon SNR $\delta\rho_h$ shown in Figs.~\ref{fig:rho_diff_iota} and~\ref{fig:rho_diff_fH}. Expanded to first order in $v_{\BNS-\BH}$, one obtains
\begin{equation}
    \delta\phi_\text{\tiny abe}^{(1)}(t) \approx
      -\frac{2\,v_{\BNS-\BH}(t)\,\cos\varphi_{\BNS-\BH}(t)}{\sin\iota}\,,
    \label{eq:dphi_abe_first}
\end{equation}
so that, comparing with Eq.~\eqref{eq:dpsi_osc}, the aberration modulation index is set by the outer orbital speed alone,
\begin{equation}
    \Theta_\text{\tiny abe} = \frac{2\,v_{\BNS-\BH}}{\sin\iota}\,,\qquad
    v_{\BNS-\BH} = \sqrt{\frac{M_\BH}{a_{\BNS-\BH}(f)}}\,,
    \label{eq:Theta_abe}
\end{equation}
independently of the NS mass and Hill fraction. 
Because the deflection is of order $v/c$, aberration produces small values of $\Theta_\text{\tiny abe}$ over the evaluated frequency interval for the reference triple and remains in the linear regime of Eq.~\eqref{eq:rho_J0_limits}; the band average then gives the closed form
\begin{equation}
    \delta\rho_h^\text{\tiny abe} \simeq \frac{8}{\sqrt2}\,
    \frac{2\,v_{\BNS-\BH}}{\sin\iota}
    = \frac{8\sqrt2}{\sin\iota}\sqrt{\frac{M_\BH}{a_{\BNS-\BH}}}\,,
    \label{eq:rho_abe}
\end{equation}
equal to $0.88$ at $\iota = 45^\circ$ for $a_{\BNS-\BH,i} = 340\,R_g$. Aberration has the opposite inclination and separation dependences to Doppler: it scales as $1/\sin\iota$ rather than $\sin\iota$, and thus dominates at small inclinations, and it weakens with the outer separation as $a_{\BNS-\BH}^{-1/2}$, whereas $\Theta_\text{\tiny Dop}$ grows linearly with $a_{\BNS-\BH}$. Being a velocity effect, it also carries no dependence on $f$, which distinguishes it from the oscillatory part of the Shapiro time delay discussed below: the two modulations repeat once per outer orbit, but probe different features of the triple geometry.

\subsubsection*{Gravitational redshift}
\label{sec:redshift}
\noindent
The gravitational potential of the central BH redshifts the
GW frequency emitted by the inner binary~\cite{Chen:2017xbi,Chen:2018axp,Kuntz:2022juv,Morton:2023wxg,Yin:2024nyz,Wu:2026pwl}:
\begin{equation}
    f(t) \;\longrightarrow\; f(t)\,\sqrt{1 - \frac{2M_\BH}{a_{\BNS-\BH}(t)}}\,,
    \label{eq:grav_redshift}
\end{equation}
with the outer separation $a_{\BNS-\BH}(t)$ evolving as in
Eq.~\eqref{eq:r_of_f}.
Expanding to linear order in $M_\BH/a_{\BNS-\BH}$, the fractional frequency shift is $\delta f/f = -M_\BH/a_{\BNS-\BH} < 0$,
and the cumulative phase correction is
\begin{equation}
    \delta\phi_\text{\tiny red}(t) = -2\pi\int_0^t
      \frac{M_\BH}{a_{\BNS-\BH}(t')}\,f(t')\,\mathrm{d} t'\,.
    \label{eq:dphi_red_full}
\end{equation}
Mapping to the frequency domain via the stationary-phase approximation
with the leading-order chirp rate $\dot f$ yields
\begin{equation}
    \delta\psi_\text{\tiny red}(f) = -\frac{5\,M_\BH}{48\,\pi^{5/3}\,\mathcal{M}_c^{5/3}}
      \int_{f_i}^{f}
      \frac{f'^{-8/3}}{a_{\BNS-\BH}(f')} \mathrm{d} f'\,.
    \label{eq:dpsi_redshift}
\end{equation}
Inserted in the perturbed waveform of Eq.~\eqref{eq:perturbed_waveform}, this
$\delta\psi_\text{\tiny red}$ enters the full numerical evaluation of the
horizon SNR $\delta\rho_h$ shown in Figs.~\ref{fig:rho_diff_iota}
and~\ref{fig:rho_diff_fH}.
Unlike the two effects above, the redshift is secular rather than oscillatory, so
Eq.~\eqref{eq:rho_J0} does not apply and no modulation index
can be assigned to it. Its size is set not by the depth of the potential, which is
small ($M_\BH/a_{\BNS-\BH,i} \ll 1$), but
by its change across the band. Indeed, a constant potential would give the
accumulated phase the same $f^{-5/3}$ dependence as the leading inspiral term, and
would be entirely reabsorbed into the chirp mass; most of the redshift phase is
absorbed by the chirp-mass fit in this way~\cite{Chen:2017xbi,Chen:2020lpq} (which must thus be maximized over in the mismatch computation), and what survives is the in-band
shrinkage of $a_{\BNS-\BH}$, fixed by the rate ratio $R_\tau$ of
Eq.~\eqref{eq:tR}. The distinguishability therefore increases rapidly with the Hill
fraction, with a scaling 
$R_\tau\propto f_\text{\tiny H}^4$: at small $f_\text{\tiny H}$ the
outer separation changes negligibly during the inner inspiral and the redshift
becomes degenerate with an unmeasurable chirp-mass shift, whereas near the disruption edge
it grows by more than two orders of magnitude. Being a scalar effect, the redshift
carries no angular factor and is independent of inclination, which --- as we show
in Sec.~\ref{sec:detectability} --- makes it the least sensitive to inclination
among the effects considered here. Its non-monotonic dependence
on the NS mass and on the outer separation is discussed in
Appendix~\ref{app:mod_redshift}.

\subsubsection*{Shapiro time delay}
\label{sec:shapiro}
\noindent
\noindent
When GWs propagate through the curved spacetime of the central BH, they experience a Shapiro time delay~\cite{Meiron:2016ipr,Sberna:2022qbn,Kuntz:2022juv}. For a source at position $\vec{r}$ relative to the lens and an observer at luminosity distance $d_\text{\tiny L}\gg|\vec{r}\,|$, the delay along the nearly straight ray reads
\begin{equation}
    \Delta t_\text{\tiny Sha}(t) = -2M_\BH\,
    \log\!\left(\frac{|\vec{r}\,| + \hat{m}\cdot\vec{r}}{2\,d_\text{\tiny L}}\right),
    \label{eq:shapiro_delay}
\end{equation}
which in the outer orbital coordinates ($|\vec{r}\,|=a_{\BNS-\BH}$, $\hat m\cdot \vec{r} = a_{\BNS-\BH}\cos\varphi_{\BNS-\BH}\sin\iota$) splits as
\begin{align}
    \Delta t_\text{\tiny Sha}(t) &= -2M_\BH\,
      \log \left(\frac{a_{\BNS-\BH}(t)}{2\,d_\text{\tiny L}} \right)
      \notag\\
    &\quad - 2M_\BH\,\log \bigl[1 + \sin\iota\,\cos\varphi_{\BNS-\BH}(t)\bigr]\,,
    \label{eq:shapiro_expand}
\end{align}
entering the phase as $\delta\phi_\text{\tiny Sha}=-2\pi f\,\Delta t_\text{\tiny Sha}$. This phase, mapped through Eqs.~\eqref{eq:dpsi_from_dphi} and~\eqref{eq:perturbed_waveform}, enters the full numerical evaluation of the horizon SNR $\delta\rho_h$ shown in Figs.~\ref{fig:rho_diff_iota} and~\ref{fig:rho_diff_fH}. The Shapiro time delay is thus a mixed effect. The first term is secular, varying only through the slowly shrinking $a_{\BNS-\BH}(t)$, and the logarithm makes it slower still, so that across the band it is nearly constant; when multiplied by $2\pi f$ it is then close to a phase linear in frequency --- a shift of the arrival time --- which the fit absorbs almost entirely. The dominant observable is therefore the oscillatory term,
\begin{equation}
  \delta\psi_\text{\tiny Sha}\simeq -4\pi M_\BH f\,
  \log \bigl(1+\sin\iota\cos\varphi_{\BNS-\BH}\bigr)\,,
  \label{eq:dpsi_sha}
\end{equation}
which repeats once per outer orbit but, unlike Doppler and aberration, is not of the single-harmonic form of Eq.~\eqref{eq:dpsi_osc}. Expanding the logarithm into orbital harmonics $\cos n\varphi_{\BNS-\BH}$ assigns to the $n$-th one the modulation index $4\pi M_\BH f\,c_n$, with $c_n = 2(-1)^{n+1}\tan^n(\iota/2)/n$. All of them are much smaller than unity, so each lies in the linear regime of Eq.~\eqref{eq:rho_J0_limits} and, since distinct harmonics are orthogonal over the orbit, their contributions add in quadrature, giving the closed form (see Appendix~\ref{app:mod_shapiro})
\begin{equation}
    \delta\rho_h^\text{\tiny Sha}=32\pi\,M_\BH\,\sqrt{\langle f^2\rangle_W}\,
    \sqrt{2\,\mathrm{Li}_2\!\bigl(\tan^2\tfrac\iota2\bigr)}
    \label{eq:rho_sha}
\end{equation}
in terms of the dilogarithm function $\mathrm{Li}_2$. The amplitude grows linearly with the lens mass $M_\BH$. The observable Shapiro modulation is independent of the outer separation, whose contribution is confined to the largely absorbed secular term [see first line of Eq.~\eqref{eq:shapiro_expand}]. The NS mass enters only indirectly through the termination frequency: a larger $m_\NS$ extends the inspiral to higher frequencies for a polytropic EoS, increasing $\langle f^2\rangle_W$ and hence the Shapiro distinguishability.

\subsubsection*{Self-lensing}
\label{sec:selflensing}
\noindent
When the GW passes close to the central BH, gravitational
lensing modifies both the amplitude and phase of the observed signal~\cite{DOrazio:2019fbq,Yu:2021dqx,Pijnenburg:2024btj,Li:2025xuh}.
For the low frequencies and small lens masses relevant here, the
wave-optics regime applies.

In the wave-optics regime, the lensed waveform is related to the
unlensed waveform by an amplification factor $F(w, y)$~\cite{Nakamura:1999uwi,Grespan:2023cpa} 
\begin{equation}
    \tilde{h}_\text{\tiny lensed}(f) = F(w, y)\,\tilde{h}(f)\,.
    \label{eq:h_lensed}
\end{equation}
Here, the amplification factor depends on two dimensionless numbers, one for the strength of the lens and one for the alignment between the source and lens. The first compares the GW frequency to the lens's gravitational time scale,
\begin{equation}
    w \equiv  8\pi M_\BH f\,,
    \label{eq:w_def}
\end{equation}
which for our subsolar-mass systems satisfies $w\ll 1$ everywhere in band, placing us deep in the wave-optics regime. The second measures how close the source comes to being directly behind the lens, in units of the Einstein radius $R_\text{\tiny E}=\sqrt{4GM_\BH D_\text{\tiny LS}}$, the natural transverse scale of lensing, set by the lens mass and the lens--source distance $D_\text{\tiny LS}$:
\begin{equation}
    y \equiv \frac{b}{R_\text{\tiny E}}\,,
    \label{eq:y_def}
\end{equation}
where $b$ (not to be confused with the argument of the confluent hypergeometric function used below) denotes the source's offset from the lens projected on the sky, i.e., the impact parameter. Both $b$ and $D_\text{\tiny LS}$ follow from the outer-orbit geometry. The inner binary sits at orbital phase $\varphi_{\BNS-\BH}$ on an orbit of radius $a_{\BNS-\BH}$ inclined by $\iota$ with respect to the line of sight:
\begin{align}
    D_\text{\tiny LS} &= a_{\BNS-\BH}\,\sin\iota\,\sin\varphi_{\BNS-\BH}\,, \notag\\
    b &= a_{\BNS-\BH}\sqrt{\sin^2\!\varphi_{\BNS-\BH}
        +\cos^2\!\varphi_{\BNS-\BH}\,\cos^2\!\iota}\,,
    \label{eq:lens_geom}
\end{align}
and lensing operates only while the source is behind the lens
($\sin\varphi_{\BNS-\BH}>0$). 
Despite the finite size of the BH lens, in the following we will make the simplifying assumption of treating it as a point-like mass, neglecting post-Minkowskian corrections associated to the finite distance between the NS binary and the central object, together with its size (see Ref.~\cite{CarrilloGonzalez:2025gqm} for related estimates).
Within this assumption, the amplification factor can be evaluated analytically to be~\cite{Nakamura:1999uwi,Takahashi:2003ix,Grespan:2023cpa,Pijnenburg:2024btj}
\begin{align}
    F(w,y) = {}& \exp\!\left[\frac{\pi w}{4}
      + \frac{\mathrm{i} w}{2}\left(\log\frac{w}{2}
      - 2\,\phi_\text{\tiny m}(y)\right)\right] \notag\\[2pt]
      &\times\,\Gamma\!\left(1 - \frac{\mathrm{i} w}{2}\right)
      {}_1F_1\!\left(\frac{\mathrm{i} w}{2},\,1,\,\frac{\mathrm{i} w y^2}{2}\right),
    \label{eq:F_analytic}
\end{align}
where $\Gamma$ is the Euler gamma function, ${}_1F_1(a,b,z)$ is the confluent hypergeometric function, and
\begin{equation}
    \phi_\text{\tiny m}(y) = \frac{\left(x_\text{\tiny m}-y\right)^2}{2}
      - \log x_\text{\tiny m}\,, \quad
    x_\text{\tiny m} \equiv \frac{y+\sqrt{y^2+4}}{2}\,.
    \label{eq:phi_m}
\end{equation}
The lensed waveform $F(w,y)\,\tilde h(f)$ of
Eq.~\eqref{eq:h_lensed} enters the full numerical evaluation of the horizon SNR
$\delta\rho_h$ shown in Figs.~\ref{fig:rho_diff_iota}
and~\ref{fig:rho_diff_fH}. Because the orbit is far larger than the Einstein
radius, $y\gg1$, and the lens is weak, $w\ll1$, the amplification factor of
Eq.~\eqref{eq:F_analytic} reduces to its first-order (Born) form, in which the
imprint is carried by the sine and cosine integrals:
\begin{equation}
    \lvert F\rvert-1 \simeq \frac{w}{2}\Bigl[\frac\pi2-\mathrm{Si}(\tfrac12 wy^2)\Bigr]\,,
    \quad
    \arg F \simeq \frac{w}{2}\,\mathrm{Ci}(\tfrac12 wy^2)\,.
    \label{eq:F_low_w}
\end{equation}
Therefore self-lensing modulates the amplitude and the phase of the signal at once. Unlike the other effects above, it cannot be represented by a single modulation index: Eq.~\eqref{eq:rho_J0} does not apply, and its distinguishability has no closed form. It does simplify in the asymptotic limit: when $\tfrac12 wy^2\gg1$ --- which covers wide orbits, high frequencies and near face-on geometries, since $\tfrac12 wy^2\propto f\,a_{\BNS-\BH}/\sin\iota$  --- the lensing power falls as $(\tfrac12 wy^2)^{-2}$, and averaging over the outer orbit leaves
\begin{equation}
    \delta\rho_h \propto \frac{\sin\iota}{a_{\BNS-\BH}}\,,
    \label{eq:lens_geo}
\end{equation}
independently of the lens mass. Self-lensing thus grows in the edge-on limit, where the source passes closest to being directly behind the lens, and weakens with the outer separation faster than any other imprint. As we quantify in Appendix~\ref{app:mod_lensing}, it is the smallest of the five modulations throughout the parameter space.

\begin{figure}[!t]
\centering
\includegraphics[width=0.99\columnwidth]{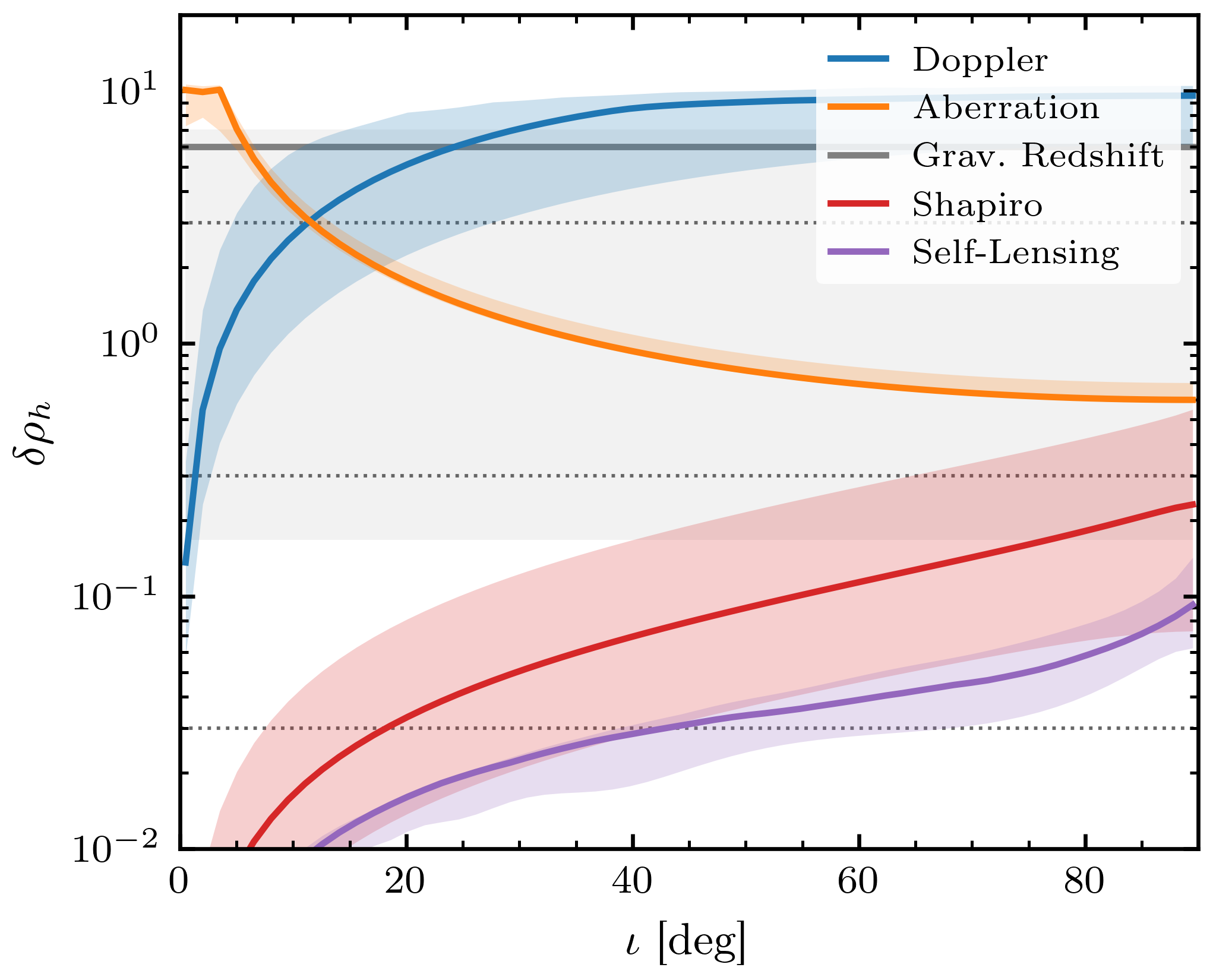}
\caption{Horizon distinguishability $\delta\rho_h$ ($\rho_0 = 8$, LIGO) versus inclination $\iota$ for the five modulation effects, assuming the polytropic EoS of Eq.~\eqref{eq:Rpoly}, at $M_\BH = 5\,\msun$, $f_\text{\tiny H} = 0.5$, $a_{\BNS-\BH,i} = 340\,R_g$. Solid lines show the reference $m_\NS = 0.1\,\msun$; the shaded bands span the NS mass range $m_\NS = 0.01 - 1\,\msun$ (the gravitational redshift band, in gray, is the most mass-sensitive). The dotted lines refer to the threshold values $\delta \rho_h = 3, 0.3, 0.03$.  The Doppler (blue) scales like $\sin\iota$ and pure aberration (orange) like $1/\sin\iota$; the gravitational redshift (gray) is exactly flat; Shapiro (red) and self-lensing (magenta) rise only weakly toward edge-on and never reach the threshold.}
\label{fig:rho_diff_iota}
\end{figure}

\subsubsection*{Detectability of modulation effects}
\label{sec:detectability}
\noindent
Having established the physics of the five modulation effects together with the detection framework, we now evaluate these predictions over the LIGO parameter range considered here.
We measure detectability of each modulation effect in terms of the distinguishability SNR $\delta\rho$ and its horizon value $\delta\rho_h$ introduced in Eqs.~\eqref{eq:rho_diff}--\eqref{eq:rho_diff_hor}, taking $\delta\rho_h>3$ (equivalently, $\mathcal{M}>9/128\approx0.07$) as the threshold for distinguishability at the horizon. All results below are exact numerical evaluations of this statistic, with no analytic approximation; the closed forms quoted alongside them serve as order-of-magnitude guides to the parameter dependences, as we elaborate in Appendix~\ref{app:modulations}.

\begin{figure*}[!t]
\centering
\includegraphics[width=\textwidth]{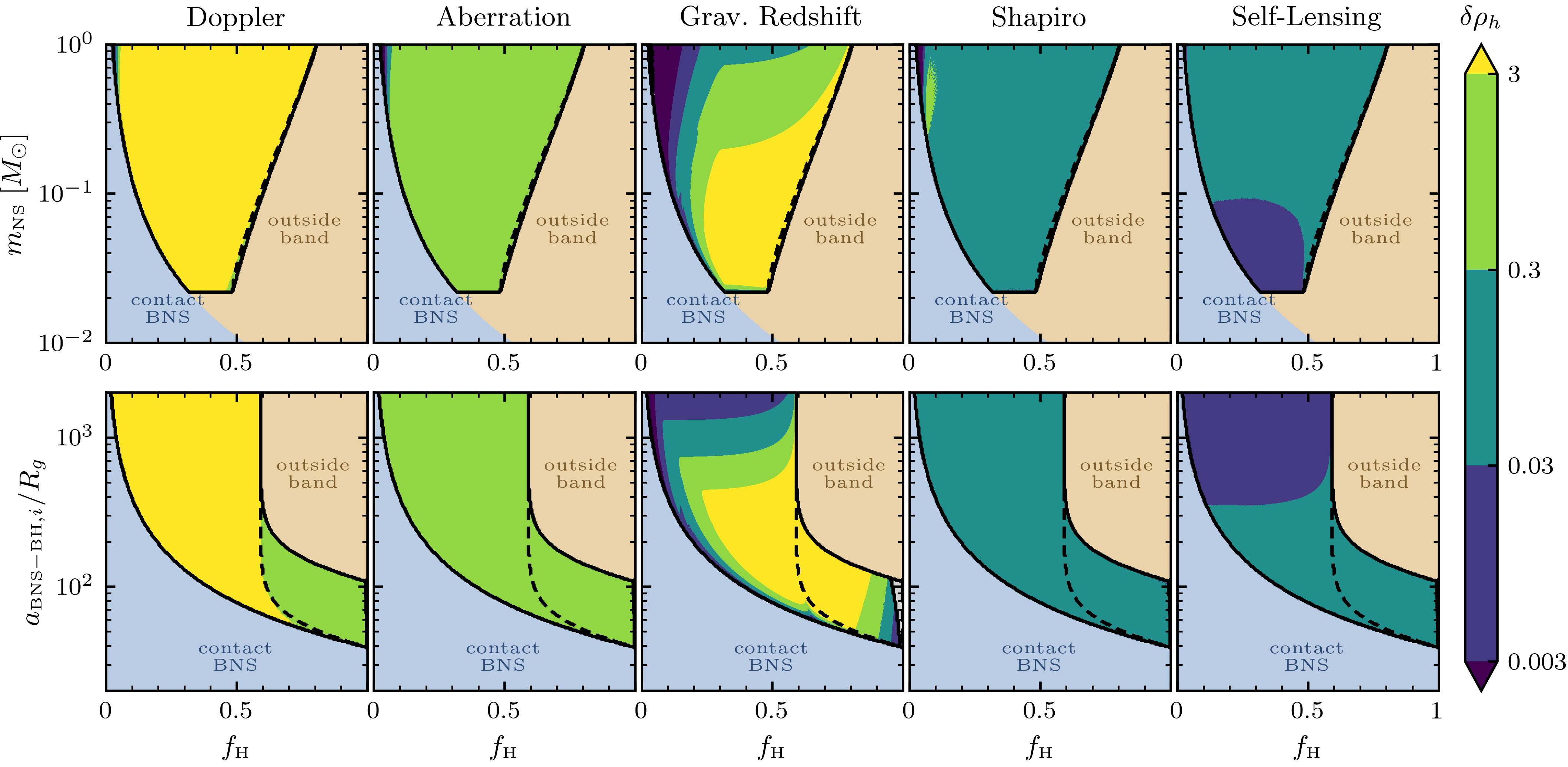}
\caption{Horizon distinguishability SNR $\delta\rho_h$ ($\rho_0 = 8$) for the five modulation effects, as a function of the Hill fraction $f_\text{\tiny H}$, in LIGO at $\iota = 45^\circ$, assuming the polytropic EoS of Eq.~\eqref{eq:Rpoly}. Columns are Doppler, aberration, gravitational redshift, Shapiro delay, and self-lensing. The top row shows the $(f_\text{\tiny H}, m_\NS)$ plane at $M_\BH = 5\,\msun$, $a_{\BNS-\BH,i} = 340\,R_g$; the bottom row shows the $(f_\text{\tiny H}, a_{\BNS-\BH,i}/R_g)$  plane at $m_\NS = 0.1\,\msun$ and $M_\BH = 5\,\msun$.  The shaded regions and boundary curves are as in Fig.~\ref{fig:hor_fH}. The solid black curves delimit the detectable region, excluding binaries born in contact and systems whose merger or disruption occurs outside the detector band.
}
\label{fig:rho_diff_fH}
\end{figure*}

The five effects have complementary coverage of the full inclination range, and the modulation indices derived above make this explicit. Doppler is the projection of the inner binary's orbital velocity onto the line of sight, so $\Theta_\text{\tiny Dop}\propto\sin\iota$ [Eq.~\eqref{eq:Theta_dop}] and the imprint vanishes face-on; aberration is the transverse, in-plane deflection of the wavefront, so $\Theta_\text{\tiny abe}\propto1/\sin\iota$ [Eq.~\eqref{eq:Theta_abe}] and it grows for face-on binaries; the gravitational redshift is a scalar potential effect carrying no angular factor at all; the Shapiro delay is a small effect and has a logarithmic dependence, $\delta\rho_h^\text{\tiny Sha}\propto M_\BH\sqrt{2\mathrm{Li}_2(\tan^2\!\tfrac\iota2)}$ [Eq.~\eqref{eq:rho_sha}]; and self-lensing grows as $\sin\iota$ toward edge-on [Eq.~\eqref{eq:lens_geo}], but would be sizable only for a source passing nearly directly behind the lens ($y\lesssim1$), a configuration never realized here because the outer orbit is far larger than the Einstein radius. 

In Fig.~\ref{fig:rho_diff_iota} we show $\delta\rho_h$ for all five effects in LIGO, as a function of $\iota$, at the reference $m_\NS = 0.1\,\msun$ (solid lines), with shaded bands spanning the NS mass range $m_\NS = 0.01 - 1\,\msun$ ($M_\BH = 5\,\msun$, $f_\text{\tiny H} = 0.5$, $a_{\BNS-\BH,i} = 340\,R_g$). The Doppler modulation (blue) increases as $\sin\iota$ from $\delta\rho_h\approx 1.4$ near face-on ($\iota = 5^\circ$) to $\approx 9.6$ edge-on, flattening toward the saturation limit of Eq.~\eqref{eq:rho_J0_limits} as $\Theta_\text{\tiny Dop}$ approaches unity. The pure-aberration curve (orange) falls as $1/\sin\iota$ from $\approx7$ at $\iota = 5^\circ$ to $\approx 0.6$ edge-on, in agreement with the closed form of Eq.~\eqref{eq:rho_abe}. The two cross near $\iota\approx11^\circ$, where the line-of-sight projection of the outer orbital velocity becomes larger than its transverse deflection. The gravitational redshift (gray) is inclination independent. Shapiro time delay (red) and self-lensing (magenta) remain below the $\delta\rho_h = 3$ threshold over the evaluated inclination range, rising only weakly toward edge-on (to $\approx0.23$ and $\approx0.09$, respectively); even there they sit at least an order of magnitude below threshold, so for these lens masses these effects remain below threshold over the evaluated range.

For the reference parameters, at least one effect exceeds the threshold at each evaluated inclination. For $\iota\lesssim 11^\circ$ aberration provides the above-threshold contribution, for $\iota\gtrsim 11^\circ$ Doppler does, and the redshift remains above threshold over the evaluated inclination range.

Figure~\ref{fig:rho_diff_fH} shows $\delta\rho_h$ for all five effects across $f_\text{\tiny H}$, in the same two planes as Fig.~\ref{fig:hor_fH}, at the intermediate inclination $\iota = 45^\circ$. Beyond the inclination complementarity outlined above, the contour plots show how each effect's distinguishability depends on the intrinsic parameters $(m_\NS, f_\text{\tiny H}, a_{\BNS-\BH,i})$. Each pattern follows from the physics of the individual modulation imprint. In all panels the colored region is bounded by the same green/orange regime curves as the horizon map; outside the inner-merger band there is no inspiral to modulate.

The panels in Fig.~\ref{fig:rho_diff_fH} (and the inclination dependence plot of Fig.~\ref{fig:rho_diff_iota}) display the horizon value $\delta\rho_h$ of Eq.~\eqref{eq:rho_diff_hor}, a direct proxy for the underlying mismatch, $\mathcal{M} = \tfrac{1}{2}(\delta\rho_h/8)^{2} = \delta\rho_h^{2}/128$, so each color scale doubles as a mismatch map: the $\delta\rho_h = 3$ contour is $\mathcal{M} = 9/128 \approx 0.07$, while $\delta\rho_h = 0.3$ is $\mathcal{M}\approx 7\times10^{-4}$. 

Because $\delta\rho \propto \rho_0 \propto d_\text{\tiny L}^{-1}$, a sub-threshold value at the horizon means that the modulation is not distinguishable at the horizon and requires a larger SNR: a modulation with $\delta\rho_h = 0.3$ becomes distinguishable once $\rho$ rises to $80$, i.e., for sources within a tenth of the horizon. However, detecting the source at all requires $\rho \ge 8$, so a modulation can never be measured beyond the detection horizon. The maximum luminosity distance out to which a modulation is distinguishable is therefore limited to the horizon,
$\min\!\left[d_\text{\tiny hor},
        \frac{\delta\rho_h}{3}\,d_\text{\tiny hor}\right]$
(up to the percent-level cosmological correction to $\rho\propto d_\text{\tiny L}^{-1}$).
This yields two regimes: effects above the threshold are distinguishable at the horizon and at smaller distances, while sub-threshold effects are distinguishable only for closer sources within $(\delta\rho_h/3)\,d_\text{\tiny hor}$. As shown in the plot, Doppler over much of the band (near the saturation limit, $\delta\rho_h \approx 9$) and the gravitational redshift at the reference point ($\delta\rho_h \approx 6$) pass the threshold for the reference-mass triples over the evaluated viewing angles. The inclination independence makes redshift less sensitive to source orientation than the other effects considered. Aberration is an order of magnitude weaker still: Eq.~\eqref{eq:rho_abe} gives
$\delta\rho_h^\text{\tiny abe}\simeq 1$ at $\iota=45^\circ$, recoverable only with an SNR of $\sim 24$ at this inclination, but its $1/\sin\iota$ scaling makes it the one effect that strengthens toward
face-on, precisely where the Doppler modulation is unobservable. Shapiro and
self-lensing ($\delta\rho_h \ll 3$) are recoverable only within a small
fraction of the detectable volume. Together, these modulations can produce
the distinguishability $\delta\rho_h\to8\sqrt2\approx11.3$, at which
point the modulated signal is essentially decorrelated from any unmodulated
template. This suggests that a collapsar triple signal could suffer significant mismatch against the point-particle waveform banks used in current subsolar searches. Since these imprints are not included in those templates, they could reduce the recovery efficiency of existing template banks.

\subsection{Environmental and tidal effects}
\label{sec:envtidalNSNS}
\noindent
As discussed in Sec.~\ref{sec:waveform:tf2}, the collapsar environment and the internal structure of the NSs both leave characteristic imprints on the GW inspiral phase~\cite{Barausse:2014tra,Barausse:2014pra,Cardoso:2019rou,Cardoso:2021wlq,Cole:2022fir,Cardoso:2022whc,Zwick:2022dih,Zwick:2025wkt}.
In this section we derive the PN phase corrections arising from (i) gas-driven dynamical friction, and (ii) the tidal deformability of the NSs. Accretion onto the system during the inspiral is discussed in Appendix~\ref{app:accretion}, where we show that its relevant timescales exceed both the GW-driven coalescence and dynamical friction times, so that it can be safely neglected in the waveform model. 

\subsubsection*{Dynamical friction}
\noindent
To assess the relevance of dynamical friction, introduced in Sec.~\ref{sec:param_NSNS}, we consider
the energy-balance equation~\cite{CanevaSantoro:2023aol}
\begin{equation}
\label{eq:balanceequation}
\dot{E}_\text{\tiny orb} = - |\dot{E}_\text{\tiny GW}|
- |\dot{E}_\text{\tiny dyn}|\,,
\end{equation}
where the GW luminosity and the dynamical-friction power
dissipation read
\begin{align}
\dot{E}_\text{\tiny GW} &= -\frac{32}{5} \eta^2 \, v_\BNS^{10} = -\frac{2}{5}  v_\BNS^{10} \,,
\nonumber\\
\dot{E}_\text{\tiny dyn} & = -\frac{4\pi\rho_\text{\tiny env}
M_\text{\tiny tot}^2}{v_\BNS} \left(\frac{1-3\eta}{\eta} \right)\,\mathcal{I} \simeq
-\frac{16\pi\rho_\text{\tiny env}
m_\NS^2}{v_\BNS} \,\mathcal{I}\,,
\end{align}
and $E_\text{\tiny orb} = - \eta M_\text{\tiny tot} v_\BNS^2/2 = -m_\NS v_\BNS^2/4$
is the total orbital energy of the binary, where in the second equalities we have focused on an equal-mass NS binary with total mass $M_\text{\tiny tot} = 2 m_\NS$ and  symmetric mass ratio $\eta \equiv m_1 m_2/(m_1+m_2)^2 = 1/4$.
Here $\rho_\text{\tiny env}$ is the post-fragmentation residual gas density introduced in Sec.~\ref{sec:param_NSNS}, the orbital velocity can be expressed as $v_\BNS = (\pi M_\text{\tiny tot} f)^{1/3}$ in terms of the GW frequency $f$, and $\mathcal{I} = \mathcal{O}(1)$ is the Coulomb logarithm characterizing the range of impact parameters for which dynamical friction is effective: see Eq.~\eqref{calIDF} and Refs.~\cite{Chandrasekhar:1943ys,Ostriker:1998fa}.

Four assumptions underlie this treatment. First, on scales $a_{\BNS} \ll h$ the binary is embedded in a medium that is approximately \emph{homogeneous}, so the standard Chandrasekhar drag formula applies directly~\cite{Chandrasekhar:1943ys}; corrections arising from density gradients across the binary orbit are suppressed by $a_{\BNS}/h \ll 1$ and can be neglected~\cite{Vicente:2019ilr}.
Second, we treat $\rho_\text{\tiny env}$ as \emph{temporally constant} throughout the inspiral. This is justified as long as $\tau_\text{\tiny acc}^{\BH}$ is larger than $\tau_\GW^{\BNS}$, i.e., as long as the gas density does not evolve appreciably over the duration of the in-band signal. Third, we assume the condition $\tau_\text{\tiny dyn} > \tau_\GW^{\BNS}$, implying that the environmental correction can be treated as a perturbative shift of the vacuum phase~\cite{CanevaSantoro:2023aol}. 
Finally, we assume \emph{circular orbits} throughout the derivation, so that the orbital velocity $v_\BNS$ is uniquely related to the GW frequency via the Keplerian relation above. Residual eccentricity sourced by disk fragmentation and gas-assisted capture can in principle modify both  the vacuum and environmental phase evolution; we assess the magnitude of this effect and its observational consequences in Sec.~\ref{sec-eccentricity}. 

Combining the relations above, the GW frequency evolution becomes  
\begin{align}
\label{eq:dotf}
    \dot{f} = \dot{f}_\GW + \dot{f}_\text{\tiny dyn} & = \frac{96}{5} \pi^{8/3}
    f^{11/3} \mathcal{M}_c^{5/3}
    +  \frac{12(1-3\eta)}{\eta^2} \, \mathcal{I}\, \rho_\text{\tiny env} \nonumber \\
    & \simeq \frac{48}{5} 2^{2/3} \pi^{8/3}
    f^{11/3} m_{\NS}^{5/3}
    + 48 \, \mathcal{I}\, \rho_\text{\tiny env}\,.
\end{align}
The accumulated time-to-coalescence and phase are then obtained by integrating the reciprocal of $\dot f$:
\begin{align}
 t(f) & = t_c - \int_f^\infty \frac{{\rm d} f'}{\dot{f}'}\,,
\nonumber \\
 \varphi(f) & = \varphi_c - \int_f^\infty
 \frac{2\pi f'\,{\rm d} f'}{\dot{f}'}\,,
\end{align} 
in terms of the coalescence time and phase reference values, $t_c$ and $\varphi_c$, respectively. 
Expanding the environmental contribution at first order in $\rho_\text{\tiny env}$, the full dynamical friction-induced phase can be written as
\begin{equation}
\delta \psi_\text{\tiny dyn}(f) = \frac{3 \, \delta_\text{\tiny dyn}}{128\,\eta\,v_\BNS^5}
\,,
\end{equation}
where the environmental contribution reads~\cite{CanevaSantoro:2023aol} 
\begin{align}
\delta_\text{\tiny dyn} &= -\frac{25\,\pi\,\mathcal{I}\,M_\text{\tiny tot}^2 \,\rho_{\text{\tiny env}}}
{304\,v_\BNS^{11}} \left(\frac{1-3\eta}{\eta^3} \right) \nonumber \\
& \simeq 
-\frac{100\,\pi\,\mathcal{I}\,m_\NS^2 \,\rho_{\text{\tiny env}}}
{19\,v_\BNS^{11}}\,.
\label{eq:deltaDF}
\end{align}
Such corrections scale as $v_\BNS^{-11}$ relative to the leading Newtonian term, i.e., they enter the waveform at $-5.5$\,PN order~\cite{Barausse:2014tra,Cardoso:2019rou}. Dynamical friction, being a negative PN correction, thus accelerates the frequency evolution with respect to vacuum, causing the binary to accumulate fewer GW cycles than a point-particle system of identical masses. From Eq.~\eqref{eq:Ncyc}, the difference in the number of cycles scales like
\begin{align}
\delta \mathcal{N}_\text{\tiny env} \simeq \frac{25 \, \mathcal{I} \, \rho_\text{\tiny env}}{512 \times 2^{1/3}\,  \pi ^{16/3} m_\NS^{10/3}} \left(f_\text{\tiny max}^{-16/3}-f_\text{\tiny min}^{-16/3}\right)\,,
\end{align}
which, when compared to the prediction including only GWs in Eq.~\eqref{eq:NGW}, shows a different dependence on the detector's frequencies.  
This effect is a potential smoking gun of a gas-rich environment~\cite{Kocsis:2011dr,Yunes:2011ws,Derdzinski:2020wlw,Speri:2022upm,Zwick:2025wkt,Tagawa:2025tfd}.

\subsubsection*{Tidal deformability of NSs}
\noindent
In addition to the environmental effects above, the finite size of the NS gives rise to a tidal contribution to the inspiral GW phase, because the tidal field of one NS induces a quadrupolar mass deformation of its companion. For a binary with individual tidal deformabilities $\Lambda_1$ and $\Lambda_2$, the leading tidal contributions to the waveform phase appear at $5$\,PN and $6$\,PN order relative to the leading Newtonian term~\cite{Flanagan:2007ix, Hinderer:2009ca, Vines:2011ud}: 
\begin{align}
\label{eq:psi_tidal}
    \delta\psi_\text{\tiny tidal}(f)
   & =  \frac{3}{128 \, \eta \, v_\BNS^5} \left[ - \frac{39}{2} \tilde{\Lambda} \,
v_\BNS^{10} \right. \nonumber \\
& \left. +
 \left(
- \frac{3115}{64}\tilde{\Lambda}
+ 
 \frac{6595}{364}\sqrt{1-4\eta} \, \delta\tilde{\Lambda}
 \right)v_\BNS^{12}
\right] 
    \,.
\end{align}
It is common to express the tidal
content of the inspiral through the mass-weighted
effective tidal deformabilities~\cite{Flanagan:2007ix,
Wade:2014vqa, LIGOScientific:2017vwq}
\begin{align}
\label{eq:Lambda_tilde}
   \tilde{\Lambda} & = 
\frac{8}{13}
\left[ \big(1+7\eta-31\eta^2\big)(\Lambda_1 + \Lambda_2) \right. 
\nonumber \\
& \left. +
\sqrt{1-4\eta}\big(1+9\eta-11\eta^2\big)
(\Lambda_1 - \Lambda_2)\right]\,, \\ 
\delta\tilde{\Lambda} & = 
\frac{1}{2}\left[
\sqrt{1-4\eta}\left(
1-\frac{13272}{
1319}\eta + 
\frac{8944}{1319}\eta^2\right)
(\Lambda_1 + \Lambda_2) \right. \nonumber \\
& \left. +
\left(
1-\frac{15910}
{1319}\eta + \frac{32850}{1319}\eta^2 + \frac{3380}{1319}\eta^3
\right)(\Lambda_1 - \Lambda_2)\right]\,.
\label{eq:deltaLambda_tilde}
\end{align}
These are the tidal parameters that enter the waveform at leading order and are therefore the primary observables accessible through GW measurements.  For an equal-mass binary ($m_1=m_2\equiv m_\NS$, $\eta = 1/4$), one has $\tilde\Lambda=\Lambda$, so that the effective tidal deformability coincides with the tidal deformability of each binary component.

The tidal phase correction of Eq.~\eqref{eq:psi_tidal} grows rapidly with increasing frequency: it scales as $(\pi\mathcal{M}_cf)^{5/3}$ at leading order, meaning that for the extremely large $\Lambda$ values characteristic of subsolar NSs, tidal effects are already significant early in the inspiral band of current and future detectors.  Combined with the environmental phase shift of Eq.~\eqref{eq:deltaDF}, the total perturbed phase entering Eq.~\eqref{eq:perturbed_waveform} reads
\begin{equation}
    \delta\psi =  \delta\psi_\text{\tiny dyn} +\delta\psi_\text{\tiny tidal} \,.
\end{equation}
The two contributions have different frequency scalings ($f^{5/3}$ for leading tides versus $f^{-16/3}$ for the dynamical friction term), which in principle allows them to be disentangled in data analysis~\cite{Cole:2022fir,Garg:2024oeu}, as we quantify in the next section.

\begin{figure*}[h!] 
    \centering
\includegraphics[width=0.9\linewidth]{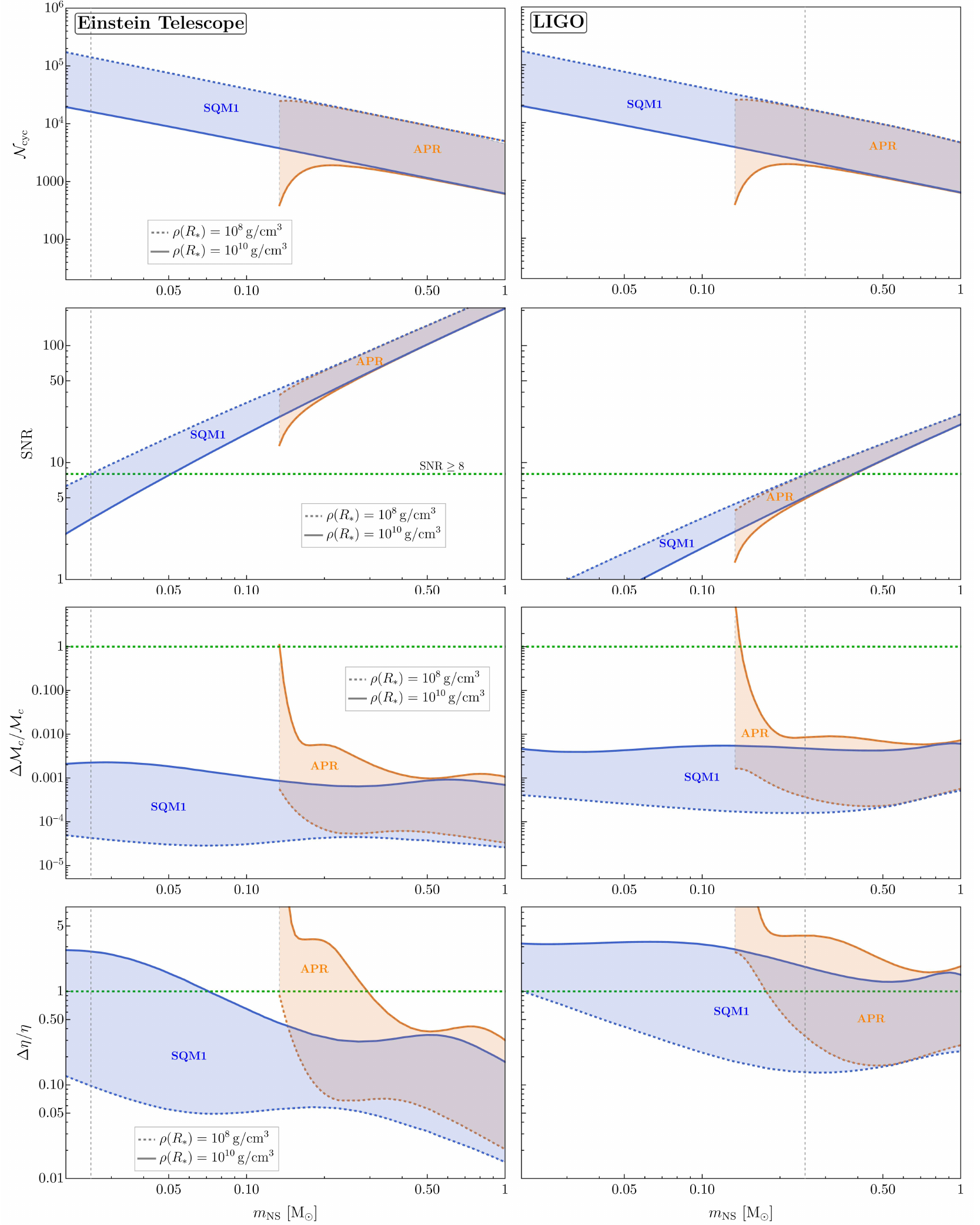}%
\caption{Measurability of model parameters in equal-mass BNS mergers at a luminosity distance of $d_\text{\tiny L} = 100 \, {\rm Mpc}$ estimated with a Fisher analysis for ET (left column) and LIGO (right column). The panels show relative errors on the number of cycles (top row), SNR (second row), chirp mass (third row) and mass ratio (bottom row), all as functions of the NS mass. Solid and dashed lines mark plausible ranges for the environmental density; blue and orange curves correspond to the QM and NM EoS, respectively. The horizontal green lines mark either regions above which the corresponding parameters are not measurable (bottom plots), or below which the binary is not detectable (SNR plot). The vertical dashed lines indicate the minimum NS mass above which binaries are detectable ($\rm SNR\geq8$), assuming the lower disk density $\rho(R_*) = 10^8 \, {\rm g/cm^3}$. }
\label{fig:measurabilityNSNS1}
\end{figure*}

\begin{figure*}[t!] 
    \centering
\includegraphics[width=0.9\linewidth]{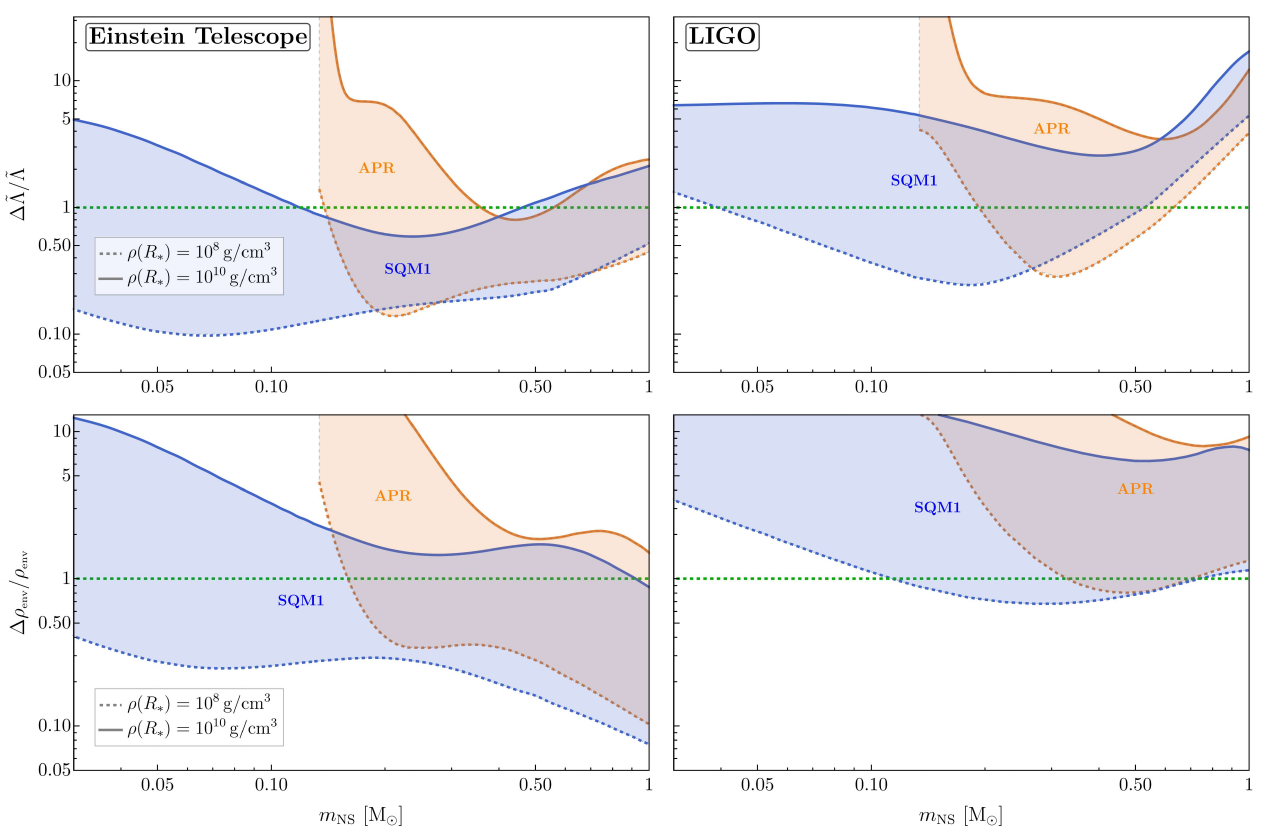}%
\caption{Measurability of model parameters in the BNS merger phase. The panels show the results of a Fisher analysis for the 5\,PN TLN (top panels) and environmental density (bottom panels) as functions of the NS mass. 
Solid and dashed lines demarcate plausible ranges for the environmental density; blue and orange curves correspond to the QM and NM EoS, respectively. The horizontal green lines mark regions above which the corresponding parameters are not measurable.}
\label{fig:measurabilityNSNS2}
\end{figure*}

\subsubsection*{Measurability of model parameters}\label{sec:FisherI}
\noindent
We assess the measurability of the BNS parameters with current and future GW interferometers using the Fisher matrix formalism of Sec.~\ref{sec:waveform:tf2}.  We restrict ourselves to the regime in which the inspiral is GW-driven. 
The corresponding region of parameter space is shown explicitly in beige in the left panel of Fig.~\ref{fig:param}. 
The initial BNS relative separation is then set by the separation below which $\tau_\text{\tiny dyn} > \tau_\GW^{\BNS}$, i.e.,
\begin{align}\label{eq:a0NSNS}
    a_{\BNS,i} &\simeq 113 \,{\rm km}
    \left(\frac{m_\NS}{0.1\,M_\odot}\right)^{7/11}\nonumber\\
    &\times\left(\frac{\rho (R_*)}{10^{8}\,{\rm g\,cm^{-3}}}\right)^{-2/11}
    \left(\frac{ f_{\text{\tiny gas}}}{10^{-4}}\right)^{-2/11}\,,
\end{align}
which depends only weakly on the disk density and on the gas fraction, and more strongly on the NS mass. The corresponding initial frequency is found via Eq.~\eqref{eq:fi}.
The inspiral continues until the two NSs touch at the frequency of Eq.~\eqref{eq:fRoche_num}, 
which depends upon the chosen EoS. For the parameters considered in the present analysis, this frequency always lies below  $\sim 3 \,{\rm kHz}$, so that the inspiral terminates within the detector band of both LIGO and ET rather than beyond it, and the signal is followed all the way to contact --- even though both instruments lose sensitivity toward the upper end of that band, above $\sim 1\,{\rm kHz}$. We perform a Fisher matrix analysis with ten parameters, $\bm{\theta} = (\mathcal{M}_c, \eta, \chi_s, \chi_a, \tilde{\Lambda}, \delta \tilde{\Lambda}, \rho_\text{\tiny env}, t_c, \varphi_c, d_\text{\tiny L})$. Assuming equal spins, we inject the symmetric and antisymmetric spin combinations $\chi_s = 0.1$ and $\chi_a = 0$, respectively. We have verified that our findings are largely insensitive to this choice.

The numerical results are presented in Fig.~\ref{fig:measurabilityNSNS1}. From top to bottom, we show the number of cycles $\mathcal{N}_{\text{\tiny cyc}}$, the SNR, and the Fisher relative errors on the chirp mass $\Delta\mathcal{M}_c/\mathcal{M}_c$ and on the symmetric mass ratio $\Delta\eta/\eta$. The left column refers to ET, which we take as the representative next-generation detector throughout this section, while the right column assumes LIGO (see Sec.~\ref{sec:waveform:tf2} for more details on the sensitivity curves). The results are reported at a benchmark luminosity distance $d_\text{\tiny L} = 100\,\rm Mpc$. We fix $f_\text{\tiny gas} = 10^{-4}$ and we consider different values for the density at the disk's outer edge $\rho(R_*)$, corresponding to different disk radii [see Eq.~\eqref{eq:Rstar}] and thus to different initial frequencies, according to Eq.~\eqref{eq:fi}. Moreover, we adopt two different EoSs, each representing one of the two  families for the matter prevailing in the NS interior.

The number of in-band cycles, shown in the top row of Fig.~\ref{fig:measurabilityNSNS1}, is essentially the same for the two detectors. Both the entry frequency $f_i$ and the contact frequency $f_\text{\tiny TD}$ lie within the detector band of LIGO and ET over the whole parameter region investigated, so the two instruments follow the same inspiral interval $[f_i,\,f_\text{\tiny TD}]$; moreover, from Eq.~\eqref{eq:NGW} the count is dominated by its lower edge, with $f_\text{\tiny TD} \gg f_\text{\tiny min}$, so the small difference in the detectors' low-frequency floors has a negligible effect. The two differ instead in the accumulated SNR (second row), set by the noise-weighted integral of Eq.~\eqref{eq:rho0}.
Moreover, we observe that a lower outer disk density $\rho(R_*)$ corresponds to a higher number of accumulated cycles. This is a direct consequence of the way collapsar disks form: according to Eq.~\eqref{eq:Rstar}, the fragmentation radius scales as $R_* \propto \rho(R_*)^{-1/3}$. Lower outer disk densities therefore correspond to larger fragmentation radii, and in turn to a larger orbital separation below which the BNS evolution is GW-driven; indeed, Eq.~\eqref{eq:a0NSNS} gives $a_{\BNS,i} \propto \rho(R_*)^{-2/11}$. A wider initial separation then translates into more in-band cycles. For the same reason, the SNR (second row of Fig.~\ref{fig:measurabilityNSNS1}) is also higher for $\rho(R_*) = 10^{8}\,{\rm g\,cm^{-3}}$ than for $\rho(R_*) = 10^{10}\,{\rm g\,cm^{-3}}$. The trend with the mass of both the number of cycles and the SNR follows from equally simple considerations: higher NS masses correspond to a higher BNS chirp mass, so that the SNR increases with $m_\NS$, while at the same time they correspond to a shorter inspiral, so that $\mathcal{N}_{\text{\tiny cyc}}$ decreases with $m_\NS$. One of the fundamental differences between QM and NM is already apparent from these first two rows. The orange curves, corresponding to the APR EoS, show a sharp feature around $m_\NS \simeq 0.1\,M_\odot$, where both $\mathcal{N}_{\text{\tiny cyc}}$ and the SNR drop rapidly as $f_{\text{\tiny TD}} \to f_i$. The SQM1 EoS, instead, always keeps $f_{\text{\tiny TD}} \gg f_i$, so that decreasing the mass below $0.1\,M_\odot$ further widens the binary, leading to increasing values of $\mathcal{N}_{\text{\tiny cyc}}$ (and decreasing SNR).
The second rows in Fig.~\ref{fig:measurabilityNSNS1} show that the detectability threshold ${\rm SNR} > 8$ is reached at different masses in the two detectors. LIGO can detect BNSs with $m_\NS \gtrsim 0.25\,M_\odot$ for $\rho(R_*) = 10^{8}\,{\rm g\,cm^{-3}}$ and $m_\NS \gtrsim 0.4\,M_\odot$ for $\rho(R_*) = 10^{10}\,{\rm g\,cm^{-3}}$, whereas ET will observe them down to $m_\NS \simeq 0.03\,M_\odot$ and $m_\NS \simeq 0.05\,M_\odot$, respectively (see dashed vertical lines in Fig.~\ref{fig:measurabilityNSNS1}). The last two rows of Fig.~\ref{fig:measurabilityNSNS1} show the relative errors on the chirp mass and on the symmetric mass ratio, respectively. Within their respective resolvable ranges, both LIGO and ET measure the chirp mass with a relative error well below unity even for $\rho(R_*) = 10^{10}\,{\rm g\,cm^{-3}}$, provided one does not approach the contact point around $m_\NS \simeq 0.1\,M_\odot$ for the APR EoS. The measurability of the symmetric mass ratio, on the other hand, depends strongly on the outer disk density: the parameter is well measured by both detectors when $\rho(R_*) = 10^{8}\,{\rm g\,cm^{-3}}$, whereas for $\rho(R_*) = 10^{10}\,{\rm g\,cm^{-3}}$ this is no longer the case, and only ET retains some sensitivity when $m_\NS \gtrsim 0.07\,M_\odot$ for the SQM1 EoS and $m_\NS \gtrsim 0.3\,M_\odot$ for the APR EoS.

In Fig.~\ref{fig:measurabilityNSNS2} we show the corresponding relative errors on the effective tidal deformability, $\Delta\tilde{\Lambda}/\tilde{\Lambda}$, and on the environmental density, $\Delta\rho_\text{\tiny env}/\rho_\text{\tiny env}$. As far as the tidal deformability is concerned, we observe that both detectors display some sensitivity for the lower value of the outer disk density, $\rho(R_*) = 10^{8}\,{\rm g\,cm^{-3}}$. In this case, ET can measure the tidal deformability regardless of the BNS mass (except, once again, close to the NM contact point around $m_\NS \simeq 0.1\,M_\odot$). In LIGO, instead, $\Delta\tilde{\Lambda}/\tilde{\Lambda}$ is below unity only for $0.05\lesssim m_\NS/M_\odot \lesssim 0.5$ with the SQM1 EoS, and $0.2\lesssim m_\NS/M_\odot \lesssim 0.6$ with the APR EoS. The prospects are more pessimistic when the outer disk density approaches the larger value $\rho(R_*) = 10^{10}\,{\rm g\,cm^{-3}}$. In that case, the tidal deformability, despite reaching $\Lambda \sim \mathcal{O}(10^9)$ for the lightest NSs, is measurable only with ET, and in the mass ranges $0.1\lesssim m_\NS/M_\odot \lesssim 0.5$ for the SQM1 EoS, and $0.35\lesssim m_\NS/M_\odot \lesssim 0.6$ for the APR EoS.
Overall, the measurability prospects for the SQM1 EoS are better than for the APR EoS. This is physically expected, as standard NM in the subsolar mass range leads to larger NS radii and therefore to a lower $f_\text{\tiny TD}$. Quark stars, on the other hand, have a longer inspiral and more signal content at high frequencies (see the left panel of Fig.~\ref{fig:param}), enabling a better measurement of the $5$\,PN term.
The two EoS predict sufficiently different values of $\Delta\tilde{\Lambda}/\tilde{\Lambda}$ in the mass range considered that a confident tidal measurement for $m_\NS \lesssim 0.2\,M_\odot$ with ET would discriminate between NM and QM, providing an independent cross-check of the mass-based EoS discrimination discussed above. Such a measurement would also correspond to a precise determination of the chirp mass and of the mass ratio (see Fig.~\ref{fig:measurabilityNSNS1}).

Concerning the environmental imprint on the waveform, for the higher outer disk density $\rho(R_*) = 10^{10}\,{\rm g\,cm^{-3}}$,
the effect appears to be too weak to be measured by either detector, independently of the EoS: the relative error on $\rho_\text{\tiny env}$ exceeds unity by a factor of $2 -10^2$ over the whole parameter region investigated (bottom row of Fig.~\ref{fig:measurabilityNSNS2}). The only exception is ET with the SQM1 EoS, which retains some sensitivity as $m_\NS$ approaches $1\,M_\odot$. For the lower value of the outer disk density, $\rho(R_*) = 10^{8}\,{\rm g\,cm^{-3}}$, the prospects are significantly better. Similarly to the tidal deformability, ET is able to measure the effect essentially over the whole resolvable mass range, while LIGO has some sensitivity only for $0.1\,M_\odot \lesssim m_\NS \lesssim 0.7\,M_\odot$ with the SQM1 EoS, and $0.4\,M_\odot \lesssim m_\NS \lesssim 0.7\,M_\odot$ with the APR EoS.

The environmental measurement improves for lower outer disk densities because of two competing effects. The environmental correction in Eq.~\eqref{eq:deltaDF} scales linearly with $\rho_\text{\tiny env} = f_{\text{\tiny gas}}\,\rho(R_*)$, so that a denser disk produces a larger dephasing per orbit. The trend with $\rho(R_*)$ is also controlled by the dependence of the initial separation, $a_{\BNS,i}\propto\rho(R_*)^{-2/11}$, given in Eq.~\eqref{eq:a0NSNS}: a denser disk fragments at a smaller radius, so that the binary enters the band closer to merger, and the resulting loss of in-band cycles exceeds the linear increase in the amplitude of the environmental correction. For the same reason, lowering $f_{\text{\tiny gas}}$ would generically improve the relative error on $\rho_\text{\tiny env}$. In practice, however, other competing effects limit the relevance of this regime. Indeed, sufficiently low values of $f_{\text{\tiny gas}}$ correspond to initial orbital separations such that $f_{\text{\tiny H}}\to 1$: a lower gas density weakens the dynamical friction, so the crossover to GW-driven evolution occurs at a wider separation --- Eq.~\eqref{eq:a0NSNS} gives $a_{\BNS,i}\propto f_{\text{\tiny gas}}^{-2/11}$, and hence $f_{\text{\tiny H}}\propto f_{\text{\tiny gas}}^{-2/11}$ at fixed Hill radius --- potentially driving the BNS into chaotic instability (see Sec.~\ref{sec:param_NSNS}).

\subsubsection*{Dynamical tidal deformability}
\noindent
The tidal phase in Eq.~\eqref{eq:psi_tidal} was computed in the adiabatic limit, where each stellar quadrupole follows the companion’s tidal field without delay. A NS instead has a finite dynamical response set by its fundamental $f$-mode, whose frequency $f_f = \bar\omega\,\sqrt{m_\NS/R_\NS^{3}}/2\pi  = \bar\omega f_\text{\tiny Roche}^\BNS$ scales with the contact frequency and the square root of the mean density, where the constant $\bar\omega$ depends on the stellar structure.  For the polytropic EoS $\bar\omega \sim 1.4$, so the $f$-mode lies above the contact frequency and the BNS never reaches resonance. The tide is nonetheless modified as the system approaches the $f$-mode frequency. This dynamical response makes the deformability in Eq.~\eqref{eq:psi_tidal} frequency-dependent:
\begin{equation}
    \Lambda_\text{\tiny eff}(f) = \frac{\Lambda}{1-\left(f/f_f\right)^{2}}\,,
    \label{eq:Lambda_dyn}
\end{equation}
which increases the induced quadrupole relative to the adiabatic model and accelerates the inspiral, reaching $\Lambda_\text{\tiny eff}/\Lambda = (1-\bar\omega^{-2})^{-1} \sim 2$ at contact. Expanding in $(f/f_f)^{2}\propto v_\BNS^{6}$, the leading dynamical correction enters at $8$\,PN order. 

This term is more readily measurable for subsolar NSs than for canonical NSs. First, the mean density scales as $m_\NS/R_\NS^{3}\propto m_\NS^{2}$ for the polytropic EoS of Eq.~\eqref{eq:Rpoly}, so that $f_f\propto m_\NS$. For subsolar NSs, the $f$-mode shifts from the kHz range characteristic of canonical binaries down to $\mathcal{O}(10^{2})$\,Hz, placing it in the most sensitive part of the detector band. Second, the dynamical response modifies a large tidal contribution, $\Lambda\propto m_\NS^{-20/3}$, so that the adiabatic tidal phase reaches $\mathcal{O}(10^{5})$\,rad over the $\sim10^{5}$ in-band cycles of Eq.~\eqref{eq:NcycBNS}, rather than the few radians available to a canonical BNS. As a result, the $8$\,PN dynamical term, which is confined to the last cycles for canonical NSs, becomes measurable for subsolar masses, as observed in Ref.~\cite{Apostolidis:2026qsg}.
For a resonance that remains above the observed frequency range, the enhanced pre-resonance tidal response remains measurable: the mismatch remains near its maximum up to $\bar\omega\simeq3$ and subsequently decreases as $1/\bar\omega^{2}$ as the increasing mode frequency produces a smaller modification to the deformability. If instead the stellar structure gives $\bar\omega < 1$, the resonance is crossed before contact: the orbit passes through the $f$-mode frequency over a short interval, during which the mode is resonantly excited, producing an observable phase shift in the binary waveform.

Throughout this section, we have neglected modulation effects in the waveform. These might bias the parameter estimation, since at the initial separation of Eq.~\eqref{eq:a0NSNS} we have $\tau_{\text{\tiny orb}} < \tau^{\text{\tiny BNS}}_{\text{\tiny GW}} < \tau_{\text{\tiny dyn}}$ (see the cross-hatched region in the left panel of Fig.~\ref{fig:param}). A Fisher or Bayesian analysis that also includes the relevant modulation effects in the waveform phase (in particular the Doppler phase shift and the gravitational redshift), and assesses their impact on parameter measurability, is left for future work.

\subsection{Eccentricity effects}
\label{sec-eccentricity}
\noindent
Eccentric pairs can generically arise as a result of disk fragmentation (see Sec.~\ref{sec:param_NSNS})~\cite{Zrake:2020zkw,Cardoso:2020iji,Lai:2022ylu}, and the hierarchical triple geometry can sustain or excite eccentricity through secular and tidal interactions~\cite{Wen:2002km,Antonini:2012ad,Naoz:2016cjb,Silsbee:2016djf,Hoang:2017fvh,Liu:2019gdc,Su:2025mdd}. In a GW-driven merger, the signature of even small initial eccentricity $e_{\BNS,i}$ is amplified by the large number of in-band cycles, which increases steeply at low masses [see Eq.~\eqref{eq:NGW}] as
\begin{equation}
\mathcal{N}_\GW^\text{\tiny BNS}\simeq 4\times10^{5}
\left(\frac{0.1\,\msun}{m_\NS}\right)^{5/3}
\left(\frac{f_\text{\tiny min}}{20\,{\rm Hz}}\right)^{-5/3}\,,
\label{eq:NcycBNS}
\end{equation} 
reaching $\approx 2\times10^{7}$ at the ET floor of $2$\,Hz. When the termination frequency enters the band --- as for NM near the minimum stable mass --- the count is further reduced by the replacement $f_\text{\tiny min}^{-5/3} \to f_\text{\tiny min}^{-5/3} - f_\text{\tiny TD}^{-5/3}$. Because $f_\text{\tiny max}\gg f_\text{\tiny min}$, this count is controlled almost entirely by the lower edge $f_\text{\tiny min}$: for the LIGO band it is the detector floor, $f_\text{\tiny min}=20$\,Hz, whenever the binary enters it from below  ($f_\text{\tiny H}\gtrsim0.3$), while for tighter orbits ($f_\text{\tiny H}\lesssim0.3$) the pair starts above the detector floor, and $f_\text{\tiny min}$ is instead dictated by the higher entry frequency $f_i$ of Eq.~\eqref{eq:fi}, increasing as $f_\text{\tiny H}$ decreases.
A small eccentricity imprints a secular phase advance that grows with the cycle count, $\delta\psi_\text{\tiny ecc} \simeq e_{\BNS,i}^2 (v_{\BNS,i}/v_\BNS)^{19/3} \psi_\text{\tiny PP} \propto e_{\BNS,i}^{2}\,\mathcal{N}^\BNS_\GW$ with an order-unity prefactor~\cite{Favata:2013rwa,Nishizawa:2016jji,Moore:2016qxz,Favata:2021vhw}, so a small $e_{\BNS,i}$ accumulates a measurable phase. We model the eccentric signal with the frequency-domain, analytic inspiral waveform of Ref.~\cite{Huerta:2014eca} (the \texttt{EccentricFD} approximant~\cite{Lower:2018seu}), which builds on the post-circular expansion of Ref.~\cite{Yunes:2009yz} by retaining eccentricity corrections to $\mathcal{O}(e^{8})$ at each PN order, and which reduces to the quasi-circular \texttt{TaylorF2} waveform of Sec.~\ref{sec:waveform:tf2} in the zero-eccentricity limit. 

\begin{figure}[!t]
\centering
\includegraphics[width=0.95\columnwidth]{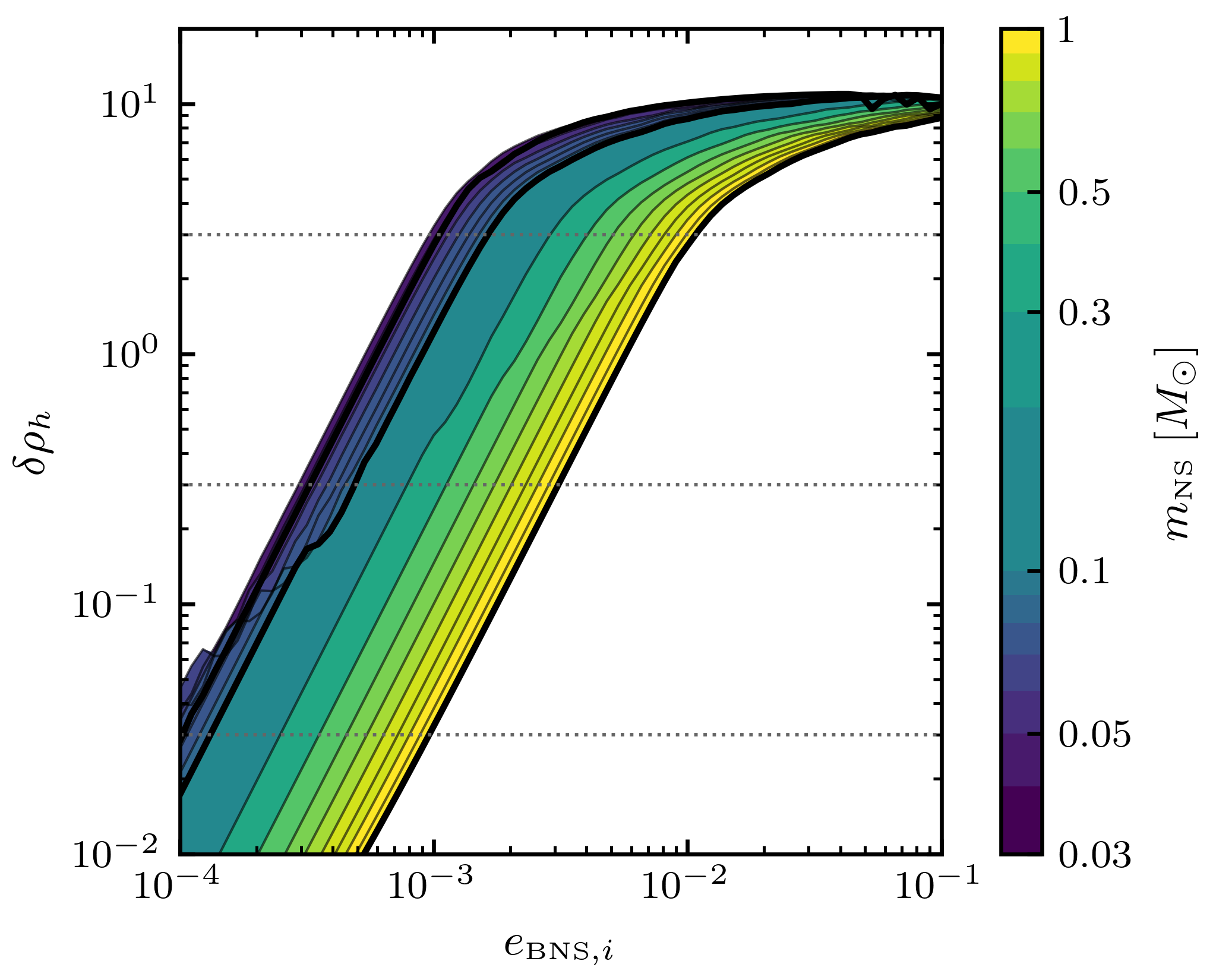}
\caption{Horizon distinguishability $\delta\rho_h$ ($\rho_0 = 8$, LIGO)  between an eccentric inner BNS and a chirp-mass-marginalized circular template, versus the initial inner eccentricity $e_{\BNS,i}$ assuming the polytropic EoS of Eq.~\eqref{eq:Rpoly}. The color scale represents the NS mass $m_\NS$, ranging from $0.03\,\msun$ (dark) to $1\,\msun$ (light). Dotted horizontal lines mark $\delta\rho_h = 3, 0.3, 0.03$. Binaries with lower $m_\NS$ cross the $\delta\rho_h = 3$ threshold at smaller $e_{\BNS,i}$ because the in-band cycle count is larger.}
\label{fig:rho_diff_e_BNS}
\end{figure}

In Fig.~\ref{fig:rho_diff_e_BNS} we show the eccentricity distinguishability in LIGO as a function of $e_{\BNS,i}$. Each curve corresponds to a different NS mass. The threshold crossing $\delta\rho = 3$ (upper dotted horizontal line) moves monotonically with mass, from $e_{\BNS,i}\sim 10^{-3}$ for the lightest fragments to $\sim 10^{-2}$ at $m_\NS = 1\,\msun$~\cite{Lenon:2020oza,Wang:2021qsu,Favata:2021vhw}. In other words, the lighter and longer-lived the BNS, the smaller the eccentricity that can be detected. At $e_{\BNS,i}\sim 10^{-3}$ the lightest subsolar triples already exceed the threshold at the detection horizon, i.e., at the largest luminosity distance at which the source is detectable at all, $\rho_0 = 8$. Since $\delta\rho \propto \rho_0 \propto d_\text{\tiny L}^{-1}$, the horizon value $\delta\rho_h$ of Eq.~\eqref{eq:rho_diff_hor} is the smallest distinguishability attainable by any detectable source, so that a curve reaching $\delta\rho_h > 3$ implies under the adopted SNR scaling that the eccentricity is recoverable for every detectable source, and not only for the closest ones. Eccentric binaries reach contact at pericenter, which increases the initial frequency and reduces $\mathcal{N}^\BNS_\GW$, but for subsolar masses this band-truncation is mild and the eccentric signature dominates across the allowed parameter region. The thresholds are nearly EoS-independent at fixed mass --- the count of Eq.~\eqref{eq:NcycBNS} is controlled by the low-frequency portion of the band --- with two exceptions: NM contributes no systems below its minimum stable mass of $m_\NS \simeq 0.1 \, M_\odot$, and the uncut QM band mildly lowers the thresholds for the lightest binaries.

\subsection{Delayed NS mergers}
\label{sec:delayed}
\noindent
Before turning to the detection prospects of the NS-BH binary merger, we comment on the multi-messenger
context that makes the timing of the subsolar BNS mergers observationally
relevant. 
Collapsars are the leading model for long gamma-ray bursts: accretion onto the
central BH powers a relativistic jet, while disk winds and the jet itself unbind
several solar masses of stellar material at high velocities, producing the
energetic, broad-lined (often hydrogen-poor) supernovae observed in coincidence
with most long gamma-ray bursts~\cite{MacFadyen:1998vz,Woosley:2006fn}. 

In the disk-fragmentation scenario considered here, this picture is enriched by the
subsolar NS mergers occurring within the disk: the neutron-rich material they
eject is rapidly processed into heavy $r$-process elements, which are mixed into
the surrounding supernova ejecta and alter its light curve and spectrum --- a superkilonova~\cite{Metzger:2024ujc,Siegel:2018zxq}.

A long gamma-ray burst or supernova temporally coincident with a subsolar GW trigger would
thus be a smoking-gun signature of the channel, and the recent
S250818k/SN\,2025ulz and S251112cm episodes have already prompted dedicated
searches along these lines~\cite{Kasliwal:2025keb}. 
Whether such a counterpart
should be sought before, during, or after the GW event depends on how the NS
merger time compares with the supernova timescale, which we now estimate (for a related discussion see also Ref.~\cite{Baibhav:2026zuy}.)

The burst itself lasts of order the disk-depletion timescale [see Eq.~\eqref{eq:taccBH}]:
\begin{equation}
    \tau_{\text{\tiny acc}}^\BH \sim 4 \, {\rm s} \left(\frac{\mathbb{H}}{0.3}\right) \left(\frac{M_\BH}{10\,M_\odot}\right) \left(\frac{\dot{M}_\BH}{M_\odot/\rm s}\right)^{-1} \,,
\end{equation}
whereas the accompanying supernova emerges on a far longer timescale,
$\sim 1 - 10\,{\rm days}$. 

A natural question is then how the merger time of
the subsolar BNS compares with this supernova timescale, since their
relative ordering determines whether the GW event would precede, accompany, or
follow the optical transient.

The merger time depends crucially on the initial orbital separation $a_{\BNS,i}$. For wide initial separations, $a_{\BNS,i}\sim r_{\text{\tiny Hill}}$, the inspiral is first driven by dynamical friction on the timescale of Eq.~\eqref{eq:dyntimescaleSSNSs}, which is inversely proportional to the environmental density. For low post-fragmentation densities $f_\text{\tiny gas} \sim 10^{-6}$, this gives
\begin{equation}
    \tau_{\text{\tiny dyn}} \sim 1 \, {\rm day} \,\left(\frac{f_\text{\tiny H}}{0.5}\right)^{-3/2}\left(\frac{f_\text{\tiny gas}}{10^{-6}}\right)^{-1} \left(\frac{\rho(R_*)}{10^{8}\,{\rm g/cm^3}}\right)^{-1/2} \,.
\end{equation}
The decrease of $\tau_\text{\tiny dyn}$ with separation $f_\text{\tiny H}$ reflects the stronger gaseous drag at the lower internal orbital speed of wider binaries.
Once dynamical friction has hardened the orbit sufficiently, the inspiral
becomes GW-driven, with the Peters timescale of
Eq.~\eqref{eq:GWtimescaleSSNSs}. Evaluated at similar fiducial parameters, this reads
\begin{equation}
    \tau_\GW^{\BNS} \sim 1 \, {\rm day} \, \left(\frac{f_\text{\tiny H}}{0.5}\right)^4\left(\frac{m_\NS}{0.1\,M_\odot}\right)^{-5/3}\left(\frac{\rho (R_*)}{10^8 \, {\rm g/cm^3}}\right)^{-4/3}\,,
\end{equation} 
which is independent of $f_\text{\tiny gas}$. For sufficiently wide initial separations and low residual densities --- both still within the collapsar disk parameter space --- the total BNS merger time can reach several days, comparable to or even exceeding the supernova timescale. How much of this time the binary spends emitting GWs will depend on the ratio $\tau_\GW^{\BNS}/\tau_{\text{\tiny dyn}}$. In particular, the higher $f_\text{\tiny gas}$, the later the binary will transition into the GW-dominated phase. A delayed merger in the collapsar scenario could therefore produce a GW-driven chirp of order seconds even if it occurs days after the gamma-ray burst.

A delayed merger could explain the observational picture emerging from current subsolar follow-up campaigns: for a S251112cm-like event, Ref.~\cite{Hall:2026gov} derives a maximum BNS delay of a few days and accordingly searches for stripped-envelope supernovae with explosion times up to $\sim 4\,{\rm days}$ before the GW alert. The same analysis stresses that the  NS-BH merger (Sec.~\ref{sec:NS-BH}) must occur after the BNS coalescence, $\tau_\GW^{\NS-\BH} > \tau^\BNS_{\GW}$ --- otherwise the central BH would tidally disrupt the NS pair before it merges --- with the NS-BH event following within $\sim 10\,{\rm days}$ of the explosion. The prospect of such a multi-day delay motivates searching for an EM counterpart not only after the merger~\cite{Vieira:2026eof}, but also in the days preceding it. 

A qualitatively distinct class of delayed mergers can arise from the
few-body dynamics of the $N_\NS\sim\mathcal{O}(10)$ fragments.
As discussed in Ref.~\cite{Lerner:2025dkd}, close scatterings among the
fragments can eject single NSs, or even bound BNSs, from the disk
altogether. Short of full ejection, such encounters can instead kick the NS onto a wide, eccentric orbit about the central BH, lengthening its subsequent merger time and further broadening the range of possible delays~\cite{Baibhav:2026zuy}}.
An ejected binary subsequently merges in a lower-density
environment than the one considered above, so that its inspiral is essentially
GW-driven from the outset and produces a near-vacuum chirp, free of the
dynamical-friction phase shift of Sec.~\ref{sec:envtidalNSNS} and of the triple
modulations of Sec.~\ref{sec:modeffects}. Such ejected systems would thus merge
on timescales decoupled from the disk dynamics and the supernova, and would lack
the environmental effects otherwise associated with the collapsar channel,
making the vacuum-like waveform itself a feature that differentiates this population. Estimates of  neutrino emission further associated with subsolar remnants in the hierarchical history are given in  Appendix~\ref{sec:postmerger}.

\section{NS-BH merger} 
\label{sec:NS-BH}
\noindent
The second dynamical stage of the collapsar scenario involves the inspiral and coalescence of the subsolar NS remnant -- either a single clump that survived disk fragmentation, or the hierarchical merger product of several such clumps -- with the central stellar-mass BH. As discussed in Sec.~\ref{sec:param_NSBH} and illustrated in the right panel of Fig.~\ref{fig:param}, the outcome depends sensitively on whether the NS reaches the inner disk before or after the surrounding gas has been depleted by BH accretion: in the former case (colored regions) the inspiral proceeds through a gas-rich environment and is migration-assisted, while in the latter (beige region) the binary evolves essentially in vacuum, driven purely by GW emission. Regardless of the environmental conditions during migration, once the NS has decoupled from the disk gas and entered the GW-driven regime the signal is that of a highly asymmetric compact binary inspiral with chirp mass set mainly by the $\sim 3 - 30\,M_\odot$ BH. This makes the NS-BH merger the loudest GW event of the disk-fragmentation scenario and its most likely channel for a first detection. In this section we characterize the detection prospects (Sec.~\ref{Sec:detNSBH}), the measurability of the binary parameters and of the tidal disruption of the NS (Sec.~\ref{sec:FisherII}), and the impact of residual eccentricity (Sec.~\ref{sec:NS-BH_ecc}).

\subsection{Detectability of NS-BH mergers}
\label{Sec:detNSBH}
\begin{figure}[t!] 
    \centering
\includegraphics[width=1\linewidth]{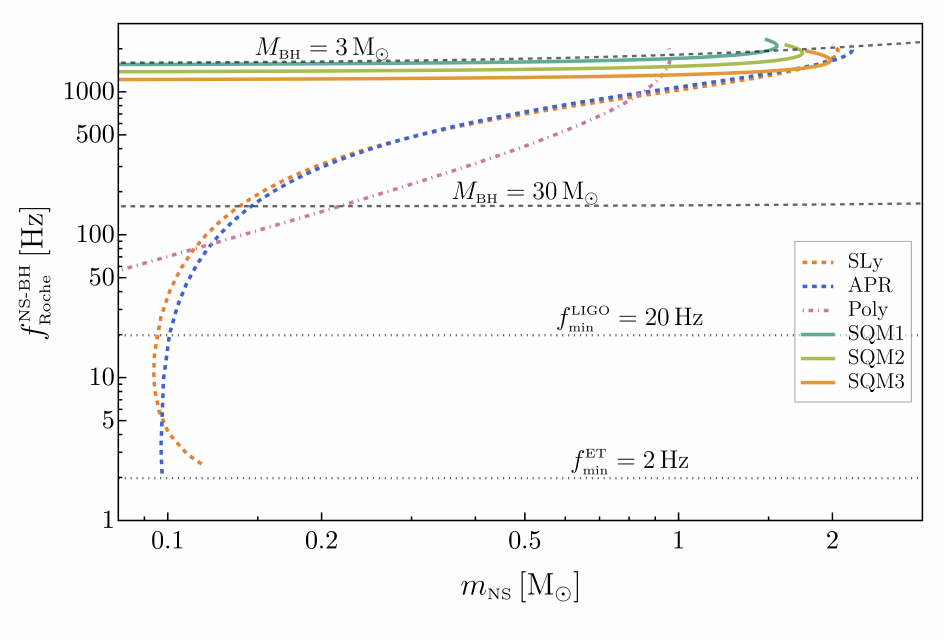}
    \caption{Tidal disruption frequency of a subsolar NS, assembled in a binary with the central BH, as a function of the NS mass, for different EoS (solid and dashed for QM and NM, respectively, dot-dashed for the non-relativistic polytrope with $Y_e = 0$). Dotted black lines indicate the minimum measurable frequency by LIGO and ET. Dashed black lines denote the ISCO frequency for a central BH with mass  $3-30\, M_\odot$, assuming a benchmark spin value of $\chi_\BH = 0.1$.
    For $M_\BH = 3\, M_\odot$ the tidal disruption frequency is mostly below the ISCO, while the opposite trend holds for $M_\BH = 30\, M_\odot$. 
    }
    \label{fig:Rochefrequency}
\end{figure}

\noindent
The NS-BH merger is the loudest GW signal of the disk-fragmentation scenario and hence the most likely to be detected. For the fiducial values of the disk parameters, the NS enters the GW-driven band at the frequency
\begin{align}
f_{\text{\tiny NS-BH},i} & = \frac{1}{\pi} \sqrt{\frac{m_\NS+ M_\BH}{a_{\NS-\BH,i}^3}}  \simeq \frac{1}{\pi} \sqrt{\frac{M_\BH}{R_*^3}} \nonumber \\
& \simeq 26\, {\rm Hz} \, \left(\frac{M_\BH}{5 \, M_\odot}\right)^{-1}
\left(\frac{\mathbb{H}}{0.3}\right)^{-3} \left(\frac{\dot{M}_\BH}{M_\odot/\rm s}\right)\,,
\label{eq:fNSBHi}
\end{align}
which is below the detectors' noise floor for heavy central BHs ($M_\BH \gtrsim 7\,\msun$ for the LIGO floor at the fiducial disk parameters) and thick disks, in which case the in-band signal begins at the detector cutoff $f_\text{\tiny min}$. The signal ends when the central BH tidally disrupts the NS at the Roche frequency $f_\text{\tiny Roche}^\text{\tiny NS-BH}$, given that unbinding is not effective in the NS-BH merger [$f_\text{\tiny TD} = f_\text{\tiny Roche}^\text{\tiny NS-BH}$, see Eq.~\eqref{eq:fTD}], which can be seen by setting $m_\text{\tiny comp} = M_\BH$ in Eq.~\eqref{eq:fRoche}. The $M_\BH$ dependence cancels except through the mass ratio $q = m_\NS/M_\BH$, leaving the EoS-agnostic form
\begin{align}
    f_\text{\tiny Roche}^\text{\tiny NS-BH} &=
    \frac{1}{\pi}\sqrt{\frac{m_\NS (1+q)}{8 R_\NS^3}}\nonumber\\
    &\simeq 65\,{\rm Hz}\,\sqrt{1+q}\left(\frac{m_\NS}{0.1\,\msun}\right)^{1/2}
    \left(\frac{R_\NS}{34\,{\rm km}}\right)^{-3/2},
    \label{eq:fRocheNSBH_gen}
\end{align}
which, for $q \ll 1$ and the mass-radius relations of Eq.~\eqref{eq:R_NS}, evaluates to
\begin{align}
{\rm Poly}:& \quad f_{\text{\tiny Roche}}^\text{\tiny NS-BH}\simeq 70\left(\frac{m_\NS}{0.1\,\msun}\right){\rm Hz}\,, \nonumber \\
{\rm NM}:&  \quad f_{\text{\tiny Roche}}^\text{\tiny NS-BH}\simeq 0.35\,{\rm kHz}\left(\frac{m_\NS}{0.1\,\msun}\right)^{1/2} \nonumber \\
& \qquad \quad \,\,\,\,\, \times \left[1+3\left(\frac{m_\NS}{0.1\,\msun}\right)^{-3}\right]^{-3/2}\,, \nonumber \\
{\rm QM}:& \quad f_{\text{\tiny Roche}}^\text{\tiny NS-BH} \simeq 1.6\,{\rm kHz}\,.
\label{eq:fNSBHRoche}
\end{align}
In Fig.~\ref{fig:Rochefrequency} we plot the Roche frequency as a function of the NS mass for different EoSs. As shown in Eq.~\eqref{eq:fNSBHRoche}, there is no dependence on the BH mass, while the NS mass dependence is determined by the EoS: for NM (dashed lines), the Roche frequency rises almost linearly with NS mass, dropping to a few~Hz for the lightest NS, $m_\NS\sim 0.1\,\msun$, and approaching the detector cutoff $f_\text{\tiny min}$ (horizontal dotted black lines); for QM (solid lines), the Roche frequency is roughly constant (with a value around the kHz) and independent of the NS mass. The polytrope (dot-dashed line), adopted here as our benchmark because it is defined over the full mass interval used here (see Sec.~\ref{sec:param_NSNS}), rises linearly with $m_\NS$ and interpolates between the two realistic families, crossing the LIGO floor near $m_\NS \simeq 0.03\,\msun$ and reaching a few thousand Hz at $m_\NS \simeq 1\,\msun$. 

In Fig.~\ref{fig:nsbh_bns_horizon} we show the horizon for circular NS-BH binaries (orange) as a function of the subsolar NS mass $m_\NS$ for LIGO, CE and ET. Since the NS mass range extends down to $m_\NS = 0.01\,\msun$, well below the minimum stable NM mass, we adopt here the polytropic mass-radius relation of Eq.~\eqref{eq:Rpoly}, which is defined across the whole range. The EoS enters only through the terminating Roche frequency, and the NM/QM alternatives are recovered by rescaling $f_\text{\tiny Roche}^\text{\tiny NS-BH}$ according to Eq.~\eqref{eq:fNSBHRoche} (see Appendix~\ref{app:EoS} for a dedicated discussion of detection prospects for the NS-BH channel considering the NM/QM EoSs). The shaded band spans BH masses in the range $M_\BH\in[3,\,30]\,\msun$, with the $M_\BH =3,\,10,\,30\,\msun$ cases shown explicitly with dotted, dashed and solid lines, respectively. Across the entire subsolar range the NS-BH horizon lies a factor $\sim 3 - 10$ above the BNS horizon, a gap set by the chirp mass through $d_\text{\tiny hor}\propto\mathcal{M}_c^{5/6}$ [Eq.~\eqref{eq:rho0}]. The NS-BH value is BH-dominated, $\mathcal{M}_c\simeq m_\NS^{3/5}M_\BH^{2/5}$, and exceeds the inner-BNS value $2^{-1/5}m_\NS$ by a factor $\sim 4 - 11$.
Both horizons increase with $m_\NS$ faster than the amplitude-only scaling ($d_\text{\tiny hor}\propto m_\NS^{1/2}$ for NS-BH, $m_\NS^{5/6}$ for the inner BNS), because the in-band window also widens with mass. The terminating frequency rises, since the polytropic radius shrinks as $R_\NS\propto m_\NS^{-1/3}$ [Eq.~\eqref{eq:Rpoly}], so that $f_\text{\tiny Roche}\propto (m_\NS/R_\NS^3)^{1/2}$ grows linearly with the NS mass, while the entry frequency stays essentially fixed. The lighter-is-larger trend is specific to the degenerate polytrope and is shared by NM but not by QM, whose nearly constant density leaves $f_\text{\tiny Roche}$ mass-independent.
The widening band increases the in-band SNR beyond the amplitude-only scaling; the effect is strongest where the band is otherwise reduced by the detector floor (LIGO at low mass).

Heavier BHs produce signals with larger SNR, so the $M_\BH=30\,\msun$ binary marks the upper edge of the horizon band, and the $3\,\msun$ binary its lower edge.
The inspiral ends at the minimum between the Roche radius and the BH ISCO. The Roche frequency terminates the signal over almost the whole plane; only in the large-$m_\NS$, large-$M_\BH$ region does $f_\text{\tiny ISCO}$ fall below $f_\text{\tiny Roche}$ and set the upper-frequency limit, reducing the upper edge of the orange band without reversing its ordering. This trend is opposite to the BNS horizon, since a heavier central BH tidally truncates the inner pair earlier, due to unbinding, lowering the ending frequency and shortening the in-band signal. The purple ordering therefore reverses, with $M_\BH=3\,\msun$ producing the largest horizon, and the band exists only where this truncation is active ($0.02\lesssim m_\NS\lesssim0.21\,\msun$); 
above $m_\NS\approx0.2\,\msun$ every mass reaches contact, the ending frequency becomes $M_\BH$-independent, and the band collapses into a single line.
  
Among the future detectors, CE has the largest horizon for $m_\NS\gtrsim0.03\,\msun$, because its noise is lowest in the tens-of-Hz band where these light binaries radiate. ET's lower cutoff is relevant for the lightest systems ($m_\NS\lesssim0.02\,\msun$), whose Roche frequency drops below the CE and LIGO cutoffs, allowing them to remain in band only for ET. In CE and ET the horizons are cosmological ($z\gtrsim1$, reaching larger values at the high-mass end): the $(1+z)$ chirp-mass increase is partly offset by the nonlinear distance--redshift relation, and the NS-BH horizon approaches this cosmological saturation before the BNS horizon, reducing the NS-BH/BNS horizon ratio toward unity.

\begin{figure*}[h!] 
    \centering
\includegraphics[width=0.92\linewidth]{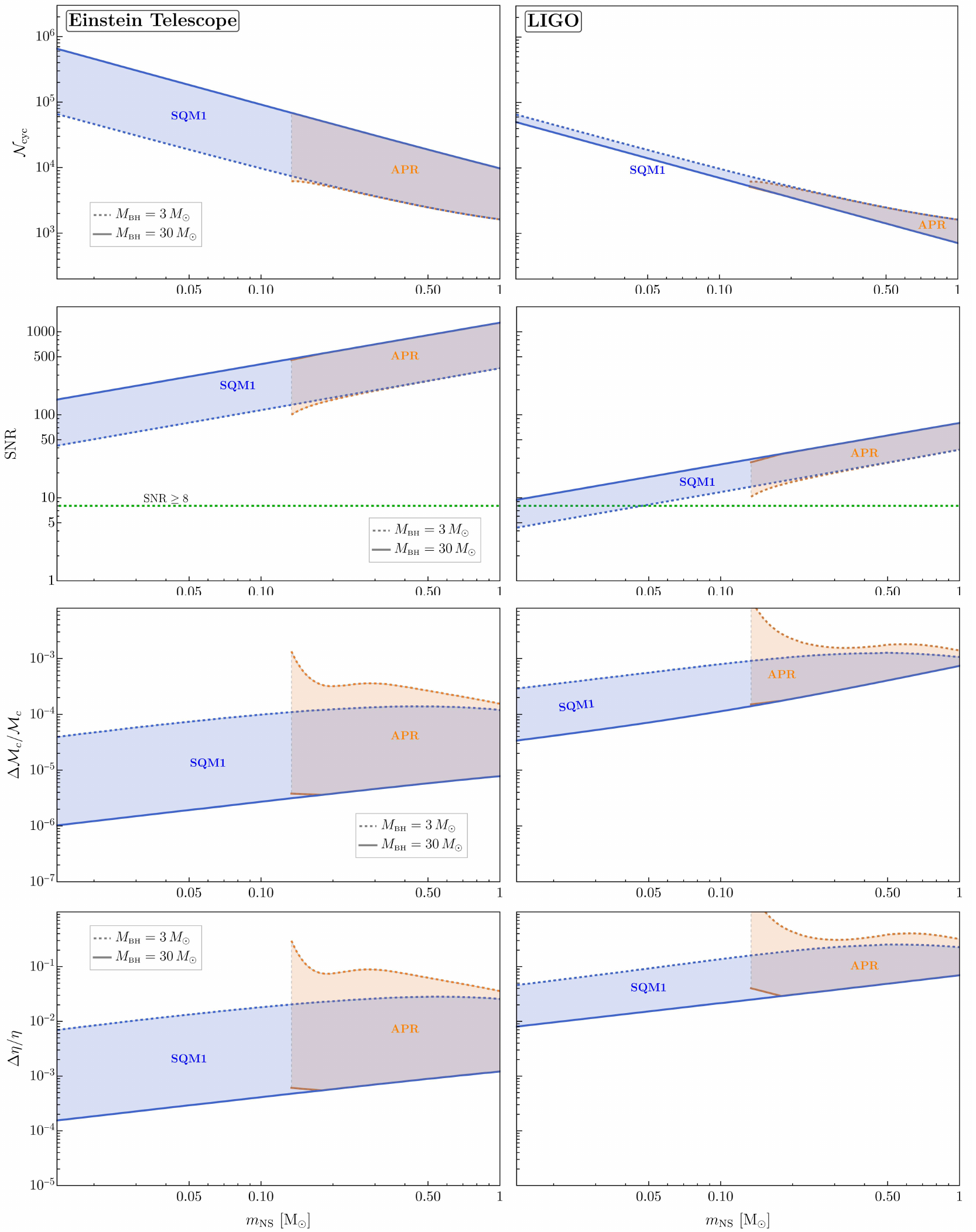}%
    \caption{Number of cycles and SNR of the NS-BH binary, as functions of the NS mass $m_\NS$, together with the relative errors on the chirp mass $\Delta\mathcal{M}_c/\mathcal{M}_c$ and on the symmetric mass ratio $\Delta\eta/\eta$, computed via a Fisher-matrix analysis for ET (left panels) and LIGO (right panels) at the benchmark luminosity distance $d_\text{\tiny L} = 100\,{\rm Mpc}$. The blue (SQM1) and orange (APR) shaded bands correspond to the QM and NM EoS, respectively. Solid and dashed boundaries bracket the range of BH masses $M_\BH = 3 - 30\,M_\odot$.}
\label{fig:measurabilityNSBH1}
\end{figure*}

\subsection{Measurability of model parameters}
\label{sec:FisherII}
\noindent
We now assess the measurability of the NS-BH binary parameters with current and future GW interferometers using the Fisher matrix formalism of Sec.~\ref{sec:waveform:tf2}, in analogy to what was done in Sec.~\ref{sec:FisherI} for the BNS case. We assume that the binary evolves in vacuum, i.e., we restrict ourselves to the beige region shown in the right panel of Fig.~\ref{fig:param}. This ensures that the binary is predominantly losing energy via GW  emission rather than dissipating energy into the environment. We consider NSs of maximum mass $m_\NS \sim 1 M_\odot$. This ensures that the highest mass ratio we encounter is $m_\NS / M_\BH \sim 1/3$, corresponding to the lightest BH compatible with the collapsar scenario. As shown in Fig.~\ref{fig:param}, even at the highest mass ratio, it is always consistent to assume vacuum domination, provided that the disk aspect ratio $\mathbb{H}$ is sufficiently large or the density suppression factor $f_\text{\tiny gas}$ is sufficiently small.

We perform a Fisher matrix analysis with ten parameters $\bm{\theta} = (\mathcal{M}_c, \eta, \chi_s, \chi_a, \tilde{\Lambda}, \delta \tilde{\Lambda}, f_{\text{\tiny TD}}, t_c, \varphi_c, d_\text{\tiny L})$, assuming the symmetric and antisymmetric spin combinations $\chi_s = 0.15$ and $\chi_a = 0.05$, respectively. Contrarily to the analysis performed in Sec.~\ref{sec:FisherI}, here we consider the tidal disruption frequency $f_{\text{\tiny TD}}$ as a model parameter, given its important role in setting the upper limit for the integration in frequency of the inspiral evolution. 

We show our results in Fig.~\ref{fig:measurabilityNSBH1}, which closely resembles  Fig.~\ref{fig:measurabilityNSNS1} for the BNS case. The number of cycles (top row) is generally higher in ET than in LIGO. In contrast to the BNS case, the initial frequency of the NS-BH merger given in Eq.~\eqref{eq:fNSBHi} is comparable to or below the LIGO floor (at $\sim \, 20 \, {\rm Hz}$), while it lies within the ET band. Therefore, the in-band number of cycles is reduced in LIGO, especially for higher BH masses ($M_\BH = 30 \, {\rm M_\odot}$), as $f_{\text{\tiny NS-BH},i} \propto 1/M_\BH$. This observation explains why the relative hierarchy of the dashed and solid curves in the top row of Fig.~\ref{fig:measurabilityNSBH1} is inverted in LIGO with respect to ET. Concerning the SNR, the NS-BH merger event is resolved in LIGO down to NS masses $\simeq 0.01 \, M_\odot$ (see the second row of Fig.~\ref{fig:measurabilityNSBH1}). This differs from the BNS case because of the higher chirp mass of the NS-BH binary.
Regarding the measurability of the model parameters, the chirp mass (third row) has a small relative uncertainty in both detectors, while the symmetric mass ratio $\eta$ (fourth row) is
recovered with relative errors that differ by about two orders of magnitude
between the two detectors: ET reaches $\Delta \eta/\eta \sim 10^{-4} - 10^{-2}$, whereas LIGO attains $\sim 10^{-2}$ to a few $\times 10^{-1}$. In both
cases the relative error remains below unity across the evaluated
subsolar range, so that $\eta$ is measurable --- albeit considerably less
precisely --- already with the current generation of detectors. The error grows
with the NS mass, since a larger chirp mass shortens the inspiral and reduces
the in-band cycle count. Superimposed on this trend, the uncertainties for the NM curves increase
at their low-mass end: the Roche frequency $f_\text{\tiny TD}$ for
standard NM decreases significantly for masses $m_\NS \lesssim 0.2 \, M_\odot$,
thus leading to a shorter signal duration and a loss of constraining power. This
effect is strong enough that $\Delta \eta/\eta$ in LIGO increases to a few $\times
10^{-1}$ for the lightest NM configurations. For the opposite reason, the
symmetric mass ratio of a binary hosting a quark star has a relative uncertainty below unity across the evaluated range, including
in LIGO. Across this parameter space, the
measurement of the symmetric mass ratio allows one to distinguish the NS-BH binary from
the BNS case, where the expectation is $\eta \simeq 1/4$: even in the configurations with the largest uncertainties, the absolute uncertainty $\Delta \eta$ remains orders of
magnitude below the separation between the NS-BH values $\eta \approx q \ll 1$
and the equal-mass one. Moreover, Fig.~\ref{fig:measurabilityNSBH1} shows that
the considerations about the distinguishability between standard NM and QM apply
also to the NS-BH binary. In particular, the parameter space for $m_\NS \lesssim
0.1 \, M_\odot$ is viable only if the NS contains QM, as signaled by the absence
of the NM bands in that region. We will discuss more concretely how to
distinguish different EoS families through tidal deformability measurements below.

\begin{figure}[t!] 
    \centering
\includegraphics[width=1\linewidth]{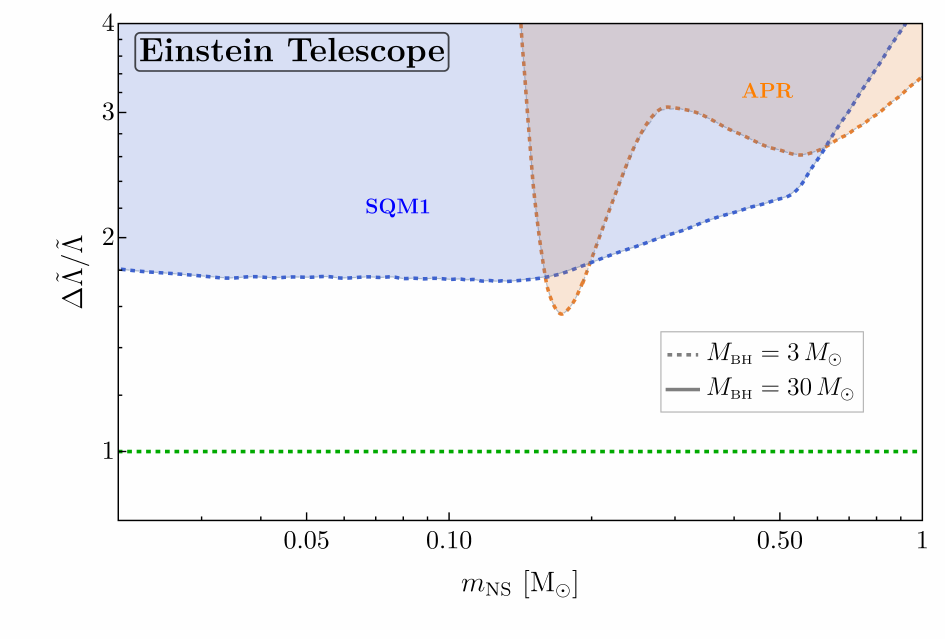}
    \caption{Same as Fig.\ref{fig:measurabilityNSBH1} for the relative uncertainty on the  tidal deformability $\Delta\tilde{\Lambda}/\tilde{\Lambda}$ 
of the NS-BH binary merger in ET. The steep $q^4$
suppression of $\tilde\Lambda_\text{\tiny NS-BH}$ [Eq.~\eqref{eq:Ltilde_NSBH}]
makes the tidal deformability largely unmeasurable through the inspiral
phase evolution across most of the parameter space.}
    \label{fig:measurabilityNSBH2}
\end{figure}

\begin{figure*}
    \centering
    \includegraphics[width=1\linewidth]{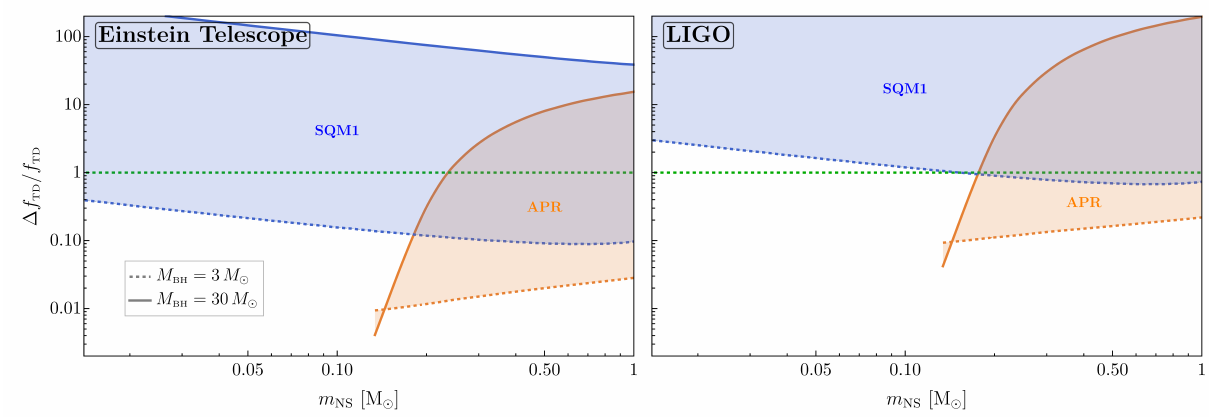}
    \caption{Measurability of the tidal disruption frequency with ET (left panel) and LIGO (right panel), for different values of the central BH mass. The color code is the same as in Figs.~\ref{fig:measurabilityNSBH1} and \ref{fig:measurabilityNSBH2}.}
    \label{fig:vacuuminner}
\end{figure*}

The large mass hierarchy $m_\NS \ll M_\BH$ limits the
measurability of the NS tidal structure through the inspiral waveform. The
tidal phase correction of Eq.~\eqref{eq:psi_tidal}, introduced in
Sec.~\ref{sec:waveform:tf2} for the symmetric BNS case, generalizes to the
NS-BH binary with $m_1 = M_\BH$, $\Lambda_1 = 0$ (the BH has no tidal
deformability~\cite{Binnington:2009bb,Damour:2009va,Hui:2020xxx}) and $m_2 = m_\NS$, $\Lambda_2 = \Lambda$.  The effective tidal
deformability of Eq.~\eqref{eq:Lambda_tilde} then reduces to
\begin{align}
\label{eq:Ltilde_NSBH}
    \tilde\Lambda_\text{\tiny NS-BH}
    & = \frac{8}{13}
\left[ \big(1+7\eta-31\eta^2\big) \right. 
\nonumber \\
& \left. -
\sqrt{1-4\eta}\big(1+9\eta-11\eta^2\big) \right]  \Lambda
    \,,
\end{align}
which, in the extreme mass-ratio limit $\eta \approx q \equiv m_\NS/M_\BH \ll 1$,
scales as $\tilde\Lambda_\text{\tiny NS-BH} \propto q^4\,\Lambda$.  The $q^4$
suppression --- a direct consequence of the mass-ratio prefactors in the
$5$\,PN tidal phase --- reduces the tidal imprint on the
waveform: even though subsolar NSs possess large individual
deformabilities $\Lambda \sim \mathcal{O}(10^6 - 10^9)$ (see Fig.~\ref{fig:mr-TLN}), the
combination $\tilde\Lambda_\text{\tiny NS-BH}$ is suppressed by four powers
of the small ratio $q \sim m_\NS/M_\BH \sim 10^{-2} - 10^{-1}$,
leaving a small phase accumulation over the in-band inspiral.  As a
consequence, the standard phase-evolution route to measuring the TLN ---
which works for the symmetric BNS stage --- is largely
ineffective here, and measuring $\Lambda$ through the inspiral alone requires
a sufficiently high SNR or a less asymmetric binary.  In Fig.~\ref{fig:measurabilityNSBH2} we quantify this statement: the
relative error $\Delta\tilde{\Lambda}/\tilde{\Lambda}$ exceeds unity across most of the
parameter space accessible to ET. Precisely, $\Delta\tilde{\Lambda}/\tilde{\Lambda} \simeq 2$ when $m_\NS \lesssim 0.5 \, {\rm M_\odot}$ and $M_\BH = 3 \, {\rm M_\odot}$, while higher BH masses, corresponding to a smaller mass ratio, lead to an even more suppressed $\tilde{\Lambda}$ and a relative error $\gtrsim 10$ (not shown in the figure).

The finite-size observable considered here in the merger stage with the central BH is therefore not the phase evolution, but the tidal disruption of the NS itself~\cite{Vallisneri:1999nq,Pannarale:2011pk,Pannarale:2015jia}.  As the orbital separation shrinks, the gravitational tidal field of the BH eventually overcomes the NS self-gravity at the Roche radius $d_\text{\tiny Roche}$ defined in Eq.~\eqref{Rocheradius}, where the companion mass is now $m_\text{\tiny comp} = M_\BH$.  At that point the NS is tidally disrupted and the inspiral terminates, so that $f_\text{\tiny TD}$ plays the role of a physical upper frequency cutoff rather than a phase correction.  
The Roche limit thus
results in a spectral cutoff whose frequency depends on the NS radius, and hence on the EoS. In Fig.~\ref{fig:vacuuminner} we quantify the measurability of $f_\text{\tiny TD}$
via a Fisher analysis. For $M_\BH = 3 \, {\rm M_\odot}$ and for the APR EoS, both detectors measure this frequency with a relative accuracy $\Delta f_\text{\tiny
TD}/f_\text{\tiny TD}$ of about $1\%$ (ET) and $10\%$ (LIGO). This is expected, as the larger radii characterizing NM lead to smaller values of $f_\text{\tiny TD}$ (which would otherwise lie closer to the corresponding ISCO frequency), as shown in Fig.~\ref{fig:Rochefrequency}. For an analogous reason, $f_\text{\tiny
TD}$ is measured less accurately for QM, with only ET retaining sensitivity in the parameter range shown. For heavier BH masses ($M_\BH = 30 \, {\rm M_\odot}$), the measurability is reduced, because the ISCO frequency in this regime is generally smaller than $f_\text{\tiny TD}$ (so that the NS is not tidally disrupted before merger). The only case where some sensitivity remains in both detectors is the APR EoS with masses $m_\NS \lesssim 0.2 \, {\rm M_\odot}$, where $f_\text{\tiny TD}$ falls below the ISCO frequency again (see Fig.~\ref{fig:Rochefrequency}). The two EoS families predict sufficiently different values of $f_\text{\tiny
TD}$ at fixed mass that a confident measurement of the cutoff frequency could
discriminate between NM and QM, complementing the
similar discriminant available from the BNS stage discussed in Sec.~\ref{sec:NS-NS}. The Fisher analyses of this section and of Sec.~\ref{sec:FisherI} adopt the NM and QM EoSs; the corresponding measurability of $\tilde\Lambda$, $\rho_\text{\tiny env}$, and $f_\text{\tiny TD}$ for the polytropic benchmark of Eq.~\eqref{eq:Rpoly}, which spans the full subsolar range, is collected in Appendix~\ref{app:EoS_meas}.

\begin{figure}[t!]
\centering
\includegraphics[width=0.95\columnwidth]{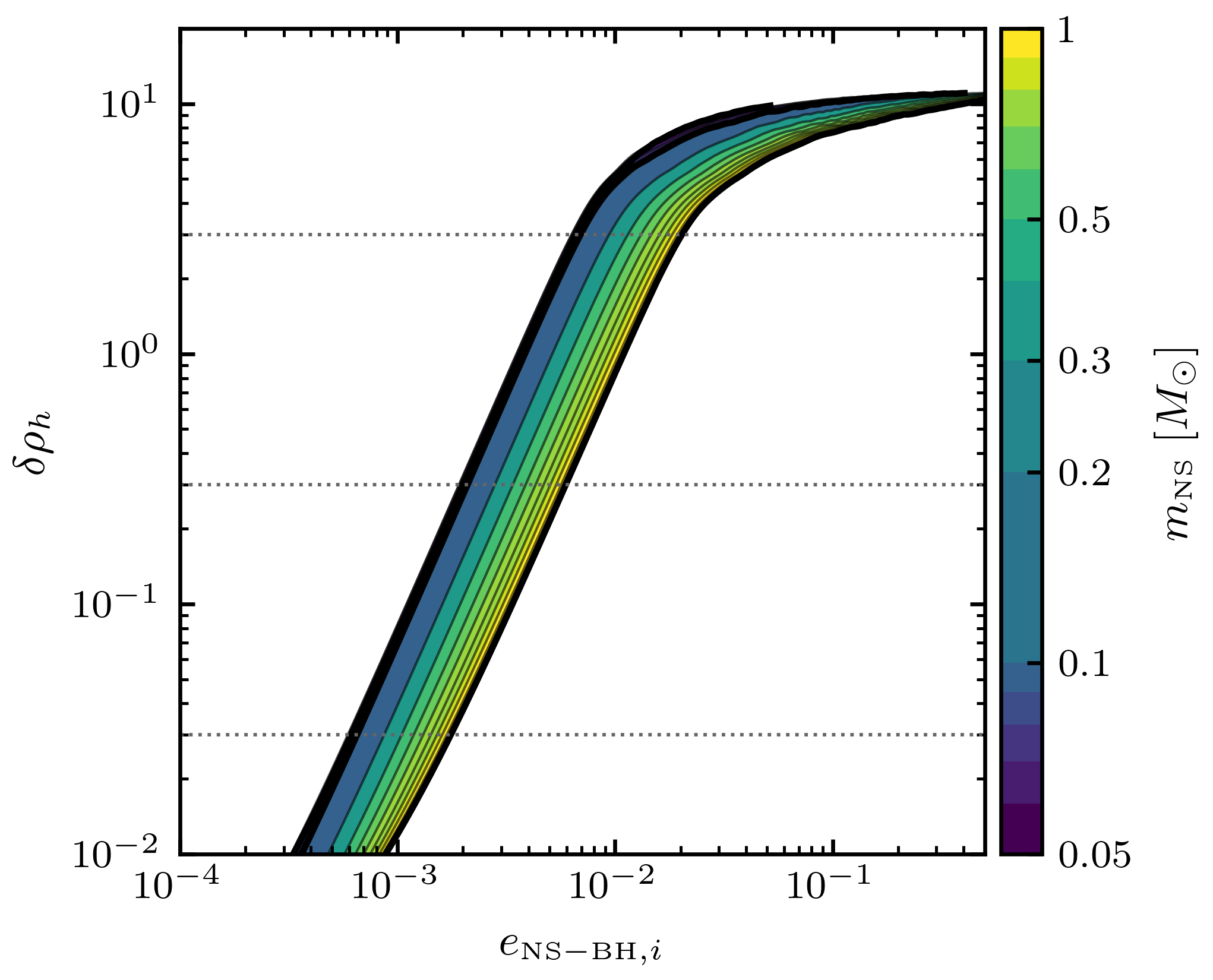}
\caption{NS-BH horizon distinguishability $\delta\rho_h$ ($\rho_0 = 8$,
LIGO) between an eccentric template and a chirp-mass-marginalized circular
reference, versus the initial eccentricity $e_{\NS-\BH,i}$. We assume $M_\BH = 5\,\msun$ and the polytropic EoS of Eq.~\ref{eq:Rpoly}. Colors correspond to values of
$m_\NS$ from $0.05\,\msun$ (dark) to $1\,\msun$ (light). The horizontal dotted lines mark
$\delta\rho_h = 3, 0.3, 0.03$. See Fig.~\ref{fig:rho_diff_e_BNS} for a comparison with the BNS merger case.}
\label{fig:rho_diff_e_NSBH}
\end{figure}

\subsection{Eccentricity effects}
\label{sec:NS-BH_ecc}
\noindent
Similarly to the BNS case, the NS clumps sourced during disk fragmentation or resulting from a NS merger hierarchy may assemble with the central BH in a binary with non-zero initial eccentricity $e_{\NS-\BH,i}$, possibly changing the features of the associated GW signal (see Ref.~\cite{Wu:2026hth} for a detailed study). The detectability of eccentricity rests on the secular phase advance it imprints on the waveform, $\delta\psi_\text{\tiny ecc}\sim e_{\NS-\BH,i}^{2}\,\mathcal{N}_\GW^{\NS-\BH}$: an eccentric signal is distinguishable from the best-fitting circular template ($\delta\rho_h>3$) only once this advance is large enough~\cite{Martel:1999tm,OShea:2021ugg,Romero-Shaw:2022fbf}. The key quantity is therefore the in-band cycle count $\mathcal{N}_\GW^{\NS-\BH}$, and this is where the NS-BH differs from the inner BNS. Its chirp mass, $\mathcal{M}_c=(m_\NS M_\BH)^{3/5}/(m_\NS+M_\BH)^{1/5}$, is far larger than the chirp mass $\mathcal{M}_c=2^{-1/5}m_\NS$ of an equal-mass subsolar BNS; and because $\mathcal{N}_\GW\propto\mathcal{M}_c^{-5/3}$, the heavier NS-BH sweeps a much shorter inspiral --- at $(m_\NS,M_\BH)=(0.1,5)\,\msun$ it accumulates roughly an order of magnitude fewer in-band cycles. A shorter inspiral builds up less phase from a given eccentricity, so a larger $e$ is needed to reach the threshold. Quantitatively, holding $\delta\psi_\text{\tiny ecc}\sim e_{\NS-\BH,i}^{2} \, \mathcal{N}_\GW^{\NS-\BH}$ at the fixed value that marks $\delta\rho_h=3$ gives a threshold eccentricity
\begin{equation}
e_\text{\tiny thr}\propto\frac{1}{\sqrt{\mathcal{N}_\GW^{\NS-\BH}}}\propto\mathcal{M}_c^{5/6}\,,
\label{eq:e_thr}
\end{equation}
which rises with the chirp mass and hence with both $m_\NS$ and $M_\BH$. Figure~\ref{fig:rho_diff_e_NSBH} confirms this trend: at $M_\BH=5\,\msun$, the $\delta\rho_h=3$ crossing moves from $e_{\NS-\BH,i}\approx7\times10^{-3}$ for the lightest NS shown ($m_\NS\sim0.05\,\msun$) to $\approx2\times10^{-2}$ at $m_\NS=1\,\msun$ --- a factor $\sim3$, consistent with $e_\text{\tiny thr}\propto m_\NS^{1/2}$ at fixed $M_\BH$ (where $\mathcal{M}_c\propto m_\NS^{3/5}$) --- and the threshold increases further with $M_\BH$. At fixed SNR, the NS-BH eccentricity threshold is approximately one order of magnitude larger than that of the inner-BNS triple (see Fig.~\ref{fig:rho_diff_e_BNS} for a comparison).

This lower per-source sensitivity is partly offset by the larger NS-BH detection horizon. The distinguishability scales as $\delta\rho_h\simeq\rho_0\sqrt{2\mathcal{M}}\sim\rho_0\,e_{\NS-\BH,i}^{2} \, \mathcal{N}_\GW^{\NS-\BH}$ so, at a fixed source distance --- where $\rho_0$ is proportional to the horizon depth --- the NS-BH's three- to ten-fold larger horizon (Sec.~\ref{Sec:detNSBH}) increases $\rho_0$ by the same factor. This increase in SNR partly offsets the roughly order-of-magnitude smaller cycle count, so, at the same distance, a NS-BH binary eccentricity is constrained about as well as a BNS triple's, and more tightly for CE and ET, where the horizon difference is largest. The thresholds quoted above are the values at the NS-BH horizon itself ($\rho_0=8$); a closer source has a smaller uncertainty.

Eccentricity is also a direct environmental discriminant for this channel: being a single binary, the NS-BH carries none of the triple modulations of Sec.~\ref{sec:modeffects} (Doppler, aberration, gravitational redshift, Shapiro, and self-lensing). Because field-formed NS-BH binaries circularize long before merger~\cite{Peters:1964zz,Kowalska:2010qg}, a subsolar NS-BH detection with a measurable $e_{\NS-\BH,i}\gtrsim10^{-2}$~\cite{Lenon:2020oza,Dhurkunde:2023qoe,Morras:2025xfu,Planas:2025plq,Kacanja:2025kpr} would favor an in-disk assembly interpretation, in line with the numerical relativity study of hierarchical fragment mergers of Ref.~\cite{Wu:2026hth}, where eccentricities as large as $e_{\NS-\BH,i}\simeq 0.1$ survive until merger. Dynamical assembly in dense stellar clusters can likewise deliver eccentric binaries to the detector band~\cite{Samsing:2017xmd,Rodriguez:2018pss,Romero-Shaw:2021ual,Zevin:2021rtf,Romero-Shaw:2022xko}, but this channel is unavailable to the objects considered here, since subsolar NSs have no formation pathway outside the collapsar disk.

\section{Interplay with PBHs}
\label{sec:PBHs}
\noindent
\noindent
A confirmed subsolar detection is frequently regarded as one of the most direct signposts of PBHs, i.e., BHs formed in the radiation-dominated era from the gravitational collapse of large primordial density perturbations, well before any star formation~\cite{Zeldovich:1967lct,Hawking:1971ei,Carr:1974nx,Carr:1975qj,Sasaki:2018dmp,Carr:2020gox,Green:2020jor,Byrnes:2025tji,LISACosmologyWorkingGroup:2023njw}. In the GW-accessible window $M_\PBH \lesssim \msun$, these subsolar sources could leave a detectable imprint in current and future searches~\cite{Prunier:2023uoo,Markin:2023fxx,Yuan:2024yyo,Miller:2020kmv,Miller:2021knj,Pujolas:2021yaw,DeLuca:2021hde,Franciolini:2023opt,Miller:2024fpo,Miller:2024jpo,Andres-Carcasona:2024wqk,LIGOScientificCollaborationtheVirgoCollaboration:2025cwh,Magaraggia:2026jhk}, while accounting for a percent-level fraction of the dark matter. For this reason, evidence for a subsolar event is often taken as potential evidence for the existence of PBHs. Because their waveforms are essentially indistinguishable from those of point-mass binaries evolving in vacuum, however, a PBH interpretation is degenerate with any compact object of astrophysical origin sharing the same masses, unless additional physical imprints --- tidal deformability, environmental effects, or an EM counterpart --- break the degeneracy~\cite{Cardoso:2017cfl,Cardoso:2019rvt,Cardoso:2019upw,Barsanti:2021ydd,Franciolini:2021xbq,Coogan:2021uqv,Cole:2022fir,Crescimbeni:2024cwh,Crescimbeni:2024qrq,DeLuca:2024uju,Roy:2024rhe,Russo:2025ivk,DeLuca:2025bph,Begnoni:2025aqc,Corman:2026lbt}. PBHs and the disk-formed subsolar NSs studied in this work are arguably the two most compelling candidates for a subsolar trigger, so disentangling them is central to the interpretation of any such event.

This interpretive landscape is summarized in the flowchart of Fig.~\ref{fig:chart}. A subsolar detection admits two broad classes of explanation: a \emph{new-physics} origin --- either a PBH binary, or a more exotic compact object with nonvanishing tidal deformability but (possibly) no in-band tidal disruption --- or an \emph{astrophysical} one, realized here by the collapsar channel. The two branches share the same anomalously low masses and are therefore separated not by mass, but by the waveform imprints listed in the previous sections. The collapsar channel produces subsolar NSs, $m_\NS \lesssim \msun$, orbiting a central BH of $M_\BH \gtrsim \mathcal{O}(5)\,\msun$, and populates the chart with two sequential merger stages: (i) mergers among the disk-formed clumps, potentially accompanied by an EM counterpart and affected by the triple geometry and residual gas (Sec.~\ref{sec:NS-NS}); and (ii) the merger of the surviving NS remnant (or remnants) with the central BH, whose EoS-dependent tidal disruption further separates NM from QM (Sec.~\ref{sec:NS-BH}). One of the central goals of this work is to provide an exhaustive list of the signatures that would point to a subsolar trigger belonging to the astrophysical branch --- and, within it, identifying which stage produced it. The remainder of this section compares the collapsar and PBH hypotheses directly through a mismatch analysis.

\begin{table}[t!]
\caption{Summary of the waveform models entering the mismatch analysis of Sec.~\ref{sec:PBHs} and of the parameters injected in the collapsar signals $h_{\text{\tiny col}}$, for the two merger stages, compared against a vacuum PBH template $h_{\text{\tiny PBH}}$ sharing the same $\mathcal{M}_c$ and $\eta$. }
\label{tab:mismatch}
\begin{ruledtabular}
\begin{tabular}{lccc}
 & $h_{\text{\tiny col}}^{\BNS}$ & $h_{\text{\tiny col}}^{\NS-\BH}$ & $h_{\text{\tiny PBH}}$ \\
\hline
\multicolumn{4}{l}{\textit{Waveform model}} \\
Tidal phase ($5+6\,${\rm PN})   & $\tilde\Lambda_\BNS$ & $\tilde\Lambda_\text{\tiny NS-BH}$ & $0$ \\
Dyn. friction ($-5.5\,${\rm PN}) & \checkmark & --- & --- \\
Tapering $S_\text{\tiny TD}$      & \checkmark & \checkmark & --- \\
Inspiral cutoff $f_\text{\tiny TD}$ & $f_\text{\tiny Roche}^\BNS$ & $f_\text{\tiny Roche}^\text{\tiny NS-BH}$ & $f_\text{\tiny ISCO}$ \\
\hline
\multicolumn{4}{l}{\textit{Injected parameters}} \\
Primary spin, $\chi_1$    & $ 0.1$ & $0.2$ & $0$ \\
Secondary spin, $\chi_2$    & $ 0.1$ & $0.1$ & $0$ \\
Mass ratio $q$                   & $1$ & $m_\NS/M_\BH$ & matched \\
$M_\BH\,[\msun]$                 & $5$ & $\{3,\,30\}$ & --- \\
$f_\text{\tiny gas}$             & $\{10^{-6},\,10^{-2}\}$ & --- & --- \\
\end{tabular}
\end{ruledtabular}
\end{table}

Following the formalism of Sec.~\ref{sec:waveform:tf2}, we denote by $h_{\text{\tiny col}}$ the waveform of a binary originating in the collapsar disk, and by $h_{\text{\tiny PBH}}$ that of a PBH binary in vacuum sharing the same chirp mass and symmetric mass ratio. We quantify their distinguishability through
\begin{align}\label{eq:rhoDiff}
\rho_{\text{\tiny Diff}}& \equiv \lVert h_{\text{\tiny col}} - h_{\text{\tiny PBH}} \rVert^{1/2}
= \Big[ \lVert h_{\text{\tiny col}} \rVert^2+\lVert h_{\text{\tiny PBH}} \rVert^2 \nonumber\\
&- 2 \lVert h_{\text{\tiny col}}\rVert \,\lVert h_{\text{\tiny PBH}}\rVert \,(1- \mathcal{M}) \Big]^{1/2}\,,
\end{align}
where $\mathcal{M}$ is the mismatch defined in Eq.~\eqref{eq:mismatch} and minimized over the coalescence time and phase~\cite{Allen:2005fk}. As in Sec.~\ref{sec:waveform:tf2}, we regard the two scenarios as distinguishable whenever $\rho_{\text{\tiny Diff}} > 3$.

\begin{figure*}[t!] 
    \centering
\includegraphics[width=1\linewidth]{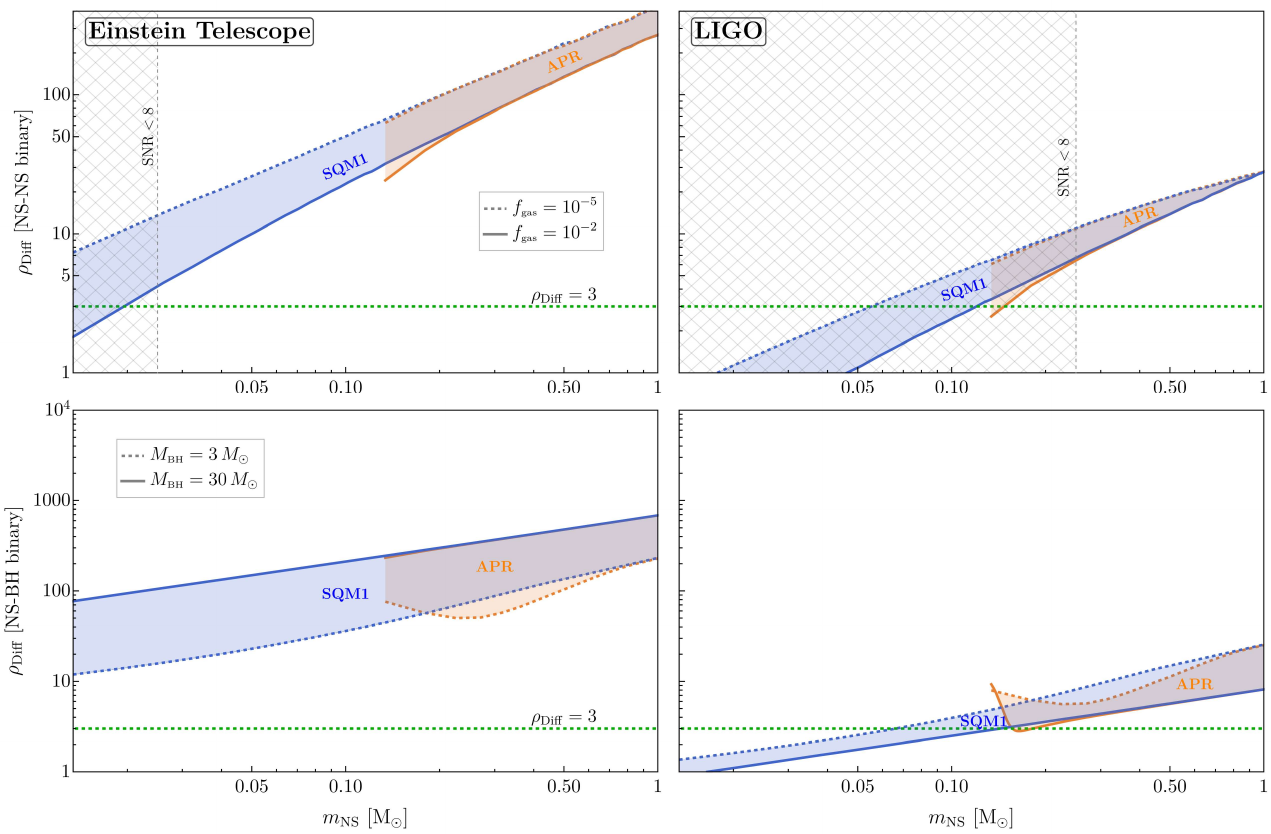}%
\caption{Distinguishability of a BNS (top panels) and of a NS-BH binary (bottom panels) from a PBH binary sharing the same chirp mass and symmetric mass ratio. The dashed green line represents the distinguishability threshold $\rho_{\text{\tiny Diff}} = 3$. Shading in the top panels denotes the regions where the SNR falls below $8$. For the chosen range of masses, the NS-BH events are always resolved.}
\label{fig:mismatch}
\end{figure*}

The two waveform models used for the BNS and NS-BH systems and the corresponding injected parameters are summarized in Table~\ref{tab:mismatch}. We model $h_{\text{\tiny col}}$ as in Eq.~\eqref{eq:perturbed_waveform}, retaining the leading and next-to-leading tidal terms ($5\,${\rm PN} and $6\,${\rm PN}) of Eq.~\eqref{eq:psi_tidal} and, for the BNS, the negative-{\rm PN} dynamical-friction correction of Eq.~\eqref{eq:deltaDF}. The end of the inspiral is imposed through the tapering function of Eq.~\eqref{eq:taperingfunction}, where the truncation frequency $f_{\text{\tiny TD}}$ is the contact frequency $f_\text{\tiny Roche}^\BNS$ at which the two NSs touch for the BNS [Eq.~\eqref{eq:fRoche_num}], and the Roche frequency $f_\text{\tiny Roche}^\text{\tiny NS-BH}$ of Eq.~\eqref{eq:fRoche} for the NS-BH binary. By contrast, $h_{\text{\tiny PBH}}$ is modeled through the same Eq.~\eqref{eq:perturbed_waveform} but with vanishing tidal deformabilities and no environmental correction, i.e., a pure-vacuum point-particle inspiral; in this case we apply no tapering and instead truncate the analysis at the Kerr ISCO frequency of Eq.~\eqref{eq:ISCOKerr}, which marks the end of the vacuum inspiral. For the PBH hypothesis, we inject $\chi_1 = \chi_2 = 0$, as the standard inflationary formation scenario favors small spin values~\cite{DeLuca:2019buf}. In the BNS case we inject equal spins $\chi_1 = \chi_2 = 0.1$, while for the NS-BH we keep $\chi_\NS = 0.1$ and $\chi_\BH = 0.2$. This corresponds to a non-zero asymmetric spin component, which is generically expected for a realistic NS-BH binary formed in a collapsar disk. In both models, we choose the minimum and maximum frequencies following the prescription of the Fisher analyses (Secs.~\ref{sec:FisherI} and~\ref{sec:FisherII}).

We present our results in Fig.~\ref{fig:mismatch}. In the top panels we compare an equal-mass BNS to an equal-mass PBH binary and plot $\rho_{\text{\tiny Diff}}$ as a function of the mass of each component, for ET (top left) and LIGO (top right) and for two representative gas density suppression factors $f_\text{\tiny gas}$. In the shaded regions, limited by masses $m_\NS \lesssim 0.025 \, \msun$ for ET and $m_\NS \lesssim 0.25 \, \msun$ for LIGO, the source is not detectable (${\rm SNR} < 8$). Two trends are visible. First, $\rho_{\text{\tiny Diff}}$ increases with the component mass because higher-mass binaries have larger SNR. Second, at fixed NS mass a smaller $f_\text{\tiny gas}$ yields larger $\rho_{\text{\tiny Diff}}$, as it corresponds to a wider BNS according to Eq.~\eqref{eq:a0NSNS}. As noted in Sec.~\ref{sec:FisherI}, widening the binary increases the relative importance of the $-5.5\,{\rm PN}$ dynamical-friction dephasing, despite $\rho_\text{\tiny env}$ decreasing linearly with $f_\text{\tiny gas}$. Even in the near-vacuum limit $f_\text{\tiny gas} < 10^{-5}$, the two hypotheses remain distinguishable ($\rho_{\text{\tiny Diff}} > 3$) across the entire mass range that yields observable events (${\rm SNR} > 8$). The environmental term therefore increases the separation of the two scenarios in the BNS channel, but it is not required. The large tidal deformabilities of subsolar NSs, $\Lambda \sim \mathcal{O}(10^6$--$10^9)$, produce a $5\,{\rm PN}$ tidal contribution that is sufficient in this comparison to exclude a point-mass PBH binary with identical $\mathcal{M}_c$ and $\eta$.

The bottom panels of Fig.~\ref{fig:mismatch} show the analogous comparison for the NS-BH binary, now for two values of the central BH mass, $M_\BH = 3$ and $30\,\msun$. 
Here distinguishability has a different physical origin, because the tidal imprint that drives the BNS separation is suppressed by four powers of the mass ratio, $\tilde\Lambda_\text{\tiny NS-BH} \propto q^4\,\Lambda$ [Eq.~\eqref{eq:Ltilde_NSBH}], leaving the Roche-frequency cutoff as the main collapsar-specific feature. A further source of distinguishability is provided by the spins, since the PBH scenario is expected to yield negligible spin, while this is not the case for binaries originating in the collapsar scenario. In ET, regardless of the BH mass, $\rho_{\text{\tiny Diff}} > 3$ holds across the full subsolar range. In LIGO, instead, the outcome depends on $M_\BH$: lighter central BHs are more easily distinguished, whereas for $M_\BH = 30\,\msun$ (for which the NS reaches the ISCO before being tidally disrupted) distinguishability is possible only for $m_\NS \gtrsim 0.15\,\msun$. 
The relative hierarchy of the dashed and solid curves is inverted in the two bottom panels. This reflects an analogous hierarchy in the number of cycles, as the accumulated mismatch scales with $\mathcal{N}_{\text{\tiny cyc}}$. As noticed in the discussion of Fig.~\ref{fig:measurabilityNSBH1}, ultimately this is because the in-band number of cycles is reduced
in LIGO, especially for higher BH masses.

\begin{figure*}[!t]
\centering
\includegraphics[width=\textwidth]{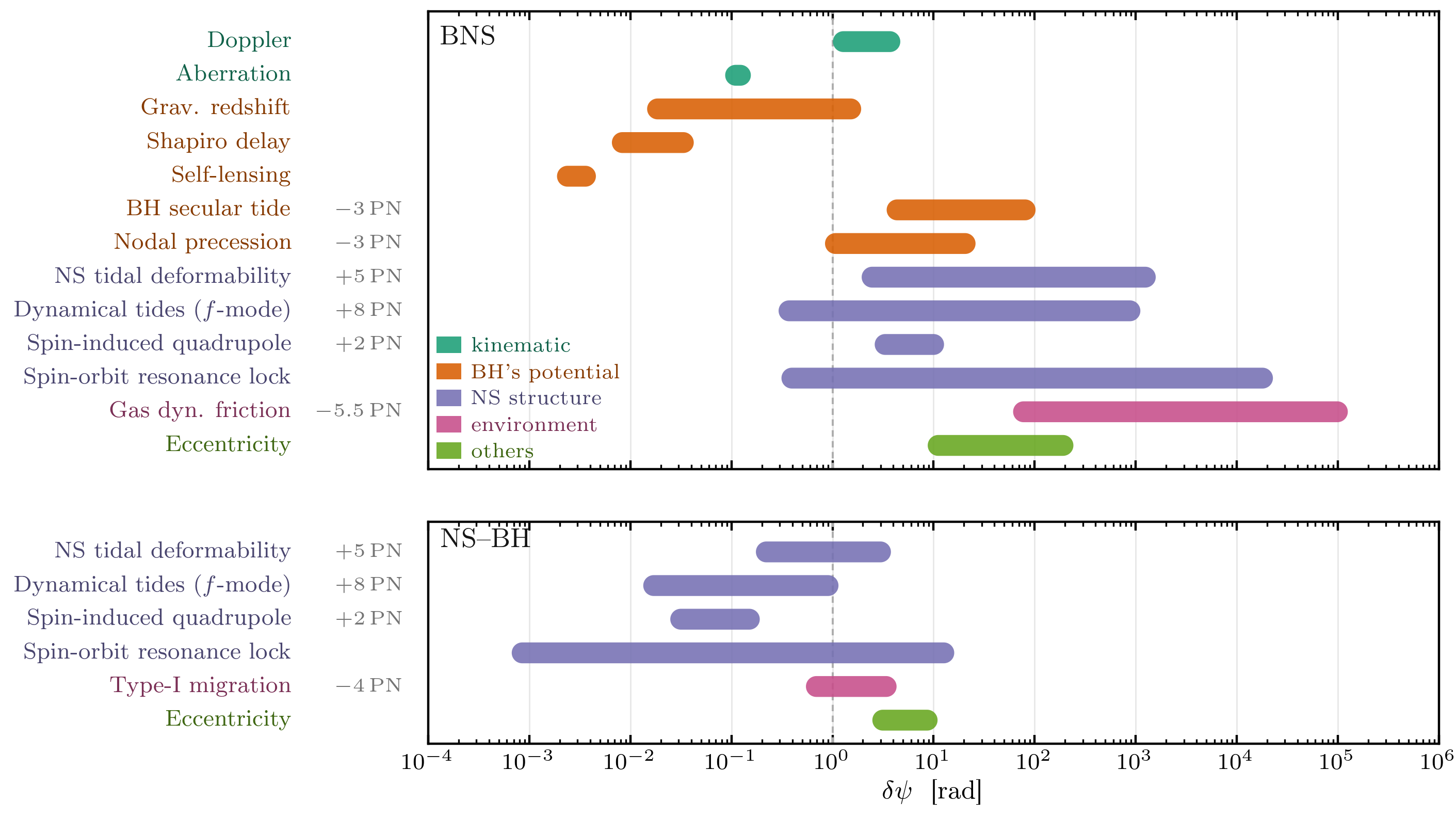}
\caption{Band-weighted root-mean-square of the unabsorbable phase residual $\delta\psi$ spanning $f_\text{\tiny H}\in[0.3,0.5]$, $m_\NS\in[0.1,1]\,\msun$ and $\iota\in[30,60]^\circ$ with $M_\BH = 5\,\msun$, $a_{\BNS-\BH,i} = 340\,R_g$ and the polytropic EoS of Eq.~\eqref{eq:Rpoly}, as observed by LIGO. The orbital velocity produces the kinematic modulations, Doppler [Eq.~\eqref{eq:dphi_doppler_explicit}] and aberration [Eq.~\eqref{eq:dphi_abe_first}]. The central BH's potential acts either on the propagating signal --- redshift [Eq.~\eqref{eq:dpsi_redshift}], Shapiro delay [Eq.~\eqref{eq:dpsi_sha}] and self-lensing [Eq.~\eqref{eq:F_low_w}] --- or, through its tidal quadrupole, on the inner orbit itself [secular tide, Eq.~\eqref{eq:dpsi_tide}, and nodal precession at an inner-plane tilt $\theta = 30^\circ$, Eq.~\eqref{eq:nodal_phase}]. The internal structure of the NSs enters through the adiabatic tidal deformability [Eq.~\eqref{eq:psi_tidal}], the finite-frequency $f$-mode response [Eq.~\eqref{eq:Lambda_dyn}], the spin-induced quadrupole [Eq.~\eqref{eq:dpsi_quadrupole}], and the spin-orbit resonance locking [Eqs.~\eqref{eq:dpsi_lock} and~\eqref{eq:dpsi_lock_nsbh}], the last two at an aligned NS spin $\chi_\NS = 0.3$, and the locking at an ellipticity $3\,\varepsilon_\text{\tiny th}$. The gas density couples to the two stages by different mechanisms --- dynamical friction on the inner BNS [Eq.~\eqref{eq:deltaDF}] at a fraction $f_\text{\tiny gas} = 10^{-5}$ of the disk density, and the Type-I disk torque on the NS-BH pair [Eqs.~\eqref{eq:dpsi_mig}]. For the eccentricity rows, we assume an initial eccentricity $e_i = 0.1$ (Sec.~\ref{sec-eccentricity}). 
}
\label{fig:rms_dpsi_ranges}
\end{figure*}

\section{Conclusions}
\label{sec:conclusions}
\noindent
\noindent
We have presented a comprehensive analysis of the GW signatures of subsolar-mass NSs formed through gravitational fragmentation in collapsar accretion disks.  In the neutron-rich, rapidly cooling environment of the outer disk, fragmentation produces NSs of mass $\sim 0.01 - 1\,\msun$ --- well below the conventional core-collapse floor.  

The collapsar scenario then predicts two sequential merger stages --- a BNS coalescence among the disk-formed clumps, followed by the inspiral of the surviving NS remnant (or other NS clumps) into the central stellar-mass BH --- each producing a distinctive GW signal that has no analog in either PBH binaries or standard BNS waveforms.  

We have characterized the parameter space, detection prospects, and measurability of source properties for both stages with current and future GW detectors, finding that the two channels are not only detectable but individually identifiable, sometimes out to cosmological distances. 

In Fig.~\ref{fig:rms_dpsi_ranges} we summarize the unabsorbable phase residual that different effects accumulate in each of the two merger stages, and in Fig.~\ref{fig:chart} we place these discriminants within the broader interpretation of a subsolar trigger. We collect our main results below, organizing the discussion by merger stage.

\subsection{BNS mergers: the signature-rich stage}
\noindent
Because the subsolar binary orbits the central BH in a hierarchical triple while embedded in residual disk gas, its waveform carries imprints that set it apart from an isolated, point-like binary. For a fiducial $0.1\,\msun$ pair (with $f_\text{\tiny H}=0.5$) the horizon reaches $\approx 36$~Mpc in LIGO, $\approx 0.57$~Gpc in ET, and $\approx 1.8$~Gpc ($z\approx0.34$) in CE, growing toward $\approx 340$~Mpc (LIGO), $z\approx1.3$ (ET), and $z\approx4.7$ (CE) at $m_\NS = 1\,\msun$; the lightest pairs ($m_\NS\lesssim0.02\,\msun$) leave the band altogether. For the fiducial disk ($\rho(R_*)=10^8\,{\rm g\,cm^{-3}}$), a Fisher analysis shows that ET can detect binaries and measure their chirp mass and mass ratio as long as $m_\NS \gtrsim 0.03\,\msun$, as opposed to $m_\NS \gtrsim 0.25\,\msun$ for LIGO. Within this reach, the following discriminants become available, in decreasing order of resolvability:
\begin{itemize}
\item \emph{Hierarchical-triple modulations.} The outer orbit carries the imprint of five relativistic modulations, whose horizon distinguishability we quantify in LIGO at the reference configuration ($M_\BH=5\,\msun$, $a_{\BNS-\BH,i}=340\,R_g$, $\iota=45^\circ$). The Doppler phase shift sits near the decorrelation ceiling over most of the accessible parameter space, vanishing only for face-on binaries ($\propto\sin\iota$). The gravitational redshift ($\delta\rho_h\approx6$) is the least sensitive to source orientation, being inclination independent; its distinguishability rises steeply with the Hill fraction ($\propto f_\text{\tiny H}^{4}$) and is largest for light, wide systems, becoming degenerate with a chirp-mass shift at small $f_\text{\tiny H}$. Relativistic aberration is complementary: although $\delta\rho_h\approx1$ at $\iota=45^\circ$, it scales as $1/\sin\iota$ and strengthens in the face-on limit (reaching $\delta\rho_h\approx7$ near $\iota=5^\circ$), precisely where the Doppler modulation is unobservable. At least one of aberration, Doppler, and redshift is above the distinguishability threshold at every inclination. The Shapiro time delay and self-lensing remain below threshold ($\delta\rho_h\ll3$) for the subsolar and central-BH masses considered, self-lensing being the smallest of the five effects. Because none of these imprints is included in the point-particle templates used in current subsolar searches, a collapsar triple can be significantly mismatched against those template banks, potentially reducing their recovery efficiency.
\item \emph{Tidal deformability.} The large deformabilities of subsolar NSs --- reaching $\Lambda \sim \mathcal{O}(10^9)$ at $m_\NS \sim 0.1\,\msun$ for an NM EoS --- produce a $5$\,PN tidal imprint. For the fiducial disk, ET measures $\tilde\Lambda$ across essentially the full resolvable mass range (except near the NM contact point at $m_\NS\simeq0.1\,\msun$), while LIGO does so over the more restricted bands $0.05\lesssim m_\NS/\msun\lesssim0.5$~(QM) and $0.2\lesssim m_\NS/\msun\lesssim0.6$~(NM). Because the two EoS families predict sufficiently different values of $\Lambda$, a confident tidal measurement for $m_\NS \lesssim 0.2\,\msun$ with ET can discriminate NM from QM, an independent cross-check of the mass-based discrimination. The finite-frequency response reinforces this conclusion: for subsolar NSs the $f$-mode shifts from the kHz range down to $\mathcal{O}(10^2)$\,Hz, so the dynamical ($8$\,PN) tide, confined to the last cycles for canonical NSs, becomes measurable here.

\item \emph{Environmental density.} Residual gas produces a $-5.5$\,PN dynamical-friction driven phase shift, so the binary accumulates fewer cycles and coalesces sooner than a point-particle system of the same masses. At the fiducial suppression factor $f_\text{\tiny gas}=10^{-4}$ (with $\rho(R_*)=10^8\,{\rm g\,cm^{-3}}$), ET can measure $\rho_\text{\tiny env}$ over essentially the whole resolvable range, while LIGO retains sensitivity in a restricted band ($0.1\lesssim m_\NS/\msun\lesssim0.7$ for QM). The measurement improves at lower disk density and lower $f_\text{\tiny gas}$, since the resulting wider orbit accumulates more in-band cycles, until $f_\text{\tiny H}\to1$ and the inner binary is driven towards chaotic instability. Note that this residual quantifies whether an effect is \emph{distinguishable} from a vacuum waveform (i.e., detectable at all), a weaker requirement than a precise measurement of its amplitude: a sizeable dynamical-friction residual can therefore coexist with a poor Fisher constraint on $\rho_\text{\tiny env}$ (Fig.~\ref{fig:measurabilityNSNS2}), the latter being diluted by the linear, low-amplitude entry of $\rho_\text{\tiny env}$.

\item \emph{Eccentricity.} Disk formation generally produces eccentric binaries, either through the fission or through the gas-assisted capture channels, and the sheer length of a subsolar inspiral --- up to $\mathcal{N}_\GW^\text{\tiny BNS} \sim 10^{6} - 10^{7}$ in-band cycles in next-generation detectors --- makes even small initial eccentricities ($e_{\BNS,i} \sim 10^{-3}$, rising to $\sim 10^{-2}$ at $m_\NS = 1\,\msun$) resolvable at the horizon. Residual eccentricity is a sensitive probe of the in-disk assembly channel.
\item \emph{Further imprints.} Besides these effects, the NS spin (through spin-induced quadrupole and spin-orbit resonance locking: see Appendix~\ref{app:NSspin}) and the tidal field of the central BH on the inner binary (secular tide, oscillatory sidebands, and nodal precession: see Appendix~\ref{sec:tidalfield}) leave additional EoS- and geometry-sensitive imprints that are summarized in Fig.~\ref{fig:rms_dpsi_ranges}.
\end{itemize}

\subsection{NS-BH merger: the loudest stage}
\noindent
The inspiral of the surviving NS remnant into the central BH is the loudest event of the scenario, with an SNR roughly an order of magnitude larger than the BNS stage because the chirp mass is set by the $\sim 3 - 30\,\msun$ BH. For a fiducial $0.1\,\msun$ NS the horizon reaches a few hundred Mpc in LIGO, several Gpc in ET, and up to $\sim 20$~Gpc in CE --- three to ten times deeper than the BNS --- and the event is resolved even by LIGO down to $m_\NS \simeq 0.05\,\msun$, making it the most likely channel for a first detection. Its discriminants differ from those of the BNS stage:
\begin{itemize}
\item \emph{Mass ratio.} The symmetric mass ratio is well measured by both detectors ($\Delta\eta/\eta \sim 10^{-4}-10^{-2}$ in ET, $\sim 10^{-2}$ to a few $\times 10^{-1}$ in LIGO), and cleanly separates the NS-BH binary from the comparable-mass BNS: even in the least favorable configurations the absolute uncertainty $\Delta\eta$ remains orders of magnitude below the gap between $\eta\approx q\ll1$ and the equal-mass value $\eta\simeq1/4$.
\item \emph{Tidal-disruption cutoff.} The tidal deformability is suppressed by four powers of the mass ratio [$\tilde\Lambda_\text{\tiny NS-BH} \propto q^4\,\Lambda$, Eq.~\eqref{eq:Ltilde_NSBH}], yielding relative errors $\Delta\tilde\Lambda/\tilde\Lambda \gtrsim 2$ even for the lightest  sources accessible by ET, so this effect is not measurable during the inspiral. The leading EoS discriminant is instead the Roche frequency at which the NS is tidally disrupted, a sharp spectral cutoff near $65$\,Hz at $0.1\,\msun$ that rises linearly with mass. For an NM EoS, the tidal disruption frequency $f_\text{\tiny TD}$ runs from tens of Hz just above the minimum stable mass to $\sim 1$~kHz at $m_\NS = 1\,\msun$, and for a light central BH ($M_\BH=3\,\msun$) it is already measured to $\sim 1\%$ (ET) and $\sim 10\%$ (LIGO). Its reach is itself EoS-limited: the QM disruption edge lies above the ISCO for essentially all collapsar-scale central BHs [Eq.~\eqref{eq:Mstar_quench}], and for heavy BHs ($M_\BH=30\,\msun$) the ISCO preempts disruption for all but the lightest NM stars, so $f_\text{\tiny TD}$ is primarily an NM-specific signature, while for QM the mass window $m_\NS < 0.1\,\msun$ --- where only quark stars exist and remain detectable --- carries the discriminating power.
\item \emph{Eccentricity.} Being a single binary, the NS-BH event carries none of the triple modulations of the BNS stage, which makes any residual eccentricity an especially clean environmental discriminant. A given source is roughly an order of magnitude less sensitive to eccentricity than the inner-BNS triple ($e_\text{\tiny thr}\propto\mathcal{M}_c^{5/6}$, from $\approx7\times10^{-3}$ at $m_\NS\sim0.05\,\msun$ to $\approx2\times10^{-2}$ at $1\,\msun$ for $M_\BH=5\,\msun$), but the larger detection horizon partly offsets this, most strongly in CE and ET. Since subsolar NSs have no formation pathway outside the collapsar disk (and field NS-BH binaries circularize long before merger), a measured $e_{\NS-\BH,i} \gtrsim 10^{-2}$ would directly favor in-disk assembly. This is consistent with the numerical relativity study of Ref.~\cite{Wu:2026hth}, where eccentricities as large as $\simeq0.1$ survive to merger.
\end{itemize}

\subsection{Distinguishing primordial black holes}
\noindent
A key question for any subsolar detection is whether it can be attributed to the collapsar channel rather than to PBHs. We address this question in Sec.~\ref{sec:PBHs} through a direct mismatch analysis against a vacuum PBH binary of identical chirp mass and symmetric mass ratio. 

For the BNS stage, the two hypotheses are distinguishable in both ET and LIGO across the entire detectable mass range, even in the near-vacuum limit $f_\text{\tiny gas} = 10^{-5}$. The enormous tidal deformabilities of subsolar equal-mass NSs alone suffice to exclude a point-mass interpretation, with the dynamical-friction induced dephasing further reinforcing the separation whenever residual gas is present. 

In the NS-BH case the tidal imprint is mass-ratio suppressed, so the leading discriminants are the Roche-frequency cutoff together with the sizeable spins expected of a collapsar origin. Distinguishability then holds across the full BH mass range in ET, and over most of the mass range in LIGO, failing only for the heaviest central BHs ($M_\BH = 30\,\msun$) combined with the lightest NSs ($m_\NS\lesssim0.15\,\msun$), where the Kerr ISCO frequency drops below the Roche frequency and the disruption cutoff is quenched. 

We caution that the very large tidal deformabilities of subsolar NSs, if neglected or mismodeled, can bias the inference of the intrinsic parameters~\cite{Corman:2026lbt}, so accurate tidal modeling is essential both to exploit the EoS information and to avoid misattributing an astrophysical source to new physics. Moreover, the boundary between the two branches of Fig.~\ref{fig:chart} is not sharp: subsolar remnants may themselves originate from PBHs captured and tidally disrupted within stars, whose fragmenting accretion disks produce subsolar compact objects in close analogy with the collapsar channel~\cite{Gottlieb:2026kpg,Cantiello:2026eur}, further motivating the waveform-level discriminants developed here.

\subsection{Multi-messenger counterparts}
\noindent
Because the disk-formed NSs are born in the neutron-rich, rapidly cooling environment of a collapsar, the material ejected in their mergers is processed into heavy $r$-process elements and mixed into the surrounding supernova ejecta, powering a ``fragmented superkilonova''~\cite{Metzger:2024ujc,Chen:2025uwd}, while the central engine may simultaneously drive a long or short gamma-ray burst and a broad-lined, stripped-envelope supernova~\cite{Kasliwal:2025keb,Baibhav:2026zuy}. As shown in Sec.~\ref{sec:delayed}, the merger of a wide, dynamical-friction hardened pair can be delayed by up to several days --- comparable to or exceeding the supernova timescale --- so that the GW event may precede, accompany, or follow the optical transient, and the search for an EM counterpart should extend to the days preceding the trigger, not only those following it~\cite{Hall:2026gov,Vieira:2026eof}. 

Recent work has argued that specific catalog events may already be collapsar-disk products preceded by a stripped-envelope supernova~\cite{Baibhav:2026zuy}, and Ref.~\cite{Tejera:2026dpr} explored the detectability of such multimessenger events with the Cross-Correlation Algorithm (CoCoA), less sensitive than matched filtering but more robust to the waveform uncertainties expected from the complex underlying astrophysics. 

The most conclusive evidence, however --- independent of any EM counterpart --- would be a spatially and temporally coincident subsolar BNS merger and NS-BH merger, the unique double-merger signature of a single collapsar disk, with no analog in either the PBH or the standard BNS scenario. 

Successful, on-axis long gamma-ray bursts (LGRBs) provide the cleanest observational signature of collapsars. Their beaming-corrected rate, $\sim\mathcal{O}(10^2)\,{\rm Gpc^{-3}\,yr^{-1}}$~\cite{Ghirlanda:2022edk}, sets a lower bound on the collapsar rate. The channel considered here, however, requires only the central engine --- a rapidly rotating BH surrounded by a massive, fragmentation-prone disk --- and does not require a successful jet. Thus, collapsars with choked, misaligned, weak, or non-breakout jets~\cite{Bromberg:2011wb}, including stripped envelope progenitors, may produce no LGRB while still undergoing disk fragmentation. The collapse itself should often produce a stripped-envelope supernova (Type Ib, Ic, Ic-BL, or IIb, depending on the degree of stripping and jet outcome), preceding the GW signals from both stages. The resulting event rate is therefore set by the collapsar rate, potentially several times the observed LGRB rate, multiplied by the fraction of disks that fragment and by the several subsolar-mass mergers that each disk may produce. Such events may plausibly be within reach of current and next-generation GW searches~\cite{Baibhav:2026zuy}.

\smallskip
Looking forward, future GW detectors will bring the full collapsar parameter space within reach.  Their enhanced sensitivity ensures an improvement over current detectors, which will enable not only detection but also precise measurements of masses, tidal deformabilities, and environmental densities for subsolar collapsar-formed systems out to cosmological distances. 

A full understanding of the BNS and NS-BH channels will ultimately require N-body and magnetohydrodynamic simulations of the fragmenting disk, which set the initial masses, separations, and eccentricities of both stages~\cite{Metzger:2024ujc,Chen:2025uwd,Lerner:2025dkd,Wu:2026hth} (and, for mildly unequal pairs, determine whether contact leads to merger or to a mass-transfer-driven outspiral~\cite{Weaver:2026qgr}). The combination of anomalously low component masses, in-band tidal disruption, environmental phase shifts, hierarchical modulation, and eccentricity nonetheless constitutes a unique waveform fingerprint of the collapsar channel --- one that next-generation detectors are well-positioned to identify.
 
\begin{table*}[t]
\caption{Notation and conventions used throughout this work.}
\label{tab:nomenclature}
\begin{ruledtabular}
\begin{tabular}{ll}
\textbf{Symbol} & \textbf{Definition / description} \\
\hline
\multicolumn{2}{l}{\textit{Model parameters}} \\
$M_\BH$ & Central BH mass \\
$R_g$ & BH gravitational radius \\
$m_\NS \,\, \& \,\, R_\NS$ & NS mass and radius \\
$N_\NS$ & Number of NSs produced by disk fragmentation \\
$\chi$ & Dimensionless spin parameter \\
$\dot{M}_\BH$ & Accretion rate onto the central BH \\
$\Omega$ & Keplerian orbital angular velocity \\
$Q$ & Toomre parameter \\
$\mathbb{H}$ & Disk aspect ratio, $h/r$ \\
$M_\text{\tiny disk} \,\, \& \,\, R_\text{\tiny disk}$ & Disk mass and outer radius \\
$R_*$ & Fragmentation radius \\
$\rho(r)$ & Disk density profile \\
$\rho_\text{\tiny env}$ & Residual post-fragmentation gas density at $R_*$ \\
$f_\text{\tiny gas}$ & Density suppression factor, $\rho_\text{\tiny env}/\rho(R_*)$ \\
\multicolumn{2}{l}{\textit{Binary parameters}} \\
$a_{\BNS}$ & Semi-major axis of the inner BNS \\
$a_{\NS-\BH}$ & Semi-major axis of the NS-BH binary \\
$a_{\BNS-\BH}$ & Outer semi-major axis of the BNS center of mass around the central BH \\
$r_\text{\tiny Hill}$ & Hill radius of the inner BNS \\
$f_\text{\tiny H}$ & Initial BNS separation in units of $r_\text{\tiny Hill}$ \\
$d_\text{\tiny Roche}$ & Roche radius \\
$f_\text{\tiny Roche}$ & Roche/contact or tidal-disruption frequency \\
$f_\text{\tiny TD}$ & Inspiral termination frequency \\
$\tau_\text{\tiny orb}$ & Orbital period of the BNS around the central BH \\
$\tau_{\GW}^{\BNS}$ & GW inspiral timescale of the BNS \\
$\tau_{\GW}^{\NS-\BH}$ & GW inspiral timescale of the NS-BH binary \\
$\tau_\text{\tiny dyn}$ & BNS dynamical-friction hardening timescale \\
$\tau_\text{\tiny mig}$ & NS migration timescale through the disk \\
$\tau_\text{\tiny acc}^\BH$ & Disk-depletion time due to accretion onto the central BH \\
$\tau_\text{\tiny J}$ & Time for a clump to migrate from $R_*$ into the stable inner disk \\
\multicolumn{2}{l}{\textit{Waveform parameters}} \\
$\mathcal{M}_c$ & Chirp mass \\
$q$ & Mass ratio; for NS-BH systems, $q=m_\NS/M_\BH$ \\
$\eta$ & Symmetric mass ratio \\
$\iota$ & Inclination angle \\
$e_{\BNS,i} \,\, \& \,\, e_{\NS-\BH,i}$ & Initial eccentricities of the BNS and NS-BH binary \\
$d_\text{\tiny L} \,\, \& \,\, d_\text{\tiny hor}$ & Luminosity and horizon distances \\
$\Lambda \,\, \& \,\, \tilde{\Lambda} $ & Dimensionless and binary effective tidal deformabilities \\
$\rho_0$ & Optimal signal-to-noise ratio \\
$\rho_{\text{\tiny Diff}}$ & Distinguishability SNR between waveform hypotheses \\
$\mathcal{N}_\text{\tiny cyc}$ & Number of GW cycles \\
$\mathcal{M}$ & Mismatch \\
\end{tabular}
\end{ruledtabular}
\end{table*}

\section*{Acknowledgments}
\noindent
V.B. acknowledges support from the NASA Hubble Fellowship grant HST-HF2-51548.001-A awarded by the Space Telescope Science Institute, which
is operated by the Association of Universities for Research
in Astronomy, Inc., for NASA, under contract NAS5-26555.
E.B., V.D.L. and L.D.G. are supported by NSF Grants No.~AST-2606672, No.~AST-2307146, No.~PHY-2513337, No.~PHY-090003, and No.~PHY-20043, by John Templeton Foundation Grant No.~62840, by the Simons Foundation [MPS-SIP-00001698, E.B.], by the Simons Foundation International [SFI-MPS-BH-00012593-02], and by Italian Ministry of Foreign Affairs and International Cooperation Grant No.~PGR01167.
This work used Delta at the National Center for Supercomputing Applications (NCSA) through allocation PHY260065 from the Advanced Cyberinfrastructure Coordination Ecosystem: Services \& Support (ACCESS) program, which is supported by U.S. National Science Foundation grants 
\#2138259, \#2138286, \#2138307, \#2137603, and \#2138296.
This work was carried out at the Advanced Research Computing at Hopkins (ARCH) core facility (\url{https://www.arch.jhu.edu/}), which is supported by the NSF Grant No.~OAC-1920103.
B.D.M. acknowledges support from the National Science Foundation (grant AST-2406637), NASA (grant 80NSSC26K0299), and the Simons Foundation
(grant 727700).


\appendix
\section{Spins of the compact objects} 
\label{app:NSspin}
\noindent
Beyond the masses and structural properties established in Sec.~\ref{sec:properties}, the spins of the compact objects formed in fragmenting collapsar disks bear directly on both their formation and the GW signals they produce. In this Appendix we estimate the natal spin of the subsolar NSs --- inherited from the angular momentum of the differentially rotating disk, and large enough to drive the fission that potentially seeds the BNSs of Sec.~\ref{sec:param_NSNS} --- and that of the central BH, set by its core-collapse origin and subsequent accretion history.

\subsection{NS spin}
\noindent
Understanding the initial spin of NSs formed through collapsar disk fragmentation is crucial for modeling both their subsequent evolution and the GW signals they may produce. Unlike NSs forged through isolated core-collapse supernovae~\cite{Miller:2014aaa}, subsolar-mass NSs emerging from disk fragmentation inherit substantial angular momentum from their progenitor region in the differentially rotating disk. At the point of collapse $r = R_*$, this spin can be estimated from the shear across the Jeans length~\cite{Metzger:2024ujc}
\begin{align}
\chi_\NS & \simeq \frac{h^2 \Omega}{m_\NS} \simeq \mathbb{H}^2 \frac{\sqrt{M_\BH R_*}}{m_\NS} \nonumber \\
& \simeq \frac{7}{Q_0^{1/3}} \left(\frac{\mathbb{H}}{0.3} \right)^2 \left( \frac{M_\BH/m_\NS}{10}\right)\,.
\end{align}
For the fiducial disk and mass parameters, this expression gives $\chi_\NS>1$. Physically, this means that the collapsing region holds more angular momentum than can be supported by a single NS --- which cannot spin faster than the critical (mass-shedding) limit, $\chi_\NS \lesssim 0.7$, largely independently of the EoS (even though quark stars can exceed unity with no universal bound~\cite{Lo:2010bj}); we note, however, that this result is established for $m_\NS \gtrsim 1\,\msun$, so for the subsolar configurations of interest it should be taken as indicative. This excess angular momentum requires redistribution during collapse. A possible outcome is fission of the collapsing fragment into two or more proto-NSs, placing the surplus angular momentum into their relative orbits rather than stellar spin. Close proto-binaries or higher-order multiples may therefore form if angular momentum is redistributed through fission during disk fragmentation~\cite{Alexander:2008wv}, with implications for the further evolution of the system.

The NS spin also leaves several distinct imprints on the waveform of the subsequent mergers. In the following we consider two such effects: 

\noindent
(i) the spin-induced quadrupole of the NS;

\noindent
(ii) the resonant locking of the spin to the orbit sustained by a non-axisymmetric deformation of the star.

\subsubsection{Spin-induced quadrupole}
\label{sec:spin_quadrupole}
\noindent
A body of mass $m$ with non-zero dimensionless spin $\chi$ acquires the mass quadrupole $\mathcal{Q} = -\bar{\mathcal{Q}}\,m^{3}\chi^{2}$. The dimensionless $\bar{\mathcal{Q}}$ quantifies the spin-induced quadrupole at fixed mass and spin. For a Kerr BH, the no-hair theorem gives $\bar{\mathcal{Q}}=1$; the same value therefore applies to a PBH. A NS is more deformable and has $\bar{\mathcal{Q}}\simeq4 - 8$ at canonical supersolar masses~\cite{Poisson:1997ha,Laarakkers:1997hb}. On the other hand, assuming a polytropic EoS with $R_\NS\propto m_\NS^{-1/3}$, subsolar stars have larger radii and lower compactness and thus deform more readily; using the approximate scaling $\bar{\mathcal{Q}} \simeq 2.5/C \propto m_\NS^{-4/3}$~\cite{Yagi:2013awa}, a $0.1\,\msun$ star has $\bar{\mathcal{Q}}\sim 500$, approximately two orders of magnitude larger than the value for a canonical NS. At fixed spin, the corresponding waveform correction is therefore larger for a subsolar binary than for a $1.4\,\msun$ binary.

The companion moves in the quadrupole field of the spinning star, so $\mathcal{Q}$ affects the orbit through the quadrupole-monopole interaction. At $2$\,PN order, the phase correction for aligned spins reads~\cite{Poisson:1997ha}
\begin{equation}
\begin{split}
  \delta\psi_\text{\tiny SIQM}^{\,\BNS}(f)
  &= -\frac{75}{32}\,\frac{(\bar{\mathcal{Q}}-1)\,\chi_\NS^{2}}{v_\BNS}\,,\\[3pt]
  \delta\psi_\text{\tiny SIQM}^{\,\NS-\BH}(f)
  &= -\frac{75}{64}\,\frac{m_\NS}{M_\BH}\,
     \frac{(\bar{\mathcal{Q}}-1)\,\chi_\NS^{2}}{v_{\NS-\BH}}\,,
\end{split}
  \label{eq:dpsi_quadrupole}
\end{equation}
where $v_\BNS = (2\pi m_\NS f)^{1/3}$, $v_{\NS-\BH} = (\pi M_\BH f)^{1/3}$, and $\bar{\mathcal{Q}}-1$ measures the deviation from the Kerr quadrupole. The $3$\,PN correction~\cite{Krishnendu:2017shb,Krishnendu:2018nqa} is $-15.5\,v_\BNS^{2}$ times Eq.~\eqref{eq:dpsi_quadrupole}.
For aligned spin, the negative sign arises because the spin-induced quadrupole reduces the orbital binding energy; the signal phase therefore advances relative to a BH template. The frequency dependence is $\delta\psi_\text{\tiny SIQM}\propto f^{-1/3}$ and is weighted toward the low-frequency part of the band. 

Detection in LIGO, $\delta\rho_h>3$, requires a minimum spin that scales approximately as $\chi_\NS^\text{\tiny min}\simeq0.1(m_\NS/0.1\,\msun)^{2/3}$ for the inner pair, while the NS-BH stage cannot be measured at its horizon and requires a source with larger SNR. The correction scales as $f^{-1/3}$, so detectors with lower low-frequency cutoffs (such as ET) provide stronger constraints.

Observability also requires the NS to retain sufficient spin between formation and merger. A rotating magnetized star, with rotational frequency $\Omega_\NS$ and magnetic field strength $B$, loses angular momentum through magnetic dipole braking. An orthogonal rotator radiates energy at the rate $\dot E_\text{\tiny MDB} = -B^{2}R_\NS^{6}\Omega_\NS^{4}/6$, and equating this to the loss of rotational energy $\frac12I\Omega_\NS^{2}$, in terms of the moment of inertia $I$, gives $\dot\Omega_\NS = -B^{2}R_\NS^{6}\Omega_\NS^{3}/6I$. By integration we have $\Omega_\NS(t) = \Omega_0(1+t/\tau_\text{\tiny sd})^{-1/2}$, with
\begin{equation}
\begin{split}
  \tau_\text{\tiny sd} &= \frac{3I}{B^{2}R_\NS^{6}\Omega_0^{2}}\\[3pt]
  &\simeq 1.6\times10^{4} \, {\rm s}
  \left(\frac{B}{10^{15}\,{\rm G}}\right)^{-2}
  \left(\frac{\chi_\NS}{0.3}\right)^{-2}
  \left(\frac{m_\NS}{0.1\,\msun}\right)^{-1}.
\end{split}
  \label{eq:tau_spindown}
\end{equation}
This gives a spin-down time of $4.4$~hr, comparable to the $\sim\!10^{4}$~s that the inner pair spends in the LIGO band. The spin-down for certain magnetic field and spin magnitudes can happen in the detector band, adding an additional observable to measure the magnetic field. Joint measurements of the tidal deformability $\Lambda$ and of the spin-induced quadrupole $\bar{\mathcal{Q}} \, \chi_\NS^{2}$ can then be compared with the compactness relation $\bar{\mathcal{Q}}\propto1/C$ to further constrain the NS EoS.  We leave a more detailed exploration of this possibility to future work.

\subsubsection{Spin-orbit resonance locking}
\label{sec:spin_lock}
\noindent
A perfectly axisymmetric rotating star in a binary experiences no torque about its spin axis. In the presence of a non-axisymmetric mass quadrupole, described by an ellipticity $\varepsilon_\text{\tiny SO}$ and potentially sustained by an internal magnetic field~\cite{Bonazzola:1995rb,Cutler:2002nw,Haskell:2007bh}, the companion can exert such a torque. The torque has a secular contribution when the stellar spin frequency matches the orbital frequency. As the orbital frequency increases during inspiral, the system may be captured into this spin-orbit resonance. During resonance, the spin and orbital frequencies evolve at the same rate. Resonance ends when the GW torque exceeds the maximum torque transmitted by the quadrupole coupling, after which the accumulated phase correction remains constant~\cite{Miao:2025tqu}.

The phase accumulated during resonance for a deformed star with moment of inertia $I$ is~\cite{Miao:2025tqu}
\begin{align}
  \delta\psi^{\,\BNS}_\text{\tiny lock}(f) &= -\frac{45}{16}\,\bar I\,g(v_\BNS)\,, \nonumber \\
  \quad
  g(v_\BNS) &= \frac{1}{v_\BNS}-\frac{4}{3v_{\BNS,i}}+\frac{v_\BNS^{3}}{3v_{\BNS,i}^{4}}\,,
  \label{eq:dpsi_lock}
\end{align}
where $v_\BNS = (2\pi m_\NS f)^{1/3}$, $v_{\BNS,i}$ is its value at the onset of the resonance (with associated frequency $f_i$), and $\bar I\equiv I/m_\NS^{3} = 0.26/C^{2}$~\cite{Yagi:2013awa}. This expression applies for $f_i\leq f\leq f_\text{\tiny br}$, in terms of the resonance-breaking frequency $f_\text{\tiny br}$, above which the phase becomes constant. The scaling $\bar I\propto C^{-2}$ substantially enhances the effect for low-compactness subsolar stars, reaching values as large as $\bar I = 1.4\times10^{4}$ at $m_\NS = 0.1\,\msun$, to be compared with $\bar I = 12$ at $1.4\,\msun$. The accumulated resonant phase is consequently $\sim 10^{4}$--$10^{7}$\,rad, rather than the order-unity correction found in Ref.~\cite{Miao:2025tqu}. It is therefore also necessary to determine the parameter region in which the resonance can be maintained.

The resonance begins when the orbital frequency reaches the spin frequency, and it ends when the GW torque exceeds the quadrupole-coupling torque. With $\bar I\propto C^{-2}\propto m_\NS^{-8/3}$ on the degenerate branch, the two frequencies are
\begin{align}
  f_i = 2f_\text{\tiny spin} &= \frac{\chi_\NS}{\pi\,\bar I\,m_\NS}
   \simeq 13.9\ {\rm Hz}
   \left(\frac{\chi_\NS}{0.3}\right)
   \left(\frac{m_\NS}{0.1\,\msun}\right)^{5/3} \,, \nonumber \\
  f_\text{\tiny br} &= \frac{1}{2\pi m_\NS}
   \left(\frac{5}{32}\right)^{3/5}\!\varepsilon_\text{\tiny SO}^{3/5} \,\nonumber \\
  &\simeq 106\ {\rm Hz}
   \left(\frac{\varepsilon_\text{\tiny SO}}{10^{-5}}\right)^{3/5}
   \left(\frac{m_\NS}{0.1\,\msun}\right)^{-1}\,.
  \label{eq:lock_freqs}
\end{align}
A sufficiently massive, rapidly rotating star does not reach its resonance within the detector band, whereas the resonance-breaking frequency decreases with mass and increases with deformation. Resonant evolution requires
$\frac{f_\text{\tiny br}}{f_i} = \left(\frac{5}{32}\right)^{3/5} \frac{\bar I\,\varepsilon_\text{\tiny SO}^{3/5}}{2\chi_\NS}>1$.
Inverting $f_\text{\tiny br} = \max(f_\text{\tiny min},f_i)$ gives the minimum deformation conditions for which the resonance persists into the detector band,
\begin{align}
  \varepsilon_\text{\tiny th}
  &= \frac{32}{5}\,\bigl[2\pi m_\NS
     \max(f_\text{\tiny min},\,2f_\text{\tiny spin})\bigr]^{5/3} \nonumber \\
  &\simeq 6.2\times10^{-7}
  \left(\frac{m_\NS}{0.1\,\msun}\right)^{5/3}\!
  \left[\frac{\max(f_\text{\tiny min},\,2f_\text{\tiny spin})}{20\ {\rm Hz}}\right]^{5/3}\,,
  \label{eq:eps_th}
\end{align}
or alternatively
\begin{equation}
  m_\NS \lesssim 0.21 \,\msun
  \left(\frac{\varepsilon_\text{\tiny SO}}{10^{-5}}\right)^{9/40}
  \left(\frac{\chi_\NS}{0.3}\right)^{-3/8}\,.
\end{equation}
The magnetic deformation scales as~\cite{Bonazzola:1995rb,Cutler:2002nw,Haskell:2007bh}
\begin{equation}
  \varepsilon_\text{\tiny SO} \simeq 5.2\times10^{-6}\,\kappa_B\!
  \left(\frac{0.1\,\msun}{m_\NS}\right)^{2}\!
  \left(\frac{R_\NS}{34\,{\rm km}}\right)^{4}\!
  \left(\frac{\bar B}{10^{13}\,{\rm G}}\right)^{2}\!,
  \label{eq:eps_field}
\end{equation}
where $\kappa_B\sim O(1)$ and $\bar B$ is the internal magnetic field. Setting $\varepsilon_\text{\tiny SO}=\varepsilon_\text{\tiny th}$ in Eq.~\eqref{eq:eps_field} and solving for the field gives the minimum internal field for which the resonance persists when the signal enters the band, $\bar B_\text{\tiny th} \propto\varepsilon_\text{\tiny th}^{1/2}\,m_\NS R_\NS^{-2}$. Using $R_\NS\propto m_\NS^{-1/3}$, this yields
\begin{align}
  \bar B_\text{\tiny th} &\simeq 3.5\times10^{12}\ {\rm G}\;\kappa_B^{-1/2}
  \left(\frac{m_\NS}{0.1\,\msun}\right)^{5/2}\, \nonumber \\
  &\quad\times
  \left[\frac{\max(f_\text{\tiny min},\,2f_\text{\tiny spin})}{20\ {\rm Hz}}\right]^{5/6}\,.
  \label{eq:B_th}
\end{align}
Whenever the resonance is present within the detector band, the BNS mismatch approaches its maximum over the evaluated frequency range, and $\delta\rho_h>3$ over the evaluated parameter range.

For the NS-BH system, Eq.~\eqref{eq:dpsi_lock} becomes~\cite{Miao:2025tqu}
\begin{equation}
  \delta\psi_\text{\tiny lock}^{\,\NS-\BH}(f)
  = -\frac{45}{64}\,\frac{m_\NS}{M_\BH}\,\bar I\,g(v_{\NS-\BH})\,,
  \label{eq:dpsi_lock_nsbh}
\end{equation}
with $v_{\NS-\BH} = (\pi M_\BH f)^{1/3}$. The resonance-breaking frequency becomes
\begin{equation}
\begin{split}
  f_\text{\tiny br}
  &= \frac{1}{\pi M_\BH}
   \left(\frac{5}{64}\,\frac{M_\BH}{m_\NS}\,
   \varepsilon_\text{\tiny SO}\right)^{\!3/5}\\[3pt]
  &\simeq 29\ {\rm Hz}
   \left(\frac{\varepsilon_\text{\tiny SO}}{10^{-5}}\right)^{3/5}
   \left(\frac{m_\NS}{0.1\,\msun}\right)^{-3/5}
   \left(\frac{M_\BH}{5\,\msun}\right)^{-2/5}.
\end{split}
  \label{eq:f_br_nsbh}
\end{equation}
The full resonance is observable only when $\max(f_\text{\tiny min},f_i) < f_\text{\tiny br} < f_\text{\tiny TD}$. The locking must be strong enough to enter the detector band but weak enough to break before the NS is disrupted: 
\begin{align}
  & 5.3\times10^{-6}
   \left(\frac{m_\NS}{0.1\,\msun}\right)
   \left(\frac{M_\BH}{5\,\msun}\right)^{2/3}
   \left[\frac{\max(f_\text{\tiny min},\,2f_\text{\tiny spin})}{20\ {\rm Hz}}\right]^{5/3}\, \nonumber \\
  &\quad< \varepsilon_\text{\tiny SO} < 3.7\times10^{-5} \,
  \left(\frac{m_\NS}{0.1\,\msun}\right)^{8/3}  \,
  \left(\frac{M_\BH}{5\,\msun}\right)^{2/3} \,.
  \label{eq:eps_window}
\end{align}
For $f_\text{\tiny br} > f_\text{\tiny TD}$, the resonance continues beyond tidal disruption, but the break in the spin-orbit resonance does not occur in the observed signal. 

Finally, the compactness that determines $\bar I$ is itself EoS dependent. A QM star is more compact, and therefore has a smaller $\bar I$. As a result, its resonance begins at a higher frequency [Eq.~\eqref{eq:lock_freqs}] and is confined to lower masses. The presence or absence of this waveform feature therefore provides additional information about the underlying EoS.

\subsection{Central BH spin}  
\noindent
The spin of the central BH in the collapsar scenario is shaped by both its formation and subsequent accretion history.  Prior to disk formation, the progenitor's angular momentum is inherited by the newborn BH, resulting in an initial spin distribution that reflects the core-collapse dynamics.  In standard supernova models, fallback and envelope angular momentum are often limited, leading to relatively modest spins for stellar-mass BHs~\cite{2019MNRAS.485.3661F}. However, collapsars represent a rare, rapidly rotating subset of massive stars~\cite{Cantiello:2007yy}, where the angular momentum at core-collapse can, in principle, be substantial~\cite{Woosley:2006fn}, despite being strongly dependent on the emission of gamma-ray bursts, which extract rotational energy from the BH. Indeed, once the BH saturates with large-scale magnetic flux in the magnetically arrested disk (MAD) state, the jet power is determined exclusively by the mass accretion rate and jet launching efficiency, the latter set solely by the BH spin, so that observations of characteristic gamma-ray burst luminosities constrain the majority of collapsar BHs to be  born slowly spinning, with a dimensionless natal spin $\chi_{\BH,i}\lesssim 0.2$~\cite{Gottlieb:2023cgm}.

The central BH formed in the collapsar scenario is then embedded in an environment distinguished by its high mass accretion rate, reaching values $\dot{M}_\BH\sim(0.1-1)\,M_\odot\,\mathrm{s}^{-1}$, and a large angular momentum reservoir.  Such sustained accretion implies that the BH can grow substantially both in mass and in spin, and tracking the spin evolution is central to understanding the details of subsequent GW emission and possible jet formation.  

\begin{figure}[t!] 
\includegraphics[width=0.99\linewidth]{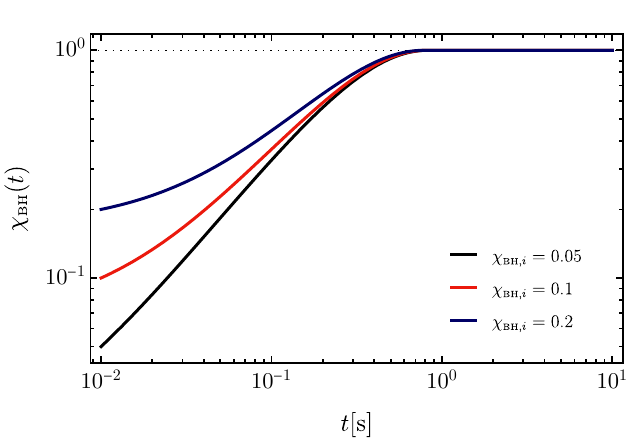}
    \caption{Evolution of the central BH spin during accretion from the surrounding disk, assumed to be thin according to the Novikov-Thorne model, for different values of the initial spin and an accretion rate of $\dot{M}_\BH = 1 \, M_\odot \, {\rm s}^{-1}$.  }
    \label{fig:BHspinup}
\end{figure}

Two qualitatively different outcomes are possible depending on the magnetic state of the disk.  If the disk remains in the standard thin-disk (Novikov-Thorne) regime~\cite{Thorne:1974ve}, the high-angular-momentum material deposited onto the BH increases $\chi_\BH$. The dimensionless spin parameter evolves as
\begin{equation}
\dot{\chi}_\BH
= \bigl(\mathcal{F}(\chi_\BH) - 2\chi_\BH\bigr)
  \frac{\dot{M}_\BH}{M_\BH}\,,
\end{equation} 
where $\mathcal{F}(\chi_\BH)=L(M_\BH,J)/M_\BH E(M_\BH,J)$ encapsulates the efficiency of angular momentum transfer through the ISCO.  The specific energy and angular momentum are determined by the relativistic geodesic structure of the ISCO as
\begin{align}\label{eq:ISCOKerr}
E &= \sqrt{1- \frac{2M_\BH}{3r_\text{\tiny ISCO}}}\,, \nonumber\\
L &= \frac{2M_\BH}{3\sqrt{3}}
    \!\left(1 + 2\sqrt{\frac{3r_\text{\tiny ISCO}}{M_\BH}-2}\,\right)\!,
    \nonumber\\
\frac{r_\text{\tiny ISCO}}{M_\BH} &= 
  \bigl[3+Z_2-\sqrt{(3-Z_1)(3+Z_1+2Z_2)}\,\bigr]\,,
\end{align}
with $Z_1=1+(1-\chi_\BH^2)^{1/3}[(1+\chi_\BH)^{1/3}+ (1-\chi_\BH)^{1/3}]$ and $Z_2=\sqrt{3\chi_\BH^2+Z_1^2}$ (for a generalization of these formulae to the initially spinning case, see~\cite{Page:1974he}). In this limit, accretion spins the BH up monotonically, approaching the theoretical maximum $\chi_\BH \lesssim 0.998$ set by radiation capture~\cite{Thorne:1974ve}, as illustrated in Fig.~\ref{fig:BHspinup} for different initial conditions.

However, this evolution changes if the disk reaches the MAD state. In collapsars, large-scale magnetic flux can accumulate near the BH on timescales of order a second~\cite{Jacquemin-Ide:2023aax}.  In a MAD, strong magnetic fields threading the BH extract its rotational energy via the Blandford--Znajek mechanism~\cite{1977MNRAS.179..433B} and simultaneously deplete the inflowing gas of its angular momentum before it reaches the horizon, so that the specific angular momentum accreted falls well below the Keplerian value.  The combined effect of these EM and hydrodynamic torques produces a spin-down more efficient than Novikov-Thorne spin-up~\cite{Lowell:2023kyu}.  Applying this model to collapsars, Ref.~\cite{Jacquemin-Ide:2023aax} found that the BH evolves toward low spin, with the final spin weakly dependent on the initial one and primarily sensitive to the fraction of mass accreted during the MAD phase. 

Overall, the collapsar scenario therefore favors a central BH with moderate-to-low spin, with the precise value depending on the magnetic history of the accretion flow. In this work, we accordingly adopt a fiducial value $\chi_\BH = 0.1$. This choice has a negligible impact on the NS-BH data analysis: the central BH spin enters the waveform only through the Kerr ISCO frequency [Eq.~\eqref{eq:ISCOKerr}], which sets the vacuum termination of the inspiral and therefore affects the results only in the regimes where the ISCO, rather than tidal disruption, limits the in-band signal.

Beyond its effect on BH spin, the MAD state also has direct implications for disk fragmentation: magnetically arrested disks are generally more stable against gravitational instability than their unmagnetized counterparts, both because magnetic pressure increases the effective sound speed and reduces the density, and because magnetic tension provides an additional restoring force against self-gravity~\cite{Begelman:2006vj}. A full magnetohydrodynamic treatment of fragmentation in strongly magnetized collapsar disks is beyond the scope of the present work.

\section{Environment-induced binary assembly} 
\label{app:capturerate}
\noindent
Here we estimate the rate at which gas dynamical friction drives close encounters between the $N_\NS\sim\mathcal{O}(10)$ NSs formed at the fragmentation radius $R_*$, leading to gravitational capture and binary formation.  The mechanism is analogous to gas-assisted capture in active galactic nuclei (AGN) disks~\cite{Tagawa:2019osr}: during a close passage within the mutual Hill sphere, the surrounding gas decelerates both NSs, dissipating enough kinetic energy to prevent the two bodies from separating after the encounter.

The NSs populate the fragmentation region of volume $\sim\pi R_*^3\mathbb{H}$, giving a number density $n_\NS\simeq N_\NS/(\pi R_*^3\mathbb{H})$.  Taking the geometric cross-section for an encounter to be set by the Hill sphere, $\sigma_\text{\tiny enc}\simeq\pi r_\text{\tiny Hill}^2$, and the typical relative velocity to be the Keplerian shear, $v_\text{\tiny rel}\simeq\sqrt{M_\BH/R_*}$, the encounter rate is $\Gamma_\text{\tiny enc} = n_\NS\,\sigma_\text{\tiny enc}\,v_\text{\tiny rel}$. 
During a single passage, the probability that gas capture forms a binary is set by the ratio of the Hill-sphere transit time to the dynamical friction timescale~\cite{Tagawa:2019osr},
\begin{equation}
    P_\text{\tiny cap}
    = \frac{r_\text{\tiny Hill}/v_\text{\tiny rel}}{m_\NS v_\text{\tiny rel}/F_\text{\tiny dyn}}
    = \frac{r_\text{\tiny Hill}\,F_\text{\tiny dyn}}{m_\NS\,v_\text{\tiny rel}^2}\,,
    \label{eq:Pcap}
\end{equation}
where $F_\text{\tiny dyn} = 4\pi m_\NS^2\rho_\text{\tiny env}\,\mathcal{I}/v_\text{\tiny rel}^2$ is the  drag force introduced in Sec.~\ref{sec:param_NSNS} and evaluated at $R_*$ [see Eq.~\eqref{eq:dyntimescaleSSNSs}]~\cite{Ostriker:1998fa}.
Defining $\Gamma_\text{\tiny cap}=\Gamma_\text{\tiny enc}\,P_\text{\tiny cap}$ and inserting the disk density profile of Eq.~\eqref{densityprofile}, the capture timescale $\tau_\text{\tiny cap}=\Gamma_\text{\tiny cap}^{-1}$ is
\begin{align}
    \tau_\text{\tiny cap}
    &\approx 10^2\,{\rm s} \, \left(\frac{f_\text{\tiny gas}}{10^{-2}} \right)^{-1} \, 
    \!\left(\frac{N_\NS}{10}\right)^{-1} \nonumber \\
    & \times 
    \!\left(\frac{\mathbb{H}}{0.3}\right)^{4}
    \!\left(\frac{M_\BH/m_\NS}{100}\right)^{2}
    \!\left(\frac{\dot{M}_\BH/M_\BH}
                 {0.1\,{\rm s}^{-1}}\right)^{-1}\,.
    \label{eq:tcap}
\end{align} 
The capture time is controlled by a few competing scalings: it shortens for heavier clumps, $\tau_\text{\tiny cap}\propto(M_\BH/m_\NS)^{2}$, since a more massive perturber feels a stronger drag ($F_\text{\tiny dyn}\propto m_\NS^2$) and subtends a larger Hill cross-section; it shortens for denser residual gas, $\tau_\text{\tiny cap}\propto f_\text{\tiny gas}^{-1}$; and it shortens for thinner disks, $\tau_\text{\tiny cap}\propto\mathbb{H}^{4}$, which are denser at fixed accretion rate.

\begin{figure}[t!]
   \centering    \includegraphics[width=\columnwidth]{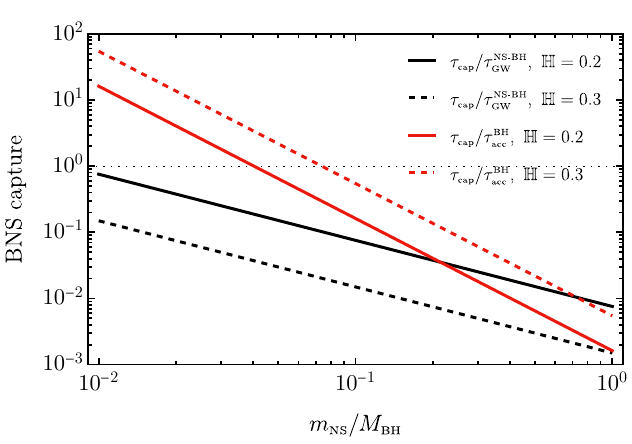}
   \caption{Comparison between the environmental-induced capture time and the timescales associated with GW merger with the central BH (black) and the environmental depletion induced by BH accretion (red), respectively. Solid and dashed lines refer to different values of the disk height $\mathbb{H}$.}
   \label{fig:capture}
\end{figure}

These trends underlie the behavior shown in Fig.~\ref{fig:capture}, where we plot the ratio of the capture timescale $\tau_\text{\tiny cap}$ to two competing timescales --- the GW-driven NS-BH coalescence time $\tau_{\GW}^{\NS-\BH}$ (black lines) and the BH accretion timescale $\tau_\text{\tiny acc}^\BH$ (red lines) --- as a function of the mass ratio $m_\NS/M_\BH$, for two representative values of the disk aspect ratio $\mathbb{H} = 0.2$ and $0.3$. Over the entire range displayed, $10^{-2}\lesssim m_\NS/M_\BH \lesssim 1$, both black curves lie below unity: gas-assisted capture always outpaces the GW-driven inspiral of an isolated clump into the central BH, so that NSs typically find a binary partner well before they can merge with the BH. The thickness dependence here is in fact the strongest of the two comparisons, $\tau_\text{\tiny cap}/\tau_{\GW}^{\NS-\BH}\propto\mathbb{H}^{-4}$, since $\tau_\text{\tiny cap}\propto\mathbb{H}^4$ while $\tau_{\GW}^{\NS-\BH}\propto R_*^4\propto\mathbb{H}^8$ when evaluated at the fragmentation radius: extrapolating below the range shown, direct inspiral would overtake capture only for $m_\NS\lesssim7\times10^{-3}\,M_\BH$ when $\mathbb{H} = 0.2$ (solid) and $m_\NS\lesssim2\times10^{-3}\,M_\BH$ when $\mathbb{H} = 0.3$ (dashed), i.e., well below the fragment masses of interest.
The comparison with $\tau_\text{\tiny acc}^\BH$ is more restrictive and reverses the ordering in $\mathbb{H}$, growing as $\tau_\text{\tiny cap}/\tau_\text{\tiny acc}^\BH\propto\mathbb{H}^{3}$ [since $\tau_\text{\tiny acc}^\BH = M_\text{\tiny disk}/\dot{M}_\BH \propto \mathbb{H}$, see Eq.~\eqref{eq:taccBH} evaluated at $R_*$]: these curves scale as $(m_\NS/M_\BH)^{-2}$ and cross unity at $m_\NS\simeq7\times10^{-2}\,M_\BH$ for the thicker disk ($\mathbb{H} = 0.3$, dashed) and at $m_\NS\simeq4\times10^{-2}\,M_\BH$ for the thinner one ($\mathbb{H} = 0.2$, solid), so that BH accretion depletes the surrounding gas faster than capture can assemble a binary only for the lightest clumps, while capture prevails for heavier ones. Taken together, the two comparisons leave a wide window, $m_\NS\gtrsim 0.1 \,M_\BH$, in which BNSs can form in a gas-rich environment before the disk is appreciably depleted.

These estimates assume a uniform density at $R_*$ 
and a relative velocity set by Keplerian shear 
alone. In practice, the stochastic nature of 
fragment creation at different radii means that 
fragments migrating inward at different rates can 
also undergo close encounters, with gas dissipation 
during such passages providing an additional 
capture channel beyond the Hill-sphere mechanism 
estimated above~\cite{Li:2022jbj}. 
Furthermore, mutual gravitational scattering and 
merger recoil kicks~\cite{Metzger:2024ujc} 
introduce additional velocity dispersion that 
would modify the capture cross-section. 
Notwithstanding these uncertainties, the analysis 
demonstrates that gas-assisted capture is a 
physically robust and, over a broad range of disk 
parameters, dominant formation channel for BNSs, complementary to the fission channel 
discussed in Sec.~\ref{sec:NS-NS}. The binaries 
formed through either capture channel are expected 
to be generally eccentric, as we quantify in 
Sec.~\ref{sec-eccentricity}.

Beyond binary capture, few-body scatterings among 
the $N_\NS\sim\mathcal{O}(10)$ fragments can eject 
single NSs or even compact BNSs entirely 
from the disk~\cite{Stone:2019, Lerner:2025dkd}. 
An ejected binary would subsequently merge in a 
much less gaseous environment, producing a 
near-vacuum GW chirp --- a 
qualitatively different signature from the 
non-vacuum waveforms discussed in 
Secs.~\ref{sec:NS-NS} and~\ref{sec:NS-BH}. 

\begin{figure}
\includegraphics[width=\linewidth]{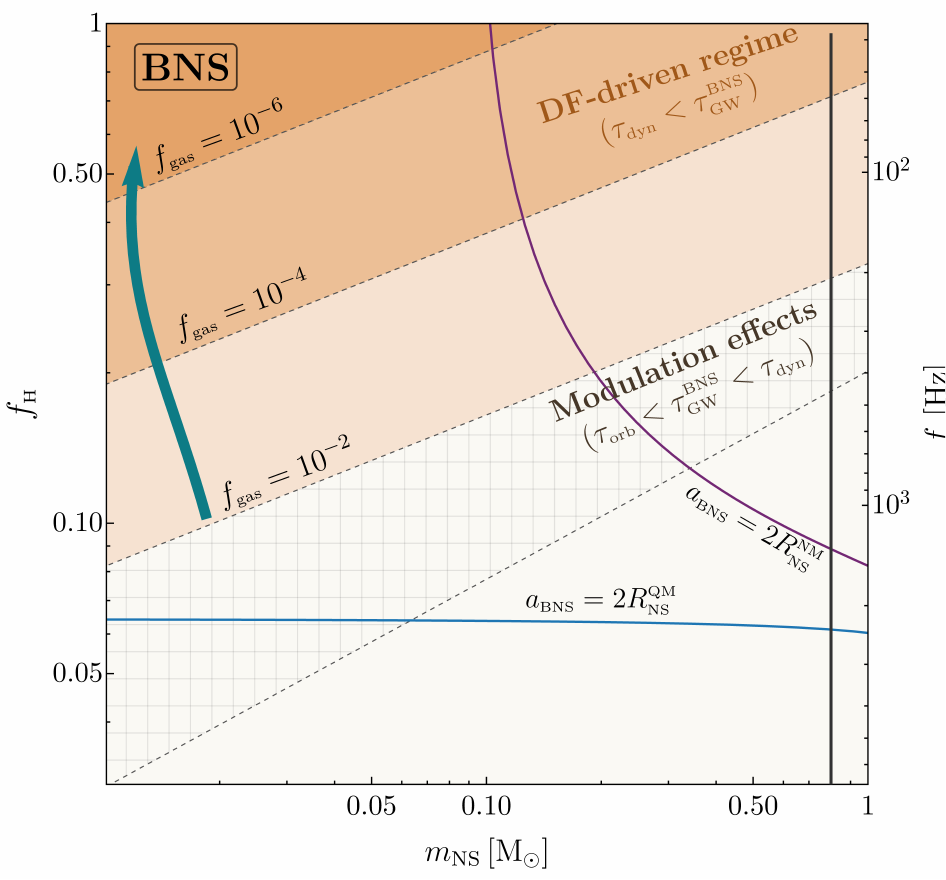}
    \caption{Parameter space for BNS mergers, but with the density at the edge of the disk before fragmentation fixed to be $\rho (R_*) = 10^{10}\,{\rm g\,cm^{-3}}$.
The orange regions identify BNS systems whose evolution is driven by dynamical friction, with smaller values of the density suppression factor $f_\text{\tiny gas}$ leading to smaller regions.
The beige region marks binaries that merge via GW-driven evolution, with the cross-hatched region corresponding to instances where the GW merger timescale exceeds the orbital period around the central BH, and the effects of Sec.~\ref{sec:modeffects} are relevant. The magenta and blue lines indicate the NS contact radius assuming an NM or QM EoS, respectively.}
\label{fig:param_app}
\end{figure}

\section{Effect of disk density on the BNS parameter space}
\label{app:NSNSdetails}
\noindent
In this appendix we collect additional material on the BNS stage. In Fig.~\ref{fig:param_app} we reconsider the parameter space of the left panel of Fig.~\ref{fig:param} for a denser disk edge, $\rho(R_*) = 10^{10}\,{\rm g\,cm^{-3}}$, two orders of magnitude above the fiducial value adopted in the main text. Raising $\rho(R_*)$ rescales the plane in three distinct ways. 

First, the fragmentation radius shrinks as $R_* \propto \rho(R_*)^{-1/3}$ [Eq.~\eqref{eq:Rstar}], and with it the Hill radius of Eq.~\eqref{eq:rHill_outer}; since the contact bound of Eq.~\eqref{eq:nstouch} scales as $f_\text{\tiny H} \propto R_\NS/R_*$, the region of binaries born already in contact moves upward by a factor $(10^{10}/10^{8})^{1/3} \simeq 4.6$, and the window of viable initial separations narrows accordingly at small $f_\text{\tiny H}$. 

Second, the entry frequency scales as $f_i \propto \rho(R_*)^{1/2}$ [Eq.~\eqref{eq:fi}] and is therefore a factor of ten larger, as displayed on the right vertical axis: these binaries begin their GW-driven inspiral well inside the sensitive band of ground-based detectors. 

Third, the residual gas density that sets the strength of dynamical friction is $\rho_\text{\tiny env} = f_\text{\tiny gas}\,\rho(R_*)$, so that the same dynamical-friction driven regions are recovered for smaller density suppression factors: the orange regions of Fig.~\ref{fig:param_app} correspond to $f_\text{\tiny gas} = 10^{-6}$ and $10^{-2}$, similarly to Fig.~\ref{fig:param}. 

The qualitative organization of the parameter space --- a dynamical-friction-dominated region at large $f_\text{\tiny H}$, a GW-dominated one at smaller $f_\text{\tiny H}$, and an intermediate strip in which the modulation effects of Sec.~\ref{sec:modeffects} are relevant --- is thus unchanged by the choice of disk density, which only shifts its location in the $(m_\NS, f_\text{\tiny H})$ plane.

\section{Accretion onto the compact objects} 
\label{app:accretion}
\noindent
In this appendix we estimate the timescales over which gas accretion modifies the masses of the compact objects involved in the two binary stages of the collapsar scenario, and compare them with the respective GW-driven coalescence times over the parameter space of interest. We consider in turn the BNS, the NS-BH binary, and accretion onto the central BH itself.  We close with a brief summary of the more detailed accretion study of Ref.~\cite{Lerner:2025dkd}, which provides complementary insight into the formation and mass growth of individual NS fragments before the binary phase begins.

\subsection{Accretion onto the BNS}
\begin{figure}[t!] 
\includegraphics[width=0.99\linewidth]{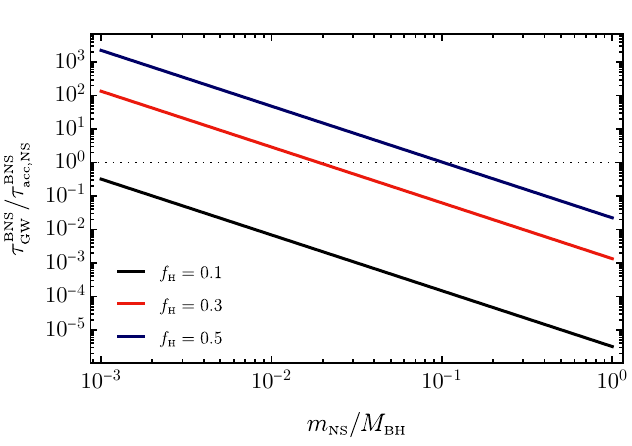}
    \caption{Ratio between the GW and NS accretion timescales during the BNS merger in terms of the mass ratio $q =m_\NS/M_\BH$, for different initial separations $f_\text{\tiny H}$, and fixing the remaining parameters as in Eq.~\eqref{tGWtaccNS-NSNS}.}
    \label{fig:accNSNS}
\end{figure}

\noindent
During the BNS inspiral, each NS moves through the residual gas of density $\rho_\text{\tiny env}= f_\text{\tiny gas}\, \rho (R_*)$ at an orbital velocity $v_\BNS\simeq\sqrt{2m_\NS/a_{\BNS}}$ set by the binary separation $a_{\BNS}$.  The relevant accretion process is Bondi-Hoyle accretion~\cite{Bondi:1944rnk,Antoni:2019pgq}, for which the accretion rate reads
\begin{equation}
    \dot{m}_\NS^{\BNS}
    = \frac{4\pi \lambda_\NS m_\NS^2 \rho_\text{\tiny env}}{v_\BNS^3}\,,
\end{equation}
where $\lambda_\NS$ is a dimensionless accretion parameter encoding the geometry of the flow, the EoS of the gas, and the gas viscosity.  We conservatively set $\lambda_\NS=1$ throughout, which provides an upper bound on the accretion rate.  We also neglect accretion feedback: in principle, the luminosity released by the accreting NS --- here carried primarily by neutrinos, since the disk is optically thick to photons --- could heat the surrounding gas and reduce the effective Bondi radius, together with the accretion efficiency~\cite{2011ApJ...739....2P, 2012ApJ...747....9P, 2013ApJ...767..163P}.  Given the short dynamical timescales and the efficient neutrino cooling of the collapsar disk, we expect this effect to be subdominant~\cite{1989ApJ...346..847C, 1991ApJ...376..234H,  Fryer:1995dr}, and our upper-bound estimate for $\dot{m}_\NS^{\BNS}$ remains conservative. 
The associated timescale, $\tau_\text{\tiny acc,NS}^{\BNS} = m_\NS/\dot{m}_\NS^{\BNS}$, reads
\begin{align}
    \tau_\text{\tiny acc,NS}^{\BNS}
    &\simeq 70 \,\text{s}\,
    \left(\frac{\mathbb{H}}{0.3}\right)^3 \left( \frac{f_\text{\tiny H}}{0.1} \right)^{-3/2} \nonumber \\
    & \times \left(\frac{f_\text{\tiny gas}}{10^{-2}}\right)^{-1}
    \left(\frac{M_\BH}{5\,M_\odot}\right) \left(\frac{\dot{M}_\BH}{M_\odot{\rm /s}}\right)^{-1}\,,
    \label{eq:taccNSNS}
\end{align}
where we have substituted the disk density profile of Eq.~\eqref{densityprofile} evaluated at the fragmentation radius $R_*$. Comparing it with the GW coalescence time $\tau_\GW^\BNS$ for an equal-mass BNS, shown in Eq.~\eqref{eq:GWtimescaleSSNSs}, one finds
\begin{equation}
    \frac{\tau_\GW^\BNS}
         {\tau_\text{\tiny acc,NS}^{\BNS}}
    \simeq \frac{f_\text{\tiny H}^{11/2}}{q^{5/3} \, Q_0^{8/3}} \left(\frac{\mathbb{H}}{0.3}\right)^5 
    \left(\frac{f_\text{\tiny gas}}{10^{-2}}\right)
    \left(\frac{\dot{M}_\BH}{M_\odot{\rm /s}}\right)^{-5/3}
    \,,
\label{tGWtaccNS-NSNS}
\end{equation}
where $q=m_\NS/M_\BH$ and $Q_0 \simeq 1$.  This ratio is plotted in Fig.~\ref{fig:accNSNS}, where we show that, for initial separations $f_\text{\tiny H} \lesssim 0.5$ and for realistic values of the NS mass ratio compared to the central BH ($m_\NS/M_\BH \gtrsim 10^{-2}$), the signal becomes GW-dominated in the detector band and NS mass growth by accretion is negligible throughout the inspiral.

It is instructive to also compare the accretion timescale of Eq.~\eqref{eq:taccNSNS} with the dynamical-friction hardening timescale derived in Sec.~\ref{sec:param_NSNS}, see Eq.~\eqref{eq:dyntimescaleSSNSs}. A direct computation gives $\tau_\text{\tiny dyn} \ll \tau_\text{\tiny acc,NS}^{\BNS}$, meaning that dynamical friction acts on the binary orbit on a timescale far shorter than the NS accretion timescale. As a consequence, the dominant environmental imprint on the GW phase is the orbital energy loss due to dynamical friction, while mass growth of the NS by accretion represents a subdominant correction that can be safely neglected in the waveform model of Sec.~\ref{sec:envtidalNSNS}.

A last comparison concerns the outer orbital motion. Evaluating the orbital period of the BNS about the central BH, $\tau_\text{\tiny orb} = 2\pi/\Omega$, at the fragmentation radius $R_*$ and dividing by the accretion time of Eq.~\eqref{eq:taccNSNS}, one finds
\begin{equation}
    \frac{\tau_\text{\tiny orb}}{\tau_\text{\tiny acc,NS}^{\BNS}}
    \simeq 10^{-3}\,
    \left(\frac{f_\text{\tiny H}}{0.1}\right)^{3/2}
    \left(\frac{f_\text{\tiny gas}}{10^{-2}}\right) \left(\frac{\dot{M}_\BH}{M_\odot \, {\rm s}^{-1}}\right)^{-1}\,.
    \label{eq:torbtaccNS}
\end{equation}
This ratio stays well below unity across the entire parameter space, so the NS mass is effectively frozen over each orbit around the central BH, and the hierarchical-triple modulations of Sec.~\ref{sec:modeffects}, which build up orbit by orbit, can be computed at fixed component masses.

Beyond accretion onto the NSs themselves, a related question is whether the residual gas that sources the dynamical-friction phase shift of Sec.~\ref{sec:envtidalNSNS} survives long enough for the inner binary to harden and merge within it. This is decided by comparing the two BNS hardening timescales --- dynamical friction, $\tau_\text{\tiny dyn}^\BNS$ [Eq.~\eqref{eq:dyntimescaleSSNSs}], and GW emission, $\tau_\GW^\BNS$ [Eq.~\eqref{eq:GWtimescaleSSNSs}] --- with the disk-depletion time $\tau_\text{\tiny acc}^\BH$ [Eq.~\eqref{eq:taccBH}], over which the central BH consumes the disk. 

\begin{figure}[t!]
   \centering
   \includegraphics[width=\columnwidth]{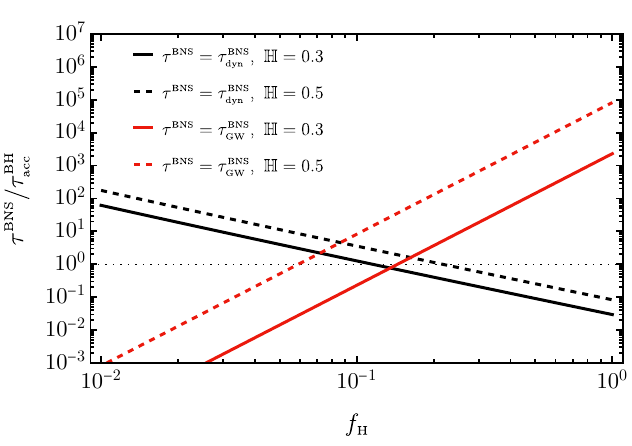}
   \caption{Ratio of the inner-BNS hardening timescales to the disk-depletion time $\tau_\text{\tiny acc}^\BH$, as a function of the Hill fraction $f_\text{\tiny H}$, for the dynamical-friction ($\tau_\text{\tiny dyn}^\BNS$, black) and GW-emission ($\tau_\GW^\BNS$, red) channels. Solid and dashed lines correspond to disk aspect ratios $\mathbb{H}=0.3$ and $\mathbb{H}=0.5$, respectively. Below the dotted line ($\tau^\BNS=\tau_\text{\tiny acc}^\BH$) the inner binary hardens before the central BH depletes the disk, so the residual gas is present during the merger and the dynamical-friction imprint applies; above it the disk is consumed first. Here, we have evaluated the accretion timescale at $R_*$ and assumed $m_\NS = 10^{-2} M_\BH$.}
   \label{fig:diskeatingBNS}
\end{figure}

In Fig.~\ref{fig:diskeatingBNS} we show the two ratios as a function of the Hill fraction $f_\text{\tiny H}$: the dynamical-friction ratio falls as $\tau_\text{\tiny dyn}^\BNS/\tau_\text{\tiny acc}^\BH \propto f_\text{\tiny H}^{-3/2}$ while the GW ratio grows as $\tau_\GW^\BNS/\tau_\text{\tiny acc}^\BH \propto f_\text{\tiny H}^{4}$, so that tight binaries harden by GW emission and wide ones by dynamical friction, and the operative --- shorter --- timescale is the lower envelope of the two curves. For thinner disks ($\mathbb{H}=0.3$, solid) this envelope lies below unity across the whole range: the inner binary always hardens before the BH depletes the disk, so the residual gas is present throughout the merger and the dynamical-friction imprint applies. For thicker disks ($\mathbb{H}=0.5$, dashed), which carry less mass and are consumed faster ($\tau_\text{\tiny acc}^\BH\propto\mathbb{H}^{-2}$) while hardening more slowly, an intermediate window $0.06\lesssim f_\text{\tiny H}\lesssim 0.2$ opens in which both hardening times exceed $\tau_\text{\tiny acc}^\BH$; there the BH clears the disk before the binary hardens, so the pair completes its inspiral by GW emission in a near-vacuum environment and the dynamical-friction phase shift is suppressed.
 
\subsection{Accretion onto the NS-BH binary}
\noindent
By the time the NS clump migrates to the inner disk and forms the NS-BH binary, the central BH may have accreted a significant fraction of the surrounding gas, as discussed in Sec.~\ref{sec:NS-BH}. Depending on the disk accretion rate $\dot{M}_\BH$ and the migration timescale, there exist regions of parameter space in which the inner disk density has been substantially depleted by the time the binary enters the GW band, rendering accretion onto the NS dynamically irrelevant. 

The prevailing regime is determined by comparing the disk-depletion time $\tau_\text{\tiny acc}^\BH$ [Eq.~\eqref{eq:taccBH}] with the Jeans crossing time $\tau_\text{\tiny J}$ [Eq.~\eqref{eq:tauJ}], the time for the clump to migrate from the fragmentation radius $R_*$ into the inner disk. 
When $\tau_\text{\tiny acc}^\BH < \tau_\text{\tiny J}$, the disk gas is substantially depleted before the NS-BH binary even forms: the density available for NS accretion upon arrival at the inner disk is strongly suppressed relative to the undepleted value, so our Bondi estimate below should be regarded as a firm upper bound on the true accretion rate. Conversely, when $\tau_\text{\tiny J} \ll \tau_\text{\tiny acc}^\BH$, the disk has not had time to drain and the full local density enters the Bondi formula.

In the latter case, for the NS-BH inspiral, the NS co-rotates with the disk material, so the relative velocity between the NS and the gas is subsonic and the relevant velocity entering the Bondi formula is the local sound speed $c_s = \mathbb{H}\,v_{\NS-\BH}$.  As in the BNS merger stage, we set $\lambda_\NS=1$ and neglect radiative feedback effects, so that the Bondi rate
\begin{equation}
    \dot{m}_\NS^{\NS-\BH}
    = \frac{4\pi \lambda_\NS m_\NS^2 \rho (R_*)}{c_s^3}\,,
\end{equation}
again provides a conservative upper bound on the accretion rate, in terms of the unsuppressed local density $\rho (R_*)$. 
The corresponding accretion timescale $\tau_\text{\tiny acc,NS}^{\NS-\BH} = m_\NS/\dot{m}_\NS^{\NS-\BH}$ evaluates to
\begin{align}
    \tau_\text{\tiny acc,NS}^{\NS-\BH} 
    \simeq 10^{-2} \, {\rm s} \,  \left(\frac{q}{10^{-2}} \right)^{-1}\left(\frac{\mathbb{H}}{0.3}\right)^6 
    \left(\frac{M_\BH}{5\,M_\odot}\right)\,.
    \label{eq:taccNSBH}
\end{align}
This timescale is very short --- of order tens of milliseconds for fiducial parameters --- meaning that, in principle, the NS could accrete significant mass during the late stages of the NS-BH inspiral. To assess whether accretion is competitive with migration, we compare $\tau_\text{\tiny acc,NS}^{\NS-\BH}$ to the Type~I migration timescale  estimated in Eq.~\eqref{eq:tau_mig},
\begin{equation}
    \frac{\tau_\text{\tiny acc,NS}^{\NS-\BH}}
         {\tau_\text{\tiny mig}}
    \approx 
   0.7 \, \mathbb{H}^2 \ll 1\,,
    \label{eq:acc_vs_mig}
\end{equation}
indicating that the NS accretes on a timescale far shorter than it takes to migrate by one orbital radius. Consequently, the NS grows rapidly by Bondi accretion as it spirals inward, gaining mass until it becomes massive enough to open a gap in the disk: see Eq.~\eqref{eq:Mgap} for the associated condition of gap opening.

Notice that this is a similar outcome of the results found in Ref.~\cite{Lerner:2025dkd}, which studies the earlier phase of Bondi-Hoyle mass growth of a single NS fragment as it migrates from the fragmentation radius $R_*$ toward the inner disk. Indeed, that work finds that the fragments grow in mass by various orders of magnitude via rapid accretion before gap opening halts further mass growth. This happens well before the fragment reaches the inner disk, so by the time binary dynamics become relevant the NS is already a cold, fully formed compact object whose mass is frozen at its gap-opening value.

\section{Tidal field of the central BH on the inner binary}
\label{sec:tidalfield}
\noindent
The modulations discussed in Sec.~\ref{sec:modeffects} of the main text arise from kinematic and propagation effects. They depend on the position of the inner binary along its orbit about the central BH and on the propagation of its radiation through the BH potential, but they do not include the dynamical influence of the BH on the inner binary. We consider this influence below.

\subsection{The secular tide}
\label{sec:tide_secular}
\noindent
Expanding the BH's potential about the inner binary's center of mass, the
monopole and dipole pieces cancel identically and the tidal quadrupole, averaged
over a circular coplanar inner orbit, is
$\langle V_\text{\tiny tid}\rangle = -M_\BH m_\NS a_\BNS^{2}/8a_{\BNS-\BH}^{3}$. Its effect on the dynamics can be parametrized by
\begin{equation}
  \tilde\varepsilon(f) \equiv
  \left(\frac{\Omega_{\BNS-\BH}}{\omega_\BNS}\right)^{2}
  = \frac{M_\BH}{a_{\BNS-\BH}^{3}\,(\pi f)^{2}}\,,
  \label{eq:epstilde}
\end{equation}
with $\omega_\BNS = \pi f$. At formation,
$\tilde\varepsilon = f_\text{\tiny H}^{3}/3$, and thereafter
$\tilde\varepsilon\propto f^{-2}$ decreases as the frequency increases during
inspiral. The perturbation is therefore largest at the greatest inner-binary
separation.

Following Ref.~\cite{Takatsy:2025bfk}, radial force balance with the outward
tidal force included gives
$\omega_\BNS^{2} = (2m_\NS/a_\BNS^{3})(1-\tilde\varepsilon/2)$. Re-expressing
the binding energy and the quadrupole luminosity at fixed frequency then gives
$E = E_\text{\tiny N}(1+\frac76\tilde\varepsilon)$ and
$\dot E = \dot E_\text{\tiny N}(1-\frac23\tilde\varepsilon)$. The tertiary therefore
increases the magnitude of the binding energy and decreases the quadrupole
luminosity at fixed frequency, where the subscript ``N'' denotes the leading-order (Newtonian) vacuum term. Energy
balance in the stationary-phase approximation gives
$\d^{2}\psi/\d f^{2} = (\d^{2}\psi_\text{\tiny N}/\d f^{2})
[1-\frac{5}{3}\tilde\varepsilon]$, and integrating twice yields
\begin{align}
  \delta\psi_\text{\tiny tert}(f)
  &= -\frac{100}{231}\,\psi_\text{\tiny N}(f)\,\tilde\varepsilon(f)\, \nonumber \\
  &= -\frac{25}{2464}\,\frac{M_\BH}{\pi^{2}a_{\BNS-\BH}^{3}}\,
     \frac{(\pi\mathcal{M}_cf)^{-5/3}}{f^{2}}\,,
  \label{eq:dpsi_tide}
\end{align}
where $\psi_\text{\tiny N} (f) = (3/128)(\pi\mathcal{M}_cf)^{-5/3}$ is the leading-order vacuum
phase. The relative correction scales as
$f^{-2}\propto v_\BNS^{-6}$ and is therefore formally a $-3$\,PN term. In addition,
$\delta\psi_\text{\tiny tert}(f)<0$ corresponds to a faster frequency evolution
and fewer accumulated cycles than in vacuum.

\begin{figure}[t]
  \centering
  \includegraphics[width=\columnwidth]{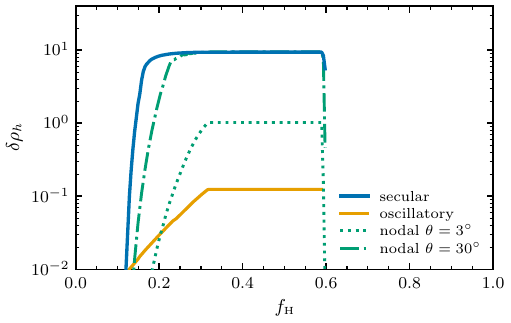}
  \caption{Detectability of the tidal field of the central BH against the Hill
  fraction at formation, at $m_\NS = 0.1\,\msun$, $M_\BH = 5\,\msun$ and
  $a_{\BNS-\BH,i} = 340\,R_g$ at LIGO. Blue: the secular term of Sec.~\ref{sec:tide_secular}; orange: the oscillatory
  term of Sec.~\ref{sec:tide_variation}; green: the nodal precession of Sec.~\ref{sec:tide_nodal}, for two different values of the tilt angle. }
  \label{fig:tide_fH}
\end{figure}

The $f^{-11/3}$ phase correction derived above is concentrated at the lowest frequencies reached by the source. Its observable contribution is therefore determined by the initial signal frequency, which depends on $f_\text{\tiny H}$ through Eq.~\eqref{eq:fi}, and by the detector's low-frequency cutoff. As shown in Fig.~\ref{fig:tide_fH}, the effect is dominant over most of the parameter region in which the triple radiates in band. For the same masses and separations, the residual is smaller in LIGO than in ET, because LIGO's minimum cutoff excludes the low-frequency part of the signal that contributes most strongly to this correction.

\begin{figure*}[t!]
\centering
\includegraphics[width=\textwidth]{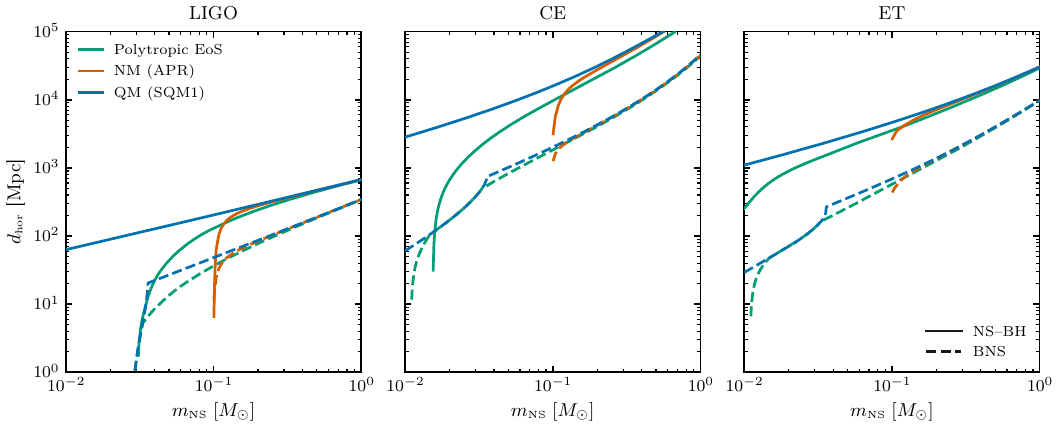}
\caption{\label{fig:eos_hor}
Horizon distance at SNR $=8$ as a function of NS mass for the NS-BH and
inner-BNS stages, for $M_\BH=5\,\msun$. The BNS has
$f_\text{\tiny H}=0.5$ and $a_{\BNS-\BH,i}=R_*=340\,R_g$, corresponding to
$\rho(R_*)=10^8\,{\rm g\,cm^{-3}}$; the NS-BH inspiral begins at $R_*$. The
three panels show LIGO, CE, and ET.}
\end{figure*}

\subsection{The oscillatory tidal sidebands}
\label{sec:tide_variation}
\noindent
The averaged potential used above discards the angle-dependent part of the tide, $-3M_\BH m_\NS r^{2}\cos[2(\varphi_\BNS-\varphi_{\BNS-\BH})]/8a_{\BNS-\BH}^{3}$, where $(r,\,\varphi_\BNS)$ locate the separation vector to the inner binary and $\varphi_{\BNS-\BH}$ the direction to the BH. For circular coplanar orbits this oscillates at $2(\omega_\BNS-\Omega_{\BNS-\BH})$, twice the synodic rate. The resulting forced motion is the classical variation in the lunar orbit about the Earth driven by its motion about the Sun~\cite{2012icm..book.....F}, of amplitude set by the same $\tilde\varepsilon$. To first order in $\tilde\varepsilon$, and in the slow-perturber limit $\Omega_{\BNS-\BH}\ll\omega_\BNS$, the radial separation oscillates with fractional amplitude $\tilde\varepsilon$, and the orbital phase acquires a variation of $11\tilde\varepsilon/8$ radians about uniform motion~\cite{2012icm..book.....F}:
\begin{align}
  r(t) &= a_\BNS\left\{1-\tilde\varepsilon
  \cos\!\left[2(\omega_\BNS-\Omega_{\BNS-\BH})t\right]\right\}\,, \nonumber \\
  \varphi_\BNS(t) &= \omega_\BNS t+\frac{11}{8}\tilde\varepsilon
  \sin\!\left[2(\omega_\BNS-\Omega_{\BNS-\BH})t\right]\,.
  \label{eq:variation_orbit}
\end{align}
So the Newtonian quadrupole $\mathcal{Q}\propto r^{2}e^{2i\varphi_\BNS}$, keeping only terms linear in $\tilde\varepsilon$ --- the \emph{slow-perturber approximation}, in which the direction to the BH changes slowly compared with the inner orbit --- splits into three contributions:
\begin{equation}
\begin{split}
  \mathcal{Q}\propto a_\BNS^{2}\Bigl[
  &e^{2i\omega_\BNS t}
  +\frac{3}{8}\tilde\varepsilon\,
   e^{i(4\omega_\BNS-2\Omega_{\BNS-\BH})t}\\
  &\;-\frac{19}{8}\tilde\varepsilon\,
   e^{2i\Omega_{\BNS-\BH}t}\Bigr]\,.
\end{split}
  \label{eq:variation_lines}
\end{equation}
The first term is the unperturbed GW. The other two are sidebands, the upper radiating at $4\omega_\BNS-2\Omega_{\BNS-\BH}\simeq4\omega_\BNS$ and the lower at $2\Omega_{\BNS-\BH}$. Defining $\hat h_\pm$ as the sideband strain amplitudes relative to the unperturbed GW, and using $h\propto\ddot {\mathcal{Q}}$,
\begin{equation}
  \hat h_{+} \simeq \frac{3}{2}\,\tilde\varepsilon\,,
  \qquad
  \hat h_{-} \simeq -\frac{19}{8}\,\tilde\varepsilon^{2}\,.
  \label{eq:variation_amp}
\end{equation}
The upper line, at $f_{+} = 2f-f_{\BNS-\BH}$ is therefore more observable than the lower line at $f_{-} = f_{\BNS-\BH}$, because two time derivatives in $h\propto\ddot{\mathcal{Q}}$ suppress its radiation by a further $\tilde\varepsilon$. 

The signal-to-noise ratio of each component is $\rho_i = \rho_0  \langle\hat h_i^{2}S_n(f)/S_n(f_i)\rangle_W$. This expression evaluates the detector noise at the frequency $f_i$ of each component within the weighted average of Eq.~\eqref{eq:weight}. Considering only the dominant upper sideband, $\delta\rho_h = \rho_+$, one can show that, for $a_{\BNS-\BH,i}=340\,R_g$, it is effectively inaccessible to LIGO near the horizon. The sidebands are strongest at formation, when the inner binary has its widest orbit and the modulation amplitude is largest, since $\hat{h}_+ \propto \tilde{\varepsilon} \propto f^{-2}$. At this outer separation, however, the formation stage falls below LIGO's low-frequency cutoff. LIGO therefore observes the binary only later in its evolution, and hence the sideband amplitude has already decreased substantially. In other words, LIGO's seismic wall removes precisely the widest-orbit stage, where the sideband signature is strongest. For more compact outer orbits, $a_{\BNS-\BH,i}\lesssim100\,R_g$, the formation stage remains above LIGO's low-frequency cutoff, allowing the stronger sidebands to be recovered. ET, by contrast, has a $\sim2\,{\rm Hz}$ low-frequency cutoff. Therefore, ET retains the formation stage and it can detect the strong sidebands at large $f_\text{\tiny H}$ even for $a_{\BNS-\BH,i}=340\,R_g$.
At $m_\NS = 0.1\,\msun$, $M_\BH = 5\,\msun$, $a_{\BNS-\BH,i} = 340\,R_g$ and $f_\text{\tiny H} = 0.3$, the lower line does not contribute in LIGO and CE because $f_{\BNS-\BH} = 2.1$\,Hz lies below both low-frequency cutoffs. Since $\tilde\varepsilon\propto f_\text{\tiny H}^{3}$, the sidebands are detectable only for binaries with the largest initial separations (Fig.~\ref{fig:tide_fH}). When detectable, their separation $2f_{\BNS-\BH}$ measures the outer orbital frequency through the conservative dynamics, independently of the Doppler modulation.

The sidebands above assume a prograde inner binary. For a retrograde inner orbit, the tide instead drives the binary at $2(\omega_{\BNS}+\Omega_{\BNS-\BH})$, shifting the upper sideband to $f_{+}=2f+f_{\BNS-\BH}$ with the same leading-order amplitude. Thus, the sign of the offset from $2f$ directly identifies whether the inner orbit is prograde or retrograde relative to the outer orbit.

Our treatment only takes into account the quadrupole of the external field. Because the octupole vanishes for an equal-mass inner binary, the leading neglected multipole is the hexadecapole, smaller by $(a_\BNS/a_{\BNS-\BH})^{2} = f_\text{\tiny H}^{2}\,(2m_\NS/3M_\BH)^{2/3}$ at formation.  In addition, we retain only the leading-order correction in $\tilde\varepsilon$. In principle, a higher-order expansion in $\tilde\varepsilon$ would provide a series of sidebands with different relative strengths.

\subsection{Nodal precession of the inner plane}
\label{sec:tide_nodal}
\noindent
Both terms above assumed the two orbits to be coplanar. If the inner plane is instead tilted by $\theta$, the same quadrupole tide torques the inner angular momentum, which regresses about the outer orbit normal at the classical lunar nodal rate~\cite{2012icm..book.....F}
\begin{equation}
  \Omega_\text{\tiny node} = \frac{3}{4}\,\tilde\varepsilon\,
  \omega_\BNS\cos\theta\,.
  \label{eq:nodal_rate}
\end{equation}
The same Eq.~\eqref{eq:nodal_rate}, when applied to the Moon, can be used to explain most of its $\sim18$-yr nodal regression period. At $m_\NS = 0.1\,\msun$, $M_\BH = 5\,\msun$, $a_{\BNS-\BH,i} = 340\,R_g$ and $f_\text{\tiny H} = 0.3$, it gives $\Omega_\text{\tiny node} = 0.48$\,rad\,s$^{-1}$: $26^\circ$ per outer orbit, a $13$\,s precession period, and $583$ cycles between band entry at $21.7$\,Hz and contact at $91.8$\,Hz.

A plane that precesses through $N_p$ cycles imprints a geometric phase on the $\ell=|m|=2$ mode,
\begin{equation}
\begin{split}
  \delta\psi(f) &= 4\pi\left[1-\cos\theta\right]N_p(f)\,,\\[3pt]
  N_p(f) &= \frac{1}{2\pi}\int_0^{t(f)}\!\Omega_\text{\tiny node}\,\d t\,.
\end{split}
  \label{eq:nodal_phase}
\end{equation}
For the Newtonian chirp the integral can be evaluated analytically, $N_p(f) \simeq (f_\text{\tiny H}^{3}f_{\BNS,i} \tau_{\BNS,i}\cos\theta/11) [1-(f_{\BNS,i}/f)^{11/3}]$, since $R_\tau \ll 1$, with $f_{\BNS,i}$ the entry frequency and $\tau_{\BNS,i}$ the time left in band there.
The residual horizon SNR crosses $\delta\rho_h = 3$ at $\theta = 6^\circ$ in LIGO, so a tilt of a few degrees is enough for the tide to leave a measurable imprint on the orientation of the inner plane as well as on the phase. Its dependence on the initial separation is included in Fig.~\ref{fig:tide_fH}. 

\section{EoS dependence of GW observables}
\label{app:EoS}
\noindent
\noindent
In Secs.~\ref{sec:detection} and~\ref{Sec:detNSBH} we determined the detectability of the inner BNS and subsequent NS-BH merger, respectively. In Sec.~\ref{sec:modeffects} we studied the five modulations induced by the outer orbit, while Secs.~\ref{sec-eccentricity} and~\ref{sec:NS-BH_ecc} considered the detectability of eccentricity in the two merger stages. Those baseline results adopted the polytropic EoS of Eq.~\eqref{eq:Rpoly}. Here we discuss how the results change when we consider instead the NM (APR) and QM (SQM1) models introduced in Sec.~\ref{sec:properties}.

The EoS affects these observables in two distinct ways. Its stable mass range determines whether a source exists at all, while its mass-radius relation determines whether a detached BNS can form and how its inspiral terminates (either through contact or tidal disruption). Unless stated otherwise, the BNS calculations use $M_\BH=5\,\msun$, $f_\text{\tiny H}=0.5$, and an outer separation $a_{\BNS-\BH,i}=R_*=340\,R_g$. From Eq.~\eqref{eq:Rstar}, this fragmentation radius corresponds to $\rho(R_*)=10^8\,{\rm g\,cm^{-3}}$. The NS-BH inspiral likewise begins at $R_*$. These choices isolate the EoS dependence at fixed disk and orbital properties. 

Note that the Fisher measurability analysis of Secs.~\ref{sec:FisherI} and~\ref{sec:FisherII} was performed for the NM and QM EoSs. In Appendix~\ref{app:EoS_meas} we complement those results by considering the polytropic EoS of Eq.~\eqref{eq:Rpoly}, which is defined over the full subsolar range, for both the BNS and NS-BH channels.

\subsection{Detectability}
\label{app:EoS_det}
\noindent
For the BNS channel, the formation-contact condition in Eq.~\eqref{eq:nstouch_num} determines whether a detached binary exists, and the contact frequency in Eq.~\eqref{eq:fRocheBNS_gen} sets the end of the inspiral. The entry frequency instead depends on the Hill geometry and is independent of the stellar radius. For the NS-BH channel, the radius enters through the Roche frequency, with the signal ending at the smaller of the tidal-disruption and ISCO frequencies. Figure~\ref{fig:eos_hor} compares the resulting horizons for both stages.

The horizons of both channels increase with $m_\NS$ because the larger chirp mass raises the signal amplitude. At fixed NS mass, the larger chirp mass of the NS-BH system makes it detectable to a greater distance than the BNS. The EoS-dependent separation is largest at low mass, where contact or disruption can remove a substantial part of the detector band.

For the BNS, the polytropic and QM curves share the same low-mass endpoint in LIGO. At this boundary the central BH unbinds the inner pair before the different stellar radii can produce distinct contact frequencies in band.
Above the boundary, the more compact QM stars reach contact later and retain more of the inspiral, whereas the radius of a low-mass polytropic star causes earlier contact. The lower frequency cutoffs of CE and ET extend both branches to smaller masses. The NM branch behaves differently: it begins near the minimum stable mass in every detector, because this is a physical boundary rather than a detector boundary. Immediately above it, the NM radius decreases rapidly with mass, moving the contact frequency through the detector band and producing the sharp rise of the horizon. At larger masses, the termination frequencies lie above the frequencies carrying most of the SNR, and the three BNS horizons converge.

\begin{figure}[t!]
\centering
\includegraphics[width=\columnwidth]{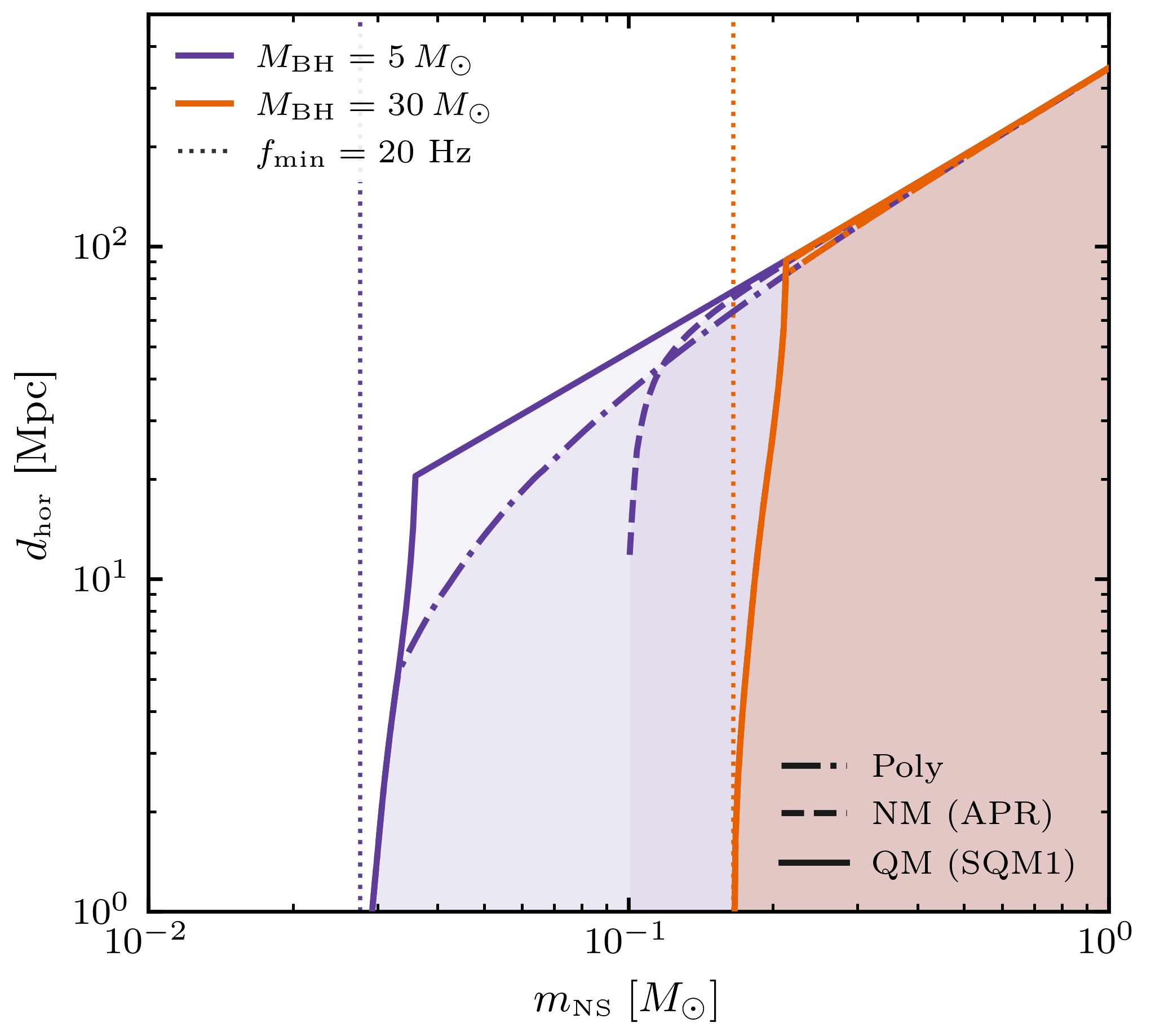}
\caption{\label{fig:eos_hor_M} 
LIGO horizon of the inner BNS for $M_\BH=5\,\msun$ and $M_\BH=30\,\msun$, with $f_\text{\tiny H}=0.5$ and the outer separation set to the fragmentation radius $R_*$ obtained from $\rho(R_*)=10^8\,{\rm g\,cm^{-3}}$. The NS masses at which the unbinding frequency crosses the LIGO low-frequency cutoff are also shown. For the heavier central BH, the same dynamical boundary controls all three EoSs. The shaded region below each horizon curve marks where a source is detectable ($d_\text{\tiny L}<d_\text{\tiny hor}$).
}
\end{figure}

Figure~\ref{fig:eos_hor_M} separates the structural and dynamical origins of the low-mass boundary. For $M_\BH=5\,\msun$, the polytropic and QM horizons begin at the common unbinding boundary, while stellar stability and contact move the NM boundary to larger mass. Increasing the central BH mass moves the unbinding boundary upward until it controls all three EoSs. Both the endpoint and the termination frequency then become independent of the stellar radius.
A detection below the NM stability limit would therefore exclude the NM branch. Detectability alone would not distinguish the QM model from the polytropic benchmark, although the latter is not a physical model of a cold star at these masses.

For the NS-BH stage, the QM model gives the largest low-mass horizon, because its small radius places the Roche frequency above the most sensitive part of the detector band. The polytropic radius grows toward lower mass, lowering the Roche frequency and shortening the observable signal. The NM horizon appears only above the minimum stable mass and rises rapidly as the stellar radius decreases. Thus, lowering the detector cutoff moves the polytropic low-mass boundary, but not the NM stability boundary. The QM branch remains in band even at lower masses because its disruption frequency remains high.

This ordering follows directly from the mass-radius relations. For the polytropic EoS, $R_\NS\propto m_\NS^{-1/3}$ and $f_\text{\tiny Roche}^\text{\tiny NS-BH}\propto m_\NS$. The NM Roche frequency likewise rises rapidly away from the minimum stable mass. For the approximately constant-density QM model, $R_\NS\propto m_\NS^{1/3}$, so the Roche frequency is nearly independent of mass. Once the Roche frequency lies above the part of the band carrying most of the SNR, however, further stellar compactness no longer increases the horizon, and the EoS curves converge.

The tidal disruption edge can be observed only if it occurs before the ISCO.
For a nonspinning BH, the critical central mass is
\begin{equation}
    M_{\BH,*} \simeq 67\,\msun
    \left(\frac{0.1\,\msun}{m_\NS}\right)^{1/2}
    \left(\frac{R_\NS}{34\,{\rm km}}\right)^{3/2},
    \label{eq:Mstar_quench}
\end{equation}
above which the ISCO terminates the signal before tidal disruption. The polytropic and NM disruption edges remain accessible at low NS mass over much of the central-BH mass range, although the ISCO preempts them for sufficiently large component masses. For QM stars, the Roche frequency is generally above the ISCO frequency, and the waveform therefore loses its radius-dependent disruption edge. Among the physical stellar models considered here, an observable disruption cutoff is consequently most useful for identifying the NM branch. A rapidly spinning central BH raises the ISCO frequency and can restore access to the QM disruption edge over a wider mass range.

\subsubsection*{Eccentricity detectability}
\begin{figure}[t!]
\centering
\includegraphics[width=\columnwidth]{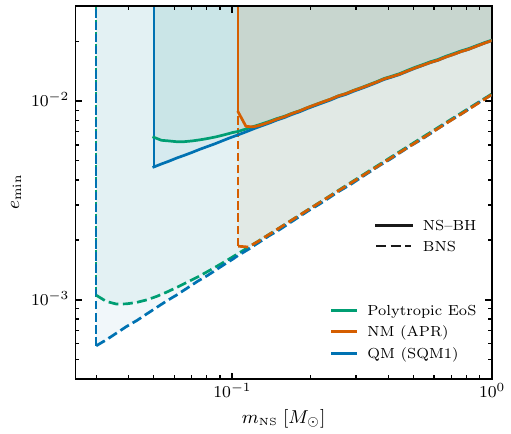}
\caption{\label{fig:eos_ecc}
Minimum resolvable initial eccentricity as a function of NS mass in LIGO, defined by $\delta\rho=3$ at the SNR $=8$ horizon. The BNS calculation uses $M_\BH=5\,\msun$, $f_\text{\tiny H}=0.5$, and $a_{\BNS-\BH,i}=R_*=340\,R_g$; the NS-BH calculation uses $M_\BH=5\,\msun$ and begins at the same fragmentation radius.}
\end{figure}

\noindent
In Sections~\ref{sec-eccentricity} and~\ref{sec:NS-BH_ecc} we showed how the eccentric contribution to the phase is accumulated over the in-band inspiral. In Fig.~\ref{fig:eos_ecc} we compare the minimum initial eccentricity $e_{\min}$ distinguishable from a circular waveform at the detection horizon for different EoS models. Both the BNS and NS-BH calculations use the same fragmentation radius adopted in the horizon comparison above.

For both merger stages, $e_{\min}$ generally increases with NS mass because a lighter binary spends more cycles in band and accumulates a larger eccentric phase correction. At fixed NS mass, the equal-mass BNS has a smaller chirp mass and more cycles than the NS-BH system, giving a lower eccentricity threshold. The EoS changes this threshold mainly near the low-mass boundaries.
The high termination frequency of the compact QM stars retains more of the inspiral and lowers $e_{\min}$ relative to the polytropic result. Near the NM stability boundary, the large radius shortens the observable interval and produces the turn in the NM curve. At larger NS masses the lower part of the band controls the eccentric phase accumulation, and the EoS curves approach one another. In summary, the estimates of eccentricity detectability presented in the main text are less sensitive to the EoS than the conditions for the existence of the lowest-mass sources and their observable horizon.

\subsection{Triple modulations}
\label{app:EoS_mod}
\begin{figure}[t!]
\centering
\includegraphics[width=\columnwidth]{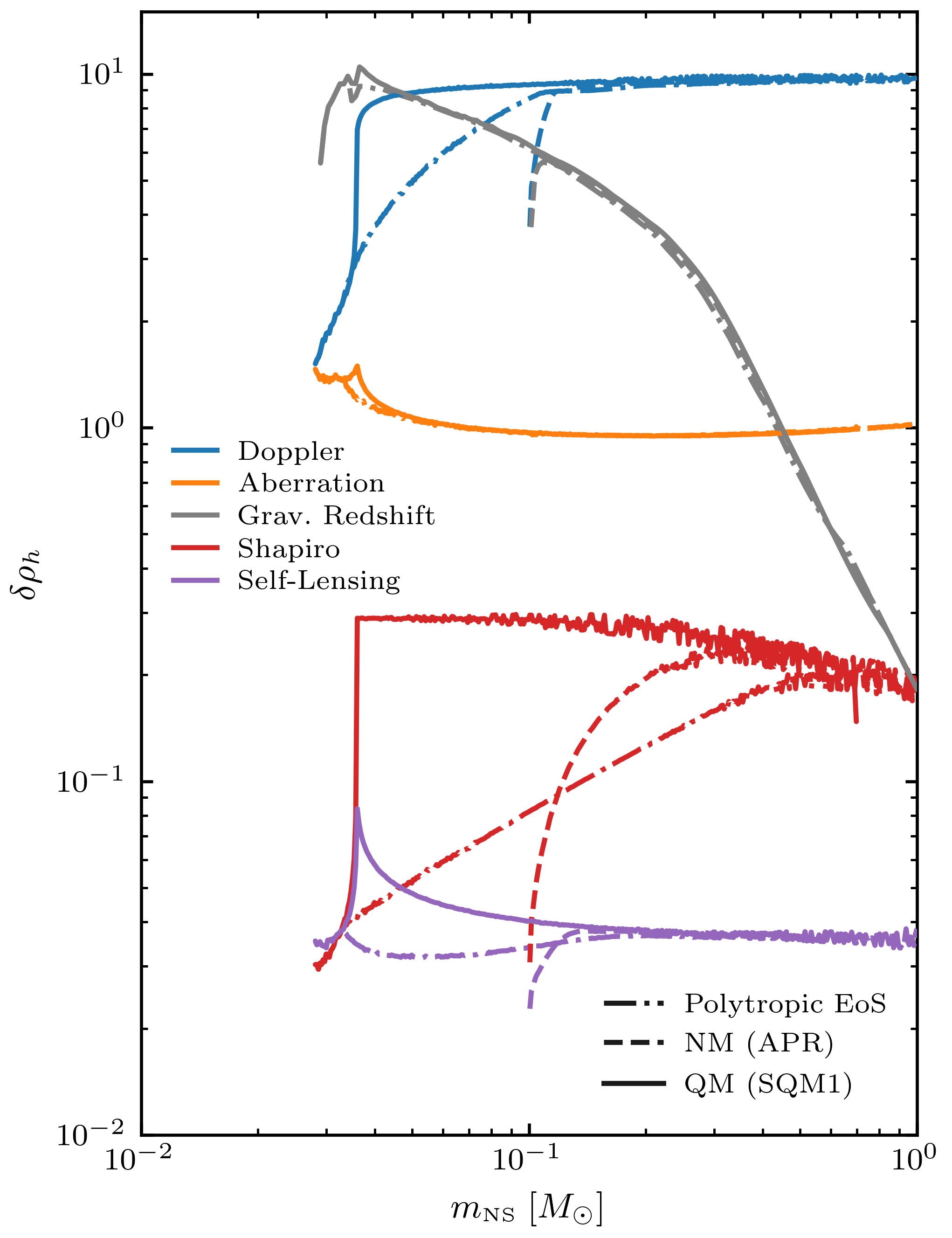}
\caption{\label{fig:eos_mod}
Horizon distinguishability $\delta\rho_h$ of the five modulation effects as a function of NS mass in LIGO, for $M_\BH=5\,\msun$, $a_{\BNS-\BH,i}=R_*=340\,R_g$, corresponding to $\rho(R_*)=10^8\,{\rm g\,cm^{-3}}$, with $f_\text{\tiny H}=0.5$ and $\iota=45^\circ$. The distinguishability threshold is $\delta\rho_h=3$.
}
\end{figure}

\noindent
The modulation indices derived in Sec.~\ref{sec:modeffects} are determined by the outer orbit and do not contain $R_\NS$. The EoS nevertheless changes the range of allowed systems and the frequency interval over which each modulation is accumulated. Figure~\ref{fig:eos_mod} isolates this dependence at fixed outer geometry.

\begin{figure*}[t!]
\centering
\includegraphics[width=0.49\linewidth]{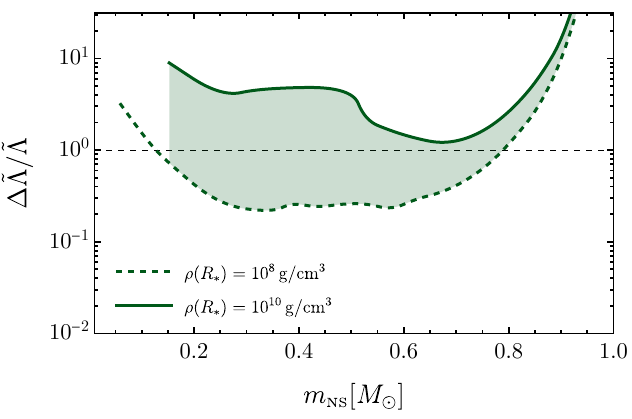}
\includegraphics[width=0.49\linewidth]{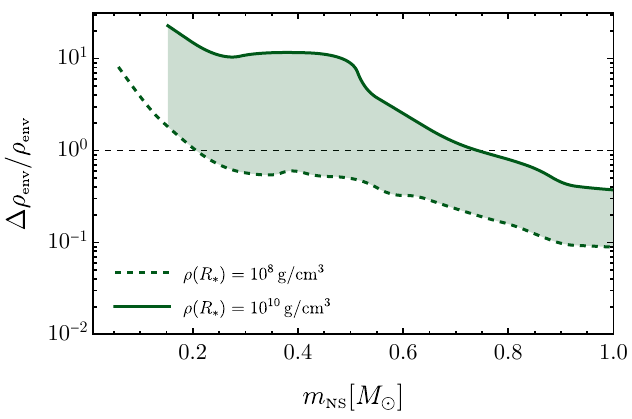}
\caption{Relative errors on the effective tidal deformability $\tilde\Lambda$ (left) and on the environmental density $\rho_\text{\tiny env}$ (right) in the BNS merger phase, for the polytropic EoS of Eq.~\eqref{eq:Rpoly}, as functions of the NS mass. The results are obtained from the Fisher analysis of Sec.~\ref{sec:FisherI} for ET at $d_\text{\tiny L}=100\,{\rm Mpc}$, with $M_\BH=5\,\msun$ and $f_\text{\tiny gas}=10^{-4}$; the initial separation follows Eq.~\eqref{eq:a0NSNS}, so that the entry frequency $f_i$ is set by the value of $f_\text{\tiny H}$ marking entry into the modulation region. Solid and dashed lines correspond to the outer-disk densities $\rho(R_*)=10^{10}$ and $10^{8}\,{\rm g\,cm^{-3}}$, respectively. The horizontal dashed line marks the threshold above which the corresponding parameter is not measurable.}
\label{fig:eos_meas_bns}
\end{figure*}

The Doppler modulation exceeds the distinguishability threshold throughout the support of every EoS in Fig.~\ref{fig:eos_mod}. It approaches the decorrelation regime of Eq.~\eqref{eq:rho_J0_limits}, so extending the upper frequency of the signal cannot substantially increase $\delta\rho_h$. The EoS therefore changes the Doppler result mainly through the low-mass support: unbinding sets the common polytropic and QM onset, whereas stellar stability sets the NM onset.
The gravitational redshift signal shows a different mass dependence. It is distinguishable for the lighter systems and decreases rapidly with mass because the secular change in the outer gravitational potential accumulates most strongly during a long, slowly chirping inspiral. The three EoSs give nearly the same result where their mass ranges overlap; their principal effect is to determine whether the low-mass systems that provide the strongest redshift signal can exist and remain in band. The choice $f_\text{\tiny H}=0.5$ does not show the strong Hill-fraction dependence derived in Sec.~\ref{sec:redshift} and Appendix~\ref{app:mod_redshift}. The observable redshift is generated by the in-band evolution of the outer potential, rather than its constant part, and rises steeply with $f_\text{\tiny H}$ through the outer-to-inner inspiral-rate ratio. This geometric dependence can change the redshift distinguishability by orders of magnitude without changing the EoS ordering shown here.

Aberration is nearly independent of both mass and EoS for these parameters because the outer orbital speed is fixed, and it remains below threshold at the chosen inclination. This conclusion applies to the displayed separation; at smaller outer separations, the compact QM branch can occupy configurations excluded by formation contact for the other EoSs, where the larger outer orbital speed can make aberration stronger.
Shapiro delay remains below threshold, but it has a strong NS-mass dependence that is distinct from aberration. Equation~\eqref{eq:rho_sha} gives $\delta\rho_h^\text{\tiny Sha}\propto\sqrt{\langle f^2\rangle_W}$, so the effect is controlled by the high-frequency end of the observable signal. For the polytropic and NM models, the large low-mass radii place the termination frequency near the detector floor; as the mass increases and the radius decreases, the growing upper frequency raises the Shapiro distinguishability substantially. The compact QM branch retains a high termination frequency and therefore reaches a larger Shapiro signal soon after its low-mass onset. Once the band extends beyond the detector's most sensitive frequencies, the curves flatten or turn over and approach one another. Thus, the Shapiro signal varies strongly with $m_\NS$, even though it never crosses the threshold in the configuration shown.

Self-lensing is the smallest of the five effects, and changes little with mass
or EoS over their common support. Its stronger dependence is on the outer
separation, as discussed in Sec.~\ref{sec:selflensing} and Appendix~\ref{app:mod_lensing}. 

Finally, the EoS does not change the angular dependences in Fig.~\ref{fig:rho_diff_iota}: Doppler weakens toward face-on, aberration strengthens, and gravitational redshift is independent of inclination. Their complementarity therefore persists for the common-mass systems in which all three EoSs are represented.

\begin{figure*}[t!]
\centering
\includegraphics[width=0.49\linewidth]{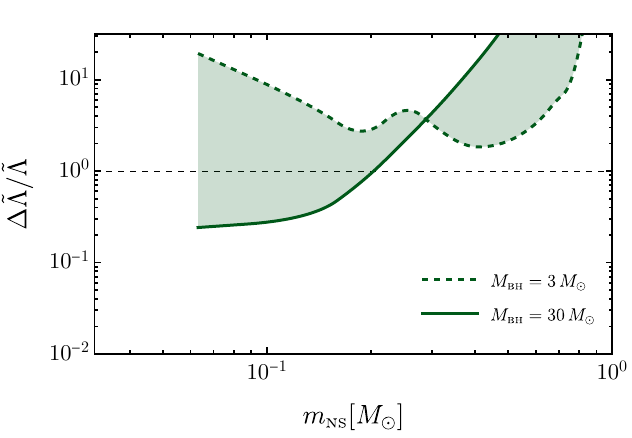}
\includegraphics[width=0.49\linewidth]{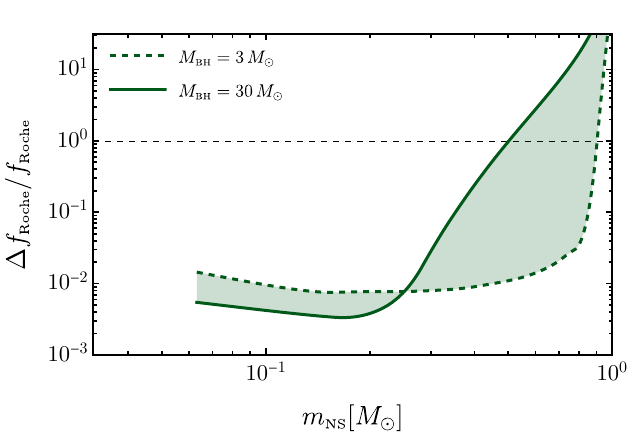}
\caption{Relative errors on the effective tidal deformability (left) and on the tidal-disruption frequency $f_\text{\tiny TD}$ (right) in the NS-BH merger phase, for the polytropic EoS of Eq.~\eqref{eq:Rpoly}, as functions of the NS mass. The results are obtained from the Fisher analysis of Sec.~\ref{sec:FisherII} for ET at $d_\text{\tiny L}=100\,{\rm Mpc}$. Solid and dashed lines correspond to central BH masses $M_\BH=30$ and $3\,\msun$, respectively. The horizontal dashed line marks the threshold above which the corresponding parameter is not measurable.}
\label{fig:eos_meas_nsbh}
\end{figure*}

\subsection{Measurability of model parameters}
\label{app:EoS_meas}
\noindent
The Fisher analysis of Secs.~\ref{sec:FisherI} and~\ref{sec:FisherII} was carried out for the NM and QM EoSs. Here we repeat it for the polytropic mass-radius relation of Eq.~\eqref{eq:Rpoly}, adopted as the fiducial benchmark throughout the main text, which extends smoothly across $m_\NS\in[0.01,1]\,\msun$. We focus on the observables most sensitive to the stellar structure (the effective tidal deformability $\tilde\Lambda$ and the environmental density $\rho_\text{\tiny env}$ for the BNS stage; $\tilde\Lambda_\text{\tiny NS-BH}$ and the tidal-disruption frequency $f_\text{\tiny TD}$ for the NS-BH stage), as the remaining parameters (chirp mass, mass ratio, SNR, and cycle count) follow the same trends discussed in Secs.~\ref{sec:FisherI} and~\ref{sec:FisherII}. All results are shown for ET at the benchmark distance $d_\text{\tiny L}=100\,{\rm Mpc}$.

In Fig.~\ref{fig:eos_meas_bns} we show the relative errors on $\tilde\Lambda$ (left) and $\rho_\text{\tiny env}$ (right) for the BNS stage, at $M_\BH=5\,\msun$ and $f_\text{\tiny gas}=10^{-4}$, for the two representative outer-disk densities $\rho(R_*)=10^8$ and $10^{10}\,{\rm g\,cm^{-3}}$. As in Sec.~\ref{sec:FisherI}, the initial separation is fixed by Eq.~\eqref{eq:a0NSNS}, i.e.\ by the value of $f_\text{\tiny H}$ at which the binary enters the modulation region, which in turn sets the entry frequency $f_i$. The curves therefore begin only above a minimum NS mass: because the contact frequency $f_\text{\tiny Roche}^\BNS\propto m_\NS$ [Eq.~\eqref{eq:fRoche_num}] falls toward low mass whereas $f_i$ does not, the two cross at a minimum mass below which $f_i>f_\text{\tiny Roche}^\BNS$, so that the pair is born already in contact and no inspiral is observed. The polytrope otherwise reproduces the qualitative picture found for the NM and QM EoSs in Fig.~\ref{fig:measurabilityNSNS2}. For the lower disk density, $\Delta\tilde\Lambda/\tilde\Lambda$ falls below unity across an intermediate mass band, reaching $\simeq0.2$ near $m_\NS\simeq0.3-0.5\,\msun$, and rising above unity for $m_\NS\gtrsim0.8\,\msun$, where the steeply decreasing polytropic deformability $\Lambda\propto m_\NS^{-20/3}$ and the shorter inspiral erode the tidal signal. For the denser disk the tidal deformability is essentially unmeasurable ($\Delta\tilde\Lambda/\tilde\Lambda\gtrsim1$ throughout), as the smaller fragmentation radius shortens the in-band inspiral. The environmental density follows the opposite mass trend: because the $-5.5$\,PN dynamical-friction dephasing of Eq.~\eqref{eq:deltaDF} grows with the component mass, $\Delta\rho_\text{\tiny env}/\rho_\text{\tiny env}$ decreases with $m_\NS$, so that for $\rho(R_*)=10^8\,{\rm g\,cm^{-3}}$ the density is measurable for $m_\NS\gtrsim0.2\,\msun$, improving to $\simeq0.1$ toward $m_\NS\simeq1\,\msun$, while for $\rho(R_*)=10^{10}\,{\rm g\,cm^{-3}}$ it is measurable only for the heaviest systems, $m_\NS\gtrsim0.75\,\msun$.

In Fig.~\ref{fig:eos_meas_nsbh} we show analogous results for the NS-BH stage, for central BH masses $M_\BH=3$ and $30\,\msun$. As in the main text, the tidal deformability is strongly suppressed by the mass ratio, $\tilde\Lambda_\text{\tiny NS-BH}\propto q^4\Lambda$ [Eq.~\eqref{eq:Ltilde_NSBH}], and is largely unmeasurable through the inspiral phase: for the lighter central BH  we find that $\Delta\tilde\Lambda/\tilde\Lambda\gtrsim2$ over the whole range, while for $M_\BH=30\,\msun$ the rapidly growing low-mass polytropic deformability partially offsets the suppression --- aided by the larger chirp mass and SNR of the heavier system --- and brings $\Delta\tilde\Lambda/\tilde\Lambda$ below unity only in a narrow window where $m_\NS\lesssim0.2\,\msun$. The finite-size information is instead carried by the tidal-disruption frequency: $\Delta f_\text{\tiny TD}/f_\text{\tiny TD}$ remains at the sub-percent to percent level over most of the mass range, confirming that the Roche cutoff is an exquisitely measurable, EoS-sensitive observable for the polytrope, and not just for the NM and QM models. Its reach is set by the interplay between the Roche and ISCO frequencies: for $M_\BH=3\,\msun$ disruption always precedes the ISCO and $f_\text{\tiny TD}$ is well measured up to $m_\NS\simeq0.7\,\msun$, whereas for $M_\BH=30\,\msun$ the ISCO preempts disruption above $m_\NS\simeq0.2\,\msun$ [Eq.~\eqref{eq:Mstar_quench}], degrading the measurement and confining sub-unity errors to lighter NSs. The polytropic analysis thus reproduces the central conclusion of the NS-BH measurability study (the tidal deformability is unmeasurable through the inspiral phase, and the tidal-disruption cutoff is the leading EoS discriminant) while extending it continuously down to the lowest subsolar masses, below the NM stability floor.

\begin{figure*}[t!]
  \centering
  \includegraphics[width=\textwidth]{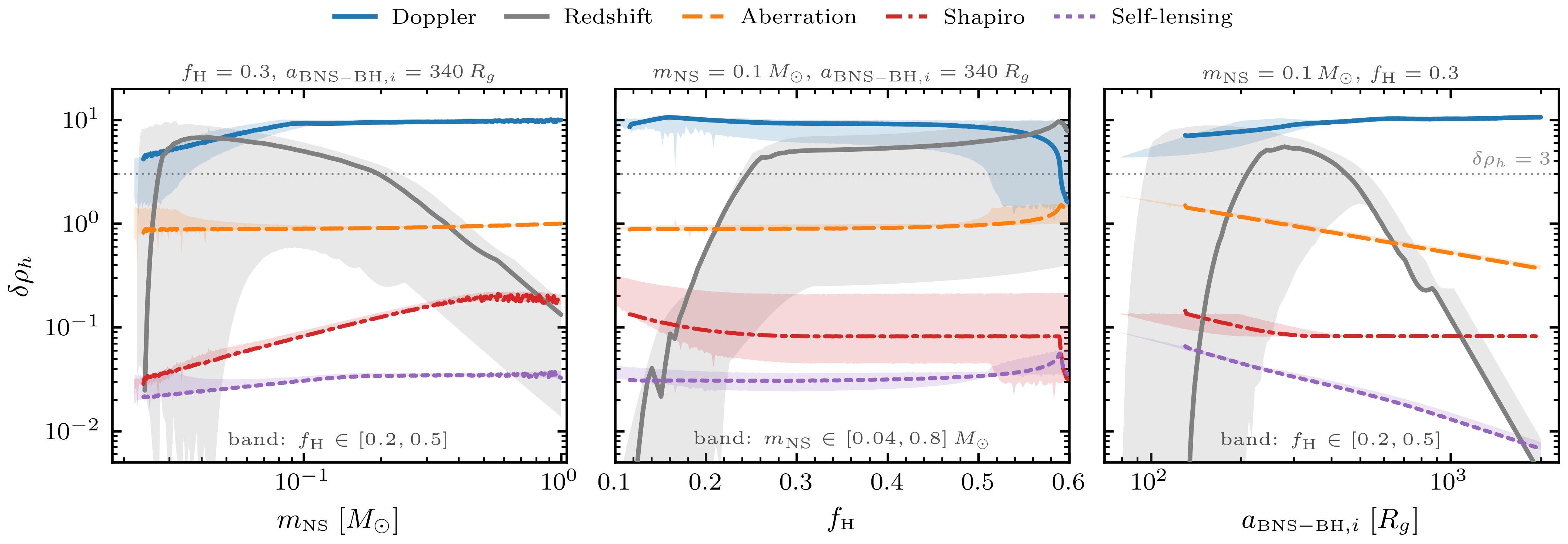}
  \caption{Horizon distinguishability $\delta\rho_h$ of the five modulation effects as a function of NS mass (left), initial Hill fraction (center), and initial outer separation (right), for LIGO with $M_\BH=5\,\msun$ and $\iota=45^\circ$. Fixed parameters are given above each panel, shaded bands span the ranges indicated within the panels, and the gray dotted line marks $\delta\rho_h=3$.}
  \label{fig:eff_1d}
\end{figure*}

\section{Derivation of modulation effects}
\label{app:modulations}
\noindent
In this appendix we derive Eq.~\eqref{eq:rho_J0} and apply it to modulations induced by Doppler, relativistic aberration, gravitational redshift, Shapiro delay, and self-lensing, computing how the outer-orbit geometry of each effect translates into its horizon distinguishability $\delta\rho_h$, and how this quantity depends on the system parameters.
These analytical calculations provide order-of-magnitude estimates but can omit subdominant ingredients: for instance, the orthogonal projection of the residual onto the re-fitted template parameters is only partially captured. The figures and numerical values quoted in Sec.~\ref{sec:modeffects} are obtained using the full numerical prescription described there.

Considering the oscillatory form of Eq.~\eqref{eq:dpsi_osc}, the outer orbital phase $\varphi_{\BNS-\BH}$ advances through many cycles within each frequency resolution element, so that $e^{i\delta\psi}$ can be averaged over one outer orbit at fixed $\Theta$ --- a WKB separation of the fast orbital phase from the slow chirp. The Jacobi--Anger identity
\begin{equation}
  \frac{1}{2\pi}\int_0^{2\pi}\! e^{-i\Theta\cos\varphi_{\BNS-\BH}}\,\d\varphi_{\BNS-\BH} = J_0(\Theta)
  \label{eq:jacobi}
\end{equation}
makes this orbit average real, so the band-weighted overlap in Eq.~\eqref{eq:overlap} reduces to $\langle J_0(\Theta)\rangle_W$. The best-fit phase $\varphi_c$ removes any leftover constant, so that the match is $\langle J_0(\Theta)\rangle_W$ and the mismatch of Eq.~\eqref{eq:mismatch} is $\mathcal{M}=1-\langle J_0(\Theta)\rangle_W$, from which Eq.~\eqref{eq:rho_J0} follows. The two limits quoted in Eq.~\eqref{eq:rho_J0_limits} of the main text follow from the expansion $J_0\simeq1-\Theta^2/4$, giving $\mathcal{M}\simeq\tfrac14\langle\Theta^2\rangle_W$ for weak modulation and, from the decorrelation of the cycle average, $\langle J_0\rangle_W\to0$ and $\mathcal{M}\to1$ for strong modulation.

As we show below, aberration and Shapiro delay remain in the linear regime over the evaluated range, while Doppler remains near the saturation limit over most of the evaluated range, so that its parameter dependence is determined primarily by the noise-weighted frequency support rather than its amplitude.

Each time-domain correction $\delta\phi(t)$ maps to $\delta\psi(f)$ through the stationary-phase relation of Eq.~\eqref{eq:dpsi_from_dphi}, using the leading chirp rate $\dot f=\tfrac{96}{5}\pi^{8/3}\mathcal{M}_c^{5/3}f^{11/3}$.
Eliminating $t$ between the inner and outer Peters inspirals gives the outer separation at inner frequency $f$, Eq.~\eqref{eq:r_of_f}, controlled by the outer-to-inner rate ratio $R_\tau$ of Eq.~\eqref{eq:tR}. We now apply these expressions to each effect in turn.

\subsection{Doppler}
\label{app:mod_doppler}

\noindent
For the reference values $m_\NS=0.1\,\msun$, $f_\text{\tiny H}=0.3$, $a_{\BNS-\BH,i}=340\,R_g$ and $M_\BH=5\,\msun$, the modulation index of Eq.~\eqref{eq:Theta_dop} increases from $\Theta_\text{\tiny Dop}\approx0.7$\,rad at the $20$\,Hz lower band edge to $\approx3.4$\,rad at the $92$\,Hz contact frequency, averaging near $2$\,rad. Since the inner binary completes $\sim\!10^4$ outer orbits while crossing the band, a modulation of this size substantially decorrelates the perturbed and vacuum waveforms, giving $\mathcal{M}\approx0.7$ and $\delta\rho_h\approx9.3$ at these values. Doppler therefore remains near the saturation limit of Eq.~\eqref{eq:rho_J0_limits} across most of the parameter space, and its curves in all three panels of Fig.~\ref{fig:eff_1d} primarily reflect the noise-weighted frequency support.
 Because $\Theta_\text{\tiny Dop}\propto f$, its band value follows the noise-weighted mean frequency $\langle f\rangle_W$, so that any parameter moving the in-band window also moves the mismatch. A heavier NS increases the contact frequency for a polytropic EoS and adds a strongly modulated high-frequency band, so $\delta\rho_h$ rises and then flattens once $m_\NS\gtrsim 0.2\,\msun$. A larger Hill fraction acts in the opposite direction, lowering the entry frequency as $f_i\propto f_\text{\tiny H}^{-3/2}$ and reducing the relative contribution of the high-frequency content, so that $\delta\rho_h$ decreases gradually along $m_\NS=0.1\,\msun$, before dropping at the disruption edge. The outer separation enters instead through the amplitude, $\Theta_\text{\tiny Dop}\propto a_{\BNS-\BH}$: widening the orbit increases the decorrelation and raises $\delta\rho_h$ from $\approx4$ at $100\,R_g$ to a plateau at $\approx10.6$. Only in the region of small $a_{\BNS-\BH}$ and large $f_\text{\tiny H}$, where both parameter dependences reduce $\Theta_\text{\tiny Dop}$ to order unity, does Doppler transition from the saturation regime to the linear regime.

\subsection{Relativistic aberration}
\label{app:mod_aberration}

\noindent
Aberration has the opposite outer-separation dependence to Doppler. Its modulation index, Eq.~\eqref{eq:Theta_abe}, is set primarily by the outer orbital speed and is therefore weakly dependent on $m_\NS$ and $f_\text{\tiny H}$; consequently, its distinguishability depends primarily on $a_{\BNS-\BH}$. The closed form of Eq.~\eqref{eq:rho_abe} decreases as $\delta\rho_h\propto v_{\BNS-\BH}\propto a_{\BNS-\BH}^{-1/2}$ (right panel of Fig.~\ref{fig:eff_1d}), whereas Doppler grows linearly with the same separation. The inclination dependence is also opposite: aberration grows as $1/\sin\iota$ toward face-on, and within a few degrees of face-on $v_{\BNS-\BH}/\sin\iota$ approaches unity, where the first-order expansion of Eq.~\eqref{eq:dphi_abe_first} breaks down and the full Lorentz boost of Eq.~\eqref{eq:lorentz_aberration} saturates the deflection. The only departure from the closed form arises in the small-$a_{\BNS-\BH}$, high-$f_\text{\tiny H}$ strip, where the outer orbit decays appreciably during the signal: there $v_{\BNS-\BH}(t)$ rises in band and $\delta\rho_h$ increases to $\approx2.4$, still below the horizon threshold. 

For the compact QM EoS, which permits configurations at the smallest outer separations, this increase can exceed the threshold and aberration becomes marginally detectable there (see Appendix~\ref{app:EoS}).

\subsection{Gravitational redshift}
\label{app:mod_redshift}

\noindent
As discussed in Sec.~\ref{sec:redshift}, the redshift distinguishability depends on the in-band shrinkage of the outer orbit, $a_{\BNS-\BH}(f)$ of Eq.~\eqref{eq:r_of_f}, set by the rate ratio $R_\tau$, and increases as $R_\tau\propto f_\text{\tiny H}^4$ (center panel of Fig.~\ref{fig:eff_1d}). Its two remaining dependences are non-monotonic because the potential amplitude competes with the time available for outer-orbit evolution.

Toward lower NS mass the residual initially increases, roughly as $m_\NS^{-5/3}$ from the $\mathcal{M}_c^{-5/3}$ prefactor of Eq.~\eqref{eq:dpsi_redshift}: a lighter binary chirps more slowly and accumulates more of the secular drift. The rate ratio has the same dependence, since at fixed $M_\BH$ a lighter NS lowers $q=m_\NS/M_\BH$, and with it $R_\tau\propto(1+2q)/q^{2/3}$ grows as $q^{-2/3}$ for $q\ll1$, so the outer orbit inspirals further relative to the inner binary and is more affected by changes in the potential. The increase does not continue indefinitely, however: the lightest NSs also have the largest radii for polytropic EoSs, $R_\NS\propto m_\NS^{-1/3}$, and so reach contact early, ending the inner inspiral before the outer orbit has drifted appreciably. Below $m_\NS\approx0.05\,\msun$ the redshift therefore decreases with decreasing mass. The separation dependence is likewise non-monotonic, peaking at a few hundred $R_g$. At large $a_{\BNS-\BH}$ the potential $\propto1/a_{\BNS-\BH}$ weakens the signal, while at small $a_{\BNS-\BH}$ the Hill radius shrinks with it, so that at fixed $f_\text{\tiny H}$ the inner binary begins at small separation, $a_{\BNS}=f_\text{\tiny H}\,r_\text{\tiny Hill}$, and merges quickly, again limiting the time available for outer-orbit evolution.
Notice also that, although the per-orbit Doppler shift ($\propto v$) is intrinsically larger than the gravitational redshift ($\propto v^2$), the redshift accumulates \emph{secularly} over the full in-band inspiral while the Doppler modulation is oscillatory and partly averages out, so the two have comparable horizon distinguishability.

\subsection{Shapiro time delay}
\label{app:mod_shapiro}

\noindent
The oscillatory Shapiro phase of Eq.~\eqref{eq:dpsi_sha} is not a single harmonic, so Eq.~\eqref{eq:rho_J0} does not apply to it directly. Averaged over one outer orbit, the logarithm separates into a constant and a sum of cosines,
\begin{align}
  \log \, \bigl(1+\sin\iota\cos\varphi\bigr)
  &=\log\left(\frac{1+\cos\iota}{2}\right)
  +\sum_{n\ge1} c_n\cos n\varphi\,,\notag\\
  c_n&=\frac{2(-1)^{n+1}}{n}\,\tan^n\!\tfrac\iota2\,,
  \label{eq:sha_fourier}
\end{align}
where the constant is the orbital mean. Multiplied by $2\pi f$, that constant is once more a phase $\propto f$, and is absorbed with the arrival time together with the secular term of Eq.~\eqref{eq:shapiro_expand}. Each harmonic $\cos n\varphi$ is instead an independent oscillation to which Eq.~\eqref{eq:rho_J0} applies, with effective index $4\pi M_\BH f\,c_n$, and because distinct harmonics are orthogonal over the orbit, their contributions add in quadrature. All of the indices lie deep in the weak-modulation limit ($M_\BH f \ll 1$), so that
\begin{equation}
  \mathcal{M}=\frac14\Bigl\langle (4\pi M_\BH f)^2\sum_{n\ge1}c_n^2\Bigr\rangle_W \,,
  \label{eq:sha_MM}
\end{equation}
and the sum of squared harmonic amplitudes resums into a dilogarithm,
\begin{equation}
  \sum_{n\ge1}c_n^2=4\sum_{n\ge1}\frac{\tan^{2n}(\iota/2)}{n^2}
  =4\,\mathrm{Li}_2\!\bigl(\tan^2\tfrac\iota2\bigr),
  \label{eq:sha_dilog}
\end{equation}
with $\mathrm{Li}_2(x)=\sum_{n\ge1}x^n/n^2$, from which the closed form of Eq.~\eqref{eq:rho_sha} follows. For the central BH masses considered here the Shapiro delay stays one to two orders of magnitude below the horizon threshold at every inclination, $\delta\rho_h\sim0.08$, and its only residual dependence on the inner binary is through the upper-frequency cutoff, as shown in the left panel of Fig.~\ref{fig:eff_1d}.

\subsection{Self-lensing}
\label{app:mod_lensing}
\noindent
As mentioned in Sec.~\ref{sec:selflensing}, self-lensing includes both an amplitude and a phase imprint, Eq.~\eqref{eq:F_low_w}, so it admits no modulation index and no closed form for $\delta\rho_h$. Its behavior nonetheless depends primarily on the single combination $\tfrac12 wy^2\propto f\,a_{\BNS-\BH}/\sin\iota$, which for the subsolar masses and low frequencies of interest is large across the accessible parameter space. The asymptotic scaling of Eq.~\eqref{eq:lens_geo} then reproduces the inverse-separation decline seen in the right panel of Fig.~\ref{fig:eff_1d}, where $\delta\rho_h\propto a_{\BNS-\BH}^{-1}$ is independent of both the inner binary and the lens mass. Self-lensing remains the smallest imprint of the five and is below the horizon threshold over the displayed range.

\section{Post-merger emission of the subsolar remnant}
\label{sec:postmerger}
\noindent
The BNS merger remnant is dominated by a quadrupolar $f$-mode oscillation at frequency $f_f$, set primarily by its mean density $\bar{\rho}_\NS$. This free oscillation of the merged remnant is distinct from the pre-merger dynamical tide of Sec.~\ref{sec:envtidalNSNS} [Eq.~\eqref{eq:Lambda_dyn}], where the same $\ell=2$ $f$-mode of each inspiraling component is driven \emph{off-resonance} by its companion's tidal field and modifies the effective deformability; here the mode instead rings freely in the remnant and radiates directly. While canonical mergers produce $f_f \simeq 2 - 4$~kHz, subsolar nuclear remnants are larger and less dense for a polytropic EoS. On the low-mass nuclear branch, $R_\NS \propto m_\NS^{-1/3}$, implying $\bar{\rho}_\NS \propto m_\NS^2$ and hence $f_f \propto m_\NS$. Thus, $f_f$ shifts from $\sim 2$~kHz at $1\,\msun$ to $\sim 200$\,Hz at $0.1\,\msun$, placing subsolar remnants near the most sensitive region of ground-based detectors with~
\cite{Andersson:1997rn,Vretinaris:2019spn,Lioutas:2017xtn,Yagi:2016bkt,Flores:2013yqa} 
\begin{equation}
  f_f \simeq 186\ {\rm Hz}
  \left(\frac{\bar\omega}{1.4}\right)
  \left(\frac{M_\text{\tiny rem}}{0.2\,\msun}\right)^{1/2}
  \left(\frac{R_\NS}{34\,{\rm km}}\right)^{-3/2}\,,
  \label{eq:f2}
\end{equation}
where $M_\text{\tiny rem}=2m_\NS$ and the dimensionless factor $\bar\omega\sim O(1)$ encodes the remnant structure. Subsolar mergers therefore shift the dominant post-merger signal from the kilohertz regime into the $\sim 100$-Hz band, near the sensitivity minima of LIGO, CE, and ET.
The mode also has a long damping time in the subsolar regime: considering gravitational radiation as the only damping mechanism, the universal relation $M_\text{\tiny rem}/\tau_\text{\tiny rem,GW} \simeq 0.112\,C^{4}$~\cite{Lioutas:2017xtn} gives
\begin{equation}
  \tau_\text{\tiny rem,GW} \simeq 1.6\times10^{3}\ {\rm s}
  \left(\frac{R_\NS}{34\,{\rm km}}\right)^{4}
  \left(\frac{M_\text{\tiny rem}}{0.2\,\msun}\right)^{-3}\,.
  \label{eq:tau_gw}
\end{equation}
This damping time is approximately half an hour, compared with tens of milliseconds for a canonical NS remnant. The post-merger signal is therefore quasi-monochromatic, with quality factor $\sim \pi f_f\tau_\text{\tiny rem,GW}\simeq10^{6}$.

\begin{figure}[t!]
  \centering
  \includegraphics[width=\columnwidth]{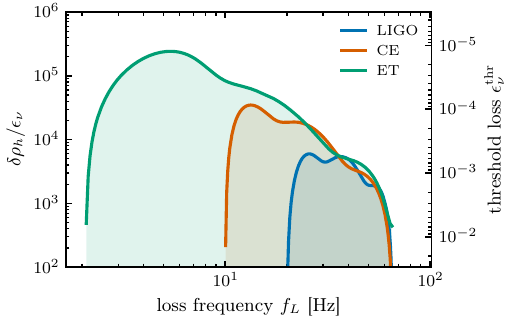}
  \caption{Response to in-band mass loss as a function of the loss frequency
  $f_L$ across the NS-BH band, at $m_\NS = 0.1\,\msun$, $M_\BH = 5\,\msun$,
  from $a_{\NS-\BH,i} = 340\,R_g$ to tidal disruption: we show $\delta\rho_h/\epsilon_\nu$ (left
  axis) and the loss that reaches the detection threshold
  $\delta\rho_h = 3$,
  $\epsilon_\nu^{\rm thr} = 3\,(\delta\rho_h/\epsilon_\nu)^{-1}$ (right axis), for
  LIGO, CE and ET.}
  \label{fig:massloss}
\end{figure}

We can adapt the ringdown SNR of Ref.~\cite{Baibhav:2018rfk}, at fixed radiated fraction $\epsilon_\text{\tiny pm}\equiv E_\text{\tiny rad}/M_\text{\tiny rem}c^{2}$. For a nuclear-branch source with inspiral SNR $\rho_0 = 8$, the post-merger SNR is $\rho_\text{\tiny pm} \sim 300\sqrt{\epsilon_\text{\tiny pm}}$. Thus, a remnant that radiates $\epsilon_\text{\tiny pm}>10^{-4}$ has $\rho_\text{\tiny pm}>3$. Depending on the radiated energy, the post-merger SNR can be comparable to or larger than that of the preceding inspiral. The subsolar nuclear post-merger may therefore be detectable independently. Its frequency $f_f$, together with the masses inferred from the inspiral, constrains the remnant's mean density through Eq.~\eqref{eq:f2} and provides an asteroseismological probe of the low-density remnant.

\subsection*{In-band mass loss due to neutrino cooling of the remnant}
\noindent
The BNS merger remnant is hot and lepton-rich. Neutrino emission reduces its gravitational mass while the GW signal is being emitted~\cite{Burrows:1986me,Pons:1998mm}.
This happens during the NS-BH stage, where the NS is the second-generation remnant. A fractional mass loss $\epsilon_\nu$ changes the leading phase coefficient $\mathcal{M}_c^{-5/3}$ by $\delta\mathcal{M}_c^{-5/3}/\mathcal{M}_c^{-5/3} = \epsilon_\nu$.

The loss is adiabatic relative to the orbital timescale: neutrino cooling occurs over seconds or longer, whereas the orbital period at band entry is much shorter ($\tau_\text{\tiny orb}^{\NS-\BH}= \frac{2\pi R_*}{v_*} \ll {\rm s}$). The orbit therefore remains approximately circular. The dephasing begins at the mass-loss frequency $f_L$: 
\begin{equation}
  \delta\psi_\nu(f) = \epsilon_\nu
    \bigl[\psi_\text{\tiny N}(f)-\psi_\text{\tiny N}(f_L)\bigr]\,\Theta (f-f_L)\,,
  \label{eq:dpsi_massloss}
\end{equation}
where $\Theta$ is the Heaviside step function.
The response is linear, $\delta\rho_h\propto\epsilon_\nu$, over the entire range of interest. 
In Fig.~\ref{fig:massloss} we show the dependence on $f_L$ across the band, from $a_{\NS-\BH,i} = 340\,R_g$ to tidal disruption at $d_\text{\tiny Roche}$ [Eq.~\eqref{Rocheradius} with $\gamma_\text{\tiny Roche}=2$]. 

Equation~\eqref{eq:dpsi_massloss} applies equally for $\epsilon_\nu<0$: in-band mass growth by accretion produces the same term with the opposite sign and the same tolerance. 


\bibliography{draft}

\end{document}